\documentclass[11pt,twoside]{book}

\usepackage{amsmath,latexsym}
\usepackage{epsf,bezier,epsfig,amssymb}
\usepackage{amsfonts}
\usepackage{theorem}
\usepackage{enumerate}

\usepackage{slashed}
\usepackage{colordvi}

\usepackage{bbm}
\usepackage{times}
\usepackage{cite}
\usepackage{graphicx,rotating}
\usepackage{lscape}
\usepackage{float}
\usepackage{tabularx}
\usepackage{fancyhdr,fancybox}
\usepackage[dutch,english]{babel}

\input{psfig}

\newcommand{\Z}{\mathbbm{Z}}
\newcommand{\R}{\mathbbm{R}}

\newcommand{\1}{\mathbbm{1}}
\newcommand{\F}{\mathcal{F}}
\newcommand{\cH}{\mathcal{H}}

\newcommand{\D}{\Delta}

\newcommand{\SU}{\mathop{\rm SU}}
\newcommand{\SO}{\mathop{\rm SO}}

\newcommand{\U}{\mathop{\rm {}U}}
\newcommand{\USp}{\mathop{\rm USp}}
\newcommand{\OSp}{\mathop{\rm OSp}}

\newcommand{\Sl}{\mathop{\rm SL}}
\newcommand{\Gl}{\mathop{\rm {}G}\ell }
\newcommand{\E}{\mathop{\rm E}}
\newcommand{\G}{\mathop{\rm G}}

\renewcommand{\l}{\lambda}
\renewcommand{\L}{\Lambda}
\newcommand{\eg}{e.g.}
\newcommand{\ie}{i.e.}

\newcommand{\br}[1]{D$#1$-brane}
\newcommand{\mbr}[1]{M$#1$-brane}
\newcommand{\re}[1]{(\ref{#1})}
\newcommand{\covder}{\mathfrak{D}}

\newcommand{\e}{\mathrm{e}}

\newsavebox{\uuunit}
\sbox{\uuunit}
    {\setlength{\unitlength}{0.825em}
     \begin{picture}(0.6,0.7)
        \thinlines
        \put(0,0){\line(1,0){0.5}}
        \put(0.15,0){\line(0,1){0.7}}
        \put(0.35,0){\line(0,1){0.8}}
       \multiput(0.3,0.8)(-0.04,-0.02){12}{\rule{0.5pt}{0.5pt}}
     \end {picture}}
\newcommand {\unity}{\mathord{\!\usebox{\uuunit}}}
\newcommand{\J}[3]{J^{#1} {}_{#2} {}^{#3}}

\def\be{\begin{equation}}
\def\ee{\end{equation}}
\def\bea{\begin{eqnarray}}
\def\eea{\end{eqnarray}}
\def\nn{\nonumber}
\def\q{\quad}
\def\qq{\quad\quad}
\def\un{\underline}

\newcommand{\fr}[2]{{\frac{#1}{#2}}}
\newcommand{\ft}[2]{{\textstyle\frac{#1}{#2}}}
\def\del{\partial}
\def\la{\label}

\newcommand{\h}[2]{h^{#1}_{#2}}

\def\half{\frac{1}{2}}
\def\su{\sqrt{-u^2}}

\def\wtd{\widetilde}

\def\sst#1{{\scriptscriptstyle #1}}
\def\fft#1#2{{#1 \over #2}}
\def\1{{\sst{(1)}}}
\def\2{{\sst{(2)}}}
\def\3{{\sst{(3)}}}

\def\a{\alpha}
\def\b{\beta}
\def\g{\gamma}

\def\d{\delta}
\def\D{\Delta}
\def\e{\epsilon}
\def\ep{{\epsilon}}
\def\ve{\varepsilon}

\def\f{\phi}
\def\F{\Phi}
\def\p{\psi}
\def\P{\Psi}
\def\vf{\varphi}
\def\k{\kappa}
\def\l{\lambda}
\def\L{\Lambda}
\def\m{\mu}
\def\n{\nu}
\def\r{\rho}
\def\s{\sigma}

\def\th{\theta}

\def\ta{\tau}
\def\o{\omega}
\def\O{\Omega}
\def\g{\gamma}
\def\x{\xi}
\def\rmi{{\,\rm i\,}}
\def\rme{{\,\rm e\,}}
\def\rmd{{\,\rm d\,}}
\def\ra{\rightarrow}
\def\lra{\longrightarrow}

\font\mybb=msbm10 at 12pt
\def\bbbb#1{\hbox{\mybb#1}}
\def\Z {\bbbb{Z}}
\def\R {\bbbb{R}}

\def\E {\bbbb{E}}

\def\cD{{\cal D}}
\def\cF{{\cal F}}
\def\cL{{\cal L}}
\def\cC{{\cal C}}

\newtheorem{theorem}{Theorem}[section]

\DeclareMathOperator{\Tr}{Tr}

\csname @addtoreset\endcsname{lemma}{chapter}

\newlength{\blength}
\numberwithin{equation}{chapter}

\numberwithin{equation}{chapter}

\begin{document}
\frontmatter

\fancypagestyle{empty}{
   \fancyfoot[CO]{} 
   \fancyhead{}
   \renewcommand{\headrulewidth}{0pt}}

\thispagestyle{empty}

\thispagestyle{empty}

\noindent
\hspace{2.6cm}
\hfill

\begin{minipage}[t]{12cm}
\begin{center}
\begin{minipage}[t]{8.5cm}
\begin{center}
\large{\sc
Katholieke Universiteit Leuven\\
Faculteit Wetenschappen\\
Instituut voor Theoretische Fysica}\end{center}
\end{minipage}
\end{center}
\par\vspace*{-2.7cm}
\begin{flushright}
\begin{minipage}[t]{2.5cm}
\hfill
\includegraphics[width=1.5cm,keepaspectratio]{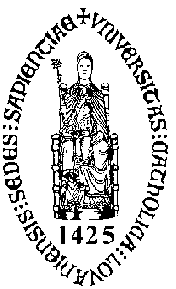}\\
\end{minipage}
\end{flushright}
\end{minipage}
\vfill

\noindent
\begin{center}
\textbf{\Large{From M-theory to}} \\[0.3cm] 
\textbf{\Large{$D=5$ supergravity and duality-symmetric theories}} 

\vspace{2cm} 

\noindent
{\Large Sorin Cucu}
\end{center}

\vfill 

\noindent
\parbox{5cm}{
  \begin{flushleft}
    Promotor: \\ Prof.~Dr.~Antoine Van Proeyen
  \end{flushleft}
}
\hfill
\parbox{7cm}{
  \begin{flushright}
    Proefschrift ingediend tot het\\
    behalen van de graad van\\
    Doctor in de Wetenschappen
  \end{flushright}
}

\vspace{0.3cm}

\noindent
\begin{center}
{\sc Leuven} 2003
\end{center}

\newpage
\thispagestyle{empty}

\begin{center}
Parin\c tilor \c si sorei mele \c si, in special, Adrianei 
\end{center} 

\
\vfill 

\newpage
\thispagestyle{empty}


\ \vfill
\ \hfill
\parbox{8cm}{   If we knew what it was we were doing, it would not be called research, would it ?
\begin{flushright}
Albert Einstein
\end{flushright}
}

\newpage
\thispagestyle{empty}

\begin{center}
{\bf {\Large Acknowledgements}}
\end{center}

The first words of thank go to my parents and my sister Raluca. It is due to your love and support that I have reached so far. Even if these simple words can hardly express my deep gratitude, I would like to thank you for your encouragements and unfailing support throughout the graduate years I have spent in Leuven.

Secondly, let me thank my advisor, Antoine Van Proeyen, who made my \mbox{Ph.D.} dream come true by accepting me, a young Romanian student, as his graduate student and by guiding me into the secrets of high-energy physics. First of all, I should thank him for helping me to transform my genuine curiosity about physics into a real passion. I am truly indebted to him for training me to become a scientist and for giving me the confidence to trust my scientific capacities. Thanks are due for his patience, for sharing his ideas with me and for his advice. I also appreciate very much that he gave me the freedom to follow my ideas and research interests.  

I would like to extend my gratitude to all other staff members of the institute for many useful discussions and for their hospitality here at the Institute for Theoretical Physics in Leuven. In particular, I benefitted from Walter Troost's critical remarks and suggestions regarding my research and didactic activities. I am also grateful to D\'esir\'e Boll\'e, Raymond Gastmans and Dieter L\"ust for their remarks concerning my thesis and for participating in the dissertation committee. Furthermore, I have enjoyed teaching under the supervision of Jos Rogiers and Andr\'e Verbeure. Thanks to Christine Detroye and Anita Raets for their constant efforts in helping me out of unsurpassable administrative problems.     

I would especially like to express my gratitude to my collaborators and close friends, Xavier Bekaert and Justin V\'azquez-Poritz: thanks for the fruitful exchange of ideas and for your friendship. I wish you success with your future scientific careers and I hope to see you back in Romania. Many thanks go out to my other collaborators during the \mbox{Ph.D.} program: Eric Bergshoeff, Martijn Derix, Tim De Wit, Jos Gheerardyn, Rein Halbersma, Hong L\"u and Stefan Vandoren, from whom I have learned so much. 

I owe such a pleasant stay in Leuven as well as many encouragements to my good friends: Annelies, Annick, Jordi, Joris, Karen, Leen, Michel, ... Their friendship enriched my social life and broadened my cultural perspectives considerably. Next, I like to acknowledge my colleagues and friends in the department: Bart, Bert, Dimitri, Geert, Ilse, Lennaert, Lieselot, Tom, Toni,... for an inspiring and enjoyable scientific environment during all those years.  

Lastly, but most significantly, I like to thank Ada who is my greatest source of inspiration. Your love is more important to me than you can possibly know.

\newpage

\pagestyle{fancy}

\fancyhf{}
\fancyhead[LE,RO]{\thepage}
\fancyhead[CE]{ \textsc{Contents}}
\fancyhead[LO]{}
\renewcommand{\headrulewidth}{0.5pt}
\renewcommand{\footrulewidth}{0pt}
\addtolength{\headheight}{3pt}
\fancypagestyle{plain}{
   \fancyfoot[CO]{\thepage} 
   \fancyhead{}
   \renewcommand{\headrulewidth}{0pt}}

\newpage
\thispagestyle{empty}

\tableofcontents

\mainmatter

\renewcommand{\chaptermark}[1]{\markboth{#1}{}}
\renewcommand{\sectionmark}[1]{\markright{\thesection\ #1}}
\fancyhf{}
\fancyhead[LE,RO]{\thepage}
\fancyhead[CO]{\textsc{\rightmark}}
\fancyhead[CE]{\textsc{\leftmark}}
\renewcommand{\headrulewidth}{0.5pt}
\renewcommand{\footrulewidth}{0pt}
\addtolength{\headheight}{0.5pt}
\fancypagestyle{plain}{
   \fancyhead{} 
   \fancyfoot[CO]{ \thepage}
   \renewcommand{\headrulewidth}{0pt}}

\chapter{Introduction \label{intro}}

\section{The story of string theory \label{story}}

During the last century, our fundamental conceptions about space and time have been radically overturned by the developing of quantum mechanics and general relativity. Einstein's theory of gravity taught us that space and time are not absolute but rather they are intertwined entities that can curl up into various  geometries. In fact, spacetime geometry is completely determined by the presence of matter and energy. In other words, gravity is nothing but the bending of spacetime and vice versa.
Shortly afterwards, we learned that the classical concept of particle trajectories has no meaning when applied to microscales. Instead, the motion of a particle between two spacetime points should be understood by considering all possible paths connecting the end points, each with a weight given by the associated classical action. The developments of the last two decades in string theory suggest that, once again, it may be necessary to modify the concept of spacetime in a fashion far from common sense.

The main goal of particle physicists since Einstein has been to encompass the four fundamental interactions (electromagnetic, weak, strong and gravitational) within the quantum mechanical framework of one theory that could, in principle, describe all matter in the universe and its interactions. Even if this is not a trivial task, by the mid 1970s the unification program for the first three interactions was completed and led to the {\it Standard Model} (SM) that describes the non-gravitational interactions of quarks and leptons. The agreement with up-to-date experimental data transformed SM into a trusted and respected theory. The main ingredient in the construction of SM was the concept of gauge symmetry. Symmetries play a crucial role in physics and, in particular, in particle physics they are believed to allow for the ultimate classification of existing particles. In fact, the idea of symmetry is a key concept also in the present work.  One distinguishes two kinds of symmetries: internal and spacetime symmetries. The first category includes for instance the $\SU(3)\times\SU(2)\times \U(1)$ gauge group of SM, while the second has the familiar Poincar\'e symmetry (spacetime translations and Lorentz transformations) appearing in General Relativity (GR). 

\begin{center}
\begin{figure}[p!]
\epsfxsize =14cm
{\epsffile{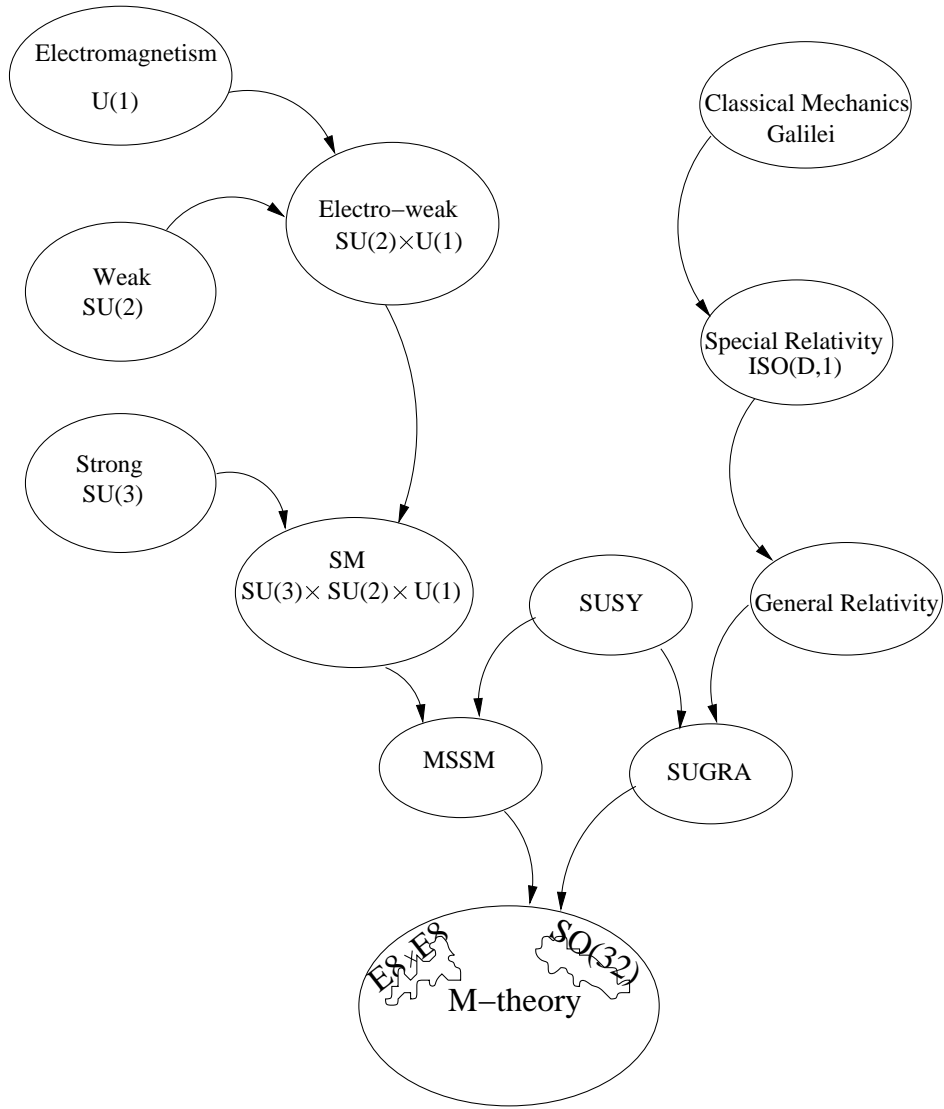}}
\caption{{\it The net of theoretical frameworks in particle physics and their symmetries.}\label{fig:sym}}\end{figure}
\end{center}

There were several hints that SM was not the full story. The unanswered questions about the number of families, the masses of the leptons and quarks and other parameters of the SM, the particular gauge group of SM, the relative strength of the fundamental forces and, most of all, the absence of gravity (i.e., the separation of internal from spacetime symmetries)  were clear signals that there should be something out there more fundamental: a theory that unifies the gauge group of SM with the spacetime symmetries of GR (see Fig.~\ref{fig:sym}).

First attempts in that direction were negative and, in fact, it was proven that under certain physical assumptions (locality, causality, etc.) the most general invariance group of a quantum field theory was, at best, a direct product of the Poincar\'e group and an internal group~\cite{Coleman:1967ad}. That was the moment string theory entered the stage. Although originally proposed with the wrong aim of describing strong interactions and then abandoned for a while after the development of Quantum Chromodynamics (QCD), rudimentary string models (dual models) led to one important discovery: {\it supersymmetry} (SUSY), a symmetry relating bosons to fermions and vice versa. The anticommuting generators of this new symmetry were initially introduced to cure the theory of unwanted tachyonic modes. However, the consequences of string theory were far richer than that.

It was immediately realized that this boson-fermion symmetry could be applied \cite{Wess:1974tw} to point-particle quantum field theory (QFT). As a result, numerous supersymmetric versions of ordinary QFT, including SM, were proposed. Besides the generic feature of predicting supersymmetric partners or superpartners for each existing particle (e.g., selectron, squarks, photino, etc.), they exhibited other interesting properties: the avoidance of the fine tuning problem in SM 
 or refinement of the running of non-gravitational force strengths at high energies (where they are believed to meet). Nevertheless, none of these particles have {\it presently} been experimentally detected\footnote{For a discussion of this issue, as well as other possible phenomenological and cosmological applications of SUSY particles, we refer to Refs.~\cite{Olive:1999ks,Schwarz:2000ew,Kane:2002tr,Olive:2003iq}.}, which implies that, at the energies accessible to our present accelerators, SUSY must be broken.
 
On top of all of this, the inclusion of fermionic generators in the hypothesis of the Coleman-Mandula no-go theorem allowed the authors of Ref.~\cite{Haag:1975qh} to show that the unification of the Poincar\'e spacetime group with internal symmetries is possible. The marriage of GR with supersymmetry was accomplished soon after that and the result was called {\it supergravity} (SUGRA). Supergravity \cite{Freedman:1976xh,Deser:1976eh} is nothing but the theory of local supersymmetry, i.e., a supersymmetric field theory  with fermionic parameters depending on spacetime (that automatically incorporates gravity). 
Even though SUGRA could solve some of the problems encountered when trying to quantize gravity (e.g., it softened the divergencies via fermionic contributions), it was still a non-renormalizable theory and, as a consequence, could not be the final answer.

In the meantime, the progress in string theory was slow but continual. Before seeing that, it is better to understand first the fundamental idea about string theory. In conventional quantum field theory \cite{Weinberg:1998s}, the elementary particles (electrons, photons, etc.) are considered as mathematical points moving through spacetime. On the other hand, string theory \cite{Green:1987sw,Polchinski:1998j} generalizes this point of view by taking as fundamental constituents  one-dimensional extended objects, i.e., tiny strings. The observed particles in nature are then just different vibrational patterns of the fundamental string. This seemingly harmless generalization has drastic consequences (extra dimensions, supersymmetry, extended objects, etc.), some of which will be the subject of the present research thesis. The obvious advantage is that there may be many elementary particles but there is only one fundamental string (as well as other extended object which we will discuss shortly) whose various excitations could reproduce the entire particle zoo. In fact, when studying the spectrum of closed strings, a spin-$2$ massless particle is found, which is the graviton. All these were indications that we may be, finally, on the right track for a unified theory of fundamental forces and matter.  

Just as the time evolution of a point particle is a worldline, a free string, either closed (loop) or open (free ends), evolves by sweeping out a 2-dimensional surface called a worldsheet (parametrized by, say, $\tau$ and $\sigma$). This surface is embedded in a higher-dimensional space where the position of the string is given by some coordinates $X^\mu (\tau, \sigma)\,.$ The free string is characterized by its tension $T = \ft{1}{2 \pi \alpha^\prime}\,,$ where $\alpha^\prime$ denotes the Regge  slope and is related to the string length $\alpha^\prime = l_S^2$. The interaction among various strings can be envisaged as the joining and splitting of these surfaces and is characterized by a coupling constant $g_S$. In this way, an interacting string reduces to the theory of two-dimensional Riemann surfaces and the interactions are determined by different topologies of free strings. The smoothness of the interaction means that strings tend to smear out the short-distance behavior  of Feynman diagrams in QFT, thus giving  the hope of a theory free of UV divergencies.  

As we already mentioned, SUSY was initially discovered through string theory. The original Ramond-Neveu-Schwarz (R-NS) supersymmetric formulation, based on worldsheet (two-dimensional) SUSY, was later modified \cite{Green:1982yb,Green:1981zg} by Green-Schwarz (GS) in the form of a target space (spacetime) SUSY. The advantage of this formulation was a manifest superymmetry (equality of bosonic and fermionic degrees of freedom at every mass level), while the main drawback was the impossibility of a covariant quantization\footnote{Recently, progress in this direction has been achieved \cite{Berkovits:2000fe,Berkovits:2002zk}.} (light-cone quantization was required). Using the latter supersymmetric formulation, Green and Schwarz were able to prove that, for some particular gauge groups, string theory\footnote{Unless otherwise specified, for simplicity, we will refer to a supersymmetric string as a string theory.} was anomaly-free \cite{Green:1984sg} and UV finite (at one-loop) \cite{Green:1985ed}. These were the proofs that theoreticians expected from a trustworthy string theory. Immediately afterwards many physicists engaged in research in this subject. Therefore, the mid 1980s was a very fruitful period for the theory, known as the {\it first superstring revolution}.   

As a result of these efforts, five {\it different} string theories emerged. Two of them, type IIA and IIB strings, are based on closed strings and have two worldsheet supersymmetries which can be translated into $N=2$ ($32$ supercharges) spacetime SUSY. The difference is that, in IIA, the two supersymmetries have opposite chirality (IIA is left-right symmetric, or non-chiral), while in IIB they have the same chirality (IIB is chiral). Next, there is type I string theory, based on closed and open strings. This theory possesses half the SUSY of type II theories. Anomaly cancellations require its symmetry group to be $\SO(32)$. Finally, there are  two heterotic string theories describing closed strings and supporting $16$ supercharges (in target space). Their consistency determines their gauge symmetries to be $\SO(32)$ and $\E_8 \times \E_8\,.$ The latter it is especially attractive from the phenomenological point of view, as we will discuss.  

However, if the world is made up of strings, why do we not detect them? The reason we do not see any strings is due to their characteristic length being of the order of {\it Planck length} $l_p\,.$ In terms of the fundamental constants (speed of\linebreak light~$c\,,$  Newton's constant $G$ and Planck's constant $\hbar$), it goes \linebreak as~$l_p = \sqrt{\hbar  G/c^3} \, \sim 10^{- 33} {\rm cm}\,,$ with a corresponding {\it Planck mass} scale
\begin{displaymath}
M_p = \sqrt{\hbar  c/G} \, = 1.2 \times 10^{19} {\rm GeV}/c^2\,.
\end{displaymath}
Hence, at the energy level reached by present day accelerators, these tiny strings look like point-particles. This explains the excellent agreement between QFT and experiment. However, once we approach the Planck energy domain, the description in terms of particles breaks down and stringy contributions become significant. This is the scale of quantum gravity. It is this energy that one deals with when studying black holes (BHs) or the first moments in the evolution of the Universe. The search for a quantum theory of gravity is thus strongly motivated by these kind of unanswered problems. String theory may be a firm step towards that aim. Even if the energy scale makes it difficult (if not impossible) to have a direct experimental test of string theory, there are hopes that some indirect measurements could confirm the string paradigm (see \cite{Witten:2002wb,Schwarz:2000ew,Dine:2000bf}).

Although our understanding of string theory is momentarily incomplete (partially due to the complications of the massive modes that appear at the Planck scale), there are certainly interesting features that can be revealed even at low energies. Indeed, considering energies far below $M_p$, the massive string states decouple from the spectrum and one is left with only massless modes. They include the metric $G_{\m\n}\,,$ the antisymmetric tensor $B_{\m\n}$ and the dilaton $\phi$ in the closed string spectrum and the gauge vector $A_\m$ in the open string spectrum. These leading contributions in $\a^\prime$ expansion (low-energy effective theories) describe nothing but \linebreak SUGRAs. Hence, string theory changed considerably the interpretation of SUGRA. Originally thought of as an elementary field theory, SUGRA is now viewed as an effective field theory describing the low-energy limit of string theory. 
SUGRAs are thus strongly motivated within the context of string theory. Nevertheless, throughout this thesis, we are going to consider them mostly in their own right.

Even before strings, SUGRAs were constructed in various spacetime dimensions. These higher dimensions were not justified theoretically but they were somehow supported by the old idea of unification proposed by Kaluza and Klein \cite{Duff:1994tn,Duff:2001jp}. Supersymmetry representation theory and field theory requirements restrict the number of spacetime dimensions (for particles with spin $\leq$ 2) to $\leq 2$. The $D=11$ SUGRA \cite{Cremmer:1978km} with $32$ conserved supercharges contains the elfbein, a gravitino and a three-index gauge field. Unfortunately, it is non-renormalizable and, at first, doesn't seem to be the low-energy limit of any string theory.

Moreover, a consistent superstring theory requires the number of spacetime dimensions in which the string moves to be ten. To make contact with the four-dimensional world we perceive, one possibility\footnote{An alternative approach to compactification is discussed in Sect.~\ref{s:rs}.} is to imagine that six of the dimensions are curled up in a compact manifold. The size of this manifold could be so small (presumably Planck scale) that it cannot be resolved by the present particle accelerators. As the spacetime geometry is determined dynamically by the equations of the theory, the form of the extra dimensions is not arbitrary. One appealing possibility is provided by a class of spaces known as Calabi-Yau (CY) manifolds \cite{Candelas:1985en}. Particular compactifications of $\E_8 \times \E_8$ heterotic string give rise in four dimensions, in the the low energy limit, to theories that resemble supersymmetric extensions of SM. The number of preserved supersymmetries is determined by the particular type of compactification. To make it phenomenologically even more attractive, the number of (lepton) families is also connected to the topology of the CY.  Given the variety of CY spaces, the main problem remains to find a dynamical mechanism that selects the correct vacuum. Furthermore, the existence of maximally supersymmetric, eleven-dimensional vacua (e.g., $AdS_4 \times S^7$, $AdS_7 \times S^4$), apparently disconnected from string theory, made things even more fuzzy.    

\section{Beyond strings: dualities and branes \label{beyond}}

The results in string theory prior to 1995 were mainly based on perturbative calculations, i.e., expansions in the dimensionless coupling constant $g_S\,.$ Similar to QFT, such approximations are reliable as long as $g_S$ is small. However, in string theory $g_S$ is a dynamical parameter, given by the expectation value of the dilaton scalar field. So, there is a priori no reason why $g_S$ should be small. It was thus expected that non-perturbative effects cannot be neglected in a realistic string theory\footnote{It is known that also in QCD non-perturbative contributions, e.g., via instantons, are relevant in some domain of the parameters.}. However, going beyond the perturbative description was a difficult task.   

Moreover, the existence of five consistent string theories was seen as a teasing wealth, considering that it was supposed to be the (unique) fundamental theory of nature. The resolution to some of these questions came together with the {\it second superstring revolution}. 
The breakthrough was based on the ideas of {\it duality} \cite{Witten:1995ex} and {\it D-branes} \cite{Polchinski:1995mt}.  

The concept of duality is not specific to string theory. The familiar electric-magnetic duality (see Ch.~\ref{ch:dual}) leaves the free Maxwell equations invariant under the interchange of electric and magnetic fields
\begin{equation}\la{EBBE}
{\vec E} \longrightarrow  {\vec B}\,, \qq {\vec B} \longrightarrow  -\, {\vec E}\,.
\end{equation}
Such a discrete symmetry can be extended to an $\SO(2)$ symmetry that rotates ${\vec E}$ and ${\vec B}$ by an arbitrary angle. In Lorentz covariant language, the set of Maxwell equations and Bianchi identities can be succintly written as
\begin{equation}\label{dualdiscr}
\partial_\m F^{\m\n} = 0 \,, \qq \partial_\m *F^{\m\n} = 0\,,
\end{equation} 
where $*F^{\m\n} = \ft 12\epsilon^{\m\n\r\s} F_{\r\s}$ and $\epsilon^{\m\n\r\s}$ denotes the antisymmetric Levi-Civita tensor. There is a $GL(2,\bbbb{R})$ duality symmetry of the form
\begin{equation}\label{dualinv}
\left(
\begin{array}{c}
F\\ *F
\end{array}\right) \longrightarrow  
\left(
\begin{array}{cc}
a&b\\ c&d
\end{array}\right)
\left(
\begin{array}{c}
F\\ *F
\end{array}\right)\,, \qq \left(
\begin{array}{cc}
a&b\\ c&d
\end{array}\right) \in GL(2,\bbbb{R})\,.
\end{equation}
Here, $F$ stands for the 2-form field strength and $*F$ for its Hodge-dual. Eq.~(\ref{dualinv}) gives an invariance of the sourceless Maxwell equations but not of the action. The implementation of this duality at the level of the action is a non-trivial issue that makes the subject of Ch.~\ref{ch:dual}.

In the presence of the electric sources $q_e$, the invariance is lost unless one accepts the existence of magnetic monopoles $q_m$, where 
\begin{equation}\label{defsource}
q_e = \frac{1}{4 \pi} \int_{S^2} *F\,, \qq q_m = \frac{1}{4 \pi} \int_{S^2} F\,,
\end{equation}  
with $S^2$ as a surface surrounding the charges. Magnetic monopoles were predicted a long time ago by Dirac \cite{Dirac:1948um}, mainly on aesthetic grounds of the Maxwell equations. To date, there is no experimental evidence for them (despite various efforts in that direction using cosmology \cite{Ambrosio:2002qq}). One possible explanation is their dilution in the inflationary era. We will adopt the standard point of view by assuming their existence in the theory. Then, the Maxwell equations with sources preserve the duality invariance, provided that the charges $(q_e, q_m)$ transform according to~(\ref{dualinv}). These considerations have been generalized for the case of several vector multiplets, in the context of $D=4$ SUGRAs with extended supersymmetry \cite{Cremmer:1979up}. Nevertheless, a consistent quantum mechanical picture requires the Dirac quantization condition
\begin{equation}\label{Diracq}
q_e q_m = 2 \pi \hbar n\,, \qq n\in \bbbb{N} \,,
\end{equation}
that spoils the invariance (\ref{dualinv}), unless one considers dyon solutions. 

Magnetic monopoles appear naturally in (supersymmetric) Yang-Mills (SYM) theories. Indeed, in addition to the perturbative excitations the $\SU(2)$ YM model admits also magnetic monopole solutions discovered by 't Hooft \cite{'tHooft:1974qc} and\linebreak Polyakov~\cite{Polyakov:1974ek}. These are solitonic solutions of the classical field equations with localized energy. Their mass goes like the inverse of the coupling constant (electric charge in this case). In other words, they are very heavy in the perturbative regime, and become relevant when one deals with non-perturbative effects. 

Dyons \cite{Julia:1975ff} are also solitonic configurations carrying both electric and magnetic charges. Taking into account the Witten effect \cite{Witten:1979ey} (characterized by the angular parameter $\theta$), the charges of the dyons form a two-dimensional lattice whose periods depend on the coupling and the $\th$ angle. The quantization condition for two dyons of charges $(q_e^i, q_m^i)_{i=1,2}$ goes under the name of Dirac-Schwinger-Zwanziger~(DSZ), and is given by 
\begin{equation}\label{DSZ}
q_e^i q_m^j - q_e^j q_m^i = 2 \pi \hbar n\,, \qq n\in \bbbb{N} \,.
\end{equation}
The Montonen-Olive duality \cite{Montonen:1977sn} or its $D=4$, $N=4$ SYM version \cite{Olive:1996sw} conjectures\footnote{Despite substantial evidence, the complete proof is still lacking.} that the $SL(2,\Z)$ symmetry of the charge lattice is an invariance of the theory. To put it differently, theories based on any pair of independent primitive vectors (stable states) of the roster describe the same physics. In particular, this implies that, in the case of $N=4$ SYM, the weak coupling description ($q_e \ra 0$) in terms of the fundamental excitations should be the same as the strong coupling description ($q_e \ra \infty$) in terms of the monopoles. This strong-weak coupling duality has an analogue in string theory, known as S-duality.

The existence of higher-dimensional objects in the context of string theory was first pointed out in \cite{Horowitz:1991cd}. Those authors put forward multi-dimensional type II supergravity solutions, baptized as $p$-branes (according to the number $p$ of spatial dimensions). They may be thought of as p-dimensional cousins of Reissner-Nordstrom BH preserving half of the $32$ supersymmetries. A significant subclass of these $p$-branes goes under the name of {\it D$p$-branes} due to the fact that they are defined as the place where open strings can end (Dirichlet boundary conditions). These exotic objects carry a complete set of electric and magnetic charges that couple to the RR gauge fields of type II string theory \cite{Polchinski:1995mt}. Consistency requires that there are D$p$-branes with even $p$ in IIA and odd $p$ in IIB. 

Another remarkable feature of the D-brane solutions is their tension (mass per unit volume), which goes like the inverse of the string coupling constant. Hence, they are not present in the effective perturbative theory but are demanded by a non-perturbative description of the type II superstring. Next to the closed string excitations, the complete type II string contains also open strings. While the former are free to move through the entire spacetime, the latter have end-points which are restricted to lie on D-branes. 
 
One important application of D-branes is to the microscopic description of BHs. By identifying and counting the number of quantum microstates corresponding to\linebreak a $D=5$ Reissner-Nordstrom BH (obtained via $K3 \times S^1$ compactification of type II string theory) the authors of \cite{Strominger:1996sh} were able to compute the statistical entropy of the BH. This was in agreement with the macroscopic entropy given by the Bekenstein-Hawking area law 
$$
S_{BH} = \frac{A_H}{4 G}\,,
$$  
where $A_H$ is the area of the horizon. Later, the result was generalized for near-extremal or rotating BHs (for a review see e.g., \cite{Maldacena:1996ky}). For the extremal case, the technique is to embed the BH into a higher-dimensional supergravity by considering superpositions of various extended objects in that theory. These D-branes are wrapped around the compact directions such that, from the lower-dimensional point of view, they appear as localized objects carrying some charge with respect to the dimensionally reduced gauge fields. The BHs obtained in this way preserve some supersymmetry that extends the perturbative estimate of the statistical entropy to strong coupling.

Some of the D-branes are electric-magnetic dual of each other, just like the magnetic monopole was dual to the electric charge. In fact, the charges of the dual branes satisfy a Dirac quantization condition of type (\ref{Diracq}). One encounters a likewise situation in $D=11$ SUGRA where the \mbr{2} \cite{Bergshoeff:1987cm} carries an electric charge and couples to the 3-form potential of the supergravity multiplet, while its magnetic counterpart \mbr{5} supports a self-dual 2-form on its worldvolume. Theories with chiral bosons need a careful treatment since they may lead to anomalies.   

Once the non-perturbative aspects started being understood, it was also realized
that the five string models and $D=11$ SUGRA were different corners of a larger theory, called {\it M-theory}.\footnote{M could stand for Magic, Membrane, Matrix, etc.} All these theories and the objects they describe are related by a web of dualities.  
The simplest of them, since it holds also perturbatively, is known as {\it T-duality}. Basically, it states the equivalence between a theory A compactified on a circle of radius $R_A$ and a second theory B compactified on another circle of radius $R_B$ provided that $R_A R_B = l_S^2\,.$ In general, the equivalence of toroidally (or K3) compactified theories falls into the same class. For example, type IIA on a circle of small radius is the same as type IIB on a circle of large radius. In the same sense, the two heterotic strings are also T-dual to each other. Moreover, $K3$ compactifications of the heterotic strings lie in the same parameter space. Another particular duality is the identification of CY threefold compactifications of type IIA with type IIB on the dual CY.

At non-perturbative level, the generalization of the electric-magnetic duality was coined {\it S duality}. In one of its forms, this duality means that the perturbative description of a given string theory coincides\footnote{ The dilaton fields of the two models are identified with opposite signs.} with the strong coupling limit of another theory with the same dimensionality. The elementary excitations on one side are identified with the solitonic ones in the dual picture. Again, these dualities are more difficult to prove but they were tested in several ways. As a first example of this duality, the heterotic $\SO(32)$ and the type I strings are S-dual in $D=10$  with the fundamental string of the former mapped into the D$1$-string of the latter. Further, type IIA compactified on $K3$ is S-dual to type I on $T^4$. It can also happen that the dilaton of one theory is mapped into another geometrical modulus of a second theory. Also, this situation relates the non-perturbative sector, on the one hand, with the perturbative regime on the other hand. Examples hereof are type II on CY threefolds being S-dual to heterotic strings on $K3 \times T^2\,.$ 

Yet another aspect of S-duality occurs  when the elementary and solitonic degrees of freedom of one and the same theory provide an alternative version of the same physics. That is, in the full quantum picture, what we call perturbative and non-perturbative treatments are just expansions of the same model around different vacua. As an illustrative example, type IIB is self-dual under strong-weak symmetry. While the corresponding SUGRA has an $SL(2,\R)$ invariance (see Sect.~\ref{s:maxsugra}), in the full IIB string quantum corrections break it to a discrete subgroup $SL(2,\Z)\,.$ In \cite{Schwarz:1995dk}, an infinite family of type IIB string solutions have been explicitly constructed. Their RR charges obey an $SL(2,\Z)$ symmetry, meaning that any of them can be interpreted as fundamental while the rest are non-perturbative contributions.

Even more exotic is the strong coupling behavior of type IIA and the heterotic $\E_8 \times \E_8$ string. In that limit, an eleventh dimensions pops up \cite{Townsend:1995kk,Witten:1995ex}. An identification of the parameters characterizing type IIA and the eleven-dimensional M-theory compactified on a circle $S^1$ of radius $R_{11}$ shows that indeed $R_{11} \sim g_S^{2/3}\,.$ It is then evident that a perturbative expansion in $g_S$ means an expansion in $R_{11}$; that is, the eleventh dimension is invisible for the perturbative string. Nevertheless, in the strong coupling regime this extra dimension decompactifies and the Kaluza-Klein (KK) modes coincide with the D-branes of type IIA. Even though poorly understood at present, this M-theory with \mbr{2}s as fundamental objects is expected to enclose all degrees of freedom of the five string theories.   

The case of the heterotic string is slightly different. The strong coupling limit leads also to M-theory but on an orbifold $S^1/\Z_2$ \cite{Horava:1996qa,Horava:1996ma}. The $\Z_2$-truncated\linebreak $D=11$~SUGRA has gravitational anomalies that are canceled by non-Abelian gauge fields living on the hyperplanes located at the two fixed points of the orbifold. Remarkably enough this cancellation takes place only if the gauge group is a copy of $\E_8$ on each of the boundaries. The relation between the string coupling  and the dimension of the orbifold resembles the one in IIA case, with the interpretation that at small couplings the two hyperplanes sit close to each-other, while at large coupling  they move far away. Such an image of a higher-dimensional space seen as brane insertions at some orbifold fixed points will be further developed in Sect.~\ref{s:rs} in connection with phenomenology. 

\section{AdS/CFT correspondence \label{s:holog}}

Another extraordinary connection between gravity and gauge theories was suggested by Maldacena \cite{Maldacena:1998re}  and made explicit in Refs.~\cite{Gubser:1998bc,Witten:1998qj}. Known as the Anti-de Sitter/conformal field theory {\it (AdS/CFT) correspondence or duality}, it conjectures the equivalence of type IIB string theory in $AdS_5 \times S^5$ background and $N=4$ SYM theory on the boundary of the $AdS_5$ space. The relevance of this duality is to give information about the YM theory when the computations in the gravity sector are tractable, and vice versa.

The first hint in that direction was the alternative description of branes as supergravity solutions or as gauge theory associated with their worldvolume. In the latter picture, a single \br{p} carries a $\U(1)$ SYM theory dimensionally reduced from $D=10$ to $D=p+1\,.$ For such $n$ coincident branes, the low-energy effective theory is again a dimensionally-reduced maximally supersymmetric YM theory but with enhanced $\U(n)$ gauge group \cite{Witten:1996im}. 

Within this class, the case of coinciding \br{3}s is of utter importance since a stack of $n$ \br{3}s has a $D=4$, $N=4$ SYM on the worldvolume. Among other nice features, this $\U(n)$ gauge theory has no UV divergences at any perturbative order and is scale invariant (vanishing $\beta (g)$ function). Moreover, the model has a conformal symmetry that combines properly with the $16$ supercharges onto a superconformal symmetry. Hence, we end up with 32 fermionic generators matching the supersymmetries on the gravity side. Also, the other symmetries agree. Indeed, the conformal group in four-dimensions $\SO(4,2)$ corresponds to the isometries\linebreak of $AdS_5$, while the R-symmetry group $\SU(4)$ of the superymmetric gauge theory agrees with the isometries of $S^5\,.$ Even more striking, the $SL(2,\Z)$ invariance\linebreak of $N=4$ SYM can be understood in terms of the $SL(2,\Z)$ S-duality of type IIB string, also preserved by \br{3}s themselves. Thus, global symmetries on both sides are identical.

Furthermore, the parameters of the dual pictures are related straightforwardly. The Yang-Mills coupling $g_{YM}$ and number of flavors $N$ determine both the string coupling as well as the $AdS_5$ and $S^5$ radii by
\begin{equation}
g_S =g_{YM}^2\,, \qq R_{AdS_5}^4 = R_{S^5}^4 = 4 \pi g_S N (\a^\prime)^2\,.
\end{equation}
Maldacena's conjecture claims the AdS/CFT duality  for all $N$ and all orders in~$g_S\,.$ Nevertheless, the full string theory quantization on (non-trivial) curved backgrounds is not yet available, such that non-trivial tests are made by looking at different approximations that are not so involved and can be handled, without being too simple. One of the possibilities is to consider $N\rightarrow \infty$ while keeping 't Hooft coupling $\l = N g_{YM}^2$ constant. In this 't Hooft limit the YM theory is defined as an $1/N$ expansion of Feynman diagrams. In the dual AdS description this limit corresponds to classical string theory since the string loop corrections are negligible ($g_S \ll 1$). When completely understood, this approximation may shed some light on the large $N$ limit of gauge theories. A further restriction is given by the large $\l$ region. In this case, the classical IIB SUGRA is a good approximation for the string theory reflected onto a $\l^{-1/2}$ expansion of the gauge sector. It is the SUGRA approximation that originally led to the formulation of the conjecture. We see once more that, even though they are just truncations of the final answer, SUGRAs can give valuable information.

However, taking AdS/CFT correspondence seriously means that the type IIB fields can be mapped onto the CFT operators. Furthermore, corresponding correlators in the two theories should give the same result. The precise recipe was given in \cite{Gubser:1998bc,Witten:1998qj} (for reviews see \cite{Aharony:1999ti,D'Hoker:2002aw}). As a result, single trace BPS operators of SYM are associated one-to-one with the canonical fields (\eg, graviton, gravitino) of supergravity compactified on $AdS_5 \times S^5\,.$ For computing the correlators it is common to use the large $N$ limit and, thus, an $1/N$ expansion in terms of Witten diagrams. They are just pictorial representations of the bulk interactions\footnote{The AdS space can be represented as a disc while its Minkowski boundary is given by the boundary circle of the disc.} generated by external sources living on the boundary space. 

It is also worth mentioning that the AdS/CFT duality is a specific realization of the Holographic Principle promoted by 't Hooft and Susskind. It conjectures that the degrees of freedom in a given region of spacetime may be completely characterized by a quantum field theory on the boundary of that region. 

\section{Randall-Sundrum braneworlds \label{s:rs}}

In parallel to their intensive application to fundamental theory, extra dimensions were also very popular among phenomenologists in recent years. This came about with the advent of {\it braneworlds}. The idea is simple and it is based on the assumption that our four-dimensional world is just an extremely thin 3-brane embedded in a higher-dimensional space. All SM fields are bound to this brane while gravity may travel in the bulk. These scenarios are, in general, oversimplified field theoretical models. Although their origin in the context of string theory is not completely clear yet, they set the stage for finding solutions to long-standing problems of particle physics (\eg, hierarchy problem, cosmological constant problem). Even more, since braneworlds allow for large or infinite extra dimensions they are potentially testable in the next generation of particle accelerators. The number as well as the size of extra dimensions is model-dependent and they are restricted also by their implications in cosmology and astrophysics.

The simplest possible scenario \cite{Arkani-Hamed:1998rs,Antoniadis:1998ig} is to neglect the tension (energy density) of the brane and to consider compact extra dimensions. The difference with the ordinary Kaluza-Klein picture is that the size $R$ of the extra dimensions needs not to be microscopic. Since at distances smaller than $R$ gravity behaves five-dimensional, it could modify Newton's attraction law (experimentally established down to $0.2\:{\rm mm}$). This fact sets an upper bound on $R$. Quantitatively, the order of magnitude of $R$ can be determined in terms of the fundamental scale of gravity\footnote{In the future we will set $c= \hbar =1\,.$} in five dimensions $M\,,$ the four-dimensional Planck scale $M_{p}$ and the number of extra dimensions $d$
\begin{equation}
R \sim M^{-1} \Big( \frac{M_p}{M}\Big) ^{2/d} \,.
\end{equation}
It is then clear that we could address the hierarchy problem, i.e., the existence of two fundamental scales $M_p$ and $M_{EW} \sim 1\: {\rm TeV}$ in four dimensions, by invoking extra dimensions. Indeed, the weakness of the gravitational force can be a consequence of the fact that $M_p$ is not fundamental in the higher-dimensional setup and its large value is an artifact of large extra dimensions. For instance, by assuming two extra dimensions and the fundamental scale $M$ slightly above the electro-weak scale $M_{EW}$, say $30\: {\rm Tev}$, the value of $R$ should be in the $1 \,-\, 10\:{\rm \mu m}$ range. However, this is merely a reformulation of the hierarchy problem because one has to explain the large size of the extra dimensions (i.e., why $1/R \gg M_{EW}$). 

A different approach \cite{Randall:1999ee} to the hierarchy problem in the context of braneworlds is to consider a setup with two $3$-branes and one finite extra dimension.  Furthermore, the gravitational contribution of the branes is accounted for by assigning them non-vanishing tensions of opposite sign. The bulk theory is AdS and the boundary conditions relate the brane tensions to the five-dimensional cosmological constant. Requiring the metric to preserve four-dimensional Poincar\'e invariance on the branes, the resulting geometry will not be factorizable as in the KK case. Instead, it is a warped Minkowski spacetime with a warp factor depending exponentially on the extra dimension. The warp factor decreases from the positive to the negative tension brane making the gravitational coupling of particles on the negative tension brane weaker than those on the positive tension companion. It is the presence of the warp factor that will finally lead to an exponential hierarchy on the brane. 

In the Randall-Sundrum I scenario (RS1) ordinary matter resides on the negative tension brane, its gravitational interactions being weak when the separation between the branes is relatively large. The interest in RS1 originates from its elegant solution to the hierarchy problem. Analogous to the results of \cite{Arkani-Hamed:1998rs,Antoniadis:1998ig} the model exhibits only one energy scale. Indeed, for an inverse AdS radius $k$ of order $1\: {\rm Tev}$, the Planck mass is  again related to the electro-weak scale by
\begin{equation}
M_p = \rme^{k z_c} M_{EW}\,,
\end{equation}
where $z_c$ is the brane separation. It is obvious that a large value of $M_p$ is achieved with only $k z_c \sim 37\,.$ Hence, in contrast to the previous model, in RS1 a large~$M_p$ is not obtained at the price of introducing another scale. On top of this at large distances the effective theory of gravity is four-dimensional. The main shortcoming of RS1 is the need of a stabilization mechanism for the relative distance between the branes (corresponding to the radion), which is not yet under control. 
 
In a subsequent work \cite{Randall:1999vf} (RS2), Randall and Sundrum came up with the concept of {\it infinite} extra dimensions (for reviews see \cite{Rubakov:2001kp,Padilla:2002tg}). The idea is to increase the brane separation from RS1 by sending the negative tension brane to infinity. The resulting setup has only one positive tension brane, on which matter is trapped, and an AdS bulk. To ensure four-dimensional Poincar\'e invariance, the brane tension is fine-tuned against the cosmological constant. The geometry of the model is again a warped spacetime corresponding to two AdS spaces glued together on the brane. Nevertheless, in this context the warp factor does not generate hierarchy.

Since the extra dimension extends to infinity, one would naively expect that gravity behaves five dimensionally. However, the novelty brought by RS2 is the localization of the gravity on the brane. The key role is played by the exponentially-dependent warp factor that suppresses the gravitational perturbations in the direction transverse to the brane. Thus, at large distances gravity looks four-dimensional to a brane observer. 
 
Even though this scenario cannot solve the hierarchy problem, it offers an ingenious `alternative to compactification'. In fact, a simple extension of it \cite{Lykken:1999nb}, by including a probe brane at finite distance and setting the matter on this brane, leads to an exponential hierarchy with an infinitely large extra dimension. 

Finally, we mention that from the phenomenological viewpoint we are more motivated to consider RS2 scenarios with dS four-dimensional braneworlds, instead of flat branes. These were indeed discussed, \eg, \cite{Padilla:2002tg}, and it turns out that no fine-tuning of the brane tension is required. In fact, to ensure a positive cosmological constant on the brane, the brane tension needs to exceed its critical value (from RS2), so that these branes are often called supercritical. Also, the warp factor changes and is vanishing at a finite distance from the brane. Thus, the damping of the metric perturbations 
reduces the distance range where Newton's law is violated, which makes this setup even more attractive if one wants to reproduce a four-dimensional gravity theory.    

\section{Thesis outline and results \label{s:outline}}

In this introductory chapter, we have sketched the theoretical and phenomenological frameworks within which the research presented in this work has been carried out. In the remainder, we briefly display an overview of the thesis together with the main results.

The thesis focuses on two central topics. The first one deals with the concept of {\it self-dual theories} and their connection to S-duality. The other part is dedicated to {\it gauged supergravity}, mainly in five dimensions, but some aspects in different dimensions will be touched on in Ch.~\ref{ch:sols}. More precisely, the conformal tensor calculus (SCTC) method will be employed as the main tool for constructing such low-energy limit supergravity theories. 

The next chapter sets the theoretical stage in which the the rest of the work was performed. We describe therein technically the theoretical tools and methods that will be used later on either for quantizing and deforming the action of duality-symmetric gauge theories or in constructing five-dimensional Poincar\'e SUGRA via SCTC. We start by briefly reviewing the idea of string theory that can be immediately extended to the worldvolume description of other multi-dimensional objects,~$p$-branes. Next, we restrict ourselves to the effective low-energy case of strings, \ie, SUGRA. In this context, p-brane solutions enter naturally as solutions to the classical equations of motion. Further, we want to introduce the method of  tensor calculus based on the superconformal group (in $D=5$). To that end we first examine the conformal symmetry and its algebra. With this knowledge, we can elaborate the conformal tensor calculus and exemplify it on a toy model. This technique will be the cornerstone of Ch.~\ref{ch:weyl} and \ref{ch:matter}. For the non-experienced reader, Sect.~\ref{confSugra}, or at least Sect.~\ref{s:sctctoy}, will be highly recommended. Finally, the Batalin-Vilkovisky (BV) formalism provides an important tool in the treatment of gauge theories, both classically and quantum-mechanically. This machinery will be introduced in Sect.~\ref{s:BV} and applied for self-dual vectors in Ch.~\ref{ch:dual}. Once again, Sect.~\ref{s:BV} represents an introduction to gauge theories and the BV method for the reader not familiar to systems with gauge degrees of freedom.

Single \br{3}s support, on their worldvolume, self-dual Abelian vector fields.\footnote{Self-duality in $D=4,\, 8$ will be further referred to as {\it duality-symmetry}.} The implementation of the duality symmetry at the level of the action is the subject of Ch.~\ref{ch:dual}. It can be done either non-covariantly \`a la Schwarz-Sen or manifestly Lorentz covariant. The former approach was the starting point for making manifest the $SL(2,\R)$ symmetry of the low-energy effective action of the heterotic string compactified on a six-dimensional torus. The latter approach was proposed by Pasti-Sorokin-Tonin (PST) and can be applied to other situations where chiral bosons are part of the spectrum: self-dual five-form field strength in type IIB or self-dual two-form living on an \mbr{5}. The price paid for maintaining both the duality symmetry and Lorentz covariance manifest in the action is the presence of an auxiliary field that enters the action non-polynomially. In Ch.~\ref{ch:dual}, we perform a BV quantization of the PST model and arrive at the conclusion that it is equivalent to the usual Maxwell theory in four dimensions.Yet another strong argument for the study of electric-magnetic duality comes from its importance in the context of extended supersymmetric field theories or vector multiplets couplings to SUGRA in $D=4$. 

A stack of \br{3}s should be characterized by a non-Abelian version of the self-dual theory discussed in the previous chapter. Such non-Abelian deformations are non-trivial in the sense that they involve non-locality or they require going beyond perturbation  theory. In fact, the goal of Ch.~\ref{ch:defo} is twofold: on one hand to look for non-Abelian extensions of duality-symmetric actions and, on the other hand, to introduce self-interactions of a single duality-symmetric gauge field. By using cohomological arguments, it will be shown that non-Abelian, local, continuous deformations of a sum of non-covariant duality-symmetric actions are NOT possible. This result points towards a non-perturbative description of non-Abelian duality. The study of self-interactions of one-form gauge potentials with manifest duality symmetry will lead to a PST-like description of the 3-brane in type IIB SUGRA. Some comments on the coupling of this duality-symmetric worldvolume description with type IIB background fields are made. 

In Ch.~\ref{ch:weyl}, we begin our program of constructing five-dimensional Poincar\'e\linebreak SUGRA with eight supercharges which include also CY compactifications\linebreak of $D=11$ SUGRA. We also want to incorporate the most general matter couplings to this theory. Our approach is based on the SCTC applied to the superconformal group $F^2(4)$ (in $D=5$). The first step is to derive the Weyl multiplet, \ie, the multiplet of gauge fields associated to the superconformal group. By either using the supercurrent method or by direct gauging, two versions of the Weyl multiplet are inferred. They go under the name of Dilatonic Weyl and Standard Weyl multiplets, depending on whether or not they incorporate the dilaton field as their scalar field. It is further shown that the Dilaton Weyl multiplet emerges from coupling the Standard Weyl multiplet to an improved vector multiplet. This was to be expected since a similar phenomenon occurs in $D=6$.  

The SCTC program is carried on in Ch.~\ref{ch:matter} where various other multiplets (linear, vector, vector-tensor and hypermultiplet) are presented. They are different representations of the superconformal group. One particularly interesting result is that requiring the realization of the supersymmetry algebra on the vector-tensor multiplet  or the hypermultiplet will automatically lead to equations of motion for the corresponding fields. In other words, we derive equations of motion from supersymmetry before having an action. The situation resembles the case of type IIB SUGRA before the PST actions were proposed. Another significant feature concerns the geometry of the scalar manifold parametrized by the hypermultiplets. This is a hypercomplex manifold, completely determined in terms of objects entering the supersymmetry transformation rules of the hypermultiplet fields. 

The next step is to derive Lagrangians for these multiplets. It turns out that the existence of an action is a restrictive requirement that implies the presence of a metric tensor on the scalar manifold. This scalar-dependent metric enters, for instance, in the kinetic term of the scalars in a similar fashion to the non-linear sigma models. In the vector-tensor multiplet sector we find, even at the rigid level, that non-trivial (off-diagonal) couplings between vectors and tensors are possible, which is a novelty with respect to the existing literature. In the hypermultiplet sector, the scalars are now parameterizing a hyperk\"ahler manifold. Finally, in the case of Abelian vectors, an embedding of the vector in the linear multiplet combined with a density formula can also be applied to determine the Lagrangian action.

Ch.~\ref{ch:sols} aims to describe different solutions of $D=5$ gauged supergravities. We will focus on black hole and string solutions of $N=2$, $D=5$ SUGRA, but we will also make some comments on $D=6,\, 7$ multicharge brane solutions. All these solutions carry either electric or magnetic charges and they are obtained using the superpotential technique. Besides these single-charged solutions, we will also devote some attention to the example known as the three-charge string. The technique used in deriving such solutions is based on first-order differential equations that can be either the BPS equations or equivalent ones expressed in terms of a superpotential. The higher-dimensional origin of some of these objects is also discussed. 

For the convenience of the reader we enclosed at the end a glossary of terms and one of symbols, which are used throughout the thesis.   


\chapter{String theory toolkit \label{ch:kit}}

The present chapter collects the essential ingredients and techniques that will be employed later on for the construction of self-dual gauge theories, SUGRA and solutions thereof. At the same time, it gives the theoretical framework within which the following chapters have to be understood. It serves as the theoretical reservoir for the rest of the thesis, which can proceed without being interrupted by messy pauses required by definitions and side explanations.

As advertised in the introductory chapter, symmetries are at the heart of string theory and ordinary gauge field theories. Therefore, throughout this chapter we will review some of the symmetries encountered in string and SUGRA theories. It will turn out that the concept of symmetry can also be used as an important tool in constructing some of these theories. Of particular interest will be the (super)conformal algebra that proved to be a key symmetry of the worldsheet description of string models.  

We start by presenting the basics of string theory, D-branes and Kaluza-Klein compactifications. The worldvolume description of D-branes will be needed in Chapt.~\ref{ch:defo}, when a similar analysis will be performed for the self-dual \br{3}. Then, we introduce the concept of rigid and local supersymmetry and construct SUGRA theories in $D=11$, $D=10$ and $D=5$. The attention goes to the five-di\-men\-sio\-nal case where some generic properties of SUGRAs are inferred. They are the backbone of later developments in the thesis. Next, the conformal algebra, its representations and application in the SUGRA construction are investigated. The gauge equivalence program on which SCTC relies, is exemplified by a simple but clarifying model. Our presentation of string theory continues with the SUGRA description of $p$-brane physics. We briefly display the standard  $p$-brane solutions and their far and near-horizon geometries. Finally, in preparation for Chapt. \ref{ch:dual}, we briefly review the formalism developed by Batalin and Vilkovisky for quantizing systems with gauge degrees of freedom. 

The reader should be aware that the knowledge related to string theory, SUGRA and D-branes is by now quite enormous and is increasing rapidly. Therefore, the selected subjects reflect a personal preference based on the relevance for the subsequent chapters. For completing the discussion that sometimes may be superficial and for further details regarding different techniques or research avenues, the interested reader is urged to consult the standard text books, reviews or original papers indicated along the way. 

\section{String theory: a guided tour\label{string}}

Even if discovered and studied in their own right, SUGRA is nowadays strongly motivated by its interpretation as low-energy limit of some string theory or compactifications hereof. Before moving on to the discussion of SUSY and SUGRA it is worthwhile to start with a brief review of the bosonic string and its symmetries. We follow here the line of \cite{Polchinski:1998j} (for reviews, see Ref.~\cite{Kiritsis:1997hj} or less technical Refs.~\cite{Schwarz:2000ew,deBoer:2002ts})

A relativistic point particle  moving in a D-dimensional spacetime sweeps out a worldline parametrized by $\tau$. The embedding of the trajectory in D-dimensional flat Minkowski space (with metric $\eta_{\mu\nu}$ `mostly plus') is given in terms of the coordinates $X^\m(\ta)$ whose evolution is encoded in the action
\begin{equation}\la{actpart}
S = - m \int d \tau \sqrt{- \eta_{\m\n} \dot X^\m \dot X^\n}\,,
\end{equation}
where $m$ is the mass of the particle and dots represent derivatives with respect\linebreak to~$\ta$. The action is invariant under local reparametrizations $\ta \ra \tilde\ta ({\ta})\,.$ This  is also referred to as diffeomorphism invariance.

\begin{center}
\begin{figure}[H]
\epsfxsize =13cm
{\hskip 0.5cm \epsffile{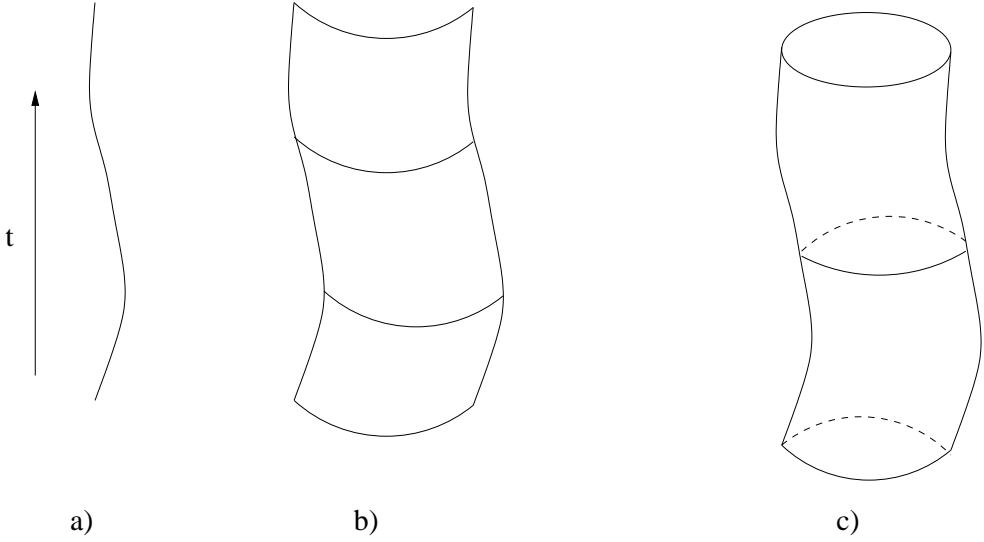}
}
\caption{{\it a) A propagating point particle (time flows in the vertical direction);
\hskip2cm 
b) The worldsheet of an open string; c) A closed string worldsheet.}\label{fig:string}}\end{figure}
\end{center}

The generalization of the action \re{actpart} for a propagating string is straightforward
\begin{equation}\la{NGact}
S = - T \int d \ta d\s \sqrt{\vert \det{\del_\a X^\m \del_\b X_\m}\vert}\,, \qq \a, \b = {\ta, \s}\,,
\end{equation}
with $T$ the tension of the string (mass per unit length). This is nothing but the area of the two-dimensional surface swept out by the string (see Fig.~\ref{fig:string}). 

A more convenient form of the action is obtained by considering the metric $h_{\a\b}$ of the string worldsheet as auxiliary fields in the action ($h^{\a\b}$ is the inverse metric and $h = \vert \det{(h_{\a\b})}\vert$). Then, one can write down the following equivalent action
\begin{equation}\la{actst}
S [X, h] = -\frac{T}{2} \int d^2 \s \sqrt{h} h^{\a\b} \eta_{\m\n} \del_\a X^\m  \del_\b X^\n\,, \qq d^2 \s = d \ta d\s\,.
\end{equation}
Indeed, since the equation for the worldsheet metric is purely algebraic one can integrate it out and recover \re{NGact}. A direct computation shows that 
both actions \re{NGact} and \re{actst} are invariant under worldsheet reparametrizations, which implies the conservation of the energy-momentum tensor $T_{\a\b}\,.$ A more important remark is the {\it conformal invariance} (Weyl invariance) of $S[X, h]\,.$ Indeed, \re{actst} possesses the following local symmetry 
\begin{eqnarray}\la{confstr}
h_{\a\b} &\ra &\rme^{\L ({\ta , \s})} h_{\a\b} \,,\nn\\
X^\m &\ra &X^\m\,.
\end{eqnarray}
It implies that the energy-momentum tensor (defined as $T^{\a\b} \sim \frac{\d S}{\d h_{\a\b}}$) is traceless
\begin{equation}\la{trac}
T_{\a\b}h^{\a\b}=0\,.
\end{equation} 
In fact, together with the diffeomorphism invariance this symmetry allows us to choose a `conformal gauge' 
\begin{equation} \la{confga}
h_{\a\b} = \rme^{\xi(\ta,\s)} \eta_{\a\b} 
\end{equation}
such that the gauge-fixed action 
describes a set of free worldsheet scalars $X^\m (\s^\a)$ (since the exponential drops out of the action).

The conformal symmetry is of utmost importance also at the quantum level. More precisely, when one performs the quantization of the bosonic string it is found that, in order to cancel the Weyl anomaly, spacetime needs to have a critical\footnote{For superstrings the critical dimension must be $D=10\,.$} dimension $D=26\,.$    

Finally, there is a D-dimensional Poincar\'e invariance 
\begin{equation}
X^\m \ra \L^\m{}_\n X^\n + a^\m\,,
\end{equation}
which leaves the worldsheet metric unchanged ($\L^\m{}_\n$ is a Lorentz matrix and $a^\m$ is a constant vector). From the worldsheet point of view this is merely an internal symmetry of the free scalars $X^\m$, but in target space it means that all particle excitations of the string form unitary representations of the Poincar\'e group. In other words, they are labeled according to their mass and spin. In principle, there are an infinite number of particle states corresponding to an infinite number of different harmonics allowed by the string oscillations, but we will subsequently focus on the massless spectrum, which is restricted to spin 2 particles. Neglecting the tachyon mode (that is absent anyway in the full superstring theory), the lowest (massless) states in the bosonic closed string sector are: the graviton spin 2 particle $G_{\m\n}\,,$ the dilaton scalar field $\Phi$ and the antisymmetric tensor field $B_{\m\n}\,.$ This is known as the universal sector. In the case of the open string one finds at the lowest level a massless photon state $A_\m$ that plays a major role in the context of D-branes. The difference between the two types of strings comes from the boundary conditions that one has to impose for each of them. For the closed string, one needs a periodicity condition
\begin{equation}\la{percond}
X^\m (\ta, \s) = X^\m (\ta, \s +  \pi)\,,
\end{equation}  
while in the open string case one may choose between Neumann or Dirichlet boundary conditions
\begin{eqnarray}
\del_\s X^m \vert_{\s = 0, \pi} =0\,, &\q m = 0,& \dots, p \qq ({\rm N.~bc.})\la{Nbc}\\
\del_\ta X^i \vert_{\s = 0, \pi} =0\,, &\q i =p+1,& \dots, D-1\qq ({\rm D.~bc.})\la{Dbc}
\end{eqnarray} 
or a mixture of them. In contrast to \re{percond}, which preserves the 26-dimensional Poincar\'e invariance, the D.~bc. are breaking this symmetry by constraining the ends of the open string to be stuck on a multidimensional hyperplane ($X^i = X^i(0)= X^i(\pi)$). This physical object goes under the name of D-brane and its dynamics will be discussed in Sect.~\ref{s:wv} (see also Sect.~\ref{s:pbra}).

We were concerned hitherto with strings propagating in a flat background. In order to see how the low-energy effective action appears, we have to investigate other backgrounds. We consider now a string in a curved spacetime characterized by $G_{\m\n}(X)$ and we turn on also a background antisymmetric tensor $B_{\m\n}(X)$ together with the dilaton field $\Phi (X)\,.$ The action that generalizes \re{actst} and is consistent with the way strings couple to these fields is a string sigma model action
\begin{equation}\la{backgr}
S = - \frac{T}{2} \int d^2 \s \sqrt{h} \Big( h^{\a\b} G_{\m\n} \del_\a X^\m \del_\b X^\n + \ve^{\a\b} B_{\m\n} \del_\a X^\m \del_\b X^\n + \a^\prime \Phi R^{(2)}\Big)\,,
\end{equation}
where $T = \ft{1}{2 \pi \alpha^\prime}$ and the components of the antisymmetric tensor $\ve^{\a\b}$ are given by $\ve^{12} = - \ve^{21} =1$ and $R^{(2)}$  is the two-dimensional scalar curvature. In general, a straightforward quantization of \re{backgr} is not possible. The standard procedure is to perform a perturbative expansion in $\a^\prime\,,$ similar to the $\hbar$ expansion in field theory. Since the dilaton term does not contain any  $\a^\prime\,,$ it represents a first-order contribution to this expansion. The action was constructed such that it is covariant under reparametrization. The first two terms preserve also Weyl symmetry but, in general, the dilaton coupling breaks it. We have seen  that conformal invariance is crucial for the consistency of the theory. We are interested to find under which circumstances this symmetry is restored. In particular, Weyl symmetry implies global scale invariance, which means the cancellation of the $\beta$-functions associated with $G_{\m\n}\,,$ $B_{\m\n}$ and $\Phi$ in the field theory. This can be expressed as a power series expansion (up to leading order) 
\begin{eqnarray}
\b^G_{\m\n} \! & \equiv &\! \a^\prime\Big( R_{\m\n} + 2 \nabla_\m \nabla_\n \Phi - \ft 14 H_{\m\s\r} H_\n{}^{\s\r}\Big) +O(\a^{\prime 2}) =0\,,\la{betaG}\\
\b^B_{\m\n} \! & \equiv &\! \a^\prime\Big( - \ft 12 \nabla^\r H_{\m\n\r} + \nabla^\r \Phi \,H_{\r\m\n}\Big) +O(\a^{\prime 2})=0\,,\nn\\
\b^\Phi  \!& \equiv &\! \a^\prime\Big( - \ft 12 \nabla^\m \nabla_\m \Phi + \nabla^\m \Phi \nabla_\m \Phi- \ft {1}{24} H_{\m\n\r} H^{\m\n\r}\Big) + O(\a^{\prime 2})=0\,,\nn
\end{eqnarray}
where the three-form field strength is defined as $H_{\m\n\r} = 3 \del_{[\m} B_{\n\r]}\,.$

In the low-energy limit $\a^\prime \ra 0 $ these constraints on the massless modes of the closed string spectrum can also be interpreted as spacetime field equations for those fields that can be inferred from the action
\begin{equation}\la{clostring}
S = \frac{1}{2 \kappa_0^2}\int d^{D} x \sqrt{G} \rme^{-2 \Phi} \Big( R + 4 \del_\m \Phi \del^\m \Phi  - \ft {1}{12} H_{\m\n\r} H^{\m\n\r} \Big)\,.
\end{equation}
This is the effective low-energy action of the universal closed string sector. The dimension D equals 26 (10) in bosonic (super)string respectively. Besides massless fermionic contributions to this action, in superstring theory one has to add specific RR terms that are model dependent (we will return to this point later).   

The constant $\k_0$ is not determined by the field equations. It is however not a relevant parameter since its value can be modified by constant shifts in the dilaton field. Then, it should be possible to relate it to the Newton constant. Obviously, the kinetic term of the scalars comes with a wrong sign in the action and there is also a non-standard Einstein term. These two problems can be solved by a metric redefinition
\begin{equation}\la{sttoE}
g_{\m\n} = \rme^{- 4 \phi/(D-2)} G_{\m\n}\,.
\end{equation}
We denoted $\phi = \Phi - \Phi_0$ for some constant $\Phi_0$ to be interpreted shortly. In terms of the new metric, the action \re{clostring} becomes
\begin{equation}\la{clostr}
S = \frac{1}{2 \kappa^2}\int d^{D} x \sqrt{g}  \Big( R - \ft 4{D-2} \del_\m \phi \del^\m \phi  - \frac {1}{12}\, \rme^{-8 \phi/(D-2)}\ H_{\m\n\r} H^{\m\n\r} \Big)\,.
\end{equation}
We see that the common Einstein term is reproduced. We can match even the coefficient of this term with the common action of General Relativity by imposing
\begin{equation}\la{genst}
\k \equiv \k_0 \rme^{\Phi_0} = \sqrt{8 \pi G} \,,
\end{equation}
with $G$ being the Newton constant. The present terminology is that the original fields determine the `string frame' while the new metric tensor $g_{\m\n}$ stands for the `Einstein frame'. All physical quantities, \eg, gravitational mass-energy, are measured in the latter frame. 

\begin{center}
\begin{figure}[H]
\epsfxsize =13cm
{\hskip 0.5cm \epsffile{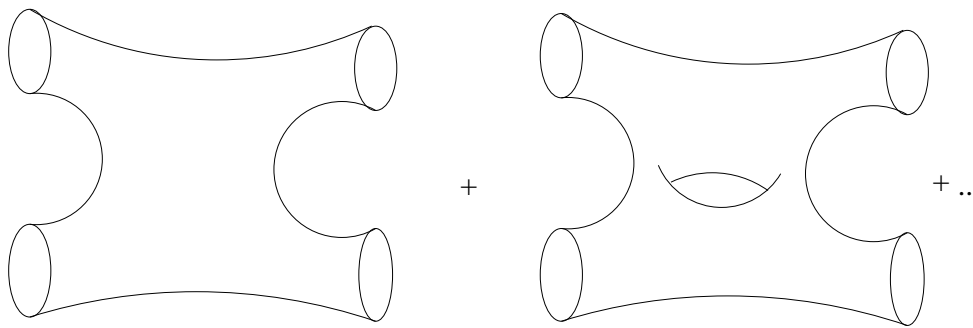}}
\caption{ {\it A perturbative expansion of string theory four-point interactions.}\label{fig:stringinter}}
\end{figure}
\end{center}

Similar to field theory, string interactions can be pictured as worldsheet joinings and splittings. The interaction is characterized by the string coupling constant $g_S$  and the computation of a scattering process is performed as a perturbation series in $g_S$ as long as $g_S\ll 1\,.$ For an example, we consider a four-point amplitude computation depicted in Fig.~\ref{fig:stringinter}. 
The four external legs correspond to incoming and outgoing strings. The first diagram shows two incoming strings joining together and later splitting apart. With that understanding, it is then obvious that such a process will contribute with a weight $g_S^2$ to the total amplitude. The one-loop interaction contains four vertices and, in general, each diagram with $n$ holes will be proportional to $g_S^{2+2n}\,.$ Hence, the total amplitude can be represented as a formal power series expansion
\begin{equation}\la{strampl}
{\cal A} = \sum_{n=0}^\infty g_S^{2+2n} {\cal A}_{(n)}\,.
\end{equation}
At this point, we remark that the power of the coupling constant is the negative of the Euler number of a two-dimensional surface with $4$ boundaries and genus $n$. More generally, for a multiparticle scattering one has to replace the power $2n +2$ by $-\chi$, where the Euler number of a two-dimensional worldsheet of genus $n$ and $b$ boundaries is defined as
\begin{equation}\la{eule}
\chi \equiv \frac{1}{4 \pi} \int d^2 \s \sqrt{h} R^{(2)} = 2 - 2 n -b\,.
\end{equation}
   
It should be noticed that, due to the particular dilatonic coupling in the action, a constant shift in the dilaton $\Phi \ra \Phi + \Phi_0$ equals a shift $S \ra S + \Phi_0 \chi$ in \re{backgr}. In the path integral this affects only the relative weighting of different topologies in the sum \re{strampl}. As a consequence, we can identify the string coupling as the exponential of the dilaton vacuum expectation value $\Phi_0$, i.e.,
\begin{equation}\la{coucon}
g_S = \rme^{\Phi_0}\,.
\end{equation}
The last relation has a very deep interpretation. It teaches us that, in contrast to QED or field theory in general, string theory does not have a theoretically arbitrary coupling constant, fixed only by experiment. Instead, the coupling constant of the theory depends on the one-parameter family of vacua of the theory. Thus, the string vacua dynamically determine the coupling of the theory. 

To summarize, in this subsection we presented some basic elements of perturbative bosonic string theory. The emphasis went to the importance of the conformal symmetry and the derivation of the low-energy effective action \re{clostr}. We now move on to the discussion of some non-perturbative aspects, namely D-branes. 

\section{Worldvolume actions\label{s:wv}}

Another hot topic in string theory that motivated part of the results presented in Ch.~\ref{ch:defo}, is the subject of D(irichlet)-branes \cite{Johnson:2000ch}. There are two main approaches to D-branes: the worldvolume picture and the SUGRA description. In this\linebreak section~\ref{s:wv}, we briefly review the former, while the later will be elaborated in Sect.~\ref{s:pbra}. The aim is to reveal the connection between D-branes and gauge theories. This is a direct consequence of the fact that D-branes are dynamical objects on which open strings can end. In particular, we will see that their dynamics is determined by a worldvolume action that is sum of a Dirac-Born-Infeld (DBI) and a Wess-Zumino~(WZ) piece. 

\begin{center}
\begin{figure}[H]
\epsfxsize =13cm
{\hskip 0.5cm \epsffile{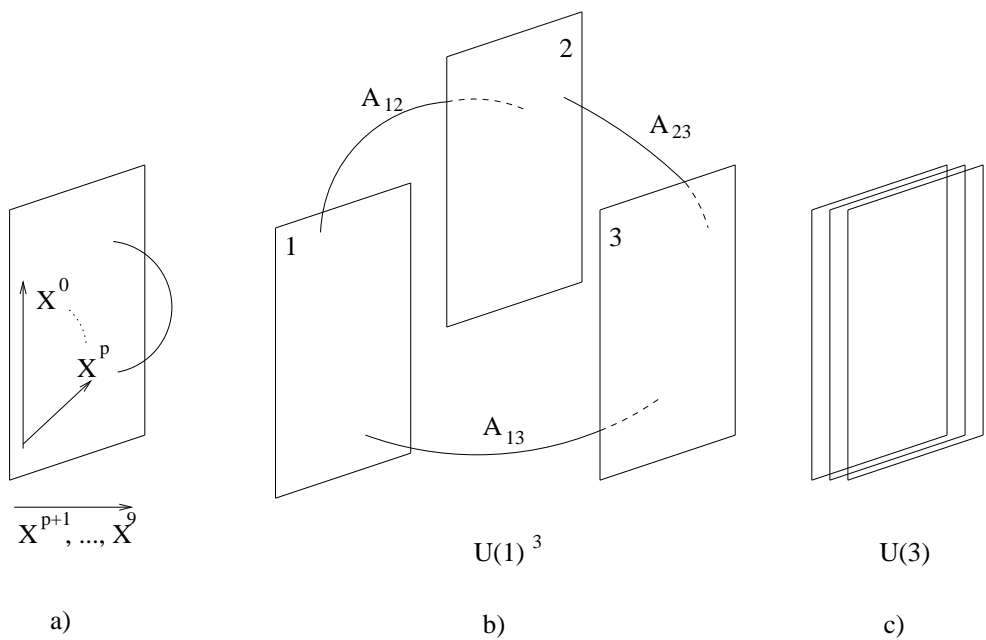}}
\caption{ {\it a) Open strings ending on a D-brane. b) N parallel D-branes and strings stretching between them ($\U(1)^N$ gauge symmetry). c) A stack of N \br{p}s with enhanced $\U(N)$ gauge group.}\label{fig:brans}}
\end{figure}
\end{center}

In string theory, \br{p}s are $p$-dimensional spatially extended objects that carry open string degrees of freedom since the ends of such strings are restricted to the worldvolume of the brane (see Fig.~\ref{fig:brans}). This can be imposed by appropriate boundary conditions \re{Nbc}-\re{Dbc}. 

We are now interested to find the field content on the worldvolume of the brane and the effective action that captures its low-energy interactions with the other string degrees of freedom. The massless spectrum of the open strings contains, in general, a $\U(1)$ gauge potential $A_\m$, which can be decomposed into a vector\linebreak field $A_m$ along the worldvolume and $9-p$ worldvolume scalars $\Phi^i$ given by the transverse components. The scalars determine fluctuations in the position of the brane. 
Together with the supersymmetric partners, these fields are expected to describe, in the low energy approximation, a $\U(1)$ SYM theory in $p+1$ dimensions. 
One way to determine the dynamics of the aforementioned degrees of freedom is to make use, once again, of the conformal invariance of the string worldsheet.\footnote{Another way advocated in \cite{Johnson:2000ch} is to use T-duality and then construct the action piece by piece.} Besides the general non-linear sigma model action \re{backgr} of the bulk string worldsheet, one has to consider boundary couplings due to the presence of open strings. They are of the form
\begin{equation}\la{boact}
S_b = - T \int_{\del \Sigma} d\s \Big( A_m  \del_\tau X^m + \Phi_i \del_\s X^i\Big)\,.
\end{equation}
Next, we require that the combined-system, bulk-boundary action is Weyl invariant in order to decouple negative norm states from the spectrum. Analogous to the previous section this leads to the dynamical equations for the ten-dimensional spacetime fields. Finally, one identifies the action that reproduces these field equations. 

The complete procedure was initially carried out in \cite{Leigh:1989jq} and the effective action describing the D-brane bosonic sector is DBI-like action (whose form is familiar from non-linear electrodynamics)  
\begin{equation}\la{DBI}
S_{DBI} = - T_p \int d^{p+1} \s \rme^{- \Phi} \sqrt{\vert \det{(G_{mn} + B_{mn} + 2 \pi \a^\prime F_{mn})}\vert}\,,
\end{equation}
where all the fields $G_{mn}$, $B_{mn}$ and $\Phi$ are pull-backs of the spacetime fields to the worldvolume of the brane (\eg, $G_{mn} = G_{\m\n} \del_m X^\m \del_n X^\nu$) and $F_{mn}$ is the field strength of the worldvolume gauge potential. The particular coupling to the dilaton field $\rme^{- \Phi}$ has its roots in the open string tree-level physics that is under consideration. The presence of the NSNS gauge two-form (the terminology will be explained in Sect.~\ref{Sugra}) was to be expected from symmetry arguments. Indeed, besides the usual worldvolume diffeomorphism invariance and the ordinary gauge symmetry of the one-form $A_m$, there is also a Kalb-Ramond symmetry
\begin{equation}\la{Kalb}
\d A_m = - \frac{1}{2 \pi \a^\prime} \,\xi_m\,, \qq \d B_{mn} = 2 \del_{[m}\xi_{n]}\, 
\end{equation}
that leaves invariant the combination $B_{mn} + 2 \pi \a^\prime F_{mn}\,.$ The geometrical interpretation of \re{DBI} is recovered by turning off $B_{mn}$, $\Phi$ and $F_{mn}$, in which case we are left with the following action
\begin{equation}\la{volDp}
S = - T_p \int d^{p+1} \s  \sqrt{\vert \det{(G_{mn} )}\vert}\,.
\end{equation}  
This gives the volume of the hypersurface swept out by the \br{p} up to a constant that is proportional to the brane tension 
\begin{equation}
\tau_p = \frac{T_p}{g_S} = \frac{1}{g_S (2\pi)^p\,\a^{\prime \, \frac{p+1}{2}}}\,.
\end{equation} 

It is instructive to see what the action \re{DBI} describes after some simplifying assumptions. We first presume that the background $G_{\m\n}$ is flat. Moreover, if the D-brane is also almost flat (static gauge) we can identify the worldvolume coordinates with the ten-dimensional ones, \ie, $\s^m = X^m\,.$ Furthermore, we consider a vanishing $B_{mn}$ and small, but of the same relative order, $\del_m X^i $ and $2 \pi \a^\prime F_{mn}\,.$ In this low-energy approximation, the DBI action takes the form \cite{Taylor:1997dy}(after subtracting the constant volume term)
\begin{equation}\la{YMU1}
S_{\U(1)-YM} = - \frac 1{4 g_{YM}^2} \int d^{p+1} \s \Big( F_{mn} F^{mn} + \frac{2}{(2 \pi \a^\prime)^2} \, \del_m X^i \del^m X^i \Big) \,,
\end{equation}
where the YM coupling constant is $g_{YM}^2 = \frac 1{4 \pi^2 \a^{\prime 2} \tau_p}\,.$ The last action describes the bosonic fields of a $\U(1)$ gauge theory on the \br{p} worldvolume with $9-p$ scalars $X^i$. After including the fermions, the low-energy theory on the worldvolume of the brane is the usual $\U(1)$ SYM in $p+1$ dimensions. This can be obtained via dimensional reduction of  the corresponding maximal supersymmetric theory\linebreak in $D=10$. 

More generally, after including the fermionic sector, the action \re{DBI} needs to be extended to a supersymmetric DBI version \cite{Cederwall:1997ri}. Furthermore, in the full superstring theory one should be aware of the RR gauge fields (to be discussed in the next section). Table~\ref{t:RR} captures the picture. 
\begin{table}[htbp]
\begin{center}
\begin{tabular}{|c|c|c|c|c|}
\hline
\rule[-1mm]{0mm}{6mm}
      &  {\rm NSNS}  & {\rm RR} & {\rm \br{p}}\, {\rm (e)} & {\rm \br{p}}\, {\rm (m)}\\
\hline
\rule[-1mm]{0mm}{6mm}
 {\rm IIA}& $G_{\m\n}$, $B_{\m\n}$, $\Phi$ & $A_\m^{(1)}$, $A_{\m\n\r}^{(3)}$ &{\rm $p = 0\,,\, 2$} & {\rm $p =  4\,,\, 6$} \\[1mm]
\hline
\rule[-1mm]{0mm}{6mm}
 {\rm IIB}& $G_{\m\n}$, $B_{\m\n}$, $\Phi$&  $A^{(0)}$, $A_{\m\n}^{(2)}$, $A_{\m\n\r\s}^{(4)}$ &{\rm $p = -1\,,\, 1\,,\, 3$}& {\rm $p = 3\,,\, 5\,,\, 7$}\\[1mm] 
\hline
\end{tabular}\caption{\it The massless bosonic sectors of type IIA and type IIB string theory, the corresponding electrically charged sources and their magnetic duals.\la{t:RR}}
\end{center}
\end{table}
For the time being, it suffices to know that, besides the same NSNS sectors, type IIA or type IIB strings have different RR fields. Thus, in IIA there is a one- and a three-form gauge field while in type IIB the zero-, two- and four-form gauge potentials are present. The last one also has a self-dual five-form field strength. 

The utility of these fields was obscure until Polchinski \cite{Polchinski:1995mt} pointed out that, similar to point particles that can be charged under a $\U(1)$ gauge field, \br{p}s carry RR charges. The coupling of the RR fields to the worldvolume of a \br{p} is given by the WZ action 
\begin{equation}\la{WZ}
S_{WZ} = \m_p \int \sum_n A^{(n) } \rme ^{2 \pi \a^\prime F}\,,
\end{equation}
where we used form notation and set $B_{mn}=0\,.$ The charge $\m_p$ they are carrying is related to the brane tension via supersymmetry. Of course, in the integrand of eq.~\re{WZ} the sum runs over all RR fields present in a given theory and the integrand picks up only a $(p+1)$-form. For example, for a \br{0} the r.h.s.~of eq.~\re{WZ} becomes simply $q\int A_m \frac{d X^m}{d \ta} d\ta\,,$ but higher-dimensional branes can play the role of charge sources also for RR potentials of rank smaller than $p+1\,.$  Indeed, in the case of a \br{3} \re{WZ} consists of two terms \cite{Tseytlin:1996it}
\begin{equation}\la{WZD3}
S_{D3-WZ} = \m_3 \int ( A_{(4) } + 2 \pi \a^\prime F_{(2)} \wedge A_{(2)} )\,.
\end{equation}

From Table~\ref{t:RR} we see that type IIA contains \br{p}s with $p=0, 2, 4, 6$ while in type IIB one encounters \br{p}s with $p=-1,1,3,5,7$. Due to Hodge-duality in $D=10$ that relates field strengths by $F_{(p+2)} =*F_{(8-p)}$ it is obvious that branes with $p+ p' =6$ are electric-magnetic dual to each other. In fact, their charges satisfy a Dirac quantization condition similar to \re{Diracq}, \ie, $\m_p \m_{6-p} = 2 \pi \,.$ Without any further details, we only notice for completeness that there is also a \br{8} in massive type IIA and filling-space \br{9} in type IIB. 
%

When we consider a system of $N$ parallel \br{p}s, besides open strings ending on the same brane there are also strings that stretch between different branes~(Fig.~\ref{fig:brans}). The latter can be labeled $A_\m^{ij}$ to indicate the orientation from the $i$th to the $j$th brane.  
The lowest mode described by such a string is massive since there is a minimal mass of the string proportional  to the brane separation. However, the massless modes corresponding to strings attached to the same brane give a total~$\U(1)^N$ gauge symmetry. Nevertheless, when the branes approach and coincide, the stretching strings become massless and the gauge symmetry is enhanced to $\U(N)$. The gauge potentials and the scalars  arrange nicely in a $\U(N)$ multiplet of SYM theory. D-branes thus offer a nice geometrical picture of spontaneous symmetry breaking. In the low-energy limit, the description of such N \br{p}s can be done by dimensionally reducing the corresponding $D=10$ $\U(N)$ SYM theory to~$D=p+1$. But for the full theory, we need a non-Abelian extension of the DBI action \re{DBI}. Such a non-Abelian generalization is, despite certain progress, still an enigma. A recent account of non-Abelian techniques applied to D-branes can be found in \cite{Myers:2003bw}.

In conclusion, the conformal invariance gives a criterion for computing the worldvolume action of a \br{p}. This is given by the sum of a DBI term and a WZ  contribution including the RR fields. For a stack of $N$ such branes, a non-Abelian generalization of \re{DBI} is required such that the low-energy limit gives a~$(p+1)$-dimensional $\U(N)$ SYM theory.  

\section{Kaluza-Klein compactification\la{s:KK}}

We have already seen that superstring theory requires higher dimensions. One way to reconcile this with our four-dimensional perspective is to assume that the extra dimensions are compact and small. We want to study how higher-dimensional fields reduce to lower dimensions. We exemplify that first on the simple case of a scalar field. Let $\hat\phi(x^\mu , z)$ be a $(D+1)$-dimensional massless scalar satisfying the flat space Klein-Gordon equation
\begin{equation}\la{KG0}
\hat \Box \,\hat \phi \equiv \partial^M \partial_M \hat \phi =0\,, \qq M= 0\,, \dots ,D\,.
\end{equation}

If one of the coordinates, here $z$, is compact, \ie, we identify  $z=z + 2 \pi R$, then we can expand $\hat \phi$ in a Fourier series 
\begin{equation}\la{KKexp}
\hat\phi(x^\mu , z) = \sum_n \phi_n (x^\mu) \rme^{\rmi n z/R}\,,
\end{equation}
where the dynamics of the lower-dimensional fields are given by 
\begin{equation}\la{KGm}
\Box \,\phi_n - \frac{n^2}{R^2} \,\phi_n = 0\,.
\end{equation}
Thus, $\phi_n$ represents a $D$-dimensional scalar field of mass $\frac{\vert n \vert}{R}$. Together with the massless zero mode, there is also an infinite tower of KK massive modes. By taking the compactification radius $R$ very small, these modes become extremely heavy and they decouple in the effective theory. Hence, when speaking of KK compactification, we have in mind not only the compactification of one or more dimensions, but also a consistent truncation to the massless sector. 

The KK compactification of other fields, like $p$-forms or graviton, happens in a similar fashion with special attention for the Lorentz index structure. Indeed, consider a $(D+1)$-dimensional pure Einstein theory described by the metric tensor~$\hat g_{MN}$. The KK idea (\eg, \cite{Duff:1994tn,Duff:2001jp,Overduin:1997pn}) is to split the Lorentz indices such that the resulting objects obey $D$-dimensional Lorentz symmetry, \ie, 
\begin{equation}
\hat g_{MN} =
\begin{pmatrix}
g_{\m\n} & \Phi A_\nu \\
\Phi A_\mu & \Phi
\end{pmatrix}, \qquad
\left\{
\begin{array}{ccl}
M,N &=& 0, \ldots ,D\,, \\
\m, \n &=& 0, \ldots ,D-1\,.
\end{array}
\right.  \label{KKmetr}
\end{equation}
It can be subsequently shown that $g_{\m\n}$, $A_\m$ and $\Phi$ satisfy $D$-dimensional field equations of a spin-2, spin-1 and dilaton fields. Note that KK reduction of a tensor fields produces in general a set of scalars in lower dimensions.

The $S^1$ compactification just presented can be directly generalized to  a torus compactification on $T^n$. Even less trivial compactifications, like on $n$-dimensional spheres $S^n$ \cite{Cvetic:2000dm} or CY \cite{Candelas:1991pi,Greene:1996cy,Aspinwall:2000fd} have been intensively studied in the context of string theory. We will apply this technique in Ch.~\ref{ch:sols}.

\section{Supergravities\label{Sugra}}

After reviewing the general features of the bosonic fields in string theory we go a step further and include fermions in our discussion. Fermionic fields are naturally incorporated into the theory by including a new symmetry transformation called supersymmetry \cite{Wess:1974kz,Wess:1974tw}. In this section, the supersymmetry algebra, its representations and dynamics in different dimensions are reviewed.

\subsection{Supersymmetry\la{SUSY}}

We start by considering rigid fermionic symmetries. Rigid supersymmetry \cite{Muller-Kirsten:1986cw,Ferrara:1984ij} refers to the spacetime-independence of the odd (anticommuting) parameters $\e_\a$ characterizing the transformation. If we denote the corresponding generators by $Q_\a\,,$ with $\a$ a spinor index, the supersymmetry operation
\begin{equation}\la{susyop}
\d(\e) = \bar \e^\a Q_\a
\end{equation} 
does not modify the parity of the field it acts upon. This property holds when the supercharges  $Q_\a$ act as follows
\begin{equation}\la{Qbf}
{\rm boson} \stackrel{Q_\a}{\longrightarrow} {\rm fermion} \stackrel{Q_\a}{\longrightarrow} {\rm translated}\: {\rm boson}\,.
\end{equation}
The last arrow expresses the most important characteristic of the super-Poincar\'e algebra, which is that the commutator of supersymmetries results in a translation. The generators of this superalgebra are labeled as $P_\m$, $M_{\m\n}$ and $Q_\a$ for translations, Lorentz rotations and supersymmetry, respectively.  Besides the usual Poincar\'e algebra, its minimal supersymmetric extension includes the commutation between the bosonic and `fermionic translations',  the spinorial character of $Q_\a$ and the anticommutator of two supercharges, \ie,
\begin{eqnarray}
&\left[P_\m   , Q_\a\right] =0\,, \qq \left[M_{\m\n}   , Q_\a \right]= - \ft 14 \g_{\m\n} Q_\a\,, &\la{suP1}\\
& \left\{ Q_{\a}, Q_{\b} \right\} = (\g^\m)_{\a\b} P_\m \,. &\la{suP2}
\end{eqnarray} 

Since supercharges square to translations, they have dimension $1/2$ ($P_\m$ has unit-mass dimension). That means that, for a scalar field $\sigma$ of dimension $(D-2)/2$ whose supersymmetry transformation commonly reads 
\begin{equation}
\d(\e) \s (x) = \bar \e \psi (x)\,,
\end{equation}
its superpartner $\psi$ carries dimension $(D-1)/2$. We thus expect that the scalar enters the supersymmetry variation of $\psi$ either via a derivative or a mass term. This general feature of supersymmetry will be encountered many times throughout this work. Due to the former contribution, the commutator of two supersymmetry operations always contains local translations\footnote{The spinor notation is explained in Appendix \ref{a:hyp}.}
\begin{equation}\la{comQQ}
\left[\d(\e_1), \d(\e_2) \right] \s (x)= \xi^\m \del_\m \s (x) + \ldots \,, \qq \xi^\m = \bar \e_1 \g^\m \e_2\,.
\end{equation}
Another direct and general valid consequence of \re{suP2} is, in the case of an invertible $P_\m$ operator, the equality between  number of bosonic and fermionic degrees of freedom in a given representation of \re{suP1}-\re{suP2}. This can be either true {\it on-shell} or {\it off-shell}, depending upon the realization of the supersymmetry algebra {\it with} or {\it without} imposing the equations of motion.  

To have the possibility of counting the fermionic degrees of freedom one needs to know the independent components of the minimal spinor in different dimensions. This is constrained by the reality conditions that can be imposed on the spinors and can be determined by analyzing the Clifford algebra of the $\g$-matrices in $D$ dimensions. For even dimensions, using  the gamma matrix $\g_* = \g_0 \cdots \g_{D-1}\,,$ left and right projection operators can be defined $P_{L,R} = \ft 12 (1 \pm \g_*)\,.$ When applied on Dirac spinors, they define {\it Weyl (chiral) spinors} (W)  $\chi_{L,R} = P_{L,R} \,\chi$, which preserve only half of their original number of complex components $2^{{\rm Int}[D/2]}$.  A further possibility for irreducible spinors is to implement a reality condition as $\chi^* = B \chi$, for some suitable matrix $B$. This also reduces by a factor of 2 the complex components and the spinors are called {\it Majorana spinors} (M). However, a Majorana condition is not always consistent with Lorentz transformations. Sometimes, one first has to double the number of spinors and then impose a pseudo-Majorana condition. If that is the case, the spinors are called {\it symplectic Majorana} (SM), and the number of independent components remains unchanged. For its simplicity, it is commonly preferred to use the formulation in terms of a doublet of SM field as we will see in $D=5$ SUGRA (see Appendix \ref{a:hyp}).  In some dimensions it is also possible to have {\it Majorana-Weyl spinors} (MW) or {\it symplectic Majorana-Weyl}~(SMW). The various possibilities for having such spinors, their minimal components and the symmetry of bispinors that appear in \re{comQQ} are displayed in Table~\ref{t:spinors} (from \cite{VanProeyen:2003zj}).\footnote{Possible spinors in spaces with various signatures were extensively discussed in \cite{VanProeyen:1999ni}.} The numbers in the last column are modulo $4$ and the interpretation for, say $1,\, 2$\linebreak in $D=4$ is $\bar \e \g^\mu \chi = - \bar \chi \g^\m \e\,, \, \bar \e \g^{\mu\nu} \chi = - \bar \chi \g^{\mu\nu} \e$ respectively.    
\begin{table}[htbp]
\begin{center}
  \begin{tabular}{|c|c|c|c|}
\hline
 Dim & Spinor & min \# components & antisymmetric \\
\hline
 $2$ & MW & $1$ & $1$ \\
 $3$ & M & $2$ & $1,2$\\
 $4$ & M & $4$ & $1,2$ \\
 $5$ & SM & $8$ & $2,3$ \\
 $6$ & SMW & $8$ & $3$ \\
 $7$ & S & $16$ & $0,3$ \\
 $8$ & M & $16$ & $0,1$ \\
 $9$ & M & $16$ & $0,1$ \\
 $10$ & MW & $16$ & $1$ \\
 $11$ & M & $32$ & $1,2$ \\
\hline
\end{tabular}\caption{\it Irreducible spinors, number of components and symmetry
  properties.\label{t:spinors}}
\end{center}
\end{table}

The minimal superalgebra \re{suP2} can be extended in several ways. A  classification of simple superalgebras and the extensions we discuss below was presented initially in \cite{Nahm:1978tg} and revised later in \cite{Strathdee:1987jr}. The first at hand generalization is to consider a collection of $N$ spinor charges denoted $(Q^i_\a)_{i=1,\dots, N}$. When that is the case, closure of the algebra might also allow an extra bosonic piece, known as R-symmetry, that basically rotates the superchages. There is thus a commutator 
\begin{equation}\la{Rrot}
\left[U^i{}_j   , Q^k_\a\right] = \d^k_i Q^i_\a\,,
\end{equation} 
with $U^i{}_j$ the R-symmetry generators. The automorphism group is not arbitrary and it is determined by the reality properties of the spinors. The corresponding groups can be found in many reviews (\eg, \cite{VanProeyen:2003zj}) and we only list them here for further reference
\begin{eqnarray}
D=2\,,10 &:& \SO(N_L)\times \SO(N_R)\,, \qq  D=3\,,9\,,11 \: : \SO(N)\,, \nn\\
D=4\,,8 \;\, &:& \U(N)\,,\label{Rgr}\\
D=6 \q \: \,&:& \USp(N_L)\times  \USp(N_R)\,, \qq D=5\,,7 \: : \USp(N)\,. \nn
\end{eqnarray}

For extended supersymmetry one has to modify slightly eq.~\re{suP2} by including a matrix $\Omega^{ij}$ on the r.h.s.~of the equation. The reality of the supercharges and the symmetry properties of the anticommutator determines the possible $\Omega^{ij}$ matrices. They reflect the symmetry properties of the R-symmetry. Moreover, the r.h.s.~of \re{suP2} can be enriched with the so-called `{\it central charges}' $Z^{ij}_{\mu_1 \dots \mu_r}$. These antisymmetric Lorentz tensors of rank $r$ commute with the supersymmetry generators.\footnote{Commonly, charges that only commute with supercharges are also coined `central charges' if they lead to the same physical consequences as real central charges.} Hence, more general we can have
\begin{equation}\la{suPc}
\left\{ Q^i_{\a}, Q^j_{\b} \right\} = \Omega^{ij}(\g^\m)_{\a\b} P_\m + \sum_r \g^{\mu_1 \dots \mu_r}_{\a\b} Z^{ij}_{\mu_1 \dots \mu_r} \,. 
\end{equation}
Central charges are associated with p-branes \cite{Townsend:1997wg}. For instance, in $D=11$ the existence of the \mbr{2} and \mbr{5} is related to the presence of two central charges $Z_{\m\n}$ and $Z_{\m\n\r\s\l}$ in \re{suPc}. A complete list of central charge extensions can be found in \cite{deWit:1997sz}.  
 
At the bosonic level, we know that in order to include a cosmological constant the Poincar\'e algebra is modified to an (A)dS version. The difference with Poincar\'e algebra is that the commutator of spatial translations no longer vanishes. Instead, 
\begin{equation}\la{adsalg}
\left[P_\m   , P_\n\right] = \pm \frac{1}{2 R^2}\, M_{\m\n}\,,
\end{equation} 
where $+$ and $-$ signs respectively correspond to AdS and dS, and $R$ is the (A)dS radius. We will return to the AdS geometry and asymptotic AdS supersymmetric solutions in Ch.~\ref{ch:sols}. This bosonic algebra can be also extended to a superalgebra \cite{Nahm:1978tg}, but we will not do it here. Finally, one can construct a superconformal generalization. We will investigate that in Sect.~\ref{s:confsym}.  

\subsection{Supergravity\la{s:maxsugra}}

One natural extension of rigid (extended) supersymmtric theories is supergravity\cite{Ferrara:1982be,Salam:1989fm,VanProeyen:2003zj}. This is a theory of local SUSY, \ie, a theory where the parameters~$\e^\a (x^\m)$ are spacetime-dependent functions. Most of the discussion from subsection~\ref{SUSY} remains valid, but there are some differences. The most significant one is the existence of a spin-$2$ particle in the representations of local SUSY \cite{Nahm:1978tg,Strathdee:1987jr}. This is the graviton, which by supersymmetry is related to a\linebreak spin-$3/2$ particle, the gravitino. As a consequence, the maximal number of independent supercharges for local SUSY is $32$, corresponding to $N=8$ in $D=4$ or $D=5$. In contrast, rigid SUSY with a highest spin-$1$ state in their supermultiplet representations allow at most $16$ supercharges, \ie, $N=4$ in $D=4$, or $D=5$. One extra subtlety is that in \re{comQQ} the local translations have to be replaced by general coordinate transformations.     

Due to the restriction to spin $\leq 2$ particles, maximal SUGRA (\eg, $N=8$, $D=4$) with $32$ supercharges in Minkowski spacetime 
are only possible up to $D=11$. Indeed, in $D=12$, the minimal spinor charge has $64$ components that requires higher-spin states. In this section we restrict to maximal SUGRAs in $D=10$ and $D=11$ keeping in the back of our mind their connection to string/M-theory \cite{Polchinski:1998j,Green:1987sw,deWit:2002vz}.    

The spectrum of $D=11$ SUGRA \cite{Cremmer:1978km} consists of a graviton $G_{MN}$, a three-form potential $A_{MNP}$ and a Majorana gravitino field $\Psi_M$. It contains no scalars, and since there are $32$ supercharges there are also no matter multiplets. These massless states transform in the $({\bf 44 + 84}) + {\bf 128}$ representations of the heliticity group $\SO(9)$. The bosonic part (for simplicity we neglect fermionic contributions) of the action describing the dynamics of pure SUGRA in $D=11$ is recast by\footnote{The form notation used here is explained in Appendix \ref{a:form}.}   
\begin{eqnarray}\la{act11}
{\cal L}_{M} &=& \hat R \hat * 1 - \ft 12\, \hat * \hat F_{(4)} \wedge \hat F_{(4)} + \ft 16\,  \hat A_{(3)} \wedge  \hat F_{(4)} \wedge \hat F_{(4)} \,,
\end{eqnarray}
where $\hat F_{(4)} = d \hat A_{(3)}$ and the hat distinguishes $D=11$ fields from type IIA dimensionally reduced objects. The action is invariant under the following rescaling of fields $G_{MN} \ra k^2 G_{MN}$ and $A_{MNP} \ra k^3 A_{MNP}$. Even if, historically, its discovery goes back to \cite{Cremmer:1978km}, a clear interpretation as the low energy-limit of what is called M-theory had to wait the until dualities became appreciated. It was then realized that a strongly coupled type IIA string theory has an extra dimension, being thus S-dual to $D=11$ SUGRA on a circle.   

Let's see now the particle content of maximal SUGRAs in $D=10$ (also referred to as $N=2$ theories). In ten dimensions the minimal spinor is MW with $16$ real components. Such an irreducible spinor transforms either under the spinor or its complex conjugate representation of $\SO(8)$, say ${\bf 8_c}$ (the difference between the two representations is important when both are present). Together with a vector transforming as ${\bf 8_v}$ of $\SO(8)$ they form the usual vector supermultiplet ${\bf 8_v} + {\bf 8_c}$ in $D=10$ (with $N=1$). It describes the ground state spectrum of the open superstring. 

As we are after the massless states of type II, we need information about closed strings. We know that the closed string spectrum is a tensor product of two copies of the open string spectrum. However, one has to pay attention to the possibility of choosing either ${\bf 8_c}$ or ${\bf 8_s}$. In fact they lead to inequivalent $N=2$ multiplets. For the first multiplet we start from two $N=1$ multiplets of opposite chirality. Then the tensor product decomposes in the following four sectors 
\begin{equation}\label{IIAsp}
{\bf (8_v + 8_s) \times (8_v + 8_c)}\longrightarrow  
\left\{ 
\begin{array}{ccccc|ccc}
 & &{{\rm \bf NSNS}}& & &     &{{\rm \bf RNS}}&\\
{\bf1}&+& {\bf  28}     &+&{\bf 35}&  {\bf 8}&    +   &{\bf 56}\\
 \phi & & B_{\m\n} &&  g_{\m\m}&  \chi^R&& \psi_\m^R\\
\hline\rule[-1mm]{0mm}{6mm}
  & &{{\rm \bf NSR}}& &   &  &{{\rm\bf RR}}&\\
    &{\bf 8} &+&{\bf 56}&   & {\bf 8}&+&{\bf 56} \\  
    &\chi^L && \psi_\m^L&  &  A_{(1)}&& A_{(3)} \\
\end{array}
 \right.
\end{equation}
giving the (non-chiral) type IIA massless spectrum. In the bosonic sector, besides the NSNS fields, we remark the presence of the RR potentials coming\linebreak from ${\bf 8_s}\times{\bf 8_c}$. They are connected to D-branes. As expected, in the fermionic sector we find massless spinors of  different handness. Since MW spinors with opposite chirality combine into one M spinor in $D=11$, we suspect that a KK reduction of $D=11$ SUGRA is nothing else than type IIA SUGRA. This is certainly true and the reduction is sketched below 
\begin{equation}\la{MIIA}
\left\{ 
\begin{array}{ccc}
{\bf 44} &\lra&{\bf 1} + {\bf 8} + {\bf 35} \\{\bf 84} &\lra& {\bf 28} + {\bf 56}\\
\hline \rule[-1mm]{0mm}{6mm}
{\bf 128}&\lra&{\bf 8} + {\bf 8} + {\bf 56} + {\bf 56}
\end{array} \right. \,.  
\end{equation}
The first line of \re{MIIA} shows the appearance of a scalar field, the dilaton $\phi$, and\linebreak a KK gauge vector $A_{(1)}$ in agreement with \re{KKmetr}. 

The dynamics of the IIA SUGRA is encapsulated in (Einstein's frame)    
\begin{eqnarray}\la{actIIA}
{\cal L}_{IIA} &=& R * 1 - \ft 12 \, * d \phi \wedge d \phi - \ft 12 \rme^{\ft 32\, \phi} \, *  F_{(2)} \wedge  F_{(2)} 
- \ft 12 \rme^{\ft 12\, \phi} \, *  F_{(4)} \wedge  F_{(4)} \nn\\
&-& \ft 12 \rme^{- \phi} \, *  H_{(3)} \wedge  H_{(3)} + \ft 12\, d  A_{(3)} \wedge d  A_{(3)} \wedge  A_{(2)}\,,
\end{eqnarray}
where
\begin{equation}\la{fIIA}
H_{(3)} = d B_{(2)}\,, \qq F_{(2)} = d A_{(1)}\,, \qq F_{(4)} = d A_{(3)} - F_{(3)} \wedge A_{(1)}\,.
\end{equation}
There exists also a massive version of IIA SUGRA proposed in \cite{Romans:1986tz}. After field redefinitions, Romans' massive IIA can be brought to a form where in the zero mass limit one recovers the usual IIA SUGRA. In other words, there are only contributions to \re{actIIA} proportional to the mass parameter \cite{Lavrinenko:1999xi}. 

The second inequivalent multiplet is the result of a tensor product between two identical $N=1$ vector multiplets ${\bf 8_v} + {\bf 8_c}$. This time the decomposition {\footnotesize 
\begin{equation}\label{IIBsp}
{\bf (8_v + 8_c) \times (8_v + 8_c)}\longrightarrow  
\left\{ 
\begin{array}{ccccc|ccccc}
 & &{{\rm \bf NSNS}}& & &    & &{{\rm \bf RNS}}& &\\
{\bf1}&+& {\bf  28}     &+&{\bf 35}&  &{\bf 8}&+&{\bf 56}&\\
 \phi & & B_{\m\n} &&  g_{\m\n}&  &\chi^L&& \psi_\m^L&\\
\hline\rule[-1mm]{0mm}{6mm}
  & &{{\rm \bf NSR}}& & &   &&{{\rm\bf RR}}&&\\
    &{\bf 8} &+&{\bf 56}&  & {\bf 1}& + & {\bf 28}& + & {\bf 35^*} \\  
    &\chi^L && \psi_\m^L& & A_{(0)}&& A_{(2)}&& A_{(4)}^* \\
\end{array}
 \right.\,.
\end{equation}
} 
produces the spectrum of type IIB SUGRA. The two changes with respect to IIA are in the RR and the fermionic sectors. In IIB the NSR and RNS spectra contain fermions with the same handness, hence type IIB is a chiral theory that cannot be obtained by KK reduction of $D=11$ SUGRA. Secondly, the RR fields are now forms of even rank $p$. They couple to odd-dimensional \br{p}s. In particular there is an axion zero-potential $A_{(0)}\equiv \chi$ and a four-form $A_{(4)}$ with a self-dual field-strength. It is precisely the self-duality that is not directly implementable into the action\footnote{Similar problems are encountered for the worldvolume description of the \br{3} or \mbr{5}. They make the subject of Chap. \ref{ch:dual}, \ref{ch:defo}.}. Here we adopt the attitude of \cite{Schwarz:1983qr,Bergshoeff:1995as} and impose this restriction by hand in addition to the equations of motion (eoms). 

The action of type IIB SUGRA is then \cite{Bergshoeff:1995as} (Einstein frame)
\begin{eqnarray}\la{actIIB}
{\cal L}_{IIB} &=&  R * 1 - \ft 12 \, * d \phi \wedge d \phi - \ft 12 \rme^{2 \phi}\, * d \chi \wedge d \chi- \ft 14  *  F_{(5)} \wedge  F_{(5)}\\
 &-& \ft 12 \rme^{- \phi} \, *  H_{(3)} \wedge  H_{(3)} - \ft 12 \rme^{ \phi} \, *  F_{(3)} \wedge  F_{(3)} - \ft 12 A_{(4)} \wedge d A_{(2)} \wedge H_{(3)}\,,\nn
\end{eqnarray}
where 
\begin{eqnarray}\la{fIIB} 
H_{(3)} &=& d B_{(2)} \,, \qq F_{(3)} = d A_{(2)} - H_{(3)} \wedge \chi \,,\nn\\
F_{(5)} &=& d A_{(4)} - \ft 12 A_{(2)} \wedge H_{(3)} + \ft 12 B_{(2)} \wedge d A_{(2)} \,, \q *F_{(5)} = F_{(5)}\,.
\end{eqnarray}

The action \re{actIIB} possesses a remarkable symmetry. It is invariant under\linebreak global $\Sl(2,\bbbb{R})$ transformations. To prove this statement, the two real scalars are combined into a complex object $\rho$ and the NSNS and RR two-form potentials into a doublet
\begin{equation}\la{IIBfiel}
\rho = \rho_1 + \rmi \rho_2 \equiv \chi + \rmi \rme ^{- \phi}\,, \qq F_{(3)}^i \equiv \left[
\begin{array}{c}
H_{(3)}\\ d A_{(2)}
\end{array} \right]\,.
\end{equation}
The $\Sl(2,\bbbb{R})$ group transforms the scalar non-linearly and it mixes the components of $F_{(3)}^i$
\begin{equation}\la{SlIIB}
\rho \ra \frac{a \rho + b}{c \rho + d}\,, \qq F_{(3)}^{i \, \prime} = \L^i{}_j F_{(3)}^j \,, \qq  \L^i{}_j = \left(
\begin{array}{cc}
d & c\\ b & a
\end{array} \right)\,,
\end{equation}
where $a,\, b,\, c, \, d \in \bbbb{R}$ with $ad - bc =1$. The rest of the fields remain unchanged. 

This symmetry of the SUGRA approximation is actually broken in the full  string theory to $\Sl(2,\bbbb{Z})$, which is an exact symmetry \cite{Hull:1995ys}. The vev of the scalar field is commonly defined as 
\begin{equation}\la{vevsc}
< \rho > = \frac {\th}{ 2 \pi} + \frac{\rmi}{g_s}\,.
\end{equation}
One particular case of the $\Sl(2,\bbbb{Z})$ invariance, namely $\rho \ra \rho +1 $ implies the periodicity of Witten's $\th$ angle. A second more important consequence appears when $\th =0$ and $\rho \ra  - \frac{1}{\rho}$. This symmetry means that type IIB string theory is invariant under S-duality $g_s \ra \frac{1}{g_s}$. Hence, the perturbative sector of IIB encodes the same physics as the strongly coupled theory. A similar S-duality (interchanging the sign of the dilaton) holds between the heterotic $\SO(32)$ and type I strings. 

\subsection{Five-dimensional supergravity \la{s:5dSU}}

Five dimensional supergravity its an old topic \cite{Cremmer:1980gs,Howe:1981ev}. Also its matter couplings were studied during that period \cite{Gunaydin:1984bi,Gunaydin:1985ak,Gunaydin:1986fg}. However, the subject was revived in the last years \cite{Gunaydin:1999zx,Gunaydin:2000xk,Gunaydin:2000ph,Ceresole:2000jd,Zucker:1999ej,Zucker:1999fn,Kugo:2000af,Kugo:2002js,Bergshoeff:2001hc,Bergshoeff:2002qk,Andrianopoli:2000fi, Chamseddine:2001hk,Gunaydin:2003yx} due to possible applications in AdS/CFT duality or to the phenomenological RS scenarios. Here we review the structure of such theories and the effective construction is done in later chapters.

Supersymmetric theories in $D=5$ \cite{Cremmer:1980gs} are possible for $Q=32,\, 24,\, 16, \, 8$ supercharges and R-symmetry group $\USp(N)$ with $N=8,\, 6,\, 4,\, 2$. The spectra of pure SUGRAs in $D=5$ are displayed in Table~\ref{t:sugra5}.  
\begin{table}[htbp] 
\begin{center}
\begin{tabular}{|c|c|c|c|c|}
\hline
\rule[-1mm]{0mm}{6mm}
  s$\backslash$ $N$    &  $N=8$  & $N=6$ & $N=4$ & $N=2$\\
\hline\hline
\rule[-1mm]{0mm}{6mm}
 2 & 1 &1 & 1 &1\\[1mm]
\hline
\rule[-1mm]{0mm}{6mm}
 3/2 & 8&  6 &4& 2\\[1mm] 
\hline
\rule[-1mm]{0mm}{6mm}
 1 & 27 &  15 &6&1\\[1mm] 
\hline
\rule[-1mm]{0mm}{6mm}
 1/2 & 48 &  20 &4& \\[1mm] 
 \hline
\rule[-1mm]{0mm}{6mm}
 0 & 42 &  14 &1& \\[1mm] 
\hline\hline
\rule[-1mm]{0mm}{6mm}
 {\rm d.o.f.} & 128+128& 64+64 &24+24& 8+8\\[1mm] 
\hline
\end{tabular}\caption{\it The spin $s$-particle content of pure SUGRA in $D=5$ for different number $N$ of extended supersymmetry. The number of degrees of freedom (bosonic + fermionic) is also indicated.\la{t:sugra5}}
\end{center}
\end{table}
Maximal SUGRA ($N=8$) is derivable via dimensional reduction on a six-torus from $D=11$ SUGRA. It contains a unique pure gravity multiplet with a graviton, 27 vector bosons and not less than 42 scalars. It worth saying that this field content is obtained upon dualizing the three-form to a scalar and all two-forms to vectors. The latter is possible in $D=5$ only in the Abelian case, to which we restrict for the time being. In addition it has also 8 gravitinos and 42 spin-$1/2$ SM fermions. They are all dimensionally reduced from the $D=11$ gravitino. The large number of moduli makes it difficult to control the scalar potential and to study its minima. Nevertheless, in some special cases, \eg, \cite{Khavaev:1998fb}, critical points that preserve some fraction of supersymmetry have been investigated. Pure gravity with lower number of supercharges can be understood as consistent truncations of maximal SUGRA.  

The situation simplifies in the case of $N=4$, when there is only one scalar field in the gravity multiplet and the potential becomes a a real, single-valued function. Supersymmetric solutions to some particular gaugings of this model were recently proposed \cite{Chamseddine:2001hk}. When there are $16$ supercharges, rigid SUSY is possible as well. One can introduce vector multiplets (Table~\ref{t:mat5}), the descendants of the $D=10$ SYM multiplet. In addition, there may be also tensors multiplets which we do not explicitly present in the table since they can be dualized to vectors in the zero limit of the YM coupling constant. Furthermore, $D=5$, $N=4$ SUGRA matter couplings are completely fixed by giving the number of vector multiplets and choosing a concrete gauging of some internal symmetry. 
\begin{table}[htbp] 
\begin{center}
\begin{tabular}{|c|c|c|}
\hline
\rule[-1mm]{0mm}{6mm}
  s$\backslash$ $N$   & $N=4$ & $N=2$\\
\hline\hline
\rule[-1mm]{0mm}{6mm}
  1  &1 & 1\\[1mm] 
\hline
\rule[-1mm]{0mm}{6mm}
 1/2  & 4 & 2\\[1mm] 
 \hline
\rule[-1mm]{0mm}{6mm}
 0  &5& 1 \\[1mm] 
\hline\hline
\rule[-1mm]{0mm}{6mm}
 {\rm d.o.f.} &8+8& 4+4\\[1mm] 
\hline
\end{tabular} 
\qq\qq \begin{tabular}{|c|c|}
\hline
\rule[-1mm]{0mm}{6mm}
  s $\backslash$ N   &  N=2\\
\hline\hline
\rule[-1mm]{0mm}{6mm}
 1/2  &  2\\[1mm] 
 \hline
\rule[-1mm]{0mm}{6mm}
 0  &4  \\[1mm] 
\hline\hline
\rule[-1mm]{0mm}{6mm}
 {\rm d.o.f.} & 4+4\\[1mm] 
\hline 
\end{tabular}\caption{\it Matter multiplets in $D=5$ and their number of degrees of freedom.\la{t:mat5}}
\end{center}
\end{table}

The by far richer and more interesting theory, $D=5$ SUGRA with eight supercharges is not fully specified by the number of matter multiplets and the chosen gauging. The models depend on functions of the matter multiplets scalars. Luckily enough, $N=2$ SUSY constrains the arbitrariness of these functions such that they define specific geometries of the moduli spaces, as it will become clear in later chapters. The same conclusion holds for $D=4$, $N=2$ or for supersymmetric theories with only 4 independent supersymmetries. However, the latter involve in their definition arbitrary functions that are less constrained by supersymmetry, and as a result, are more difficult to handle.  

$D=5$, $N=2$ SUGRA can be obtained by dimensionally reducing $D=11$ SUGRA \re{act11} on a CY threefold  \cite{Cadavid:1995bk, Papadopoulos:1995da} or in a dual formulation from heterotic string on $K3 \times S^1$ \cite{Antoniadis:1996cy}. It contains a graviton $e_\m^a$, an $\USp(2)$-doublet of SM gravitini $\psi_\m^i$ (for notations see Appendix \ref{a:hyp}) and a spin-$1$ graviphoton $A_\m^0$. There are no scalars in the gravity multiplet. They appear only after coupling matter multiplets. We are going to discuss later on vector multiplets, hypermultiplets (see Table~\ref{t:mat5}), tensor and linear multiplets. 
Without being complete, we will reveal in the remainder of the section some generic features of Einstein-Maxwell theories by looking at $N=2$ SUGRA coupled to a set of $n$ vector multiplets (gauge vectors $A_\m^{\widetilde x}$, gauginos $\l^{i \,{\widetilde a}}$ and scalars $\s^{\widetilde x}$). The moduli $(\s^{\widetilde x})_{{\widetilde x} = 1, \dots, n}$ can be seen as coordinates parameterizing a manifold whose tangent space directions are labeled by the index ${\widetilde a}$. It is common to combine the $n$ vectors with the graviphoton into an $(A_\m^I)_{I=0,\dots, n+1}$ object. The generic form of the transformation rules is 
\begin{eqnarray}\la{genertr}
\d e_\mu^a &=& \ft 12 \bar \e \g^a \psi_\mu\,,\nn \\
\d \psi_\mu^i &=& \nabla_\mu \e^i + \rmi g S^{ij} (\s) \g_\mu \e_j + \dots \,,\nn\\
\d A^I_\mu &=& \frac 12 \, h^I_{\widetilde a} (\s) \bar \e \g_\mu \l^{\widetilde a} - \frac 12 \,\rmi h^{I } (\s) \, \bar \e \psi_{\mu }  \nn\\
\d \l^{i\,\widetilde a} &=& g N^{i\,\widetilde a}_j (\s) \e^j - \frac 12  \rmi f_{\widetilde x}^{\widetilde a} \,{\slashed{\cal D}} \s^{\widetilde x} \e_i + \dots \,, \nn\\
\d \s^{\widetilde x} &=&  \frac 12\,\rmi f^{\widetilde x}_{\widetilde a}\bar \e^i \l^{\widetilde a} \,, 
\end{eqnarray}   
where $f_{\widetilde x}^{\widetilde a}$ and $f^{\widetilde x}_{\widetilde a}$ denote the $n$-bein and its inverse of the target (scalar) \linebreak space, \ie, $f_{\widetilde x}^{\widetilde b} f^{\widetilde x}_{\widetilde a} = \d ^{\widetilde b}_{\widetilde a}$ and $f_{\widetilde y}^{\widetilde a} f^{\widetilde x}_{\widetilde a} = \d ^{\widetilde x}_{\widetilde y}$. There appear also some covariant derivatives: $\nabla_\m$ is Lorentz covariant, whilst ${\cal D}_\m$ may include gaugings of some internal symmetries. The dots stand for extra derivative terms that are not relevant here. The significant pieces are the gaugino shifts $N^{i\,\widetilde a}_j (\s)$ and the gravitino (symmetric) mass matrix $S^{ij} (\s)$, which appear proportional with the gauge coupling constant~$g$.

Schematically, the supersymmetric Lagrangian can be summarized as follows
\begin{eqnarray}\label{genaction}
e^{-1}{\cal L} \!&=&\! -\ft 12 {\cal R} -\ft 12 \bar\psi_\mu \g^{\mu\nu\rho} \nabla_{\nu} \psi_{\rho} - \ft 12 g_{{\widetilde x}{\widetilde y}} (\s){\cal D}_{\mu}  \s^{\widetilde x} {\cal D}^\m\s^{\widetilde y} \nn\\
&-& \ft 14 a_{{ I}{ J}} (\s) F_{\mu\nu}^{ I} F^{\mu\nu \, J} + {\cal L}_{CS}  - \ft 12 \d_{{\widetilde a}{\widetilde b}}  \bar\l^{\widetilde a} \slashed{\cal D} \l^{\widetilde b} \nn\\
&-& \ft 12\, \rmi  f_{\widetilde x} ^{\widetilde a}  \bar\l_{\widetilde a}  \slashed{\cal D}  \s^{\widetilde x}\g^\n \psi_{\nu } - g  N^{i \,{\widetilde a}}_j (\s) \bar \l_{i \,{\widetilde a}} \g^\mu \psi^j_{\mu } \nn\\
& -& \ft 32 \rmi g  S^{ij} (\s) \bar \psi_{\mu i } \g^{\mu\nu} \psi_{\nu j} + \rmi g {\cal M}_{{\widetilde a}{\widetilde b}} (\s) \bar\l^{\widetilde a}  \l^{\widetilde b} - g^2 V(\s) + \dots \,.
\end{eqnarray}
In the first two lines we have the usual Einstein term, the Rarita-Schwinger contribution from the gravitini kinematic terms for the matter fields and a Chern-Simon term ${\cal L}_{CS}$ specific to odd dimensions. Moving lower, we remark some vertex couplings of the scalars with two fermions and, the last line, includes fermionic mass terms and the scalar potential. The metric on the target space $g_{{\widetilde x}{\widetilde y}}$ is connected to the $n$-beins by $g_{{\widetilde x}{\widetilde y}}= f_{{\widetilde x}}^{{\widetilde a}} \d_{{\widetilde a}{\widetilde b}}f^{{\widetilde b}}_{{\widetilde y}}$.

The closure of the SUSY algebra and the SUSY invariance of the action yield the following important relations
\begin{eqnarray}
\d^i_k V(\s) & = & 2 N^{j,\,i \,{\widetilde a}}  N_{j,\,k \,{\widetilde a}} - 24 S^{ij} S_{jk}\,, \la{potentge} \\
\del_{\widetilde a} S_{ij} (\s) & = &  \ft 16 \Big( N_{i, \,j\, {\widetilde a}}+ N_{j, \,i\, {\widetilde a}}\Big) \,,\la{massmatr} \\
\ft 12 \d_i^j f^{\widetilde x}_{\widetilde a}\del_{\widetilde x} V(\s) & \equiv & \ft 12 \d_i^j \del_{\widetilde a} V(\s) = 5 \d_{{\widetilde a}{\widetilde b}} N^{k, \,j\, {\widetilde b}} S_{ki} - 2 {\cal M}_{{\widetilde a}{\widetilde b}} N_i^{j\, {\widetilde b}}\,.\la{derpot} 
\end{eqnarray}
The first two equations give the scalar potential as a linear combination of a square term and the square of its derivatives. They are computed from the variation of the action to gravitinos. However, the last relation expressing the derivative of the potential in terms of the fermionic mass matrices and shifts also holds in rigid SUSY. If we consider the particular solution $S_{ij} = Q_{ij} W(\s)$, $N_{i, \,j\, {\widetilde a}} = 3 Q_{ij}\del_{{\widetilde a}} W (\s)$ of \re{massmatr} (with normalization $Q_{ij} Q^{ij} = 2$) the potential acquires the familiar form 
\begin{equation}\la{superPot}
V (\s) = 12 \left[3 \del_{{\widetilde x}} W (\s) \del^{{\widetilde x}} W (\s) - 4 W(\s)^2 \right] \,.
\end{equation}
The function $W(\s)$ is referred to as superpotential.

The results \re{potentge}-\re{derpot} are not accidental to these models. They are generic features of extended SUGRAs and they appear in other dimensions as well. For a complete discussion see \cite{D'Auria:2001kv,Fre:2001jd}. Thus, in this section we reviewed the SUGRA framework in $D=11\,,\, 10$ and $D=5$. In the latter case, the general structure \re{genertr} and \re{genaction} will  reappear in the construction of Ch.~\ref{ch:matter}. Also the generic expression of the potential \re{superPot} is the key ingredient for obtaining solutions in~Ch.~\ref{ch:sols}. 

\section{Superconformal tensor calculus\label{confSugra}}

Among various techniques for constructing SUGRAs our preference goes throughout this thesis to superconformal tensor calculus \cite{Kaku:1977pa,VanProeyen:1983wk}. It is based on the superconformal group and it was successfully applied in $D=4\,,\, 6\,,$ and now in $D=5$. Here, we only explain the essence of the method and the effective construction is presented later in Ch.~\ref{ch:weyl} and Ch.~\ref{ch:matter}.

\subsection{Conformal symmetry and conformal algebra\label{s:confsym}}

Before presenting the superconformal method for constructing SUGRA we make a detour through conformal symmetry and its algebra. The full fermionic complications will be added in Ch.~\ref{ch:weyl}. Here, we only focus on the bosonic subalgebra in presenting the core of the conformal tensor calculus.

Conformal transformations are defined as general coordinate transformations $x \ra x^\prime$ that leave the geometry invariant up to a scale factor
\begin{equation}\la{confTr}
g_{\m\n} (x) \lra g^\prime_{\m\n} (x) = \Omega (x) g_{\m\n} (x)\,, 
\end{equation} 
\ie, they are angle-preserving transformations. When the coordinate transformation is infinitesimal $x^{\mu \prime} \ra x^{\mu } + k^\mu (x)$ the infinitesimal displacements $k^\mu (x)$ satisfy the conformal Killing equation
\begin{equation}\la{cve}
\d_{\rm g.c.t.} (k) g_{\mu\nu}(x) \equiv \nabla_\mu k_\nu(x) +
\nabla_\nu  k_\mu (x) = \Omega(x) g_{\mu\nu}(x)\, , 
\end{equation}
for some arbitrary function\footnote{The vector $k^\mu (x)$ satisfying \re{cve} for $\Omega(x)=0$  (Killing equation) is called a Killing vector.} $\Omega(x)$. The covariant derivative $\nabla_\m$ is defined with respect to the Levi-Civita connection. Restricting to a D-dimensional flat space, eq.~\re{cve} implies (symmetrization is done with weight one, see Appendix~\ref{a:form})
\begin{equation}\label{eq:conf_killing_eqn}
\partial_{(\mu}k_{\nu)}(x) - \frac 1D \eta_{\mu\nu}\partial_\rho k^\rho (x)
= 0\,. 
\end{equation}
Excluding the special case $D=2$, this equation has the following general solution
\begin{equation}\label{ximu}
k^\mu(x)=\xi^\mu +\lambda_M^{\mu\nu}x_\nu+\lambda_D x^\mu
+\left(x^2\Lambda_K^\mu-2x^\mu x\cdot \Lambda_K\right). 
\end{equation}
The terms in the r.h.s.~describe translations, Lorentz rotations, dilatations and conformal boosts, respectively. The corresponding generators are denoted by $P_\m$, $M_{\m\n}$, $D$ and $K_\m$ such that a conformal transformation can be expressed as\footnote{We emphasize that the scale transformation present in the conformal group does not preserve the mass-shell condition $p^2 + m^2 =0$ for massive particles. As a result, the conformal symmetry can only describe massless physical states.}
\begin{equation}\la{confTra}
\delta_C= \xi^\mu  P_\mu + \lambda_M^{\mu\nu}M_{\mu\nu}+ \lambda_D D +
\Lambda_K^\mu K_\mu \,.
\end{equation}

The conformal generators in $D$ dimensions possess a simple representation
\begin{eqnarray}\label{confRep}
P_\m = \del_\m \,, &\qq & M_{\m\n} = x_{[\m} \del_{\n]}\,, \nn\\
D  = x^\m \del_\m\,, &\qq& K_\m = x^2 \del_\m - 2 x_\m x^\n \del_\n\,.
\end{eqnarray}
It follows that they form an $\SO (D,2)$ algebra
\begin{eqnarray}\label{confAlg}
\left[P_\m , M_{\n\r} \right] = \eta_{\m[\n} P_{\r]} \,, & \qq& \left[M_{\m\n}, M^{\r\s}\right] = - 2 \d_{[\m}^{[\r}  M_{\n]}{}^{\s]} \,, \nn\\
\left[K_\m, M_{\n\r}\right] = \eta_{\m[\n} K_{\r]} \,, & \qq& \left[P_\m, K_\n \right] = 2(\eta_{\m\n} D + 2 M_{\m\n})\,, \nn\\
\left[D , P_\m\right] = P_\m \,, & \qq&  \left[D , K_\m \right] = - K_\m\,,
\end{eqnarray}
including the Poincar\'e symmetry as a subalgebra. An explicit embedding of\linebreak the $D$-dimensional generators into $(D+2)$-dimensional objects satisfying $\SO (D,2)$ can be found easily (\eg, \cite{VanProeyen:1999ni})
\begin{equation}\la{SOD2em}
M_{\hat \m \hat \n} = 
\left(\begin{array}{ccc}
M_{\m\n}& \ft 14 \,(P_\m - K_\m)&\ft 14 \,(P_\m + K_\m)\\
- \ft 14 \,(P_\m - K_\m)& 0 & - \ft 12\, D\\
- \ft 14 \, (P_\m + K_\m)& \ft 12\, D & 0
\end{array}\right)\,.
\end{equation}
Without being aware of this embedding, an immediate check can be made by counting the number of generators: $D + D(D-1)/2 + 1 + D = (D+1)(D+2)/2$.

Besides \re{confRep} the conformal symmetry admits also non-trivial representations connected to the stability algebra of $x^\m =0$. They are
\begin{eqnarray}\label{xdepP}
 \d_M \phi
(x) &=&{\rm derivatives} + \d_\Sigma(\lambda_M)\phi(x)\, ,\nonumber\\
 \d_D
\phi(x)&=& {\rm derivatives} + \d_\Delta
(\lambda_D)\phi(x)\, , \\ 
\d_K
\phi(x) &=& \!\! {\rm derivatives} + \! 
\Bigl (\d_\D (-2
x\cdot \L_K) + \d_\Sigma (-4x_{[\mu}\Lambda_{K\nu]}) + \d_\kappa
(\Lambda_K)\Bigr ) \phi(x) \,,\nonumber
\end{eqnarray}
for some generic field $\phi(x)$. The `{\rm derivatives}' are given by \re{confRep}. The concrete form of the other contributions in \re{xdepP} depends on the type of field they act upon. Excepting some fields in the Weyl multiplet whose $\d_\kappa$-transformation will be explicitly given in Ch.~\ref{ch:weyl}, all our fields are invariant under internal special conformal transformations $\delta_\kappa
\phi^\alpha = 0$. For Lorentz representations we distinguish between antisymmetric tensors $\phi_{a_1\cdots
a_n}(x)\ (n=0,1,2, \ldots)$ and spinors $\psi_\alpha(x)$ 
\begin{eqnarray}\la{Lorrep}
\d_\Sigma (\l_M) \phi_{a_1\cdots a_n}(x) &=& -n (\l_M)_{[a_1}{}^b
\phi_{|b|a_2 \cdots a_n]}(x) \,, \nonumber\\
\d_\Sigma (\l_M) \psi(x) &=& -\frac 14 \l^{ab}_M \g_{ab} \psi(x) \,.
\end{eqnarray}

Internal dilatations are, in general, characterized by means of a number $w$, the Weyl weight of the field
\begin{equation}\label{simplediltr}
\d_\Delta(\l_D) \phi^\alpha(x) = w \l_D \phi^\alpha(x) \,.
\end{equation}
One exception hereof are scalars parameterizing an internal manifold (see\linebreak Sect.~\ref{s:5dSU}). In that case we generalize \re{simplediltr} to
\begin{equation}\label{diltr}
\delta_\Delta(\l_D) \phi^\alpha = \lambda_D k^\alpha (\phi) \,.
\end{equation} 
But then, the dilational invariance of the corresponding non-linear sigma model
\begin{equation} \la{nonsig}
S=-\frac 12 \int d^Dx \; g_{\alpha\beta}(\phi) \partial_\mu  \phi^\alpha
\partial^\mu  \phi^\beta
\end{equation}
implies that $k^\alpha (\phi)$ are {\it homothetic Killing vectors}, \ie, they are solutions of internal equations, similar to  \re{cve}, with constant $\O(x) = D-2$. Further, it can be proven \cite{Bergshoeff:2002qk} that the invariance of the action \re{nonsig} under the special conformal transformations restricts $k_\alpha = \del_\a \chi (\phi)$. This offers an example of {\it exact} homothetic Killing vector. In fact, one can alternatively introduce an exact homothetic Killing vector $k^\a$ via 
\begin{equation}\la{exhomK}
\covder_\alpha k^\beta \equiv \partial_\alpha k^\beta +
\Gamma_{\alpha\gamma}{}^\beta  k^\gamma  = \frac{D-2}{2}\, \delta_\alpha{}^\beta\,,
\end{equation} 
which is a metric-independent relation. When the target space has vanishing Levi-Civita connection, the vector $k^\alpha  = \phi^\a \,(D-2)/2 $ and, eq.~\re{diltr} reduces to eq.~\re{simplediltr} for $w = (D-2)/2$. 

After this primer on conformal symmetry we are ready to explain (super)con\-for\-mal tensor calculus (SCTC).

\subsection{The method\label{s:SCTC}}

The aim of SCTC is to construct super-Poincar\'e invariant (SUGRAs) theories with the help of superconformal symmetry \cite{Ferrara:1984ij,VanProeyen:1983wk}. Even if not realized in nature, this superconformal symmetry can be a convenient instrument for theoretical constructions if it is finally broken. The advantage of a higher symmetry is that it constrains the structure of the theory and avoids certain complications present in a direct construction based on the lower symmetry. In SUGRA context for instance, it considerably reduces the non-linearities encountered in super-Poincar\'e. In order to preserve the number of degrees of freedom one has to introduce extra fields, called compensators, that account for the extra symmetries. They are used in the end just to gauge-fix the unwanted symmetries. In this way, one actually deals continuously with the same theory (degrees of freedom), but one splits in the intermediary steps the original reducible representations into irreducible ones, easier to handle. 

SCTC is thus a particular application of the gauge equivalence program. This method consists in the following algorithm:
\begin{itemize}
\item Introduce larger symmetry and compensators;
\item Construct invariant action or eoms;
\item Break the unwanted symmetries;
\item Rewrite the action or eoms.
\end{itemize}  
In the framework of SCTC, it is the superconformal group that extends the super-Poincar\'e symmetry. In the first step, different multiplets (representations) of the former are defined together with their transformation rules. The first one to construct is the Weyl multiplet, which contains the gauge fields associated to the superconformal symmetry. Then, other representations are elaborated and coupled to the Weyl multiplet. In $D=5$, $N=2$ we will consider vector, linear and hypermultiplets. Some of them are realized off-shell, but for others the closure of the algebra already yields dynamical equations (before having an action). 

In the next move, conformal invariant actions are obtained. This can be achieved by means of either density formulae or by integrating the eoms from the previous step. After determining the action and the superconformal transformations, we want to return to the Poincar\'e formulation. To that end, we choose appropriate gauge conditions that break the extra symmetries present in the conformal case. And finally, we reduce the SUSY transformations and recast the action in a super-Poincar\'e invariant form. It must be stressed that using different compensating multiplets may lead to different (off-shell) formulations of SUGRA theories. Hence, SCTC explains the connection between these off-shell models and allows the construction of all of them without setting up each time the whole machinery. A simple choice of the compensating multiplet will suffice. It is nevertheless believed that, upon eliminating the auxiliary fields, the off-shell formulations should yield the same on-shell theory. The procedure described above will be worked out in detail for the $D=5$, $N=2$ SUGRA. However, we consider it important to illustrate the method on an oversimplified example that avoids the complications of SUGRA but encodes the goal, the advantages and the full algorithm.

\subsection{A toy model\label{s:sctctoy}}

Consider a massive vector (Proca) field $V_\m$ whose Lagrangian reads
\begin{equation}\la{Proc}
L_{{\rm Proc}} = - \ft 14 F_{\m\n} (V) F^{\m\n} (V)- \ft 12 \, m^2 V_\m V^\m\,.
\end{equation}
Obviously, the mass term forbids a local $\U (1)$ symmetry of the action. We aim to show that the theory can be formulated in a $\U (1)$-invariant manner. Hence, in our toy model, the $\U (1)$ group represents the extra symmetry. To balance the new degree of symmetry we have to introduce a compensating scalar field $\phi$. Redefining the Proca field   
\begin{equation}\la{redProc}
V_\m = A_\m - m^{-1} \del_\m \phi\,,
\end{equation}
it is then invariant under a local symmetry
\begin{equation}\la{locU1}
\d_\L A_\m = \partial_\m \L \,, \qq \d_\L \phi = m\L\,.
\end{equation}
The transformation of the scalar is a pure shift meaning that the scalar itself is a pure gauge field. It can be set to any value, as we will shortly do.
In terms of the new fields the $\U (1)$-invariant action 
\begin{equation}\la{U1Proc}
L_{\U (1)} = - \ft 14 F_{\m\n} (A) F^{\m\n} (A) - \ft 12 \, D_\m \phi D^\m \phi \,,
\end{equation}
where the covariant derivative is defined as always
$D_\m \phi \equiv \partial_\m \phi - \d_\L (A_\m) \phi =\partial_\m \phi - m A_\m$.

The action \re{U1Proc} with a $\U (1)$ symmetry is certainly  equivalent to the initial Lagrangian \re{Proc}. One way to see it is to break $\U (1)$ by a gauge-fixing condition~$\phi = 0$ that reduces $L_{{\rm Proc}}$ to $L_{\U (1)}$.
Another, but equivalent, way is to make the field redefinition \re{redProc} and repeat the above argument backwards. Thus, we can reformulate the massive vector degrees of freedom in terms of a massless vector and a scalar. Of course, deriving such a simple and well-known theory with this method is meaningless. The purpose was only to illustrate that the gauge equivalence program is not restricted to conformal symmetry or SUSY. Further illuminating examples are provided in  \cite{deWit:1981gj}.


\section{$p$-brane solutions\label{s:pbra}}

In preparation of Ch.~\ref{ch:sols} where new charged brane solutions of SUGRA theories are presented, we review here the standard supergravity charged \br{p} solutions of Ref.~\cite{Lu:1995cs}.

In general, $p$-branes appear in string theory as solutions to the low-energy limit of the theory. They are generalizations of black holes solutions of general relativity and, similarly, they are characterized by a mass parameter $M$. However, we saw that SUGRAs include various $p$-forms that can supply $p$-branes with a charge~$Q$. To avoid naked singularities, these parameters are constrained by the inequality $Q \leq M$. We shall be interested in the extremal situation $Q=M$ when the electromagnetic and gravitational forces between such objects balance out. Such cases are called BPS-solutions and they preserve half of the number of supersymmetries.

It is familiar that in $D=4$, the Maxwell field can support an electrically and/or a magnetically charged black-hole. Analogously, in generic $D$ dimensions an $n$-form field strength $F_{(n)}$ will support either electric $(n-2)$-brane or its magnetic dual $(D-n-2)$-brane. For example, $D=11$ SUGRA \re{act11} has a $4$-form  $\hat F_{(4)}$ in its spectrum suggesting that we should find $2$- and $5$-brane solutions. Indeed, there exists an electrically charged \mbr{2} \cite{Duff:1991xz}
\begin{eqnarray}\la{M2br}
ds_{11}^2 &=& H^{-2/3} d x_{(3)}^2 + H^{1/3} d y_{(8)}^2 \,, \nonumber\\
\hat F_{(4)} &=& d^3 x \wedge d H^{-1} \,, \qq H (r) = 1 + R^6/r^6\,,
\end{eqnarray}
and a magnetic \mbr{5} \cite{Gueven:1992hh}
\begin{eqnarray}\la{M5br}
ds_{11}^2 &=& H^{-1/3} d x_{(6)}^2 + H^{2/3} d y_{(5)}^2 \,, \nonumber\\
\hat F_{(4)} &=& * (d^3 x \wedge d H^{-1}) \,, \qq H (r) = 1 + R^3/r^3\,
\end{eqnarray}
with $r = \sqrt{\sum y^m y^m}$ the radial coordinate in the transverse space and the constant~$R$ related to the charge.

Going one dimension lower to type II SUGRA, the dilaton enters the stage and couples to both NSNS and RR forms \re{actIIA}, \re{actIIB}. In the NSNS sector, the $2$-form couples to the fundamental string
\begin{eqnarray}\la{F1st}
ds_{10}^2 &=& H^{-3/4} d x_{(2)}^2 + H^{1/4} d y_{(8)}^2 \,, \qq \rme^{- 2\phi} = H\,, \nonumber\\
H_{(3)} &=& d^2 x \wedge d H^{-1} \,, \qq H (r) = 1 + R^6/r^6\,
\end{eqnarray}
or its dual NS$5$-brane. In the RR sector the dilaton coupling to the RR fields can be summarized as
\begin{eqnarray}\la{dilcoupl}
\rme ^{\frac{3-p}{2}\, \phi} * F_{(p+2)} \wedge F_{(p+2)}\,.
\end{eqnarray}
Then the corresponding \br{p} solutions have the following form \cite{Horowitz:1991cd}
\begin{eqnarray}\la{Dpbr}
ds_{10}^2 &=& H^{(p-7)/8} d x_{(p+1)}^2 + H^{(p+1)/8} d y_{(9-p)}^2 \,, \qq \rme^{\phi} = H^{(3-p)/4}\,,\nonumber\\
 F_{(p+2)} &=& d^{p+1} x \wedge d H^{-1} \,, \qq H (r) = 1 + (R/r)^{7-p}\,,
\end{eqnarray}
where the cases $p>3$ are described by the magnetic field strength $F$.

As noticed when constructing the action, the \br{3} deserves special attention. For instance, there is no dilatonic coupling to the kinetic term \re{dilcoupl} of the self-dual field-strength. Consequently, the dilaton can be set to a constant value. The \br {3} solution carrying both electric and magnetic charges is \cite{Horowitz:1991cd}
\begin{eqnarray}\la{D3br}
ds_{10}^2 &=& H^{-1/2} d x_{(4)}^2 + H^{1/2} d y_{(6)}^2 \,, \nonumber\\
F_{(5)} &=& d^4 x \wedge d H^{-1} + *(d^4 x \wedge d H^{-1}) \,, \qq H (r) = 1 + R^4/r^4\,.
\end{eqnarray}

It is clear that the presented $p$-brane solutions approach\linebreak asymptotically ($r \ra \infty$) a Minkowski spacetime. In the near region, when $r\ra 0$, the function $H$ diverges suggesting a horizon limit $g_{00} \ra 0$. Nevertheless, the presence of the dilaton in the solution, behaving as $\del_r \phi \sim 1/r \ra \infty$, implies that the metric should have a singularity for these solutions (because the Ricci scalar is proportional to the square of $\del_r \phi$). This conclusion can be escaped in the absence of a dilaton, \ie, for M$2$-, M$5$- and D$3$-branes. For example, the $r\ra 0$ limit of\linebreak the \br{3} is
\begin{eqnarray}\la{D3brlim}
ds_{10}^2 &=& \frac{r^2}{R^2} d x_{(4)}^2 + \frac{R^2}{r^2} d r^2 +R^2  d \O_{(5)}^2 \,. 
\end{eqnarray}
It describes a product space of an $ AdS_5$ and a five-sphere $S^5$. Similarly, the near horizon limit of the \mbr{2} and the \mbr{5} yields $AdS_4 \times S^7$, respectively $AdS_7 \times S^4$. All these are maximally symmetric spaces that preserve the full supersymmetry. Thus, in the asymptotic and near-horizon regimes we encounter an enhancement of supersymmetry. This concludes our examination of standard SUGRA $p$-brane solutions in $D=10$ and $D=11$.

\section{Batalin-Vilkovisky formalism\label{s:BV}}

Here we give only some of the main ideas underlying the BV quantization method. For more details we refer the reader to~\cite{Batalin:1983jr,Henneaux:1990jq,Henneaux:1992ig,Gomis:1995he}. This formalism is very useful when one wants to derive consistently the path integral of a system with gauge degrees of freedom. We choose here to work in the Lagrangian framework even though an equivalent Hamiltonian approach will be referred to later on. 

Let $S_0[\phi^i]$ be an action with the following bosonic gauge
transformations\footnote{We use the DeWitt notation. We neglect here trivial transformations $\d_{\rm triv} \phi^i \sim \ve \eta^{ij} \frac{\d S_0}{\d \phi^j}\,,$ with $\eta^{ij} = - \eta^{ji}$ for bosonic $\phi^i$.}
\begin{equation}
\label{e:GT}
\delta_\varepsilon \phi^i = R^i_\alpha\epsilon^\alpha\,,
\end{equation}
which (for simplicity) are supposed to be irreducible.
In the first step one has to enlarge the ``field'' content to
\begin{equation}\la{e:fieldsgh}
\{\Phi^A\}=\{\phi^i,C^{\alpha}\}\,.
\end{equation}
The fermionic ghosts $C^{\alpha}$ correspond to the parameters
$\varepsilon^\a$ of the gauge transformations~(\ref{e:GT}).  To each
field $\Phi^A$ we associate an antifield $\Phi_A^*$ of opposite
parity. The set of associated antifields is then
\begin{equation}
\{\Phi_A^*\}=\{\phi^{*}_i,C^*_\alpha\}\,.
\end{equation}

Subsequently, we introduce various gradings on the space of fields and antifields. The fields \re{e:fieldsgh} possess a vanishing antighost number (antigh) and a
non-vanishing pure\-ghost number (pgh)
\begin{equation}
\hbox{pgh}(\phi^i)=0\,,\qquad \hbox{pgh}(C^\alpha)=1\,.
\end{equation}
The pgh number of the antifields vanishes, but their respective
antigh number is equal to
\begin{eqnarray}
\hbox{antigh}(\Phi_A^*)=1+\hbox{pgh}(\Phi^A)\,.
\end{eqnarray}
The total ghost number (gh) equals the difference between the pgh
number and the antigh number.  The antibracket of two functionals
$X[\Phi^A,\Phi_A^*]$ and $Y[\Phi^A,\Phi_A^*]$ is defined as
\begin{equation}
(X,Y)=\int d^nx\left( \frac{\delta^RX}{\delta\Phi^A(x)}
\frac{\delta^LY}{\delta\Phi_A^*(x)}
-\frac{\delta^RX}{\delta\Phi_A^*(x)}\frac{\delta^LY}{\delta\Phi^A(x)}\right)\,,
\end{equation}
where $\delta^R/\delta Z(x)$ and $\delta^L/\delta Z(x)$ denote
functional right- and left-derivatives.

The \emph{extended action} $S$ is defined by adding to the classical
action $S_0$ terms containing the antifields in such a way that the
classical \emph{master equation},
\begin{equation}
(S,S)=0\,,
\label{e:master}
\end{equation}
is satisfied, with the following boundary condition
\begin{equation}
S=S_0+\phi^*_iR^i_\alpha C^{\alpha}+\dots \,.
\end{equation}
The dots denote terms of higher antigh number. This imposes the value of terms quadratic in ghosts and
antifields. The extended action has also to be of vanishing gh number.
If the algebra is non-Abelian, we know that we have to add other
pieces of antigh number two in the extended action with the general
form (due to structure functions)
\begin{equation}
S_2 =\frac{1}{2}C_\alpha^* f_{\beta \gamma}^\alpha C^\beta
C^\gamma\,.
\end{equation}
If the algebra is open\footnote{An algebra is called {\it open} if its closure involves trivial symmetries $\left[ \d(\ve_1), \d(\ve_2)\right] \phi^i = \dots + \eta^{ij} (\ve_1, \ve_2)\frac{\d S_0}{\d \phi^j}$. When that is the case, the algebra determines also the dynamics of the system.}, other terms in antigh number must be added,
quadratic\linebreak in $\phi^*_i$'s.  Furthermore, other terms in higher antigh
number could be necessary, e.g., when the structure functions depend on
the fields $\phi^i$.

The extended action captures all the information about the gauge
structure of the theory: the Noether identities, the (on-shell)
closure of the gauge transformations and the higher order gauge
identities are contained in the master equation.

The BRST (Becchi-Rouet-Stora-Tyutin) transformation $s$ \cite{Becchi:1974xu,Tyutin:1975qk} in the antifield formalism is a canonical
transformation, i.e., $sA=(A,S)$.  It is a differential: $s^2=0$, its
nilpotency being equivalent to the master
equation~(\ref{e:master}). The BRST differential decomposes according
to the antigh number as
$$
s=\d + \gamma + \hbox{"more"}\,, \quad \hbox{antigh}(\d) = -1\,, \quad \hbox{antigh}(\g) = 0
$$
and provides the gauge invariant functions on the stationary surface,
through its cohomology group at gh number zero $H_0(s)$.  The
Koszul-Tate differential $\delta$
\begin{equation}
\label {e:KT}
\d \F^*_i=(\F^*_i,S)\vert_{\F_A^*=0}
\end{equation}
implements the restriction on the stationary surface, and the exterior
derivative along the gauge orbits $\gamma$
\begin{equation}
\label {e:gammaop}
\gamma\F^i=(\F^i,S)\vert_{\F^*_A=0}
\end{equation}
picks out the gauge-invariant functions.

The solution $S$ of the master equation possesses gauge invariance,
and thus, cannot be used directly in a path integral.  There is one
gauge symmetry for each field-antifield pair. The standard procedure
to get rid of these gauge degrees of freedom is to use the
\emph{gauged-fixed action} $S_\P$ defined by
\begin{equation}
S_\P=S_{\rm non-min}\left[ \F^A,\F^*_A=\fr{\d\P[\F^A]}{\d\F^A}\right].
\label{e:GAUGEFIXED}
\end{equation}
The functional $\P[\F^A]$ is known as the \emph{gauge-fixing fermion}
and must be such that~$S_\P[\Phi]$ is non-degenerate, i.e.,  the
equations of motion derived from the gauge-fixed action $\d
S_\P[\F^A]/\d\F^A=0$ have a unique solution for arbitrary initial
conditions, which means that all gauge degrees of freedom have been
eliminated.  It also has to be local in order that the antifields are
given by local functions of the fields.

The generating functional of the theory is
\begin{equation}
\label{e:PATHINT}
Z=\int[\cD\F^A] \exp iS_\P \,.
\end{equation}
The value of the path integral is independent of the choice of the
gauge-fixing fermion $\P$.  The notation $[\cD\Phi]$ stands for
$\cD\Phi\,\mu [\Phi]$, where $\mu[\F]$ is the measure of the path
integral.  It is important to notice that the expression of the
measure $\mu[\Phi]$ in this path integral is not completely determined
by the Lagrangian approach.  A correct way to determine it, would be
to start from the Hamiltonian approach for which the choice of measure
is trivial. Indeed, it is known to be $\cD\F\cD\Pi$, that is the
product over time of the Liouville measure $d\Phi^Ad\Pi_A$.

It can be proved that, when correctly handled, the two approaches are
equivalent (see~\cite{Henneaux:1990jq,Henneaux:1992ig} and references therein).  This justifies a
posteriori the choice of the measure $\mu [\Phi]$
in~(\ref{e:PATHINT}). This method will be worked out explicitly in the next chapter for the case of duality-symmetric theories.

To conclude this chapter, we reviewed special topics in string and gauge field theories. For future use the reader should remember the worldvolume description
of D-branes in terms of the BI and WZ actions from Sect.~\ref{s:wv}. Another essential ingredient is conformal symmetry and the SCTC given in Sect.~\ref{confSugra}. Also, when constructing SUGRA solutions in Ch.~\ref{ch:sols}
we will rely on the property of the scalar potential that can be expressed in terms of a superpotential \re{superPot}. And for immediate use, the BV method is going to serve as quantization procedure for the PST model.

\chapter{Free duality-symmetric theories \label{ch:dual}}


\section{Introduction}

Chiral bosons represent a respectable subject in string and gauge theories. They are mostly related to long-standing problems regarding the complete action and quantization of such theories. To start with an example, we recall the duality transformation \re{EBBE} of the Maxwell equations that interchanges ${\vec E}$
with ${\vec B}$, without changing the physics these equations describe. We also remarked that such a symmetry does not hold directly at the level of the action. The way to circumvent that shortcoming will be explained in this chapter.

{\it Chiral bosons}, in general, are defined as $p$-forms gauge potentials with\linebreak a $(p+1)$-form field-strength $F_{(p+1)}$ satisfying a self-duality requirement
\begin{equation}\label{selfdual}
F_{(p+1)} = * F_{(p+1)}\,.
\end{equation}
Obviously, this is only possible in $2(p+1)$ dimensions. Furthermore, in spacetimes with Minkowskian signature (where $** F_{(p+1)}= (-)^{p} F_{(p+1)}$)
the self-duality condition is non-trivial only in twice odd dimensions. In twice even dimensions, we have to take $F_{(p+1)}$
to be a complex field and redefine the dual operator to be imaginary
by $*\rightarrow {\rm i}*$.  The complexification of the fields is
also equivalent to the dualization of a pair of real $p$-forms gauge
fields, in which case we will call the theory {\it duality-symmetric}. This will be for instance the case of Maxwell theory~($p=1$).

The study of chiral bosons is nowadays motivated by their applications to\linebreak SUGRA and brane physics. The $4$-form with a self-dual field-strength shows up as a component of the RR sector in type IIB SUGRA \cite{Dall'Agata:1997ju}. Another example is the \br{3} that supports a duality-symmetric vector \cite{Tseytlin:1996it,Nurmagambetov:1998gp} in its worldvolume description.\footnote{Similar features of other types of branes will be mentioned in Ch.~\ref{ch:defo}.} Electric-magnetic duality, in particular, is encountered in $D=4$ SUGRA coupled to vector multiplets or just rigid supersymmetric theories, like SYM. 

The difficulty with chiral bosons theories is the construction of an action that encodes the self-duality condition \re{selfdual} as a field equation. Indeed, this is a first-order differential equation, in contrast to the usual bosonic eoms that are second order. There are several ways to overcome this problem \cite{McClain:1990sx,Schwarz:1994vs,Pasti:1995ii,Pasti:1995tn}, but in each of them the manifest self-duality (duality-symmetry) of the action is gained at the expense of the some other symmetry or natural feature of the action. To be specific, for the free duality-symmetric Maxwell field\footnote{We choose this particular case just for the sake of clarity and further relevance, since analogous descriptions are applied also for other chiral bosons.} one can give up manifest Lorentz covariance to obtain manifest duality-symmetry \cite{Schwarz:1994vs}. We will explain how this can be done in Sect.~\ref{s:ssnon}. For quantization reasons \cite{Devecchi:1996cp}, it is more convenient to have a Lorentz-covariant version of the duality-symmetric action. Also this can be achieved \cite{Pasti:1995ii,Pasti:1995tn} with the help of an auxiliary scalar field that couples the Maxwell field strength in a non-polynomial manner. We will clarify the PST method in Sect.~\ref{s:freePST}. The polynomiality can be restored by exchanging the PST auxiliary scalar with with an infinite tower of auxiliary fields, which now enter the action polynomially \cite{McClain:1990sx,Pasti:1997vs}. 

It can be shown directly that all these formulations are classically equivalent and, actually, they describe a (free) Maxwell field. However, there is a priori no guarantee that the same conclusion holds for the associated quantum theories. Sect.~\ref{s:BVPST} will deal precisely with this issue and we will prove that the above mentioned approaches remain equivalent at the quantum level. The BV method is used to quantize the covariant duality-symmetric action. This method was also successfully applied for chiral $2$-forms in six dimensions \cite{VanDenBroeck:1999xw}. We conclude the chapter by briefly reminding the reader how the PST approach tackles the problem of self-dual $4$-form of type IIB SUGRA. 

\section{Non-covariant formulation}\la{s:ssnon}

One possible solution for reformulating Maxwell theory such that it incorporates the duality symmetry directly into the action was put forward by Schwarz and Sen~(SS) \cite{Schwarz:1994vs}. The price they payed for doing that is giving up manifest Lorentz covariance, even if the theory is still Lorentz invariant.

In short, the idea is to consider, besides the usual Maxwell potential, its dual vector potential. The role of the latter is to implement the duality as a field equation. Hence, our theory contains two gauge potentials.\footnote{In the context of duality-symmetric models the $\a$-label denotes an $\SO (2)$ index. Summation over repeated indices is understood.} $(A_\m^\a)_{\a =1,2}$ Schwarz and Sen proposed that the dynamics of the free Maxwell field is encoded in 
\begin{equation}\la{SSact}
S_0^{\rm SS} = - \frac 12 \int d^4x \Big( B^{\a\, i} \cL^{\a\b} E^{\b\, i} + B^{\a\, i} B^{\a\, i} \Big)\,,
\end{equation}
with $\cL^{\a\b}$ the antisymmetric unit matrix of $\SO(2)$ ($\cL^{12} =1= - \cL^{21}$) and 
\begin{eqnarray}
E^{\a\, i}&\equiv& \del_0 A_i^\a - \del_i A_0^\a\,, \nn\\
B^{\a\, i} &\equiv& \ft 12 \ve^{ijk} F^\a_{jk} = \ve^{ijk} \del_j A_k^\a\,. \la{SSnot}
\end{eqnarray}

In order to understand the equivalence to ordinary electrodynamics. we need to make use of the invariances this action possesses.  The non-vanishing transformations leaving the theory invariant are
\begin{eqnarray}
\d A_0^\a &=& \L^\a\,, \la{pureSS}\\
\d A_i^\a &=& \del_i \varphi^\a \,, \la{usualSS}
\end{eqnarray}
where the latter is the usual gauge transformation of the Maxwell field. However, due to the former invariance, the $A_0^\a$ component can be interpreted as pure gauge and set to $A_0^\a =0$. Note that, in contrast with Maxwell theory, we lose no physical degree of freedom by this choice. Then, the field equation for $A_i^2$ 
\begin{equation}\la{SSeq}
\ve^{ijk} \del_j (B_k^2 - E_k^1) = 0 
\end{equation}
contains no time derivative of $A_i^2$. This allows us to treat it as an auxiliary field and eliminate it using the solution
\begin{equation}\la{SSsol}
B^{2\,i} = E^{1\, i} + \del_i \theta
\end{equation}
for some arbitrary function $\theta(x)$. With the help of the \re{usualSS} symmetry the $\theta$ function can be chosen zero, \ie, $B^{2\,i} = E^{1\, i}$. We thus recover Maxwell theory (in the $A_0^1 =0$ gauge)
\begin{equation}\la{SS2M}
S_0^{\rm M} = - \frac 12 \int d^4x \Big( B^{1\, i}  B^{1\, i} - E^{1\, i} E^{1\, i} \Big)\,,
\end{equation} 
with the normal gauge invariance $\d A_i^1 = \del_i \varphi^1$.

The reason to consider the SS action \re{SSact} is its manifest $\SO(2)$ invariance
\begin{equation}\la{SSdual}
A_\m^\a \ra \cL^{\a\b} A_\m^\b\,.
\end{equation}
Note that in the SS formulation, unlike the Maxwell theory, the realization of the duality symmetry only depends on the gauge potentials and not on the field strengths. The SS action is invariant under rotations, but it lacks Lorentz covariance. Nevertheless, it is also invariant under (for some constant $\vec{V}$) 
\begin{equation}
\d A_i^\a = x^0 V^k \del_k A_i^\a + \vec{V} \cdot \vec{x}\, \cL^{\a\b} \ve^{ijk} \del_j A_k^\b
\end{equation}
that upon using the solution \re{SSsol} gives the Lorentz transformation of the physical~$A_i^1$ field. Thus, the SS model is Lorentz invariant. 

The SS formulation can be made supersymmetric. More appealingly, it can be generalized to obtain manifestly $\Sl(2, \bbbb{R})\times \mathop{\rm O} (6,22)$ invariant action for toroidally reduced $D=10$, $N=1$ SUGRA \cite{Schwarz:1994vs}. Of course, it will be more interesting and useful to dispose of a Lorentz covariant action with manifest duality symmetry. This is discussed in Sect.~\ref{s:freePST}.

\section{The PST model}\la{s:freePST}

In order to achieve a Lorentz covariant and manifest duality-symmetric Maxwell theory it suffices not only to work with a pair of dual vectors $(A_\m^\a)_{\a =1,2}$, but one needs extra auxiliary fields and symmetries. The possibility advertised here was proposed by Pasti, Sorokin and Tonin (PST) \cite{Pasti:1995ii,Pasti:1995tn}. It is based on a single auxiliary scalar field $a(x)$ that enters the action non-polynomially. 

The PST model is defined as follows
\begin{equation}\label{e:pst}
S_0= \frac 12 \int d^4x\, v^m F^{*\a}_{mn}\left( \cL^{\b\a }F^{\beta pn} - F^{*\a \,pn}\right) v_p \,,
\end{equation}
where the $m,n,\dots\,$ stand for Lorentz indices in $4$ dimensional
space-time. The common notation for the PST action reads
\begin{equation}
\begin{array}[b]{rclcrcl}
u^2&=&u^mu_m \,,
&\qquad&
v_m&=&\displaystyle \fr{u_m}{\su}\,,
\\[8pt]
F^\a_{mn}&=&2\partial_{[m}A^\a_{n]}\,,
&\qquad&
 F^{*\a}_{mn}&=&\displaystyle \half\e_{mnpq}F^{\a pq} \,,
\\ 
\cH^\a_{m} &=& \cL^{\b\a}v^n F^\b_{nm}\,,
&\qquad&
\tilde \cH^\a_{m}  &=& v^n F^{*\a}_{nm}\,,\\
\cF^\a_{mn}&=&\cL^{\b\a}F^\b_{mn}-F^{*\a}_{mn}\,,&  &h^{\a}_{m}&=& \cH^\a_{m} -  \tilde \cH^\a_{m} \,.
\end{array}
\label{e:notSS}
\end{equation}
The scalar field $a$ is present in the action only via its gradient $u_m = \del_m a$, which is a time-like vector. 

The claim that \re{e:pst} describes on-shell a single Maxwell field can be sustained by inspecting its dynamical equations and symmetries. The equations of motion associated to~(\ref{e:pst}) take the form
\begin{eqnarray}
\d A_m^\a &:\qquad &  \e^{mnqp}\partial_n(v_p \h{\a}{q})=0\,,
\label{e:eomA}\\ 
\d u_m & :\qquad & \fr{1}{2\sqrt{-u^2}}\left(\h{\a}{n}
 \cF^{\a mn} -\h{\a}{n} h^{\a\, n}v^m\right)=0\,.
\label{e:eomu} 
\end{eqnarray}
A first remark concerns the scalar $a(x)$, which is auxiliary because its field equation~\re{e:eomA} is a consequence of the equation corresponding to $A_m^\a$. This remark is also supported by the gauge invariances of the action
\begin{eqnarray}
\d_IA_m^\a&=&\partial_m\varphi^\a\,, \qquad \d_I a=0 \,,
\label{e:ginv1}\\
\d_{II}A_m^\a&=&-\cL^{\a\b}\h{\b}{m}\fr{\f}{\sqrt{-u^2}}\,, \qquad \d_{II}a=\f \,,
\label{e:ginv2}\\
\d_{III}A_m^\a&=&u_m\l^\a\,, \qquad \d_{III}a=0 \,,
\label{e:ginv3}
\end{eqnarray}
that are irreducible. Indeed, the second symmetry \re{e:ginv2} states that the transformation of $a(x)$ is a pure shift. It is in this sense that $a(x)$ is pure gauge. Yet, its equation of motion is not purely algebraic and one could not solve it for $a$ and then eliminate this field from the action. However, it can be easily fixed away using a suitable gauge condition (avoiding the singularity $u^2 =0$). One could choose, for instance, $u_m = \d_m^0$ and recover the SS action.

The third symmetry \re{e:ginv3}, also new with respect to Maxwell model, is crucial for enforcing the self-duality condition. Similar to the SS case, we can solve the field equation \re{e:eomA} as
\begin{equation}\la{solpst}
v_{[p} h_{q]}^\a = \del_{[p} \theta^\a u_{q]}
\end{equation}  
for some arbitrary function $\theta^\a (x)$. Because $\d_{III} (v_{[p} h_{q]}^\a) = \del_{[p} \l^\a u_{q]}$ we can adjust~$\l^a$ such that $\theta^\a (x) =0$. This approach yields the self-duality condition
\begin{equation}
\cF^{\a}_{mn}=0
\label{e:duality}
\end{equation}
as consequence of the field equations and not as an extra condition, which is usually imposed by hand. The self-duality requirement can be employed to express one of the gauge fields $A_\m^\a$ in terms of the other one, reducing the model to an ordinary Maxwell system.

So, the field $u_m$ as well as one of the two $A_m^\a$ are auxiliary, in the sense that one needs them only to lift self-duality and Lorentz invariance at the rank of manifest symmetries of the action. But, they can be removed on the mass-shell taking into account the gauge invariances of the new system. Nevertheless, the way we gauge fix the last invariance~(\ref{e:ginv3}) can be applied only at the classical level since we make explicit use of the field equations, which cannot be done in a BRST path integral approach. The manner of fixing the
unphysical degrees of freedom in the BV formalism will be clarified
in Sect.~\ref{s:g-fixing}.

Another key ingredient that will be needed shortly is the gauge algebra 
\begin{eqnarray}
[\d_{II}(\f_1),\d_{II}(\f_2)]&=&\d_{III}\left(\fr{\cL^{\a\b}\h{\b}{p}}
{(-u^2)^{3/2}}(\f_1\partial^p\f_2-\f_2\partial^p\f_1)\right)\,,
\label{e:galg1}\\
\left[\d_{II}(\f),\d_{III}(\l^\a)\right]&=&\d_I(\f\l^\a)+\d_{III}
\left(\fr{u^p\f}{(-u^2)} \partial_p\l^\a\right)\,.
\label{e:galg2}
\end{eqnarray}
These relations reflect the non-Abelian character of the PST system, with the structure constants replaced by non-polynomial structure \emph{functions}.

Inspired by the non-covariant version, the PST model was supersymmetrized by properly adding neutral fermions to the spectrum. A Lorentz covariant version of the manifestly $\Sl(2, \bbbb{R})\times \mathop{\rm O} (6,22)$ invariant action for the heterotic string low-energy approximation has also been proposed \cite{Pasti:1995tn}. 

We mention that the PST non-polynomial formulation can be traded for an equivalent model with infinitely many auxiliary fields, entering the action in a polynomial manner \cite{Devecchi:1996cp,Pasti:1997vs}. The latter are also relevant as effective actions of the massless RR type IIB fields compactified on a a CY\cite{Berkovits:1996tn}. 
They will not be discussed here.
 
Thus, we indicated several ways to implement the duality symmetry of the Maxwell equations into the action. Irrespective of the chosen path, SS, PST or with infinitely many auxiliary fields there is always a price to pay: the loss of Lorentz covariance, non-polynomiality or the infinity of fields. An interesting issue is to see if the classical equivalence among different formulations is taken over at the quantum level. We investigate it right away. 
 
\section{BV quantization of the PST action}\la{s:BVPST}

In this section we plan to prove the quantum equivalence between the free Maxwell theory on one hand and the SS and PST formulations on the other hand. This has been partially carried out in \cite{Girotti:1997kn} where the SS-Maxwell equivalence was achieved with the help of the Hamiltonian Batalin-Fradkin-Vilkoviski formalism. The novelty of this section is the proof (along the lines of \cite{Bekaert:2000rh}) that the same conclusion holds in the case of the Lorentz covariant formulation (PST) described in the previous section. To that end, we will apply the Lagrangian BV technique to determine the corresponding path integral.\footnote{Similar considerations for the SS model can be found in \cite{Bekaert:2001wa}.} Finally, the propagator of the PST model will be computed using the example of gravity-coupled PST system.  

\subsection{Minimal solution of the master equation}\label{s:BRST}

Building on the classical analysis of the model (Sect.~\ref{s:freePST}), we can now start the
standard BV procedure.\footnote{For some generalities on the BV method, see Sect.~\ref{s:BV}.} The first step is to construct the
minimal solution of the master equation with the help of the gauge
algebra. In order to reach that goal we will introduce some new fields, called ghosts, and their antibracket conjugates, known as antifields.

The minimal sector of fields and antifields dictated by the gauge\linebreak invariances (\ref{e:ginv1})-(\ref{e:ginv3}) as well as their ghost numbers and statistics are displayed in Table~\ref{t:mingh}.

\begin{table}[htbp] 
\begin{center}
\begin{tabular}{|c|c|c|c|c|c|c|c|c|c|c|}\hline\rule[-1mm]{0mm}{6mm}
$\Phi$&$A_m^\a$&$a$&$A_m^{\a *}$&$a^*$&$c^\a$&$c$&$c'^\a$&$c^{\a *}$&$c^*$&$c'^{\a *}$\\
\hline
\rule[-1mm]{0mm}{6mm}
$gh(\Phi)$&$0$&$0$&$-1$&$-1$&$1$&$1$&$1$&$-2$&$-2$&$-2$\\
\hline
\rule[-1mm]{0mm}{6mm}
${\rm antigh}(\Phi)$&$0$&$0$&$1$&$1$&$0$&$0$&$0$&$2$&$2$&$2$\\ 
\hline
\rule[-1mm]{0mm}{6mm}
\mbox{stat($\Phi$)}&$+$&$+$&$-$&$-$&$-$&$-$&$-$&$+$&$+$&$+$\\
\hline
\end{tabular}
\caption{{\it Ghost number, antighost number and statistics of the minimal fields and the associated antifields.}\label{t:mingh}}
\end{center}
\end{table}

The transformations~(\ref{e:ginv1})-(\ref{e:ginv3}) determine directly
the antigh number one piece of the extended action, i.e.,
\begin{equation}
S_1=\int d^4x\left[A_m^{\a *}\left(\partial^m c^\a- \cL^{\a\b} h^{\b \,m}
\fr{c}{\sqrt{-u^2}}+u^mc'^\a\right)+ a^* c\right].
\label{e:ext1}
\end{equation}
In order to take into account the structure functions one has to
insert in the solution of the master equation a contribution with
antigh number two of the form
\begin{equation}
\label{e:ext2}
S_2 = \int d^4x \left[c'^{\a *}\left( \fr{\cL^{\a\b}\h{\b}{p}}
{(-u^2)^{3/2}}c\,\partial^pc +\fr{v_p}{\su}c\partial^p c'^{\a}\right) +c^{\a *}c\,c'^\a \right].
\end{equation} 
Due to the field dependence of the structure functions one should
expect that $S_1$ and $S_2$ are not enough to completely specify the
extended action and one will need an extra piece of antigh number
three to do the job. Indeed, that was already the case for chiral
$2$-forms in $6$ dimensions discussed in~\cite{VanDenBroeck:1999xw}. Nevertheless, one
can readily check that, in the present situation, $S_{\rm
min}=S_0+S_1+S_2$ determines the minimal solution of the classical master
equation $(S_{\rm min},S_{\rm min})=0$, i.e.,
\begin{eqnarray}
(S_1,S_1)_1 + 2(S_1,S_2)_1 &=&0\,,
\nonumber \\
(S_2,S_2)_2 + 2(S_1,S_2)_2 &=&0\,.
\end{eqnarray}
This follows also as a consequence of the irreducibility of our
model. 

Once $S_{\rm min}$ has been derived, we can infer the BRST operator
$s$, which is the sum of three operators of increasing antigh
number
\begin{equation}
s=\delta+\gamma+\rho\,.
\end{equation}
For instance, using the general relation \re{e:KT}, the non-trivial action of the Koszul-Tate differential,
of antigh number $-1$, reads in our case 
\begin{eqnarray}
\d A_m^{\a *} &=& \e^{mnqp}\partial_n (v_p \h{\a}{q})\,,\\
\d a^* &=& \partial_m\left(\fr{1}{2\sqrt{-u^2}}\left( \h{\a}{n}\cF^{\a mn}
- \h{\a}{n} h^{\a\, n} v^m\right)\right)\,,\\
\d c^{\a *}&=&-\partial^m A^{\a *}_m\,,
\label{e:KT1} \\
\d c^* &=&-\fr{\cL^{\a\b} \h{\b}{p}}{\su} A^{\a *\, p} +a^*\,,
\label{e:KT2} \\
\d c'^{\a *}&=&u^m A^{\a *}_m\,.
\label{e:KT3}
\end{eqnarray}
The operator $\g$ can be as well directly computed by applying \re{e:gammaop}. For exemplification
\begin{eqnarray}\la{e:gammaop3}
\g A_m^\a &=& \partial_m c^\a- \cL^{\a\b} \h{\b}{m}
\fr{c}{\sqrt{-u^2}}+u_m c'^\a\,,\\
\g a&=& c\,.
\end{eqnarray}

The third piece, $\rho$, of antigh number $+1$ is present also
because the structure functions determined
by~(\ref{e:ginv1})-(\ref{e:ginv3}) depend explicitly on the fields.

In this way the goal of this section, i.e., the construction of the
minimal solution for the master equation, was achieved.

\subsection{The gauge-fixed action}\label{s:g-fixing}

The minimal solution $S_{\rm min}$ will not suffice in fixing all the
gauge invariances of the system and, before fixing the gauge, one
needs a non-minimal solution\linebreak for $(S,S)=0$ that takes into
account the trivial gauge transformations.  In this subsection, we initially 
construct such a non-minimal solution and, afterwards, we propose two
possible gauge-fixing conditions, which will yield two versions for the
gauge-fixed action: a covariant and a non-covariant one.

\subsubsection{Non-minimal sector}

Inspired by the gauge transformations~(\ref{e:ginv1})-(\ref{e:ginv3})
and their irreducibility we propose a non-minimal sector given in
Table~\ref{t:non-minsect}.

\begin{table}[b]
\begin{center}
\begin{tabular}{|c|c|c|c|c|c|c|c|c|c|c|c|c|} \hline
\rule[-1mm]{0mm}{6mm}
$\Phi$&$B^\a$&$B$&$B'^\a$&${\bar C}^\a$&${\bar C}$&${\bar C}'^\a$&
$B^{\a *}$&$B^*$&$B'^{\a *}$&${\bar C}^{\a *}$&${\bar C}^*$&${\bar
C}'^{\a *}$\\ 
\hline\rule[-1mm]{0mm}{6mm}
$gh(\Phi)$&$0$&$0$&$0$&$-1$&$-1$&$-1$&$-1$&$-1$&$-1$&$0$&$0$&$0$\\
\hline\rule[-1mm]{0mm}{6mm}
\mbox{stat($\Phi$)}&$+$&$+$&$+$&$-$&$-$&$-$&$-$&$-$&$-$&$+$&$+$&$+$\\
\hline%
\end{tabular}\caption{{\it Ghost number and statistics of the non-minimal fields and
their antifields.}\label{t:non-minsect}}
\end{center}
\end{table}

They satisfy the following equations
\begin{eqnarray}
s{\bar C}^{\cdots} &=& B^{\cdots}\,,
\nonumber\\
\qquad sB^{\cdots}&=&0\,, 
\nonumber\\
sB^{\cdots *}&=&{\bar C}^{\cdots *}\,,
\nonumber\\
s{\bar C}^{\cdots *} &=&0\,.
\end{eqnarray}
The dots are there to express that these relations are valid for the
corresponding three kinds of non-minimal fields.  We immediately see
that $\bar{C}^{\cdots}$'s and $B^{\cdots}$'s constitute trivial pairs,
as well as their respective antifields, in such a way that they do not
enter in the cohomology of $s$. Hence, they are called non-minimal. A
satisfactory explanation for the necessity of the presence of a
non-minimal sector is provided by BRST-anti-BRST formalism.  Their
contribution to the solution of the master equation is
\begin{equation}
\label{e:non-min}
S_{\rm non-min}=S_{\rm min} +\int d^4x \left({\bar C}^{\a *}B^\a +
{\bar C}^* B +{\bar C}'^{\a *}B'^\a \right).
\end{equation}

\subsubsection{Covariant gauge fixing}

We will first try a covariant gauge fixing, which in principle should
yield a covariant gauge-fixed action, and we will see what is the main
problem that occurs. One can consider the following \emph{covariant}
gauge choices
\begin{eqnarray}
\d_{I} &\ra & \partial^m A^\a _m =0\,,
\label{e:gc1} \\
\d_{II} &\ra & u^2 +1=0\,,
\label{e:gc2} \\
\d_{III} &\ra & u^m A^\a _m =0\,.
\label{e:gc3} 
\end{eqnarray}
The gauge choice~(\ref{e:gc1}) is analogous to the Lorentz gauge. In
its turn~(\ref{e:gc2}) allows to take a particular Lorentz frame in
which $u^m(x)$ is the unit time-like vector at the point $x$. In such a
case, at the point $x$,~(\ref{e:gc3}) is the temporal gauge condition
for the two potentials.

A gauge-fixing fermion corresponding to the gauge
choices~(\ref{e:gc1})-(\ref{e:gc3}) is
\begin{equation}
\label{e:gaugeferm}
\P[\F^A]=-\int d^4x \left[{\bar C}^\a \partial^m A^\a _m  
+{\bar C} (u^2 +1) +{\bar C}'^\a u^m A^\a _m \right]. 
\end{equation}
One next expresses all the antifields with the help of $\P[\F]$ and
\begin{equation}
\label{e:antifields}
\F^*_A =\fr{\d\P[\F^A]}{\d\Phi^A}\,,
\end{equation}
which for our model becomes
\begin{eqnarray}
\label{e:antifield1}
A_m^{\a *} &=& \partial_m {\bar C}^\a -u_m {\bar C}'^\a\,, 
\qquad a^*= 2\partial_m (u^m {\bar C}) +\partial^m(A_m^\a {\bar C}'^\a) \,, 
\nonumber\\
\label{antifield2}
c^{\cdots *}&=& 0\,, \qquad B^{\cdots *}= 0\,, 
\nonumber\\
{\bar C}^{\a *} &=& -\partial^m A^\a _m \,, \qquad 
{\bar C}^* = -(u^2 +1)\,, \qquad {\bar C}'^{\a *} = -u^m A^\a _m  \,.\label{e:antifield3}
\nonumber
\end{eqnarray}
Using the last relations one can find the gauge-fixed action as
in~(\ref{e:GAUGEFIXED}). Explicitly, this is
\begin{eqnarray}
\label{e:fixedaction}
S_\P &=& S_0 + \int d^4x \Biggl[-{\bar C}^\a \square c^\a -
\fr{\cL^{\a\b}\h{\b}{m}}{\sqrt{-u^2}}\partial^m {\bar C}^\a \cdot c
+u^m \partial_m {\bar C}^\a \cdot c'^\a  
\nonumber\\&&
	\hphantom{S_0 + \int d^4x \Biggl[}\!
-u_m {\bar C}'^\a \partial^m c^\a - u^2{\bar C}'^\a c'^\a -(2 u_m
{\bar C} + A_m^\a {\bar C}'^\a)\partial^m c 
\nonumber\\ &&
	\hphantom{S_0 + \int d^4x \Biggl[}\!
-\,(\partial^m
A^\a _m )B^\a -(u^2+1)B -(u^m A^\a _m )B'^\a \Biggr]\,.
\end{eqnarray}
Writing down the path integral~(\ref{e:PATHINT}), one can integrate
directly the fields $B^{\cdots}$ producing the gauge
conditions~(\ref{e:gc1})-(\ref{e:gc3}). A further integration of
${\bar C}$, ${\bar C}'^\a$\linebreak and $c'^\a$ (in this order) leads to
\begin{equation}
\label{e:fixed1}
Z = \int\left[\cD A\cD a\cD c^\a \cD c\cD {\bar C}^\a \right] \,\d
(\partial^m A_m^\a) \,\d (u^2+1) \,\d (u^mA_m^\a) \,\d (u^m \partial_m
c) \exp iS'_\Psi\,,
\end{equation}
where
\begin{equation} 
\label{e:covgfa}
S'_\Psi =S_0 + \int d^4x\left[ -{\bar C}^\a \square c^\a +\left(
\fr{\cL^{\a\b}\h{\b}{m}}{\sqrt{-u^2}} c +\fr{u_m}{u^2} (u^p\partial_p
c^\a +A_p^\a\partial^p c )\right)\partial^m {\bar C}^\a\right].
\end{equation}
Of course, the next step in deriving a covariant generating functional
from which we should read out the \emph{covariant} propagator for the
fields $A_m^\a$ would be the elimination of $c$ and $a$
in~(\ref{e:fixed1}). Due to the `gauge condition' for the ghost $c$
(i.e., $u^m \partial_m c =0$) and the way it enters the gauge-fixed
action $S'_\Psi$, this integration is technically difficult. One
could try to integrate both $c$ and $a$ at the same time. This is
also not straightforwardly possible as a consequence of the gauge
condition~(\ref{e:gc2}). This requirement was necessary to
\emph{covariantly} fix the symmetry~(\ref{e:ginv2}). Nevertheless, one
can attempt to find the general solution to this
equation~(\ref{e:gc2}), which reduces to the integration of
$\partial^m a=\L^m{}_p(x)n^p $ (with $\L^m{}_p(x)$ a point-dependent
Lorentz boost and $n_p$ a constant time-like vector, i.e., $n_p n^p
=-1$). Such a solution is still inconvenient due to $x$-dependence of
the Lorentz transformation matrix $\L^m{}_p(x)$.

A way to overcome this sort of complication is to choose a particular
form for this matrix, breaking Lorentz symmetry. It is precisely this
price that we have to pay in order to explicitly derive the propagator
of $A_m^\a$ fields. As it will be explained shortly, by
taking a particular solution for~(\ref{e:gc2}), i.e., by giving up
Lorentz invariance, we will be able to express the gauged-fixed action
in a more convenient form for our purposes.

\subsubsection{Non-covariant gauge fixing}

As it was remarked in the previous subsection, in order to explicitly
derive the Feynman rules for the PST model one has to break up its
Lorentz symmetry by taking a specific solution of the
equation~(\ref{e:gc2}). In this subsection we present a non-covariant
gauge of the theory and the advantages for such a choice will become
clear in the forthcoming sections. A possible \emph{non-covariant} gauge
fixing is
\begin{eqnarray}
\d_{I} &\ra & \partial^m A^\a _m =0\,,
\label{e:gnc1} \\
\d_{II} &\ra & a-n_m x^m=0\,,\qquad n_m n^m =-1 \,,
\label{e:gnc2} \\
\d_{III} &\ra & n^m A^\a _m =0\,.
\label{e:gnc3} 
\end{eqnarray}
By~(\ref{e:gnc2}), the gradient $\partial^m a$ becomes equal to the
vector $n^m$ introduced above.  In a Lorentz frame where
$n^m=(1,0,0,0)$ the requirement~(\ref{e:gnc3}) is the temporal gauge
condition, while~(\ref{e:gnc1}) is the Coulomb gauge condition for the two
potentials $A^\a_m$.

Then, the gauge-fixing fermion assumes the following form
\begin{equation}
\label{e:gaugeferm-non}
\P[\F^A]=-\int d^4x \left[{\bar C}^\a \partial^m A^\a _m +{\bar C}
(a-n_m x^m) +{\bar C}'^\a n^m A^\a _m \right].
\end{equation}

Using the same non-minimal contribution $S_{non-min}$ as before, the
non-covariant gauge-fixed action is
\begin{eqnarray}
\label{e:gauge-fixedaction}
S_\P &=& S_0 + \int d^4x \Biggl[\left(\partial^m {\bar C}^\a -u^m {\bar
C}'^\a \right)\left(\partial_mc^\a-
\cL^{\a\b}\h{\b}{m}\fr{c}{\sqrt{-u^2}}+u_m c'^\a\right) 
\nonumber\\ &&
	\hphantom{S_0  }\!
+\left(-{\bar C}+\partial^m(A_m^\a {\bar C}'^\a)\right) c-\partial^m
A^\a _mB^\a - (a-n_m x^m)B \,
\nonumber\\ &&
	\hphantom{S_0  }\!
-u^m A^\a _mB'^\a \Biggr]\,.
\end{eqnarray}
This action is by far more convenient in deriving the propagator of
the gauge fields than its covariant expression~(\ref{e:covgfa})
because one can completely integrate the ghost sector and, as a bonus, the
bosonic part takes a more familiar form. It represents also the cornerstone in proving the quantum mechanical equivalence of the PST model to ordinary Maxwell theory.

\subsection{PST-Maxwell quantum equivalence\label{s:qmequiv}}

The gauge-fixed action corresponding to the non-covariant gauge choice
can be used to recover the Schwarz-Sen theory, which is itself
equivalent to the Maxwell theory.  Our starting point is the generating functional (see Sect.~\ref{s:BV} and Appendix~\ref{a:Max})
\begin{equation}
Z= \int {\cal D} A^\a_m \,\cD a \,\cD c^{\cdots} \,\cD B^{\cdots}
\,\cD \bar{C}^{\cdots} \, {\rm det}(\square)\,{\rm det} ^{-1}({\rm
curl})\, \exp iS_\P\,,
\end{equation}
with $S_\P$ given by (\ref{e:gauge-fixedaction}).

After integrating out some fields, in the following order
($B^{\cdots},$ $\bar{C},$ $c,$ $\bar{C}'^\a,$ $c'^\a ,$ $a$), we
obtain a simplified path integral
\begin{equation}
Z= \int {\cal D} A^\a_m \,\cD {\bar C}^\a\,\cD c^\a \,{\rm
det}(\square)\, \,{\rm det} ^{-1}({\rm curl})\,\delta(\partial^m
A^\a_m)\,\delta(n^m A^\a_m)\, \exp iS'_\P\,,
\end{equation}
where the gauge-fixed action reduces now to
\begin{equation}
\label{e:non-fixedaction1}
S'_\P =\int d^4x \left[ -\fr12 n^mF^{*\a}_{mn}\cF^{\a np}n_p -{\bar
C}^\a \square c^\a -{\bar C}^\a n^pn^q\partial_p\partial_q c^\a \right].
\end{equation}
If we place ourselves in a Lorentz frame where $n^m=(1,0,0,0)$, the
functional~$S'_\P$ assumes the form of the sum of Schwarz-Sen
gauge-fixed action~(\ref{e:ScSe}) and a ghost term
\begin{equation}
\label{e:ghosts}
-\int d^4 x\, {\bar C}^\a \bigtriangleup c^\a \,. 
\end{equation}
At this point we can integrate the ghosts $\bar{C}^\a$ and $c^\a$, and
the two fields $A^\a_0$, obtaining exactly the generating
functional~(\ref{e:genfunc}) of the Maxwell theory in the
non-covariant formulation (The quantum equivalence of Maxwell and
Schwarz-Sen actions is briefly reviewed in appendix~\ref{a:Max}).

This proves the quantum equivalence (see also footnote~\ref{f:PSTq}) of the PST action~(\ref{e:pst})
to the Maxwell theory, which was already known at the classical
level. The quantum equivalence was not obvious because the PST formulation is not quadratic (and so the path integral
is not Gaussian) and the scalar field $a$ is not, strictly
speaking, an auxiliary field (its equation of motion is not an
\emph{algebraic} relation which allows its elimination from the
action).


\subsection{Coupling to gravity} \label{s:gravity}

The gravitational couplings of the PST model can be treated along the same line of thought, a fact that allows an immediate and explicit computation of the effective propagator for the PST fields $A^\a_\m$.


%
We consider thus the same PST action~(\ref{e:pst}) but in a
gravitational background characterized by a metric $g_{\mu\nu}$,
i.e., we take\footnote{The Greek letters $\m$, $\n$, $\r$ etc. label
curved indices, while $\a$ and $\b$ denote $\SO(2)$ indices.}
\begin{equation}
\label{e:pstgrav}
S^g_0=-\frac{1}{2}\int \sqrt{-g} \, v_\mu
\h{\a}{\nu}g^{\m\r}g^{\n\sigma}F^{*\a}_{\r\sigma}\,,
\end{equation}
where $g=\det{g_{\m\n}}$.

Next, we apply the same BRST formalism as before, following precisely
the same steps (merely replacing the flat indices by curved
ones). Moreover, using the same non-covariant gauge, one infers for
the ghost sector a similar contribution of the type~(\ref{e:ghosts}),
which becomes here
\begin{equation}
-\int d^4x \sqrt{-g}\,{\bar C}^\a (\square_{cov}
 +\nabla^{cov}_N\nabla^{cov}_N )c^\a \,.
\end{equation}
The only difference resides in replacing the ordinary derivatives by
covariant quantities. In any case, the fermionic ghosts decouple from
the bosonic fields and can be handled as explained
following eq.~(\ref{e:ghosts}) (by integrating over them in the path
integral). This is the reason why we focus our attention on the
bosonic part of the gauge-fixed action arising from the original
action $S^g_0$.

As we are only interested in the first-order interaction of the model
with the background $g_{\m\n}$ it is natural to try to expand this
metric around the flat one. In other words, we consider
\begin{equation}
g^{\m\n}=\eta^{\m\n} + h^{\m\n}\,,
\end{equation}
where the fluctuation $h^{\m\n}$ is parametrized in terms of inverse $e_m^\m$ of the
orthogonal vectors $e_\m^m$, i.e.,
\begin{equation}
h^{\m\n} = e_m^\m  e_n^\n \eta^{mn}\,.
\end{equation}
Our next move consists in developing the bosonic part of the
gauge-fixed action to the first-order in the perturbation $h^{\m\n}\,$. A straightforward computation yields\footnote{The indices are from now on raised and lowered with the flat
metric $\eta^{\m\n}$.}
\begin{eqnarray}
S^g_\P &=& \int d^4x \Biggl\{-\frac{1}{2}
A_\m^\a[\d^{\a\b}\eta^{\m\n}(-\square-\partial^2_N)
+\cL^{\a\b}T^{\m\n}\partial_N ]A_\n^\b  
\nn\\&&
	\hphantom{\int d^4x \Biggl\{}\!
+ \frac{1}{2} (T^{\m\n}A^\a_\n)\Biggl[\d^{\a\b}\eta_{\m\s}(\frac{1}{2}{\tilde
h}-(n^\tau e_\tau)^2) +\d^{\a\b}e_\m e_\s  
\nn\\&&
	\hphantom{+ \frac{1}{2} (T^{\m\n}A^\a_\n)\Biggl[}\!
-\frac{1}{2} \cL^{\a\b}(n^\zeta e_\zeta)\e_{\m\kappa\tau\s}e^\kappa
 n^\tau\Biggr](T^{\s\r}A^\b_\r)\Biggr\} \,,
\label{e:orders}
\end{eqnarray}
where we neglected the second order in $h^{\m\n}$ or higher.\footnote{The flat indices $m$ of the vierbeins have been dropped out for simplicity.} In the
meantime we employed the notation ${\tilde h} =
h^{\m\n}\eta_{\m\n}$ and also
\begin{equation}
\label{e:T}
T^{\m\n}A_\n^\a = \e^{\m\n\r\s}n_\r \partial_\s A_\n^\a \,.
\end{equation}
The object $T^{\m\n}$, defined in this way, is a differential operator
transforming one-forms into one-forms. It is antisymmetric under the
interchange of its indices and it is characterized by a very important
feature, namely
\begin{equation}
\label{e:propT}
T^{\m\r}T_{\r\s}T^{\s\n} = - (\square +\partial_N^2) T^{\m\n}\,.
\end{equation}
This property allows one to transform any series expansion in
$T^{\m\n}$ into a polynomial containing only $1$, $T$ and $T^2$.

Let us return to the interpretation of the
expansion~(\ref{e:orders}). The first remark is that
the zeroth-order, given in the first line, coincides with the one from the
flat space discussion.

This term delivers the gauge-fixed kinetic operator 
\begin{equation}
K_{\m\n}{}^{\a\b} = \d^{\a\b}\eta_{\m\n}(-\square-\partial^2_N)
+\cL^{\a\b}T_{\m\n}\partial_N\,,
\end{equation}
whose inverse is nothing but the propagator $P_{\m\n}{}^{\a\b}$ of the
vector fields $A_\m^\a$. A simple computation based on the
property~(\ref{e:propT}) of $T^{\m\n}$ gives then the explicit form of the
propagator
\begin{equation}
\label{e:propag}
P_{\m\n}{}^{\a\b} = -\fr{1}{\square +\partial_N^2}\left[
\d_{\m\n}\d^{\a\b} + \fr{\cL^{\a\b}T_{\m\n}\partial_N}{\square}
-\fr{\d^{\a\b} T_{\m\r}T^\r{}_\n \partial_N^2}{\square(\square
+\partial_N^2)}\right].
\end{equation}

If we consider also the first-order
interaction with a gravitational background, \ie, the last two lines of~(\ref{e:orders}), we notice that in such an interaction the gauge fields $A_\m^\a$ couple to the perturbation $h^{\m\n}$ only as
$T^{\m\n}A_\n^\a$. Therefore, we conclude that the effective
propagator in the presence of gravity must be
\begin{equation}
\label{e:effpropag}
T^{\m\r}P_{\r\s}{}^{\a\b}T^{\s\n} = [\cL^{\a\b}\d^\m{}_\s \partial_N
-\d^{\a\b} T^\m{}_\s ]T^{\s\n}\fr{1}{\square}\,,
\end{equation}
where we see that the apparent pole $-{1}/({\partial^2_N +\square})$ has
been replaced by an expected massless propagator ${ 1/\square}$. This
should not be understood as a result of the specific gravitational
coupling, but as a characteristic of Feynman computations for the PST
model.
 
The expression of the effective propagator together with the
interaction terms in~$S^g_\P$ can further be used to determine the
building blocks of the one-loop Feynman diagrams for the coupling of
the PST model to a gravitational background. A similar method was
carried out in~\cite{Lechner:1998ga} for evaluating the gravitational anomalies of chiral bosons in $4p+2$ dimensions. 
\begin{center}
\begin{figure}[ht]
\epsfxsize =10cm
{\hskip 2cm \epsffile{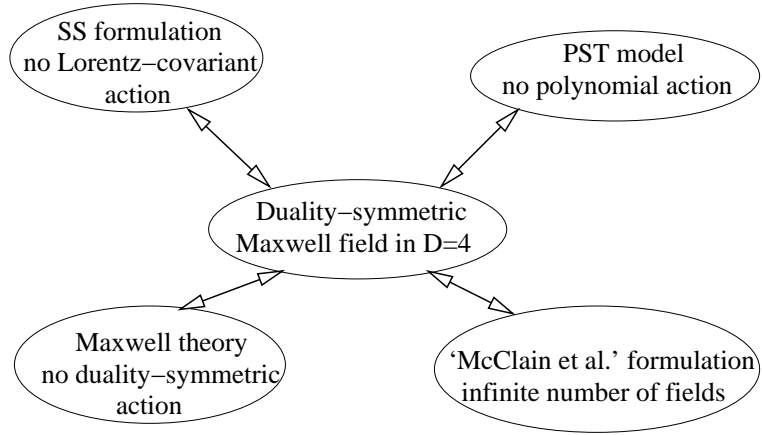}
}
\caption{{\it Various off-shell formulations describing (on-shell), both classically and quantum mechanically, one Maxwell gauge potential.}\label{fig:dualmax}}
\end{figure}
\end{center}

\subsection{Conclusions}\label{s:concl}

Summarizing, we proved the equivalence of the PST \re{e:pst} and
Schwarz-Sen \re{SSact} formulations of duality-symmetric Maxwell theory at the
quantum level.\footnote{The quantization of the formulation with infinitely many auxiliary fields has been carried out in \cite{Martin:1994wf}.} The latter, Schwarz-Sen, is quantum mechanically also
equivalent to the ordinary Maxwell theory~\cite{Girotti:1997kn} such that all
these models are physically related (on-shell) at the classical and
quantum level, even if their off-shell descriptions are different. Figure~\ref{fig:dualmax} captures the actual picture.

To prove the equivalence, we have adopted a formal path integral quantization (`formal' in the sense that possible UV divergencies due to the non-Gaussian character of the the integral were not considered\footnote{The absence of gauge anomalies 
together with the finiteness 
of the PST model have been recently established in \cite{DelCima:2000nr} 
.\la{f:PSTq}}) based on the
BV quantization formalism. This approach resides in
compensating all the gauge symmetries of the original system by some
fermionic ghosts and their antibracket conjugates - called
antifields. After extending the action to a suitable chosen
non-minimal sector, we had to fix the gauge. We were able to perform
two different gauge-fixings. The covariant one \re{e:gaugeferm} preserves Lorentz
covariance but it has the disadvantage of an intricate form in the
ghost sector which makes its integration difficult.  On the other
hand, giving up Lorentz covariance we presented also a non-covariant
gauge~\re{e:gaugeferm-non} that has a simple structure in its fermionic part leading us to
favorable results. Firstly, it was the cornerstone in proving the
quantum equivalence of the studied PST model and the Schwarz-Sen
action. 
Secondly, the same gauge choice~\re{e:gaugeferm-non} was employed in the example of gravitational interactions of the PST system. As a result, we were able to determine the effective propagator \re{e:effpropag} of the PST fields, which involves the massless pole of the Maxwell theory. Such a
first-order expansion in the perturbed metric was performed also by
Lechner~\cite{Lechner:1998ga} in studying the gravitational anomalies of the
self-dual tensors in $4p+2$ dimensions. However, as discussed
in~\cite{Alvarez-Gaume:1984ig}, the self-dual vector field in $4$ dimensions should be anomaly free.

\section{The PST formulation of Type IIB SUGRA\la{s:pstIIB}}

The PST approach to duality-symmetry is not restricted to $D=4$. It can be directly extended to duality symmetry in $D=4 p$ and self-dual $2p$-forms in $D=4 p+2$. One particular relevant application in the context of string theory is the $4$-form $A^{(4)}$ (with self-dual field strength) of type IIB SUGRA. Since it represents the background to which the deformed $D=4$ PST theory (self-dual \br{3}) will be coupled in the next chapter and also due to its intrinsic importance, we take the chance to briefly comment on the PST-like action for type IIB. Further details can be found in \cite{Dall'Agata:1997ju}.

In Sect.~\ref{s:maxsugra}, we pointed out that the self-duality condition $* F_{(5)} = F_{(5)}$ constitutes an extra requirement~\re{fIIB} in addition to the field equations. It is needed to reduce the number of physical components of a ten-dimensional $4$-form from ${\bf 70}$ down to ${\bf 35}$. However, following the PST suggestion, this condition can be lifted to the rank of an equation of motion. In order to preserve also the general coordinate invariance, we introduce again the auxiliary scalar $a(x)$ and we modify the kinetic term of $A^{(4)}$ in~\re{actIIB} {\it \`a la} PST
\begin{eqnarray}\la{actIIBpst}
{\cal L}^{PST}_{IIB} &=&  R * 1 - \ft 12 \, * d \phi \wedge d \phi - \ft 12 \rme^{2 \phi}\, * d \chi \wedge d \chi \nn\\
 &-& \ft 12 \rme^{- \phi} \, *  H_{(3)} \wedge  H_{(3)} - \ft 12 \rme^{ \phi} \, *  F_{(3)} \wedge  F_{(3)} - \ft 12 A_{(4)} \wedge d A_{(2)} \wedge H_{(3)}\nn\\
&-& 4  * (i_v * F_{(5)}) \wedge  (i_v {\cal F}_{(5)})\,, 
\end{eqnarray}
where the $5$-form field strength and its antisymmetric part denote 
\begin{eqnarray}\la{fIIBpst} 
F_{(5)} &=& d A_{(4)} - \ft 12 A_{(2)} \wedge H_{(3)} + \ft 12 B_{(2)} \wedge d A_{(2)} \,, \nn\\
{\cal F}_{(5)} &=& F_{(5)} - *F_{(5)} \,.
\end{eqnarray}

Besides the global $\Sl(2,\bbbb{R})$ symmetry discussed in Sect.~\ref{s:maxsugra}, general covariance and usual gauge invariances
\begin{eqnarray}\la{usuagIIB}
\d B_{(2)} &=& d \a_{(1)}\,, \qq \d A_{(2)} = d \tilde\a_{(1)}\,,\nn\\
\d A_{(4)} &=& d \varphi_{(3)}  - \ft 12\, d \a_{(1)} \wedge A_{(2)} + \ft 12 \, d \tilde\a_{(1)} \wedge B_{(2)}\,, 
\end{eqnarray}
this type IIB action \re{actIIBpst} has two extra symmetries 
\begin{eqnarray}
\d A_{(4)} &=& \frac{\d a}{\sqrt{- u^2}} i_v {\cal F}_{(5)}\,, \qq \d a = \phi\,, \la{ginv2IIB}\\
\d A_{(4)} &=& d a \wedge \l_{(3)}\,, \qq \q \:\,\d a =0\,.\la{ginv3IIB}
\end{eqnarray}
In analogy with the Maxwell theory, the first transformations \re{ginv2IIB} state the unphysical character of the scalar $a(x)$ that can be gauge fixed. For instance, one could choose locally $d a =n$, for some $n^2 = -1$, at the expense of general coordinate invariance. The second new transformation \re{ginv3IIB} is responsible for generating the self-duality condition
\begin{eqnarray}\la{selfdIIB}
{\cal F}_{(5)} &\equiv & F_{(5)} - *F_{(5)} = 0 \,
\end{eqnarray}
as a consequence of the field equations. The dynamics described by \re{actIIBpst} is identical to the one inferred from \re{actIIB} supplemented with \re{fIIB}. The reason why the standard formulation of Sect.~\ref{s:maxsugra} is still preferred to the PST approach, is the absence in the former formulation of the non-polynomialities caused by the auxiliary field. One equivalent alternative is to reformulate the PST-like action \re{actIIBpst} with the help of infinitely many auxiliary (anti-)self-dual $5$-forms \cite{Dall'Agata:1997ju}. This latter approach has also its own shortcomings and it won't be discussed here.

Through this chapter, we examined free theories of chiral bosons with emphasis on the duality-symmetric Maxwell field and the self-dual $4$-form. In the former case we presented three different\footnote{We mentioned even a fourth one, but it will play no role in our subsequent considerations.}, but (classically) equivalent descriptions of the same physics: Maxwell theory \re{SS2M}, SS \re{SSact} and PST \re{e:pst}. In each of these models one sacrifices some symmetry or nice feature of the action: manifest duality-symmetry, manifest Lorentz covariance or polynomiality respectively. Furthermore, we demonstrated that the three formulations remain equivalent also at the quantum mechanical level. As a bonus, we computed the physical propagator of the PST model \re{e:effpropag}. And, finally, we illustrated how the PST approach can meet the challenge of another long-standing problem of string theory, namely writing down a complete action of type IIB SUGRA. 

\chapter{Interacting duality-symmetric theories\label{ch:defo}}

\section{Introduction}

Theories of chiral bosons appear not only as free models as in the cases presented in the previous chapter, but they are equally relevant as interacting theories. The necessity for  rigorous study of consistent interactions of such chiral systems comes again from M-theory. We already met several times the \br{3}, the description of which requires the presence of a duality-symmetric worldvolume gauge vector. On the other hand, the \br{3} action must be of DBI type \re{DBI} (plus some WZ term). Combining these two remarks, the need of a duality-symmetric formulation of the DBI action becomes evident. This goal can be reached by either deforming directly \cite{Gibbons:1995cv,Bekaert:2001wa,Bekaert:2002cz,Deser:1998gq} the free models of Ch.~\ref{ch:dual} or by dimensionally reducing from higher dimensions \cite{Berman:1997iz,Nurmagambetov:1998gp}.

A similar situation occurs for the \mbr{5} \cite{Pasti:1997gx} and the type IIA NS$5$-bra\-ne~\cite{Callan:1991ky}. They both admit a self-dual second-rank tensor on their worldvolume. The corresponding actions have been obtained as well using similar methods to the ones discussed hereafter \cite{Perry:1997mk,Schwarz:1997mc,Pasti:1997gx}. In \cite{VanHoof:1999xi,Claus:1998cq} tensor multiplets containing these self-dual $2$-forms were supersymmetrized both in rigid and local superconformal background.\footnote{Analogous superconformal theories for $D=5$ are elaborated in detailed in Ch.~\ref{ch:weyl} and \ref{ch:matter}.} 

With this motivation in the back of our mind, we would like in this chapter to generalize the free actions \re{SSact} and \re{e:pst} to interacting theories. We start gradually by deforming the theory with one (on-shell) free Maxwell field and the step to possible interactions of $N$ such fields is postponed to Sect.~\ref{s:nogo}. For the case of only one vector field, the two situations (non-covariant and manifestly Lorentz covariant) are handled separately (in Sect.~\ref{s:defSS} and \ref{s:defPST}, respectively) with emphasis on the technique of deforming a free theory and on the requirements that need to be imposed in both cases. One of the novelties presented in Sect.~\ref{s:solC-H} concerns the family of possible solutions for the equation enforcing duality-symmetry. One of these solutions leads to a manifest duality-symmetric action for the \br{3} and constitutes the subject of Sect.~\ref{s:D3br}. The duality-symmetry property of DBI Lagrangian was recently analyzed \cite{Kamani:2002za} in the context of non-commutative field theory.

From the algebraic point of view, the deformations of $N$ free duality-symmetric actions are even more challenging. The resulting theory would, of course, describe a stack of \br{3}s in $D=4$ or an \mbr{5} if one prefers to work in $D=6$. Despite some progress on the theory of such systems \cite{Myers:1999ps,Taylor:1999pr}, these models are yet poorly understood. As we have learned in the case of non-chiral formulations of Maxwell or $2$-form theories, consistent interactions are derivable in the perturbative regime using the cohomological approach \cite{Berends:1986xx,Barnich:1994pa,Henneaux:1998bm}. Unfortunately, the results~\cite{Bekaert:1998yp,Bekaert:2001wa,Bekaert:2002cz} discussed in Sect.~\ref{s:nogo} exclude simple non-Abelian extensions for the chiral corresponding models.    




\section{Self-interacting SS\label{s:defSS}}

Similar to the free theories, we open the discussion of duality-symmetric interactions with the non-covariant formulation. We aim to introduce consistent self-couplings for SS action \re{SSact} describing on-shell one free Maxwell gauge potential. In Ref.~\cite{Deser:1998gq}, it was proposed to tackle this problem using the Hamiltonian formulation, which is appropriate for dealing with first-order actions like \re{SSact}. Let us adopt here that line of thought, trying to prepare the reader for the developments of Sect.~\ref{s:nogo}. In fact, the ansatz made hereafter is better justified in the light of the results obtained in Sect.~\ref{s:nogo}.

A deformation of a theory is called {\it consistent} if the deformed theory possesses the same number (not necessarily the form) of independent gauge symmetries, reducibility identities, etc.~, as the free system we started from. This guarantees that the unphysical degrees of freedom decouple from the theory.\footnote{There is no theorem enforcing this assumption as the unique way of deriving a consistent deformation, but we will adopt it as a sine qua non condition through this work.} In other words, the number of physical degrees of freedom is unchanged. As an example, the Born-Infeld theory is a deformation that modifies the action but not the gauge symmetries of the free model.

From the very beginning, we state three basic requirements on the non-covariant interacting model: (i)  the deformation of \re{SSact} remains a first-order action, (ii) manifest rotation invariance, (iii)  manifest duality-symmetry.
The first assumption comes from the fact that we work in Hamiltonian formalism. 
The second and third requirements simply extend the properties of the free model. Under these assumptions, the non-linear action generalizing \re{SSact} is
\begin{equation}
S=-{1\over 2}\int d^4x\,B^{i\alpha}{\mathcal L}^{\alpha\beta}E^\beta_i - \int dx^0 H,
\end{equation}
where
\begin{equation}\la{defSSan}
H = \int d^3x \left({\mathcal H}(\partial_{i_1\dots i_{k-1}}B^{\a}_{i_k})+\zeta A_i^\a B^{i\a}\right)
\end{equation}
stands for the Hamiltonian of the model. In Sect.~\ref{s:nogo}, when we will deform the theory of $N$ Maxwell vectors, the last term in \re{defSSan} can include also some time derivatives. However, since we are dealing here with only one (on-shell) vector, this kind of extensions are prohibited here by our first requirement~(i). 
The factor $\zeta$ of the Chern-Simons like term is constant. 
If we now restrict ourself to (iv) Hamiltonian densities that do not depend explicitly on the derivatives of the magnetic
field\footnote{In other words, we assume a slowly varying field strengths.}: ${\mathcal H}={\mathcal H}(B^{\a A}_i)$,
we can deduce from (ii) and (iii) that the Hamiltonian density ${\mathcal H}$ only depends on two independent space scalars that are
manifestly duality invariant
\begin{equation}
y_1 = \frac12 B^{\a}_iB^{\a i}\,, \qq
y_2 = \frac{1}{4} B^{\a}_iB^{\b i}B^{\a}_jB^{\b j}\,.
\end{equation}
We set ${\mathcal H}= f(y_1,y_2)$.

All the expected symmetries are now manifest, except Lorentz symmetry. With the help of tensor calculus, it is straightforward to construct interactions that preserve Lorentz invariance. But, there is an alternative way to control Lorentz invariance. It is through the commutation relations of the energy-momentum tensor components. Dirac and Schwinger gave a sufficient condition for a
manifestly rotation and translation invariant theory (in space) to be also Lorentz-invariant. The condition is necessary when one turns to gravitation. 
The Dirac-Schwinger criterion yields in our case $\zeta=0$ and a non-linear first-order differential equation\linebreak for $f$, namely
\begin{equation}\label{e:C-H2}
f_1^2 +2 y_1 f_1 f_2 + 2 (y_1^2 -y_2 )f_2^2 = 1\,,
\end{equation}
where $f_\a = \frac{\partial f}{\partial y_\a}$ for  $\a=1,2.$
The equation \re{e:C-H2} can be simplified after the change of variables
\begin{eqnarray}\label{e:vartrans}\left\{\begin{array}{lll}
y_1 &=& u_+ + u_- \\
y_2 &=& u^2_+ + u^2_-
\end{array}\right.\,.
\end{eqnarray}
Denoting the function derivatives by $f_\pm \equiv {\partial f\over \partial u_\pm}$, one has the remarkably simple first-order differential equation \cite{Gibbons:1995cv,Perry:1997mk,Schwarz:1997mc,Deser:1998gq,Bekaert:1998yp}
\begin{equation}\label{duality}
f_+ f_-=1.
\end{equation}
We will refer to this equation enforcing duality symmetry as {\it Courant-Hilbert equation} \cite{Courant:1962xx}. 

Another way to derive the equation \re{duality} (as it was initially obtained \cite{Gibbons:1995cv}), is to start from the usual one-potential Lagrangian formulation of non-linear electrodynamics, manifestly gauge and Lorentz invariant, with action $S[A_\m]=\int d^4x\, L(x,y)\,.$
The function $L$ depends only on the two independent Lorentz scalars constructed from the curvature $F_{\m\n}$, namely $x = -\fr{1}{4}F_{\m\n}F^{\m\n}$ and $y =\fr{1}{64}(F_{\m\n}F^{*\m\n})^2.$
An important physical requirement is that we want to recover Maxwell theory in the weak field limit. 
For this purpose, we require that $L(x,y)$ is \emph{analytic} in the neighborhood of $x=y=0$ and 
\begin{equation}\label{e:weakfield}
L(x,y)=x+O(x^2,y).
\end{equation}
To make the link with equation \re{e:duality}, we perform a change of variables
\begin{equation}\label{e:xy}
x = u_+ + u_- \,,\qq
y = u_+ u_-.
\end{equation}
If we ask for (*) duality invariance of the eoms, and simultaneously for (**) the weak field limit \re{e:weakfield},
we obtain exactly the equation \re{duality} for the function $f(u_+,u_-):=L(x,y)$.
It is not surprising to find the same equation for the Hamiltonian density and the Lagrangian density because they are
related by a Legendre transformation, and a Legendre transformation relates a model and its dual \cite{Gaillard:1981rj}.

The same constraint \re{duality} was inferred for a chiral two-form in six dimensions, by choosing as special either the fifth direction
\cite{Perry:1997mk,Schwarz:1997mc} or the time direction \cite{Bekaert:1998yp} (in order to use the Dirac-Schwinger condition to enforce Lorentz invariance).
This confirms the fact that dimensional reduction of a chiral two-form from six to four dimensions
gives duality-symmetric electrodynamics, at the linear \cite{Verlinde:1995mz} and non-linear \cite{Berman:1997iz} level. We finally mention that this equation was also obtained for a self-interacting massless scalar
field in four dimensions \cite{Deser:1998gq}.

\section{Self-couplings of the PST model\label{s:defPST}}

Having constructed consistent self-couplings of SS action, we would like in this section to do something similar for the PST model. We thus look for an interacting, manifestly-covariant and duality-symmetric theory describing consistent self-couplings of one Maxwell field (on-shell). Our model has to be Lorentz covariant, generalizing the discussion of the previous section and the free PST model. Since Lorentz invariance is a built-in characteristic, a constraint of type \re{duality} will not follow any more from requiring such a symmetry. Just as in the case of the free system, the covariantization involves an auxiliary field and an extra gauge invariance. As we will see below, it is the preservation of the self-duality condition under these deformed gauge symmetries that will finally lead us to the same Courant-Hilbert equation. 

For reasons explained above, we require that the interacting action satisfies the same kind of symmetries as the free one. In other words, we should expect besides the Lorentz invariance also a manifest duality-symmetry. If on top of that we ask that the interaction should depend only on the field strengths this reduces  the number of invariants to only two, namely
\begin{equation}\la{defineinv}
z_1 = \frac{1}{2}\tilde\cH^{\a}_p \tilde\cH^{\a p}\,, \qq
z_2 = \frac{1}{4}\tilde\cH^{\a}_p \tilde\cH^{\b p}\tilde\cH^{\a}_q \tilde\cH^{\b q}\,.
\end{equation}

Thus, we propose as action for a self-interacting gauge vector 
\begin{eqnarray}
S^{I}&=& \int d^4x \, \left( \frac12 \cH^\a _m \tilde \cH^{\a m}  - f(z_1 ,z_2) \right) \,
\label{e:interpst}
\end{eqnarray}
where, similar to the non-covariant situation, we kept the ``kinetic" term and added an interaction term $f(z_1 ,z_2)$ depending only on  the invariants of the theory. The function $f$ is up to now arbitrary, but the connection with the free theory \re{e:pst} imposes its analyticity at the origin and its reduction to $f\ra z_1$ in the weak field limit. We are going to restrict the class of possible interactions by demanding the field equations as well as the action to remain invariant under some modified gauge transformations of type \re{e:ginv1}-\re{e:ginv3}. In fact, we deform only the gauge symmetry\footnote{The other two gauge symmetries remain the same and play the same  role as in the free model.} \re{e:ginv2} as 
\begin{eqnarray}
\delta A_m^\a = \cL^{\b\a} (\cH_m^\b - J_m^\b )\fr{\f}{\sqrt{-u^2}}\,, &\,\delta a=\f \,,\label{e:defginv2}
\end{eqnarray}
where we denoted the deforming contribution by $J_m^\b = \frac{\d f}{\d \tilde \cH^{\b m}}\,.$ One can observe that, in the free limit case, the gauge transformation \re{e:defginv2} reduces to \re{e:ginv2}.

Using the same approach as for the free theory, i.e., gauge-fixing \re{e:ginv3}, the solution to the equations of motion determined by  $S^I$ reads
\begin{eqnarray}
\cH^\a_{p} = J^\a_{p} = f_1 \tilde \cH^\a_{p} + f_2 (\tilde \cH^3)_p^\a \,,\label{e:defduality}
\end{eqnarray}
which should be understood as a generalization of the self-duality condition \re{e:duality}. We denote here $f_1 = \frac{\partial f}{\partial z_1}$, $f_2 = \frac{\partial f}{\partial z_2}$, $f_{12} = \frac{\partial^2 f}{\partial z_1\partial z_2}$, etc.  and $(\tilde \cH^3)_p^\a = \tilde \cH^\b_{p} \tilde \cH^{\b q} \tilde \cH^\a_{q}$.

We aim subsequently to find the implications on the function $f$ of the invariance of the deformed self-duality condition \re{e:defduality} under \re{e:defginv2}. To see that, one takes first its general variation, i.e.,
\begin{eqnarray}
\d \cH^\a_{p} &=& f_1 \d  \tilde \cH^\a_{p} + f_2 \left(\d \tilde \cH^\b_{p} \tilde \cH^{\b q} \tilde \cH^\a_{q} + \tilde \cH^\b_{p} \d \tilde \cH^{\b q} \tilde \cH^\a_{q} + \tilde \cH^\b_{p} \tilde \cH^{\b q} \d \tilde \cH^\a_{q} \right)  \nn\\
&& + \tilde \cH^\a_{p} \left( f_{11} \d \cH^\b_{q} \tilde \cH^{\b q} + f_{12} \d \cH^\b_{q} (\tilde \cH^3)^{\b q} \right)  \nn\\
&& + (\tilde \cH^3)_p^\a \left( f_{12}\d \cH^\b_{q} \tilde \cH^{\b q} + f_{22} \d \cH^\b_{q} (\tilde \cH^3)^{\b q} \right) \,.\label{e:genvar}
\end{eqnarray} 
Instead of plugging in directly the transformation \re{e:defginv2}, we first use the other symmetries of the system (Lorentz and SO(2) rotation invariance) to choose a basis in which the vector $u_p = \d_p^0$ (i.e., it is time-like) and the only non-vanishing components of the tensor $\tilde \cH^\a_{p}$ are 
\begin{equation}
\tilde \cH^1_{1} = \l_+ \,, \qq  \tilde \cH^2_{2} = \l_-\,.
\end{equation}
With this special choice, the equation of motion reduces to only
\begin{eqnarray}
\g_{\pm} = \l_{\pm} (f_1 + f_2 \l_\pm^2) \,,
\end{eqnarray}
where $\g_+ =\cH^1_{1}$, $\g_- =\cH^2_{2}$ are the non-zero components of $\cH^\a_{p}$ in this basis.

This simplifies considerably the analysis of \re{e:genvar}. In fact, there remain only two non-trivial possibilities for the choice of the $\a$, $p$ indices. The first one\linebreak is $\a = 1$, $p=3$ (the situation $\a = 2$, $p=3$ is similar). It leads (up to some factors that cancel in \re{e:genvar} and after using the field equations \re{e:defduality}) to the following variations $\d \cH^1_{3} = \l_-$, $\d \tilde\cH^1_{3} = \g_-$. The variation \re{e:genvar} becomes then
\begin{displaymath}
\l_- = f_1 \g_-  + f_2 \g_- \l_+^2 \,,
\end{displaymath}
which upon using once more the equations of motion gives
\begin{equation}\label{e:C-H1}
(f_1 + f_2 \l^2_+)(f_1 + f_2 \l^2_-) = 1\,.
\end{equation}
One can reformulate it in terms of the two invariants $z_1 = \frac{1}{2} (\l^2_+ + \l^2_- )$\linebreak and $z_2 = \frac{1}{4}(\l^4_+ + \l^4_-)$ and, after performing the transformation \re{e:vartrans}, one ends up with the Courant-Hilbert equation. Thus, we re-derived the Courant-Hilbert equation \re{duality} by imposing the invariance of the equations of motion under the gauge transformation \re{e:defginv2}. 

The second non-trivial possibility resides in taking $\a =1$, $p=1$\linebreak or $\a =2$, $p=2$. One should a priori expect that the two conditions derived in this way  from \re{e:genvar} will lead to further constraints on the second-order derivatives of $f$. Nevertheless, one can show that, taking the sum and the difference of the two relations, one arrives to some consequences of \re{e:C-H2} (linear combinations of the derivatives of this equation). In other words, eq.~\re{duality} is the only restriction imposed by the invariance of \re{e:defduality} under the deformed gauge symmetries \re{e:defginv2}. 

Next, we would like to show explicitly that the action \re{e:interpst} is also invariant (up to total derivatives) under the same gauge transformations \re{e:defginv2}. Inserting \re{e:defginv2} in the general variation of the action
\begin{eqnarray}
\d S^I &=&  \int d^4x \, \left[ ( \cH^\a_{p} - J^\a_{p}) \d_{A,a} \tilde \cH^{\a p} +\frac{1}{2} \d_a \cH^\a_{p} \tilde \cH^{\a p} - \frac{1}{2} \cH^\a_{p} \d_a \tilde \cH^{\a p} \right]
\end{eqnarray} 
and canceling the contributions of  first-order in the derivatives of $f$ coming from the variation with respect to $A_m^\a$, respectively $a$, one deduces
\begin{eqnarray}\label{e:JJ}
\d S^I &=& \frac{1}{2} \int d^4x \, \e^{mnpq} \cL^{\b\a} \d v_m v_n \left( \tilde\cH^\a_{p} \tilde\cH^\b_{q} - J^\a_{p} J^\b_{q} \right)\,.
\end{eqnarray} 
The idea is to prove that the `second-order' (in  the derivatives) term $JJ$ gives the same contribution as in the free theory. Thus, it cancels the first term. The simplest way to achieve that, is to evaluate it in the special basis mentioned above. Indeed, in such a basis \re{e:JJ} becomes  
\begin{equation}
\d S^I = - 2 \int d^4x\, \d v_3 \,\l_+ \l_- \,[(f_1 + f_2 \l^2_+)(f_1 + f_2 \l^2_-)-1]\,.
\end{equation}
Therefore, upon applying the Courant-Hilbert equation \re{duality} (or its \re{e:C-H1} version) one gets $\d S^I =0$, i.e., the invariance of the self-interacting action under the modified gauge transformations \re{e:defginv2}. Hence, the restriction \re{e:C-H1} is sufficient to guarantee the invariance of both field equations and action.  

In conclusion, we have constructed a modified Lorentz-covariant theory for the Maxwell field that possess also an electric-magnetic duality. The allowed self-interactions are restricted to a class of functions of two variables that must satisfy the Courant-Hilbert equation. The physical solutions of this equation are discussed in the next section. Moreover, it was shown that the deformed model has some modified gauge invariance, similar to the free case.

\begin{center}
\begin{figure}[h]
\epsfxsize =12cm
{\hskip 0.5cm \epsffile{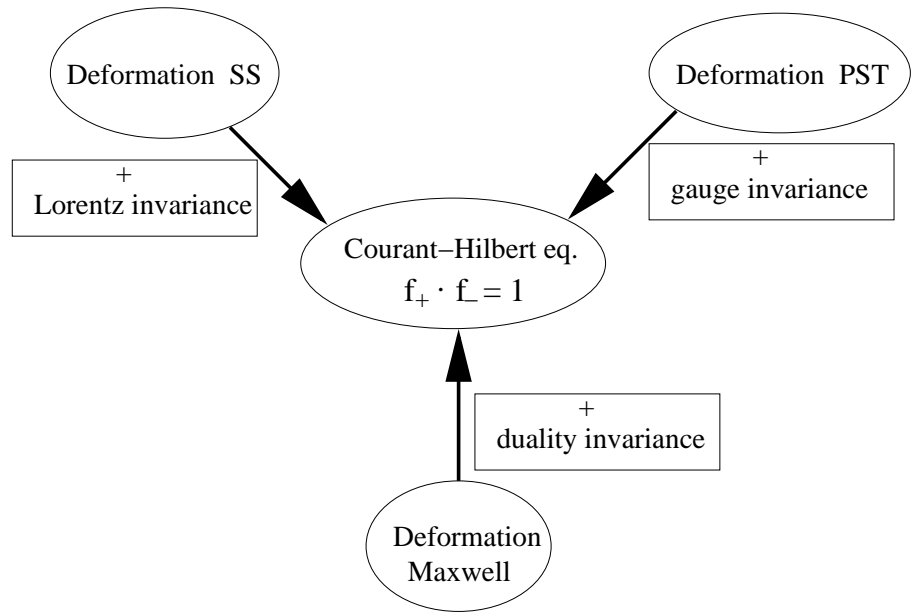}
}
\caption{{\it Various ways to deform different formulations of Maxwell theory lead (under certain requirements) to the Courant-Hilbert equation.}\label{fig:deform}}
\end{figure}
\end{center}

Before trying to find explicit solutions to the Courant-Hilbert equation and to connect them with familiar models in particle physics, we recollect the various paths to derive this equation 
in Fig.~\ref{fig:deform}. One context in which this equations showed up, was when deforming the SS theory. There, it was the Lorentz invariance requirement that led us to \re{duality}. We mentioned next that the duality invariance of non-linear electrodynamics produces the same condition. And finally, gauge invariance of a deformed PST model is guaranteed, provided the same restriction on the deforming function holds. 

\section{Solutions of the Courant-Hilbert equation\label{s:solC-H}}

We have already motivated the necessity of a solution for the Courant-Hilbert equation. This constitutes the subject of the present section. As pointed out in \cite{Perry:1997mk,Schwarz:1997mc}, the general solution of \re{duality} has been given long ago by Courant and Hilbert \cite{Courant:1962xx}.\footnote{That is why we refer it as the Courant-Hilbert equation. Let us mention that the authors of
\cite{Hatsuda:1999ys} gave an interesting alternative form of the general solutions.}
Their solution is expressed implicitly, in terms of an arbitrary function $z(t)$
\begin{eqnarray}
f &=& {2u_+\over \dot z(t)} + z(t)\,,\nonumber \\
u_- &=& {u_+\over (\dot z(t))^2} + t,\label{e:general}
\end{eqnarray} 
where the dot means here the derivative of the function with respect to its argument. In principle, the second equation determines $t$ in terms of $u_+$ and $u_-$, which can then be substituted into the first one to give $f$ in terms of $u_+$ and $u_-$. Unfortunately, in practice this method to generate solution is not tractable for arbitrary $z(t)$.

Furthermore, we should not forget the analyticity requirement at the origin on~$L(x,y)$, as well as \re{e:weakfield}.
The theorem presented below states that these two requirements can be equivalently translated into a precise condition on the
generating function $z(t)$.
\begin{theorem}\label{t:analyticity}
Let $f(u_+,u_-)$ be a solution of $f_+ f_-=1$.
The function $L(x,y)\equiv f\left(u_+(x,y),u_-(x,y)\right)$ 
\begin{itemize}
\item is analytic near $(x,y)=(0,0)$ and 
\item satisfies $L(x,y)=x+O(x^2,y)$,
\end{itemize}
if and only if the boundary condition $L(t,0) \equiv z(t)$ is such that the\linebreak function $\Psi(t)\equiv
-t\dot{z}^2(t)$
\begin{itemize}
\item[(i)] is equal to its inverse: $\Psi\left(\Psi(t)\right)=t$, 
\item[(ii)] is distinct from the identity: $\Psi(t)\neq t$, 
\item[(iii)] is analytic near the origin $t=0$ and 
\item[(iv)] vanishes at the origin: $\Psi(0)=0$.
\end{itemize}
\end{theorem}
It has been shown by Perry and Schwarz~\cite{Perry:1997mk,Schwarz:1997mc} that $(i)-(iv)$ are necessary, while the proof that they are also sufficient is given in~\cite{Bekaert:2001wa}.

Theorem \ref{t:analyticity} is relevant for understanding the existence of \emph{an infinite class of physically relevant duality-symmetric theories}, even if only one explicit example is known.\footnote{Notice that the
Lagrangians given in \cite{Hatsuda:1999ys} are not analytic at the origin in the variables $x$ and $y$. This point was brougth to
our attention by \"{O}. Sar{\i}o\~{g}lu.} 

To see that there exists indeed an infinite class of them, we follow the procedure presented by Perry and Schwarz to generate
a large class of solutions for
\begin{equation}
\Psi\left(\Psi(t)\right)=t.
\end{equation}
Let $F(s,t)$ be an analytic function near the origin such that $F(s,t)=s+t+O(|(s,t)|^2)$.
Then the implicit equation $F(s,t)=0$ defines a function $s=\Psi(t)=-t+O(t^2)$ analytic near the origin (application of
implicit function theorem). 
An interesting point is that if the function $F$ is symmetric, then the implicit function $\Psi(t)$ is equal to
its inverse because $F(s,t)=F(t,s)=0$
implies $t=\Psi(s)=\Psi\left(\Psi(t)\right)$. 

The simplest non-trivial example $F(s,t) = s + t +  \alpha s t$ generates (for some constant $\a$) the Born-Infeld electrodynamics (DBI action) at the end of the whole procedure ($\a=0$ corresponds to Maxwell theory, which is not considered here as a distinct example of solution). 
Unfortunately, this entire procedure becomes rapidly cumbersome and no other explicit example of duality-symmetric theory is known.
Anyway, our theorem shows that duality invariance together with analyticity is not enough to single out uniquely DBI theory, contrary to what could have been conjectured from the fact that only one explicit example is known. The theorem~\ref{t:analyticity} ensures that we can generate implicitly an infinite class of analytic solutions at the origin.

We will shortly analyze in detail the DBI solution. However, the method advertised here for solving the equation \re{duality} opens new avenues for finding unknown duality-symmetric models that may equally play a role in understanding the symmetries of string/M-theory.

\section{Duality-symmetric D$3$-brane \la{s:D3br}}

Even though we have just learned that the DBI theory is not the unique theory describing one self-coupled duality-symmetric gauge vector, it is  nevertheless remarkable that the DBI action emerges when deforming the covariant PST formulation of Maxwell theory. It is therefore worthwhile, before trying to deform the theory of $N$ duality-symmetric vectors, to investigate explicitly the form of the deformed theory and its physical interpretation. As one expects, one self-interacting gauge vector describes nothing than the duality-symmetric worldvolume of a \br{3} in type IIB SUGRA. The deformation of the free PST model just leads us to its manifestly duality-symmetric and Lorentz-covariant worldvolume action.  

This section should be understood as a logical continuation of the first part of Sect.~\ref{s:defPST} where we will implement the knowledge of one particular solution given in Sect.~\ref{s:solC-H}. To explicitly determine the solution, we repeat the construction of Sect.~\ref{s:solC-H} in reverse order. 

Since we are interested in the DBI solution, our starting point is given by the symmetric function
\begin{equation}\la{e:Func}
F(s,t) = s + t +  \alpha s t\,,
\end{equation}  
with $\a$ a non-vanishing constant. The equation $F(s,t) = 0$ determines the\linebreak function $\Psi(t) = - \, \frac{t}{1 + \a t}$ and the generating function
\begin{equation}\la{e:genFunc}
z(t) = \frac{2}{\a}\, \sqrt{1 + \a t}\,.
\end{equation} 
It satisfies indeed the conditions $(i)-(iv)$ of Theorem \ref{t:analyticity}. Introducing this expression of $z(t)$ into the solution~\re{e:general}, we infer the deforming function
\begin{equation}\la{e:defFunc}
f(u_+, u_-) = \frac{2}{\a}\, \sqrt{(1 + \a u_+) (1 + \a u_-)}\,.
\end{equation} 
Expressed in terms of the invariants $z_1$ and $z_2$, this gives a DBI-like Lagrangian
\begin{equation}\la{e:defInv}
f(z_1, z_2) = \frac{2}{\a}\, \sqrt{1 + \a z_1 + \frac 12 \,\a^2 (z_1^2 - z_2)}\,.
\end{equation}

We insert in the last formula the defining relations~\re{defineinv} of the two invariants and, for making the connection with known results in the literature, we set $\a =2$. This allows us to write down an explicit expression for the self-interacting PST action
\begin{equation}\la{DBIPST}
S^{I} = \int d^4x \, \left( \frac12 \cH^\a _m \tilde \cH^{\a m}  - \sqrt{1 + \tilde\cH^{\a}_p \tilde\cH^{\a p} 
+ \frac{1}{2}\Big( (\tilde\cH^{\a}_p \tilde\cH^{\a p})^2 - \tilde\cH^{\a}_p \tilde\cH^{\b p}\tilde\cH^{\a}_q \tilde\cH^{\b q}\Big)} \right) \,.
\end{equation} 
The last action is nothing else than a duality-symmetric reformulation of the DBI theory \cite{Berman:1997iz}. To understand that, we recall the symmetries of the theory. Besides worldvolume diffeomorphism, the action is gauge invariant under \re{e:ginv1}, \re{e:ginv3} and the modified transformation \re{e:defginv2}, with $f$ given in \re{e:defInv}. The presence of these symmetries determines the on-shell self-duality condition \re{e:defduality} that allows, just like for the free case, the elimination of one of the vectors $A_m^\a$. Upon this elimination, the action \re{DBIPST} reduces to the corresponding DBI action for a Maxwell field.  

Even if the way in which we presented the action \re{DBIPST} seems to indicate that it holds only for a flat worldvolume metric, the generalization to include also curved worldvolume metrics is immediate. We will further work with this generalization that amounts to the introduction in \re{DBIPST} of the square root of the metric determinant and the replacement of flat indices by curved ones. We also mention that the same action \re{DBIPST} was inferred in \cite{Nurmagambetov:1998gp} via dimensionally reducing the self-dual formulation of the M-theory five-brane \cite{Pasti:1997gx}.

Additionally, the self-interacting gauge vector can be coupled to the background fields of type IIB SUGRA in an $\Sl(2,\R)$-invariant manner. We follow here \cite{Nurmagambetov:1998gp} to explain how such a coupling can be achieved.\footnote{The $\Sl(2,\R)$ duality symmetry was also investigated in \cite{Kamani:2002za} in the context of non-commutative formulation of the DBI action.} Our first step is to rearrange the complex scalar $\rho$ from \re{IIBfiel} as a matrix-valued scalar field
\begin{equation}
{\cal S}= \frac{1}{\rho_2(X)}\left(\begin{array}{cc}
1 & \rho_1(X)\\
\rho_1(X)& \rho_1^2(X) + \rho_2^2(X)
\end{array}\right)\,, \q \phi = \phi (X(x))\,,\q \chi = \chi (X(x))\,,
\end{equation}
where $X$ denote the coordinates of the ten-dimensional space.In fact, the just introduced scalar matrix is symmetric and satisfies ${\cal S}\,{\cal L}\,{\cal S}^T = {\cal L}$.
 
In this dilaton-axion background the following generalized action
\begin{eqnarray}\la{dilaPST}
S^{I}_{dil-axi} &=& \int d^4x \, \sqrt{g}\,\Big( \frac12 \cH^\a _\m \tilde \cH^{\a \m}  \\
&-& \sqrt{1 + \tilde\cH^{\a}_ \m ({\cal L}^T{\cal S}\,{\cal L})_{\a\b} \tilde\cH^{\b \m} 
+ \frac{1}{2}\Big( (\tilde\cH^{\a}_\m \tilde\cH^{\a \m})^2 - \tilde\cH^{\a}_\m \tilde\cH^{\b \m}\tilde\cH^{\a}_\n \tilde\cH^{\b \n}\Big)} \Big) \nn
\end{eqnarray} 
is invariant under global $\Sl(2,\R)$ transformations $\o$
\begin{equation}
{\cal S} \ra \o^T\, {\cal S} \,\o\,, \qq \o\, {\cal L} \,\o^T = {\cal L}\,, \qq A_\m^\a = (\o^T A)_\m^\a\,.
\end{equation}

The antisymmetric gauge potentials of type IIB (see Sect.~\ref{s:maxsugra}) can be taken into account by combining them as 
\begin{equation}\la{2forpot}
A^\a_{\m\n} = \left(\begin{array}{c} B_{\m\n}\\ A_{\m\n} \end{array}\right)\,,
\end{equation} 
and then, replacing every $F^\a_{\m\n}$ appearing in the action and transformations rules by 
\begin{equation}\la{IIBrepla}
C^\a_{\m\n}= F^\a_{\m\n} - A^\a_{\m\n}\,.
\end{equation}
This behaves as a vector under $\Sl(2,\R)$ 
\begin{equation}
C^\a_{\m\n} \ra (\o^T C)^\a_{\m\n}\,.
\end{equation}
The final ingredient for obtaining the complete worldvolume theory for the duality-symmetric \br{3} is the WZ action. As we discussed in Sect.~\ref{s:wv}, \br{p}s are in general sources for the RR fields. For the \br{3} in particular, this implies the presence of a WZ term \re{WZD3} in its worldvolume description. Because now we look for a duality-symmetric formulation, the form of this piece of the action is going to be adapted to make this symmetry manifest.\footnote{We set here $2 \pi \a^\prime = 1$.}  

The final answer giving the duality-symmetric version of the \br{3} action can be presented in the form
\begin{eqnarray}\la{3brPST}
S^{D3}_{IIB} &=& -\int d^4x \, \sqrt{g}\,\Big(  \sqrt{1 + \tilde\cC^{\a}_ \m ({\cal L}^T{\cal S}\,{\cal L})_{\a\b} \tilde\cC^{\b \m} 
+ \frac{1}{2}\Big( (\tilde\cC^{\a}_\m \tilde\cC^{\a \m})^2 - \tilde\cC^{\a}_\m \tilde\cC^{\b \m}\tilde\cC^{\a}_\n \tilde\cC^{\b \n}\Big)} \nn\\
&-&  \frac12 \cC^\a _\m \tilde \cC^{\a \m}\Big) + \int \Big( A_{(4)} + \frac 12 {\cal L}^{\a\b} F^\a_{(2)} \wedge A^\b_{(2)}\Big)\,.
\end{eqnarray} 
Of course, in the full type IIB superstring theory, the duality group is\linebreak expected \cite{Tseytlin:1996it,Berman:1997iz} to be broken to $\Sl(2,\Z)$. This $\Sl(2,\Z)$ symmetry of the \br{3}, generalizing the strong-weak duality, finds its roots in the M-theory picture (see also \cite{Bekaert:2001vq}). Similarly to the $\Sl(2,\Z)$ symmetry of type IIB generated by a toric reduction of M-theory, one can start with an \mbr{5} and wrap two of its directions around the torus. The worldvolume of the \mbr{5} supports a self-interacting chiral
two-form potential which couples minimally to dyonic strings. 
Upon dimensional reduction one ends up with a \br{3} reduced on a transverse circle. 
The procedure not only reduces the self-dual $B_{\m\n}$ to the duality-symmetric\linebreak pair $A_\m^\a$, but also a dilaton-axion scalar combination becomes identified with the complex structure of the torus. The $\Sl(2,\Z)$ modular group of the compact space offers hence a geometrical interpretation of the duality-symmetry property of the \br{3}. In this light, the action \re{3brPST} simply confirms the expectations of an~$\Sl(2,\R)$-invariant formulation of the \br{3} effective action.

We finish our trip through the theory of consistent deformations of one duality-symmetric vector by mentioning that the authors of \cite{Perry:1997mk} proposed a string solitonic solution for the self-duality condition of a chiral $2$-form in $D=6$. Their string solution acts as a source for both the electric and magnetic fields and is non-singular. It will be interesting, of course, to investigate if our condition \re{e:defduality} for duality-symmetry admits similar solutions for the DBI particular form of $f$. It would be expected, due to dimensional reduction considerations, that a so\-li\-to\-nic~$0$-brane carrying a self-dual charge 
will solve equation \re{e:defduality}.

\section{No-go theorem\la{s:nogo}}

In the previous sections of this chapter we computed possible self-interactions of a single duality-symmetric Maxwell field. Hereafter, we inquire the more general question of deforming the theory of $N$ such objects. 

We want thus to analyze the consistent deformations of a system of $N$ free Abelian $1$-forms (on-shell) described by the sum of $N$
non-covariant actions, manifestly invariant under duality. We are looking for a Yang-Mills extension of this system. 

We know that for a set of $N$ Abelian $1$-forms described by a sum of  covariant Maxwell actions, the Yang-Mills theory is the unique
Lorentz invariant consistent deformation of the model deforming the gauge symmetry\footnote{There exist also interactions that only deform the action and leave the transformations untouched. As an example hereof we mention the DBI action.} (e.g., \cite{Berends:1986xx,Barnich:1994pa}).
But, non-Abelian duality seems to involve intrinsically non-local or non-perturbative features. 
For example, the standard Hamiltonian procedure for usual Maxwell theory gives $\pi_i=F_{0i}={\dot A}_i-\partial_iA_0$.
Hence the potential $A_0$ can be expressed non-locally in terms of the other variables (e.g., by a line integral).
Thus, the Yang-Mills cubic vertex containing explicitly $A_0$ would be non-local if we express it in terms of the other variables
to make the link with the duality-symmetric formulation. In any case (even in temporal gauge), non-locality will arise from
solving perturbatively
the Gauss law $D_i\pi^i=0$ for $\pi^i$, and inserting its non-local expression in the cubic vertex.
Hence it could be expected that such a non-Abelian, local, continuous deformation of a sum of manifestly invariant actions
under duality is not possible at all.
The results presented in this section will confirm that expectation.

\subsection{Cohomological approach}

The issue of deforming a gauge theory consistently proves to be generally cumbersome. This is a consequence of the non-linearities produced by the equations enforcing a consistent deformation, as illustrated for the simple case of a single Maxwell field by \re{e:C-H1}. This is not difficult to understand if we think that, in the first place, we look for interactions modifying simultaneously the actions and its gauge invariances. Further complications come from the necessity of sorting out  trivial deformations that are equivalent to field redefinitions. In practice, the celebrated Noether method determines these deformations perturbatively, \ie, as a power series in the coupling constant $g$. However, the method becomes easily intricate and we need another approach that organizes better the equations we have to handle. 

Consistent deformations of an action are equivalently achieved by deforming the solution $S$ of the
master equation $(S_0,S_0)=0$ into a solution $S^I$ of
the deformed master equation $(S^I,S^I)=0$ (In fact, this is due to the fact the master equation encodes completely the gauge structure of a theory). 
We can treat also this last problem perturbatively, 
that is, assume that $S^I$ can be expressed as an expansion $S^I = S_0 + g S_1 + g^2 S_2 + \dots$, 
the zeroth-order term in the expansion of~$S^I$ being equal to the solution of the free theory $S_0$. 
We try then to construct the deformations order by order in $g$.
It can be shown \cite{Barnich:1993vg}that the first-order non-trivial deformations of $S$ are elements of the {\it BRST cohomology group} in vanishing ghost number\footnote{The $H_0(s)$ group consists of objects $A$ of vanishing ghost number that are BRST closed \ie, $s A=0$, with the remark that objects that differ by a BRST-exact quantity, say $A$ and $A + s B$, are identified.} $H_0(s)$. This is nothing else than an elegant reformulation of the well-known Noether method.

If we now require locality, the non-trivial deformations of the theory will be even more constrained, because
in that case we restrict the deformation to be a local functional, i.e., $S^I=\int \alpha^n$, where the integrand
$\alpha^n$ is a {\em local $n$-form} in $n$ dimensions, that is a differential $n$-form with local functions as
coefficients.
{\em Local functions} depend polynomially on the fields
(including the ghosts and the antifields) and their derivatives up to a finite order. 
The non-trivial deformations $\a^n$ are elements of the {\it local} BRST cohomology group in gh number zero
\cite{Barnich:1993vg}, that is $H^n_0(s|d)$ 
where $d$ is the spacetime differential.\footnote{The elements $\a^n\equiv \a$ of $H^n_0(s|d)$ are BRST closed modulo the $d$-exact form \ie, $s \a + d \b =0$, and they are identified as follows $\a \sim \a + s \b + d \eta$. We neglect further the form degree.}

The construction of first-order consistent deformations is hence controlled in the cohomological approach by the ghost-zero cohomology group $H_0(s|d)$. But, one has to verify that these interactions are consistent with higher-order equations in $g$ and to try to compute higher-order deformations. The BRST cohomology comes once more to the rescue since, it turns out  that it is $H_1(s|d)$ which determines the obstruction to the existence of higher-order deformation terms. Without going into details, the antibracket of $S_1$ with itself must be a trivial element in $H_1(s|d)$ in order for this deformation not to be obstructed at the second order. It follows that, for the particular situation of vanishing $H_1(s|d)$, no first-order deformation can be obstructed and, from $(S_1, S_1)$ one can read off the second order deformation. As an exemplification, the familiar YM cubic vertex 
\begin{equation}\la{YMver}
S_1 \sim g f_{ABC} F^{A\, \m\n} A_\m^B A_\n^C
\end{equation}  
is consistent at the second order if and only if the functions $f_{ABC}$ satisfy the Jacobi identity (form a Lie algebra). In fact, using the powerful algebraic tools of cohomology one can prove that the YM vertex is the only possible one that modifies also the gauge algebra.\footnote{Other deformations of Maxwell theory (in general a dimension) like Freedman-Townsend in $D=3$ or Chern-Simons in odd dimensions deform the transformations or/and the action but leave the algebra Abelian.} 

We conclude this survey of the cohomological approach to consistent deformations by saying that, in certain cases, it gives not only possible interactions, but it offers as a bonus also their uniqueness. It is also useful in proving no-go theorems that forbid certain deformations (as we will do shortly), and it can be applied outside the context of $p$-form gauge theories, \eg, in SUGRA \cite{Bautier:1997yp}.

\subsection{The model and results}

The cohomological technique for constructing consistent deformations is in the remaining used on a system of $N$ SS actions. We aim to classify all consistent interactions of that theory.  

After carrying out the BV formalism \cite{Batalin:1983jr} for a sum of $N$ non-covariant actions~(\ref{SSact}) in the
temporal gauge ($A^{\a}_0=0$) we find the following minimal solution to the master equation
\begin{equation}
S= \sum_{A=1}^N  \int d^4 x\,\left[\fr12 (\cL^{\a\b}\dot{A}^{\a A}_i -\delta^{\a\b}B^{\a A}_i)B^{\b A\,i}+ A^{\a
A*}_i\partial^iC^{\a A}\right]\,.
\end{equation}
The field content of the theory discussed in Sect.~\ref{s:ssnon} was appended throughout the BV method with some ghosts $C^{\a A}$ 
and the associated antifields $C^{\a A *}$,\linebreak and $A^{\a A *}_i$. Their respective statistics, ghost numbers and antighost numbers are listed in Table~\ref{t:SSN}. 

Because the theory is Abelian, the BRST operator of this theory is simply the sum of the Koszul-Tate differential $\delta$
and the differential $\gamma$. They act on the fields and antifields in the following way
\begin{eqnarray}\la{dgSS}
\delta A^{\a A}_i&=&\delta C^{\a A}=0,\\
\delta A^{\a A*}_i&=& \partial^jF_{ij}^{\a A}+\fr12\cL^{\a\b}\epsilon_{ijk}\dot{F}^{\b A\,jk},\\
\delta C^{\a A*}&=& - \partial^i A^{\a A*}_i,\\
\gamma A^{\a A}_{i}&=&\partial_i C^{\a A},\\
\gamma C^{\a A}_i&=&0,\\
\gamma A^{\a A*}_i&=&\gamma C^{\a A*}=0.
\end{eqnarray}
Furthermore, $\delta$ and $\gamma$ commute with the partial derivatives $\partial_m$ and act trivially on the coordinates
$x^m$, thus their action on all the generators of the algebra of local forms can be easily determined using the Leibnitz rule.
Recall that since $S$ is solution to the master equation, the BRST operator must be nilpotent $s^2=0$ implying that $\delta^2=\delta\gamma+\gamma\delta=\gamma^2=0$, as can be verified directly from \re{dgSS}.
\begin{table}[htbp]
\begin{center}
\begin{tabular}{|c|c|c|c|c|}
\hline
 & statistics & antigh & pgh & gh\\
\hline\rule[-1mm]{0mm}{6mm}
$A_i^{\a A}$ & $+$ & $0$ & $0$ & $0$\\ \hline\rule[-1mm]{0mm}{6mm}
$C^{\a A}$ & $-$ & $0$ & $1$ & $1$\\ \hline\rule[-1mm]{0mm}{6mm}
$A_i^{\a A *}$ & $-$ & $1$ & $0$ & $-1$ \\ \hline\rule[-1mm]{0mm}{6mm}
$C^{\a A *}$ & $+$ & $2$ & $0$ & $-2$ \\ \hline\rule[-1mm]{0mm}{6mm}
$s$ & $-$ & $-1$ & $-1$ & $1$ \\ \hline
$\delta$ & $-$ & $-1$ & $0$ & $1$ \\ \hline\rule[-1mm]{0mm}{6mm}
$\gamma$ & $-$ & $0$ & $1$ & $1$ \\ \hline
\end{tabular}
\caption{{\it Respective statistics, ghost and antighost numbers of the variables and various differential operators.}\label{t:SSN}}
\end{center}
\end{table}

The consistent deformations of this model were investigated in \cite{Bekaert:2001wa} with the help of the aforementioned cohomological approach. 
After a long but well-established procedure, we derived the allowable interactions. The results are contained in the following theorems. 
\begin{theorem}\label{t:classification}
All consistent, continuous, local deformations of a system of free Abelian vector fields ($A=1,\cdots,N$)
described by a sum of $N$ free, duality-symmetric, non-covariant actions as the coupling constant goes to zero, are only of two
types:
\begin{itemize}
\item[I.]{those which are strictly invariant under the original gauge transformations; they are polynomials in curvatures
and their partial derivatives, i.e.
of the form 
\begin{equation}\la{polcurv}\int d^4x\,f(\partial_{m_1\dots m_k}F^{\a A}_{ij})\,.\end{equation}} 
\item[II.]{those which are invariant only up to a boundary term;
they are linear combinations of Chern-Simons like terms, i.e., 
\begin{equation}\la{CSlike}\int d^4x\,\lambda_{\a\b\,ab\,AB}\,(\partial_0)^aA_i^{\a A}(\partial_0)^bB^{\b B\, i} \,,\end{equation}
where $\lambda_{ab\,\a\b\,AB}$ are constants such that 
$\lambda_{ba\,\b\a\,BA}=\lambda_{ab\,\a\b\,AB}\,$.}
\end{itemize} 
\end{theorem}
For instance, the Hamiltonian of the free theory (second term in \re{SSact}) is a term of type I, as well as the Hamiltonian for DBI theory (\re{defSSan} with $\zeta =0$ and the associated $\cH$). 
The kinetic term is of type II (first term in \re{SSact}). Note that theorem~\ref{t:classification} provides the complete list of possible deformations of a sum of SS actions. The reader should notice that the ansatz \re{defSSan} becomes completely justified in the light of the results \re{polcurv} and \re{CSlike}, when only one SS gauge field is present.

As a corollary, the following {\it no-go theorem} holds
\begin{theorem}\label{t:nogo}
No consistent, local interactions of a set of free Abelian vector fields can deform the Abelian gauge transformations
if the local deformed action (free action + interaction terms) continuously reduces to a sum of free, duality-symmetric,
non-covariant actions in the zero limit for the coupling constant. 
\end{theorem}

As argued also in the introductory discussion, we see that simple YM-like deformations do not appear among the local, perturbative interactions of the self-dual SS formulation. We want to stress that an analogous theorem holds for any duality-symmetric theories in twice even dimensions. 
This is a consequence of the fact that, in general, the gauge symmetries of higher-order forms have a high degree of rigidity, as explained for instance in \cite{Henneaux:1998bm}. 

The result of theorem \ref{t:nogo} could have been anticipated in the M-theory context. Indeed, following the spirit our discussion at the end of Sect.~\ref{s:D3br}, a stack of $N$ \br{3}s whose worldvolume theory should be described by such an YM deformation is nothing but a toric reduction of $N$ coinciding, wrapped \mbr{5}s. The effective theory describing the latter system should be a non-Abelian extension for chiral two forms. This kind of deformations made the subject of \cite{Bekaert:1999dp}, where it was shown (using the same cohomological technique) that no such local interactions are allowed in the perturbative approach. Thus, YM-like deformations of the dimensionally reduced theory should also fall outside the reach of conventional perturbative and local field theory. 

\section{Conclusions\label{s:conclusio}}

The consistent deformations of \re{SSact} and \re{e:pst} derived in the present chapter fall under two categories: self-interactions of one gauge field and interactions of $N$ such copies. Although the methods used to compute these generalizations of the free theories were slightly different, they all have a common denominator: the interacting models must ``preserve" (modulo possible deformations) the same symmetries as the original theories.

We pointed out, either in Lagrangian or Hamiltonian formalism, various approaches leading to the Courant-Hilbert equation \re{duality}. This equation enforces duality symmetry for sourceless non-linear electrodynamics. Besides the general assumption of slowly varying field strengths, for the non-covariant SS deformations, the equation \re{duality} was a consequence of demanding Lorentz invariance. Self-interactions of the PST model on the other hand, had to satisfy some deformed gauge symmetries \re{e:ginv2}, that ultimately produced the same equation \re{duality}. This has been done for simplicity in a flat Minkowski space but, due to the covariance, it can be directly generalized to a curved background, as it was discussed for a DBI interaction in $6$ dimensions by the authors of \cite{Pasti:1997gx}. Of course, it would be interesting to generalize these last results to higher dimensions. 
Finally, the Courant-Hilbert equation results also when deforming the usual Maxwell action if one one imposes duality symmetry. The complete philosophy is visualized in Figure~\ref{fig:deform}.

The general solutions to Courant-Hilbert equation were examined in the spirit of \cite{Perry:1997mk,Bekaert:2001wa}. It turned out to exist a whole class of such solutions, thus an infinite set of physically relevant theories of duality-invariant non-linear electrodynamics. Additionally, we constructed  and throughly analyzed one of these solutions, namely the duality-symmetric \br{3}. We emphasized in particular the $\Sl(2,\Z)$ invariance of the latter theory, which could be the M-theory origin of the electric-magnetic duality in $D=4$.  

When dealing with more than one duality-symmetric vectors, we completely classified all consistent local interactions of a system of $p$-form gauge fields (with~$p$ odd)
described at the free level by a sum of non-covariant duality-symmetric actions in $2p+2$ dimensions. To handle this problem, we made use of the powerful tool of homological perturbation theory, by reformulating the problem in
terms of the local BRST cohomology. 

We found that no deformation of the gauge transformations is allowed, the only consistent interactions being of two types: either polynomials in the curvatures
and their spatial derivatives \re{polcurv}, or Chern-Simons like terms \re{CSlike}.
Of course, the strength of a no-go theorem is directly proportional to the weakness of its hypothesis, in this case: 
(i) continuous and (ii) local deformations of a sum of duality-symmetric actions.
Hence, the absence of the analogue of a Yang-Mills type deformation suggests that non-Abelian duality requires to take into account unavoidable non-perturbative or non-local features (or even more exotic properties). For instance, the authors of \cite{Chan:1999bj} proposed a generalized duality symmetry for non-Abelian Yang-Mills fields (Their use of loop space formalism is presumably responsible for non-locality.). However, the fundamental action principles describing a stack of \br{3}s or its six-dimensional cousin for a stack of \mbr{5}s remain unsettled problems.  
%
\chapter{Conformal supergravity in $D=5$: Weyl multiplets \label{ch:weyl}}

The connection of (matter-coupled) five-dimensional SUGRA with various fashionable topics of string theory, like $AdS_6/CFT_5$~\cite{Nishimura:2000wj,D'Auria:2000ad} and
$AdS_5/CFT_4$~\cite{Balasubramanian:2000pq} correspondences or RS scenario~\cite{Randall:1999vf,Randall:1999ee}, leades nowadays to the necessity of a more complete understanding of the subject. Besides its importance to M-theory physics, this research area is valuable on its own due to the relation with quaternionic-like and very special geometries \cite{deWit:1992cr,VanProeyen:2001wr,Bergshoeff:2002qk} or Jordan algebras \cite{Gunaydin:1984bi,Gunaydin:1985ak}.

The construction of such matter-couplings was recently tackled with different tools. Nishimura \cite{Nishimura:2000wj} for instance based his computation on the $AdS/CFT$ conjecture, while in \cite{Zucker:1999ej,Zucker:1999fn,Gunaydin:1999zx,Gunaydin:2000ph} the Noether method was preferred. At the same time, using the rheonomic approach in superspace, the authors of \cite{Ceresole:2000jd} claimed to provide the most general matter couplings of minimal $D=5$ SUGRA. However, that result asks for an independent confirmation, preferably using a different method. Therefore, another attempt was undertaken in~\cite{Kugo:2000hn} by reducing the
known six-dimensional result~\cite{Bergshoeff:1986mz} to five
dimensions. The authors of~\cite{Kugo:2000hn} already gauge-fixed some
symmetries of the superconformal algebra during the reduction process
in order to simplify the matter multiplet coupling. In this way, they
found a multiplet that is larger than the Weyl multiplet that we will
construct in this work, because they do not aim to obtain
superconformal symmetry in five dimensions. Our strategy is to start from
the basic building blocks of superconformal symmetry in five dimensions. 

Our derivation of matter-coupled SUGRA is built on the superconformal approach (see Sect.~\ref{s:SCTC}). SCTC is an elegant way to construct general couplings of Poincar\'e-supergravities to matter~\cite{Kaku:1977pa,Kaku:1978nz}. The advantages and the algorithm of the method were elucidated in Sect.~\ref{s:SCTC}. This chapter and the next one illustrate the effective construction for the case of $D=5$ SUGRA with $8$ real supercharges. 

In the present chapter, after completing the introduction to superconformal symmetry started in Sect.~\ref{s:confsym}, we initiate our matter-couplings investigation by considering the building block of SCTC, \ie, the Weyl multiplet. This is the multiplet of gauge fields associated to the superconformal algebra appended with matter fields whose role is to balance the bosonic and fermionic degrees of freedom. We will see that in five dimensions there are two
possible sets of matter fields one can add, yielding two ($32+32$) versions of
the Weyl multiplet: the Standard Weyl multiplet and the Dilaton Weyl
multiplet. This result is similar to what was found for~$(1,0)\ D=6$
conformal supergravity theory~\cite{Bergshoeff:1986mz}. Also in that
case, two versions were found: a multiplet containing a dilaton and
one without a dilaton.

There exist two methods of deriving the Weyl multiplets. The first one resides in coupling the supercurrent multiplet associated to the $D=5$ vector multiplet to either version of the Weyl multiplet. That turns out (Sect.~\ref{s:curr}) to be easier in the case of the Dilaton formulation, because the latter contains a dilaton-like field that can compensate for the lack of scale invariance of the original vector multiplet. The second manner (Sect.~\ref{s:gaug}), a direct gauging of the superconformal algebra $F^2(4)$ combined with suitable curvature constraints and matter fields, will allow us to determine the full (non-linear) transformation rules of the two Weyl multiplets. They are collected in Sect.~\ref{s:wmul}. We briefly mention the connection between the two Weyl multiplets.   

This chapter is based on \cite{Bergshoeff:2001hc}, which is further reviewed in \cite{Cucu:2001cs}. A related paper is \cite{Fujita:2001kv}.

\section{Conformal supersymmetry \la{s:csusy}}

In Sect.~\ref{s:confsym}, we introduced the notion of conformal symmetry. Following the same strategy (defining equation, algebra and representations), we extend it here to conformal supersymmetry. The parameters
of these supersymmetries define a conformal Killing spinor
$\epsilon^i(x)$, \ie, soltutions of the equation
\begin{equation}
 \nabla_\mu \e^i(x) -\frac 1D
\gamma_\mu \gamma^\nu\nabla_\nu \e^i(x)=0 \, .
\end{equation}
In $D$-dimensional Minkowski spacetime this equation implies
\begin{equation}
\partial_\mu \epsilon^i(x) - \frac 1D \gamma_\mu \slashed{\partial}\epsilon^i(x) = 0\, .
\end{equation}
The solution to this equation is given by
\begin{equation}
\epsilon^i(x) = \epsilon^i + \rmi x^\mu \gamma_\mu \eta^i\, ,
\label{susyc}
\end{equation}
where the (constant) parameters $\epsilon^i$  correspond to ``ordinary''
supersymmetry transformations $Q_\alpha^i$ and the parameters $\eta^i$
define special conformal supersymmetries generated by $S_\alpha^i$. The
conformal transformation~(\ref{ximu}) and the
supersymmetries~(\ref{susyc}) do not form a closed algebra. To obtain
closure, one must introduce additional R-symmetry generators. In
particular, in the case of 8 supercharges $Q_\alpha^i$ in $D=5$, there is
an additional $\USp(2) \equiv\SU(2)$ R-symmetry (automorphism) group with generators $U_{ij}= U_{ji}\
(i=1,2)$. Thus, the full set of superconformal transformations $\delta_{SC}$
reads\footnote{The $\rmi$ coefficients appear due to the reality properties explained in Appendix~\ref{a:hyp}.} 
\begin{equation}
\delta_{SC}= \xi ^\mu  P_\mu + \lambda_M^{\mu\nu}M_{\mu\nu}+\lambda_D D +
\Lambda_K^\mu K_\mu + \Lambda^{ij} U_{ij} + \rmi \bar{\e} Q + \rmi
\bar{\eta} S \,.
\end{equation}

The special supersymmetries appear naturally in the commutator of special conformal transformations and supersymmetries, justifying  the nomenclature. And similarly, the commutation relations of spacetime translations and special supersymmetries produce ordinary supercharges. Moreover, the anti-commutator of usual and special supersymmetries yields, besides some conformal transformation, the R-symmetry generators. As a result, the superconformal algebra can be expressed schematically as 
\begin{equation}
\left(
\begin{array}{cc}
\SO(D,2) & Q - S\\
Q + S & R
\end{array}
\right)\,.
\end{equation} 
We refer to Appendix~\ref{a:alg} for the full superconformal algebra
$F^2(4)$ formed by (anti-)commutators between the (bosonic and fermionic)
generators.


We end our general introduction to conformal supersymmetry with some remarks about its field realizations that are going to be used in the subsequent developments. In the supersymmetric case, we must specify the $\SU(2)$-properties of the
different fields as well as the behavior under $S$-supersymmetry.
Concerning the $\SU(2)$, we will only encounter scalars $\phi$, doublets
$\psi^i$ and triplets $\phi^{(ij)}$ whose transformations are given by
\begin{eqnarray}\la{SUtranfo}
\d_{\rm SU(2)} (\Lambda^{ij}) \phi &=& 0\, , \nonumber\\
\d_{\rm SU(2)} (\Lambda^{ij}) \psi^i(x) &=& -\Lambda^i{}_j \psi^j (x) \,,
\nonumber\\ \d_{\rm SU(2)} (\Lambda^{ij}) \phi^{ij} (x) &=& -2
\Lambda^{(i}{}_k\phi^{j)k}(x)\, .
\end{eqnarray}
The scalars of the hypermultiplet will also have an $\SU(2)$
transformation despite the absence of an $i$ index. We refer for that to
Sect.~\ref{s:hyper}.

This leaves us with specifying how a given field transforms under the
special supersymmetries generated by $S_\alpha^i$. In superfield language
the full $S$-transformation is given by a combination of an $x$-dependent
translation in superspace, with parameter $\epsilon^i(x) = \rmi x^\mu
\gamma_\mu \eta^i$, and an internal $S$-transformation. This is in
perfect analogy to the bosonic case. In terms of component fields, the
same is true. The~$x$-dependent contribution is obtained by making the
substitution
\begin{equation}
  \epsilon^i \rightarrow \rmi\slashed{x}\eta^i
 \label{xdepQ}
\end{equation}
in the $Q$-supersymmetry rules. The internal $S$-transformations can be
deduced by imposing the superconformal algebra. The explicit form of these internal $S$-transformations will become evident, case by case, in the forthcoming sections. 

\section{The supercurrent method\la{s:curr}}

The general algorithm of SCTC method was explained in Sect.~\ref{s:SCTC} where it was stressed that the first step resides in constructing the Weyl multiplet. We will see that there exist two versions of this multiplet. Both of them can be constructed in two different ways: either with the supercurrent method or by gauging the superconformal algebra. With the help of the former, we obtain in this section the two linearized Weyl multiplets. After discussing the supercurrent method we will construct the currents of a
rigid on-shell vector multiplet (Sect.~\ref{s:currm}), and
define a Weyl multiplet as the fields that couple to the currents
(Sect.~\ref{s:lweyl}). The comparison with known Weyl multiplets
in $4$ and $6$ dimensions tells us that there is also another Weyl
multiplet, and we point out that it can be obtained from the first one
by redefining some fields.

\pagebreak[3]
{\bf Noether method}

The Noether procedure of transforming a rigid symmetry into a local one is the backbone of the supercurrent method. We exemplify it by considering a Lie-valued scalar field $\Phi$, in the adjoint of $\U(N)$, whose dynamics is governed by
\begin{equation}\la{scrigU}
{\cal L}_{\rm scalar} = - \ft 12 \Tr{\del_\m \Phi \del^\m \Phi}\,.
\end{equation}
This action is invariant under rigid $\U(N)$ transformations $\d (\L) \Phi = - \left[ \L, \Phi\right]$, where the commutator is computed with respect with the structure constants of $\U(N)$. However, the invariance is lost when the symmetry is made local
\begin{equation}\la{locUN}
\d (\L (x)) {\cal L}_{\rm scalar} = -\Tr{J^\m \del_\m \L (x) }\,,
\end{equation}
where the Noether current 
\begin{equation}\la{Noecu}
J_\m \equiv \left[ \Phi, \del_\m \Phi\right]
\end{equation} 
transforms covariantly (for a definition see Sect.~\ref{s:gaug}) and has (on-shell) vanishing divergence, \ie,
\begin{equation}
\d (\L (x))J_\m =  - \left[ \L (x), J_\m\right]\,, \qq \del^\m J_\m \vert_{\Box \Phi =0 } = 0\,.
\end{equation} 

To recover the invariance of the theory in its local form, a (Lie-valued) gauge field $h_\m$ is introduced, with gauge transformation
\begin{equation}\la{gtransfor}
\d (\L (x))h_\m =  \del_\m \L (x)- \left[ \L (x), h_\m\right]\,.
\end{equation}
To compensate the non-vanishing variation \re{locUN}, we add to the scalar kinetic term a Noether term
\begin{equation}\la{Noethert}
{\cal L}_{\rm Noether} = \Tr{h^\m J_\m}
\end{equation}
such that the total action is now gauge invariant under $\U(N)$. In fact, in terms a suitable covariant derivative $D_\m$ (see also Sect.~\ref{s:gaug}), the action can be expressed in a manifestly gauge-invariant form 
\begin{equation}\la{sclocU}
{\cal L}_{\rm local\, U(N)} =  - \ft 12 \Tr{D_\m  \Phi D^\m \Phi}\,, \qq D_\m \Phi \equiv \del_\m \Phi + \left[ h_\m ,\Phi \right]\,.
\end{equation}

We intend to reproduce the same procedure for the $F^2(4)$ superconformal algebra. It will turn out that the currents associated to the superconformal algebra are not invariant but they transform among themselves, thus building up a complete superconformal current.

The multiplet of currents in a superconformal context has been
discussed before in the literature, e.g., the current multiplet
corresponding to the $N=1$, $D=4$~\cite{Kaku:1978nz}, the $N=2$,
$D=4$~\cite{Ferrara:1975pz,Sohnius:1979pk} and the $N=4$, $D=4$ vector
multiplets~\cite{Bergshoeff:1981is} and to the (self-dual) $(2,0)\
D=6$ tensor multiplet~\cite{Bergshoeff:1999db}.

After adding local improvement terms (which are terms constructed from the trace of the corresponding objects) one obtains a supercurrent multiplet containing an energy-momentum tensor $\th_{\m\n} = \th_{\n\m}$
and a supercurrent
 $J^i_\m$ which are both conserved and (gamma-)traceless
\be
\partial^\m \th_{\m\n} = \th_\m^{~\m} = \partial^\m J^i_\m = \g^\m J^i_\m = 0\,.
 \ee
These improved current multiplets were used in the past to construct
the linearized transformation rules for the Weyl
multiplet 
since a traceless
energy-momentum tensor is equivalent to scale-invariance of the
kinetic terms in the action.

However, the standard kinetic term of the $D=5$ vector field
\begin{equation}
\label{kinetic}
{\cal L} = - \ft14 F_{\mu\nu}F^{\mu\nu}
\end{equation}
is not scale invariant, i.e., the energy-momentum tensor is not traceless
\begin{equation}
\th_{\m\n} = - F_{\m\l} F_\n^{~\l} + \ft14 \eta_{\m\n} F_{\rho\sigma}
F^{\rho\sigma}\, ,\hskip 2truecm \theta_\mu{}^\mu = \ft14 F_{\m\n} F^{\m\n}
\ne 0\, .
\end{equation}
Moreover, there do not exist gauge-invariant local improvement terms. 


There is a remedy for this problem. Whenever there is a compensating
scalar field present, i.e., a scalar with mass dimension zero but
non-zero Weyl weight, then the kinetic term (\ref{kinetic}) can be
made scale invariant by introducing a scalar coupling of the form
\begin{equation}
{\cal L} = - \ft14 e^\phi F_{\mu\nu}F^{\mu\nu}\, .
\end{equation}
This compensating scalar is called the dilaton (a reminiscent of the string theory dilaton). In general, there are
three possible origins for a dilaton coupling to a non-conformal
matter multiplet: the dilaton is part of
\begin{enumerate}
\item the matter multiplet itself (the multiplet is then called an
`improved' multiplet); 
\item the conformal supergravity multiplet;
\item another matter multiplet.
\end{enumerate}
The $N=2$, $D=5$ vector multiplet contains precisely such a scalar. We
could therefore use it to compensate the broken scale invariance of
the kinetic terms. This leads to the so-called improved vector
multiplet, further
investigated in Chapt.~\ref{ch:matter}.

The second possibility will be considered here (the third possibility
is included for completeness). This possibility thus occurs when the
Weyl multiplet itself contains a dilaton. We will see that there
indeed exists a version of the Weyl multiplet containing a
dilaton. This version is called the Dilaton Weyl multiplet. It turns
out that there exists another version of the Weyl multiplet without a
dilaton. This other version will be called the Standard Weyl
multiplet.

The coupling of the supercurrent multiplet to the Dilaton Weyl multiplet
can happen without adding improvement terms to the former (see below). That is a consequence of the existence of a dilaton field in the Dilaton Weyl multiplet that compensates for the lack of scale invariance. In particular, the dilaton will couple directly to the trace of the energy-momentum tensor. 


However, when coupling to the Standard Weyl multiplet one needs to add
\emph{non-local} improvement terms to the current multiplet. This was
done for $D=5$ in Ref.~\cite{Halbersma:2002th} where the linear Standard Weyl multiplet was inferred using the same method that we are shortly presenting for its Dilaton version.\footnote{The same analysis was carried out for the current multiplet coming from the $D=10$ vector multiplet~\cite{Bergshoeff:1982av}. In both cases the non-local improvement terms that were added, required the use of auxiliary fields satisfying differential constraints in order to make the
transformation rules local.} 
In this section we adopt a different strategy and construct the linearized transformation rules of the Standard Weyl multiplet via field redefinition from those of the Dilaton Weyl multiplet.

Thus, for matter multiplets having a traceless energy-momentum tensor, no
compensating scalar is needed. To see the difference between the
various cases it is instructive to consider $(1,0)\ D=6$ conformal
supergravity theory~\cite{Bergshoeff:1986mz} which was constructed
without the supercurrent method. In that case, two versions were
found: a multiplet containing a dilaton and one without a dilaton. We
expect that both versions can be constructed using the supercurrent
method: the one without a dilaton starting from the conformal $(1,0)$
tensor multiplet (being a truncation of the $(2,0)$ case), and the
version containing the dilaton by starting from the non-conformal
$D=6$ vector multiplet (which upon reduction should produce our
results in $D=5$).

\subsection{Supercurrent multiplet\la{s:currm}}

Our starting point is the on-shell $D=5$ vector multiplet. Its field
content is given by a massless vector
$A_\m$, a symplectic Majorana spinor $\p^i$ in the fundamental of $\SU(2)$
and a real scalar $\s$. See Table~\ref{tbl:nAvm}
\begin{table}[htbp]
\begin{center}
\begin{tabular}{|c|c|c|c|c|}
\hline
\rule[-1mm]{0mm}{6mm}
Field       & Equation of motion   & {$\SU(2)$} & $w$ & \# d.o.f.  \\
\hline
\rule[-1mm]{0mm}{6mm}
$A_\m$      & $\partial_\m F^{\m\n} = 0$ & 1 & 0             & 3 \\
\rule[-1mm]{0mm}{6mm}
$\s$        & $\Box \s = 0$          & 1 & 1             & 1 \\
\hline
\rule[-1mm]{0mm}{6mm}
$\p^i$      & $\slashed{\partial} \p^i = 0$   & 2 & ${3/ 2}$ &
4 \\[1mm]
\hline
\end{tabular}
\caption{{\it The $4+4$ on-shell Abelian vector multiplet.}\label{tbl:nAvm}}
\end{center}
\end{table}
for additional information. Our conventions are presented in
Appendix~\ref{a:hyp}.

The action for the $D=5$ Maxwell multiplet is given by
\be 
{\cal L} = -\ft14 F_{\m\n}F^{\m\n} - \ft12 \bar{\p} \slashed{\partial}
\p - \ft12 (\partial \s)^2\,. 
\label{eq:onshell_maxwell}
\ee
This action is invariant under the following (rigid) supersymmetries
\bea
\d_Q A_\m &=& \ft12 \bar{\e} \g_\m \p\, , 
\nonumber\\
\d_Q \p^i &=& -\ft14 \g \cdot F \e^i - \ft12\rmi \slashed{\partial}
\s \e^i\, , 
\nonumber\\
\d_Q \s   &=& \ft12\rmi \bar{\e} \p\, ,
 \eea
as well as under the standard gauge transformation
\be 
\d_\Lambda  A_\m = \partial_\m \L\, . 
\ee

The various symmetries of the Lagrangian (\ref{eq:onshell_maxwell})
lead to a number of Noether currents: the energy-momentum tensor
$\theta_{\mu\nu}$, the supercurrent $J_\mu^i$ and the $\SU(2)$-current
$v_\mu^{ij}$. The supersymmetry variations of these currents lead to a
closed multiplet of $32+32$ degrees of freedom (see
Table~\ref{tbl:currentmult}). As explained above, an
unconventional feature, compared to the currents corresponding to a
$D=4$ vector multiplet or a $D=6$ tensor multiplet, is that the
current multiplet cannot be improved by local gauge-invariant terms,
i.e., $\theta_{\mu}{}^\mu \ne 0$ and $\gamma^\mu J_\mu^i \ne 0$. It is
convenient to include these trace parts as separate currents since, as
it turns out, they couple to independent fields of the Weyl multiplet.

\begin{table}[htbp]
\begin{center}
\begin{tabular}{|c|c|c|c|c|c|c|}
\hline
\rule[-1mm]{0mm}{6mm}
Current        & Noether & {$\SU(2)$} & $w$ &
\# d.of. \\
\hline
\rule[-1mm]{0mm}{6mm}
$\th_{(\m\n)}$ & $\partial^\m \th_{\m\n} = 0$  & 1 & 2              & 9 \\
\rule[-1mm]{0mm}{6mm}
$\th_{\m}{}^{\m}$       &  & 1 & 4              & 1 \\
\rule[-1mm]{0mm}{6mm}
$v^{(ij)}_\m$  & $\partial^\m v^{ij}_\m =0$    & 3 & 2              & 12 \\
\rule[-1mm]{0mm}{6mm}
$a_\m$       & $\partial^\m a_\m = 0$       & 1 &  3             & 4 \\
\rule[-1mm]{0mm}{6mm}
$b_{[\m\n]}$   & $\partial^\m b_{\m\n} = 0$   & 1 &  2             & 6 \\
\hline
\rule[-1mm]{0mm}{6mm}
$J^i_\m$     & $\partial^\m J^i_\m = 0$     & 2 & ${5 / 2}$  & 24  \\
\rule[-1mm]{0mm}{6mm}
$\Sigma^i \equiv \rmi \g \cdot J^i$     & & 2 & ${7 / 2}$  & 8 \\[1mm]
\hline
\end{tabular}
\caption{{\it The $32+32$ current multiplet. The trace  $\theta_\mu {}^\mu
$ and the gamma-trace of $J^i_\mu $ constitute separate currents. The latter is
denoted by $\Sigma^i$.}\label{tbl:currentmult}}
\end{center}
\end{table}
\vspace{-0.27cm}

We find the following expressions for the Noether currents and their
supersymmetric partners (particular cases of \re{Noecu}) in terms of bilinears of the vector multiplet fields
\bea \la{currmvec}
\th_{\m\n} &=& - \partial_\m \s \partial_\n \s +\ft12 \eta_{\m\n}
\left(\partial \s\right)^2 - F_{\m\l} F_\n^{~\l} + \ft14 \eta_{\m\n}
F^2 - \ft12\bar{\p} \g_{(\mu }\partial_{\nu )} \p\,,
\nonumber\\
J^i_\m &=& -\ft14 \rmi \g \cdot F \g_\m \p^i - \ft12
(\slashed{\partial} \s) \g_\m \p^i\, ,
\nonumber\\
v^{ij}_{\m} &=& \ft12 \bar{\p}^i \g_{\m} \p^j\, ,
\nonumber\\
a_\m &=& \ft18 \ve_{\m\n\l\r\s} F^{\n\l} F^{\r\s} + (\partial^\n \s)
F_{\n\m}\, ,
\nonumber\\
b_{\m\n} &=&\ft12 \ve_{\m\n\l\r\s} (\partial^\l \s) F^{\r\s} +\ft12
\bar{\p} \g_{[\mu }\partial_{\nu ]} \p\, , 
\nonumber\\ 
\Sigma^i & \equiv & \rmi \g \cdot J^i = \ft14 \g \cdot F \p^i + \ft32 \rmi
\slashed{\partial} \s \p^i\, ,
\nonumber\\
\th_\m^{~\m} &=& \ft32 \left(\partial \s\right)^2 + \ft14 F^2\,. 
\eea
From these expressions, using the Bianchi identities and equations
of motion of the vector multiplet fields, one can calculate the
supersymmetry transformations of the currents. A straightforward
calculation yields
\bea 
\d_Q \th_{\m\n} &=& \ft12\rmi \bar{\e} \g_{\l(\m}\partial^\l J_{\n)}\,,
\nonumber\\
\d_Q J^i_\m &=& - \ft12 \rmi \g^\n \th_{\m\n} \e^i - \rmi
\g_{[\l}\partial^\l v^{ij}_{\m]} \e_j - \ft12 a_\m \e^i + \ft12\rmi
\g^\n b_{\m\n} \e^i\,,
\nonumber\\
\d_Q v^{ij}_\m
&=& \rmi \bar{\e}^{(i} J^{j)}_\m\,,
\nonumber\\
\d_Q a_\m &=& -\bar{\e} \partial^\l \g_{[\l} J_{\m]} + \ft14 \bar{\e}
\g_{\r\m} \g^\s \partial^\r J_\s + \ft14\rmi \bar{\e} \g_{\r\m}
\partial^\r \Sigma\,,
\nonumber\\
\d_Q b_{\m\n} &=& \ft34\rmi \bar{\e} \g_{[\l\m}\partial^\l J_{\n]} -
\ft18\rmi \bar{\e} \g_{\r\m\n} \g^\l \partial^\r J_\l + \ft18 \bar{\e}
\g_{\r\m\n} \partial^\r \Sigma\, ,
\nonumber\\
\d_Q \Sigma^i &=& \ft12 \th_{\m}{}^{\m} \e^i - \ft12 \slashed{\partial}
\slashed{v}^{ij} \e_j -\ft12 \rmi \slashed{a} \e^i - \ft12 \g \cdot b
\e^i\, ,
\nonumber\\
\d_Q \th_\m^{~\m} &=& \ft12 \bar{\e} \slashed{\partial} \Sigma\, .
\label{lintrcurr} 
\eea
Note that we have added to the transformation rules for $a_\mu$ and
$b_{\mu\nu}$ terms that are identically zero: the first term at the
r.h.s.~contains the divergence of the supercurrent and the last two
terms are proportional to the combination $(i \g \cdot J - \Sigma)$
which is zero. Similarly, the second term in the variation of the
supercurrent contains a term that is proportional to the divergence of
the $\SU(2)$ current.

The reason why we added these terms is that in this way we obtain
below the linearized Weyl multiplet in a conventional
form. Alternatively, we could not have added these terms and have brought the Weyl multiplet into the same conventional form later on by
redefining the $Q$-transformations via a field-dependent $S$- and
$\SU(2)$-transformation.

\subsection{Linearized Weyl multiplets\la{s:lweyl}}

We are now prepared to couple the supercurrent to one of the Weyl multiplets and determine the linear transformation rules of the latter. For reasons debated above, it is easier to do that for the Dilaton Weyl multiplet. Combining some of the Dilaton Weyl components into the matter fields of the Standard Weyl multiplet the transformation rules of the latter are inferred without introducing non-local counter terms.

{\bf Linearized Dilaton Weyl multiplet}

The linearized $Q$-supersymmetry transformations of the Weyl multiplet
are determined by coupling every current to a field, and demanding
invariance of the corresponding action. The field-current action is
given by
\be
\label{eq:cur2lin}
S = \int d^5 x \bigg(\ft12 h_{\m\n} \th^{\m\n} + \rmi
\bar{\p}_\m J^\m + V_\m^{ij}v_{ij}^\m + A_{\m} a^{\m} + B_{\m\n} b^{\m\n}
+ \rmi \bar{\p} \Sigma + \vf\ \th_\m^{~\m} \bigg)\,. 
\ee 
In Table~\ref{tbl:fieldsWeyls}, we give some properties of the Weyl
multiplets. In particular of the one just derived, which we call the
Dilaton Weyl multiplet.\footnote{Note that the Dilaton Weyl multiplet
contains a vector $A_\mu$, a spinor $\psi^i$ and a scalar $\sigma$
which, on purpose, we have given the same names as the fields of the
vector multiplet as they can be identified with each-other.}
A similar Weyl multiplet containing a
dilaton exists in $D=6$~\cite{Bergshoeff:1986mz}.

\begin{table}[htbp]
\begin{center}$\begin{array}{|c|cccc||c|cccc|} \hline
\mbox{Field} & \#  & \mbox{Gauge}&\SU(2)& w & \mbox{Field} & \#  &
\mbox{Gauge}&\SU(2)& w\\
\hline \rule[-1mm]{0mm}{6mm}& \multicolumn{4}{c||}{\mbox{Elementary gauge
fields}} && \multicolumn{4}{c|}{\mbox{Dependent gauge fields}}\\
\rule[-1mm]{0mm}{6mm} e_\mu{} ^a & 9 & P^a &1&-1 & \omega _\mu
^{[ab]}&- & M^{[ab]}&1&0   \\ \rule[-1mm]{0mm}{6mm}
b_\mu   & 0 & D &1&0&f_\mu{} ^a &-& K^a &1&1   \\
\rule[-1mm]{0mm}{6mm}
V_\mu ^{(ij)} &12  & \SU(2)\,&3&0&& & &  &   \\
\hline \rule[-1mm]{0mm}{6mm}
\psi _\mu ^i  & 24 & Q^i_\alpha
&2&-1/2 & \phi _\mu ^i & - &
 S^i_\alpha&2&1/2 \\ [1mm]
\hline\hline \rule[-1mm]{0mm}{6mm} & \multicolumn{4}{c||}{\mbox{Dilaton
Weyl multiplet}} && \multicolumn{4}{c|}{\mbox{Standard Weyl
multiplet}}\\ \rule[-1mm]{0mm}{6mm} A_\mu  & 4 & \d A_\m =
\partial_\m \L &1&0 & T_{[ab]}&  10& & 1&1  \\ \rule[-1mm]{0mm}{6mm}
B_{[\mu \nu] }  & 6 & \d B_{\m\n} = 2 \partial_{[\m}
\L_{\n]}\hspace{-2pt} &1&0 &&& &  &  \\
\rule[-1mm]{0mm}{6mm} \varphi & 1 & &1&1 & D&1 &  & 1&2 \\ \hline
\rule[-1mm]{0mm}{6mm} \p ^i  & 8 & &2&3/2 & \chi ^i & 8& & 2&3/2
\\ [1mm]
\hline
\end{array}$
\caption{{\it Fields of the Weyl multiplets, and their roles. The upper
half contains the fields that are present in all versions. They are
the gauge fields of the superconformal algebra (see Sect.
~\ref{s:gaug}). The fields at the right-hand side of the upper
half are dependent fields, and are not visible in the linearized
theories. The symbol $\#$ indicates the off-shell degrees of
freedom. The gauge degrees of freedom corresponding to the gauge
invariances of the right half are subtracted from the fields at the
left on the same row. In the lower half are the extra matter fields
that appear in the two versions of the Weyl multiplet. In the left
half are those of the Dilaton Weyl multiplet, at the right are those
of the Standard Weyl multiplet. We also indicated the (generalized)
gauge symmetries of the fields $A_\m$ and $B_{\m\n}$. (The linearized
fields, corresponding to $e_\mu{}^a$ and $\s \equiv e^\f$ are denoted
by $h_\mu{}^a$ and $\vf$, respectively.)}\label{tbl:fieldsWeyls}}
\end{center}
\end{table}

Using the supersymmetry rules for the current multiplet, we find that the
following transformations leave the action (\ref{eq:cur2lin}) invariant
{\allowdisplaybreaks
\bea 
\d_Q h_{\m\n} &=& \bar{\e} \g_{(\m} \p_{\n)}\,,
\nonumber\\
\d_Q \p^i_\m &=& - \ft14 \g^{\l\n} \partial_\l h_{\n\m} \e^i -
V_\m^{ij} \e_j + \ft18\rmi \bigg(\g \cdot F + \ft13\rmi \g \cdot
H\bigg) \g_\m \e^i\, ,
\nonumber\\
\d_Q V^{ij}_\m &=& - \ft12 \bar{\e}^{(i} \g^\l \p^{j)}_{\l\m} \nn +
 \ft12\rmi \bar{\e}^{(i} \g_\m \slashed{\partial} \p^{j)}\,,
\nonumber\\
\d_Q A_\m &=& -\ft12\rmi \bar{\e} \p_\m + \ft12 \bar{\e} \g_\m \p\, ,
\nonumber\\
\d_Q B_{\m\n} &=& \ft12 \bar{\e} \g_{[\m} \p_{\n]} + \ft12\rmi
\bar{\e} \g_{\m\n} \p\,,
\nonumber\\
\d_Q \p^i &=& -\ft18 \g \cdot F \e^i -\ft12\rmi \slashed{\partial}
\vf \e^i + \ft1{24} \rmi \g \cdot H \e^i\, ,
\nonumber\\
\d_Q \vf &=& \ft12 \rmi \bar{\e} \p\,, 
\label{lintrDilW}
\eea
}
where we have defined
\be 
F_{\m\n} = 2 \partial_{[\m} A_{\n]}, \quad H_{\m\n\l} = 3
\partial_{[\m} B_{\n\l]}, \quad \p_{\m\n} = 2 \partial_{[\m}
\p_{\n]}\, .
\ee

{\bf Linearized Standard Weyl multiplet}

It turns out that there exists a second formulation of the Weyl
multiplet in which the fields $A_\m$ and $B_{\m\n}$ are replaced by an
anti-symmetric tensor $T_{ab}$ and where also the spinor and the
scalar are redefined. It is the multiplet we should have expected if
we compare it with the Weyl multiplets of the $D=4$ and $D=6$ theories
with 8 supercharges. This can be seen in Table~\ref{tbl:countWeyl}.

This second Weyl multiplet is called the Standard Weyl multiplet. More
information about the component fields can be found in
Table~\ref{tbl:fieldsWeyls}. The Standard Weyl multiplet cannot be
obtained from the same current multiplet procedure we applied to get
the Dilaton Weyl multiplet, unless we would consider an `improved'
current multiplet \cite{Halbersma:2002th}. The reason is that the Standard Weyl multiplet
contains no dilaton scalar with a zero mass dimension that can be used
as a compensating scalar. Therefore it can not define a conformal
coupling to a non-improved current multiplet.


The connection between the two versions of the Weyl multiplet at the
linearized level is given by algebraic relations. First of all we
denote some particular terms in the transformations of $\p^i_\m$ and
$V^{ij}_\m$ by $T_{ab}$ and $\chi^i$. Then we compute the variations
of these expressions under supersymmetry, finding one more object
called $D$. We find
{\allowdisplaybreaks
\bea
T_{ab} &=& \ft18 \bigg(F_{ab} - \ft16 \ve_{abcde} H^{edc} \bigg)\,,
\nonumber\\
\chi^i &=& \ft18\rmi \slashed{\partial} \p^i + {1 \over 64} \g^{ab}
\p_{ab}\, ,
\nonumber\\
D &=& \ft14 \Box \vf - \ft1{32} \partial^\m \partial^\n h_{\m\n} +
\ft1{32} \Box h^\m_\m\,.
\label{linrelW}
\eea
}
\begin{table}[htb]
\begin{center}$ \begin{array}{|l|rrr|} \hline
 \mbox{Field}  & $D=4$ & $D=5$ & $D=6$ \\
\hline\hline
e_\mu {}^a     & 5 & 9 & 14 \\
b_\mu    & 0 &0  &0 \\
\omega _\mu {}^{ab}&- &- & - \\
f_\mu {}^a &- &-&- \\
V_{\mu i}{}^j   & 9 & 12 & 15  \\
A_\mu &3 & - & - \\ \hline &&&\\[-4mm]
\psi _\mu {}^i  & 16 & 24 & 32 \\[1mm]
\phi _\mu {}^i &- &- &- \\[1mm] \hline\hline
T_{ab}, T_{abc}^-  & 6 & 10 & 10 \\
D          & 1 &  1 &  1\\
\hline &&&\\[-4mm]
\chi ^i  & 8 & 8 & 8  \\[1mm]
\hline\hline
\mbox{TOTAL}  & 24+24 & 32+32 & 40+40\\
\hline
\end{array}$
\caption{{\it Number of components in the fields of the Standard Weyl
multiplet.  The dependent fields have no number. The field $T$ is a
two rank tensor in $4$ dimensions and a self-dual three rank tensor in $6$
dimensions. In $5$ dimensions we can choose between a two-rank or a
three-rank tensor as these are dual to each
other.}\label{tbl:countWeyl}}
\end{center}
\end{table}

The resulting supersymmetry transformations are those of what we call
the linearized Standard Weyl multiplet. They are given by
{\allowdisplaybreaks
\bea 
\d_Q h_{\m\n} &=& \bar{\e} \g_{(\m} \p_{\n)}\,,
\nonumber\\
\d_Q \p^i_\m &=& - \ft14 \g^{\l\n} \partial_\l h_{\n\m} \e^i -
V_\m^{ij} \e_j + \rmi \g \cdot T \g_\m \e^i\,,
\nonumber\\
\d_Q V^{ij}_\m &=& -\ft18 \bar{\e}^{(i} \bigg(\g^{ab} \g_\m - \ft12
\g_\m \g^{ab}\bigg) \p^{j)}_{ab} + 4 \bar{\e}^{(i} \g_\m \chi^{j)}\, ,
\nonumber\\
\d_Q T_{ab} &=& \ft12\rmi \bar{\e} \g_{ab} \chi -\ft3{32}\rmi \bar{\e}
\bigg(\p_{ab} -\ft1{12} \g_{ab} \g^{cd} \p_{cd} + \ft23 \g_{[a} \g^c
\p_{b]c} \bigg)\,,
\nonumber\\
\d_Q \chi^i &=& \ft14 D \e^i - {1 \over 32} \g^{\m\n} \partial_{\m} V^{ij}_{\n}
\e_j + \ft3{32}\rmi \g \cdot T \overleftarrow{\slashed{\partial}} \e^i
+ \ft1{32}\rmi \slashed{\partial} \g \cdot T \e^i\, ,
\nonumber\\
\d_Q D &=& \bar{\e} \slashed{\partial}
\chi\,, 
\label{lintrStW}
\eea}

This concludes our discussion of the linearized Weyl multiplets.

\section{Gauging superconformal algebra\la{s:gaug}}

An alternative manner to construct the two Weyl multiplets is to implement the general technique of gauging some generic group symmetry for the specific case of the superconformal group.  The method was successfully applied to other dimensions as well. It was initially developed for $N=1$ in four dimensions \cite{Kaku:1978nz,Ferrara:1977ij} and later on exploited also for $N=2$ \cite{deWit:1980ug,VanProeyen:1983wk}, $N=4$ \cite{Bergshoeff:1981is} and $D=6$ \cite{Bergshoeff:1986mz}. We first discuss the general method of gauging and then elaborate it for constructing the full (non-linear) Weyl
multiplets that we found at the linearized level in
Sect.~\ref{s:lweyl}. For convenience, we have collected the final
results in Sect.~\ref{s:wmul}.

To explain how one gauges a generic group $G$, we consider its algebra 
\begin{equation}\la{galgebr}
\left[ \d (\L_1^A) ,\d (\L_2^B)\right] = \d (\L_2^B\L_1^A f_{AB}{}^C)\,,
\end{equation}
where $f_{AB}{}^C$ are the structure constants (and can be promoted to structure functions as it will be explained in Sect.~\ref{s:wmul}). For each parameter $\L^A$ we associate a gauge field $h_\m^A$ (the analogue of $h_\m$ in~\re{gtransfor}) transforming according to \re{gtransfor}. 

{\it Covariant derivatives} defined as normal derivatives modulo (all) gauge transformations with parameters replaced by the corresponding gauge fields
\begin{equation}\la{covderg}
D_\m = \del_\m - \d_A (h^A_\m)
\end{equation}
have the property (when applied on fields) of transforming covariantly, \ie, without any derivatives on the parameters. The commutator of two such covariant derivatives determines the {\it curvatures}
\begin{equation}\la{comcovd}
\left[ D_\m , D_\n\right] = - \d_A (R_{\m\n}^A)\,, \qq R_{\m\n}^A = 2 \del_{[\m} h^A_{\n]} + h^C_\n h^B_\m f_{BC}{}^A
\end{equation}
that are covariant objects
\begin{equation}\la{covob}
\d R_{\m\n}^A = \L^C R_{\m\n}^B f_{BC}{}^A
\end{equation}
and satisfy the Bianchi identity
\begin{equation}\la{BIg}
D_{[\m} R_{\n\r]}^A = 0\,.
\end{equation}
Subsequently, we will extend this procedure to the superconformal algebra.

\subsection{Fields \& curvatures\la{s:fic}}

The $D=5$ conformal supergravity theory is based on the superconformal
algebra $F^2(4)$ whose generators were discussed in Sect.~\ref{s:confsym} and \ref{s:csusy}. They are repeated in Table~\ref{tbl:superconformal5} together with the corresponding generators and the associated gauge fields. For clarity, $Q_{\a i}$ are symplectic Majorana spinors.
\begin{table}[htbp]
\begin{center}
\begin{tabular}[htbp]{|c|c|c|c|c|c|c|c|}
\hline 
Generators & $P_a$ & $M_{ab} $& $D$ & $K_a$ & $U_{ij}$ & $Q_{\a i}$ &
$S_{\a i}$\\
\hline
Fields & $e_\m{}^a$ & $\o_\m^{ab}$ & $b_\m$ & $f_\m{}^a$ & $V_\m^{ij}$ &
$\p_\m^i$ & $\phi_\m^i$\\
\hline
Parameters & $\xi^a$ & $\l^{ab}$ & $\L_D$ & $\L_K^a$ & $\L^{ij}$ & ${\e}^i$ &
${\eta}^i$\\
\hline
\end{tabular}\caption{ \it The gauge fields  and parameters of the superconformal
algebra $F^2(4)$.\label{tbl:superconformal5}}
\end{center}
\end{table}

We can read off the transformation rules for the gauge fields from the
algebra~(\ref{suf2(4)}) using the general rules for gauge theories. We
find
{\allowdisplaybreaks
\bea
\label{ftr}
\d e_\m{}^a & = & {\cal D}_\m \x^a -\l^{ab}e_{\m b} -\L_De_\m{}^a +\ft
12{\bar\e} \g^a\p_\m \, , \nn\\
\d \o_\m{}^{ab} & = & {\cal D}_\m\l^{ab} - 4\x^{[a}f_\m{}^{b]} - 4\L_K^{[a}
e_\m{}^{b]}
{+\ft 12}\rmi{\bar\e} \g^{ab} \f_\m
{-\ft 12}\rmi{\bar\eta} \g^{ab} \p_\m\, , \nn \\
\d b_\m & = & \partial_\m \L_D -2 \x^a f_{\m a} +2\L_K^a e_{\m a}
{+\ft 12}\rmi {\bar\e} \f_\m + {\ft 12}\rmi {\bar\eta} \p_\m\, , \nn\\
\d f_\m{}^a & = & {\cal D}_\m \L_K^a - \l^{ab}
f_{\m b} + \L_D f_\m{}^a {+\ft 12}{\bar\eta} \g^a \f_\m\, ,\nn\\
\d V_\m^{ij} & = & \partial_\m\L^{ij} -2\L^{(i}{}_\ell V_\m^{j)\ell} -{\ft
{3}2} \rmi{\bar\e}^{(i}\f_\m^{j)} +{\ft {3}2}\rmi{\bar\eta}^{(i}\p_\m^{j)}
  \,,\\
\d \p_\m^i & = & {\cal D}_\m \e^i {+\rmi}\x^a\g_a\f_\m^i -\ft14 \l^{ab}
\g_{ab}\p_\m^i -\ft12\L_D\p_\m^i
-\L^i{}_j\p_\m^j
{-\rmi}e_\m^a\g_a\eta^i \,,\nn\\
\d \f_\m^i & = & {\cal D}_\m \eta^i -\ft14 \l^{ab}\g_{ab}\f_\m^i +\ft12\L_D
\f_\m^i
-\L^i{}_j\f_\m^j
-{\rmi}\L_K^a\g_a\psi_\m^i +{\rmi}f_\m^a\g_a\e^i  \nonumber \,,
\eea
}
where ${\cal D}_\m$ is the covariant derivative with respect to
dilatations, Lorentz rotations and $\SU(2)$ transformations, \eg,
\bea
{\cal D}_\m \x^a & = & \partial_\m \x^a + b_\m \x^a + \o_\m{}^{ab} \x_b \, ,
\nn \\
{\cal D}_\m \e^i & = & \partial_\m \e^i+\ft 12 b_\m\e^i +\ft 14 \o_\m^{ab}
\g_{ab}\e^i - V^{ij}_\m\epsilon _j\,.
\label{covder2}
 \eea
Applying the general relation~\re{comcovd} for~(\ref{suf2(4)}), we obtain the expressions for the curvatures (terms proportional to
vielbeins are underlined for later use), \eg,
\bea
\label{fcurv} 
R^{~~~a}_{\m\n}(P) & = &
2\del_{[\m} e_{\n]}{}^a + \underline{2\o_{[\m}{}^{ab} e_{\n]b}} +
\underline{2b_{[\m} e_{\n]}{}^a } {-\ft 12}{\bar\p}_{[\m}\g^a\p_{\n ]}\,,
 \\
R_{\m\n}{}^{ab}(M) & = & 2\del_{[\m} \o_{\n]}{}^{ab} + 2\o_{[\m}{}^{ac}
\o_{\n]c}{}^b + \underline{8 f_{[\m}{}^{[a} e_{\n]}{}^{b]}}
+\rmi{\bar\f}_{[\m}\g^{ab} \p_{\n ]}
\, , \nn \\
R_{\m\n}{}^i(Q) & = & 2\del_{[\m}\p^i_{\n ]} +\ft12
\o_{[\m}{}^{ab} \g_{ab}\p^i_{\n ]} +b_{[\m}\p^i_{\n ]}
-2V_{[\m}{}^{ij}\p_{\n ]\,j}
+ \underline{2\rmi\g_a\f^i_{[\mu }e_{\nu ]}{}^a}\,.\nn
\eea

Our direct construction yields a gauge theory of $F^2(4)$, yet not a
gauge theory of diffeomorphisms of spacetime. So far, the the spin connection is perceived as an independent
field. 
To make the connection with spacetime symmetries, we have to consider it as a composite field depending on the vielbein.


Furthermore, the mismatch between the bosonic and fermionic degrees of freedom ($96 + 64$) provides a different argument for the failure of the present status of the theory to describe invertible general coordinate transformations generated by the square of supersymmetry operations.

\subsection{Constraints \la{s:cons}}

The solution to the problems described above is well known. In order
to convert the $P$-gauge transformations into general coordinate
transformations and to obtain irreducibility we need to impose
curvature constraints and we have to introduce extra matter fields in
the multiplet.

The constraints will define some gauge fields as dependent fields. The
extra matter fields will also change the transformations of the gauge
fields. In fact, we will have for the transformation (apart from the
general coordinate transformations) of a general gauge field $h_\mu
^I$
\begin{equation}
\delta _J(\epsilon ^J)h_\mu ^I= \partial _\mu \epsilon ^I
+\epsilon ^J h_\mu{}^Af_{AJ}{}^I+\epsilon ^JM_{\mu J}{} ^I\,,
\label{transfogauge}
\end{equation}
where we use the index $I$ to denote all gauge transformations apart
from general coordinate transformations, and an index $A$ includes the
translations.

The last term depends on the matter fields, and its explicit form has
to be determined below. But also the second term has contributions
from matter fields. This is due to the fact that the structure
`functions' of the final algebra $f_{IJ}{}^K$ are modified from those
of the $F^2(4)$ algebra which were used for~(\ref{ftr}). These extra
terms lead also to modified curvatures $\widehat{R}_{\mu \nu }{}^I$ including also matter contributions.

Our next steps can be justified by remarking that the general coordinate transformation (g.c.t.) of the vielbein can be brought to a convenient form
\begin{eqnarray}\label{gcte}
\d_{g.c.t.} (\xi) e_\m{}^a & \equiv & \xi^\n (x) \del_\n e_\m{}^a + e_\n{}^a \del_\m \xi^\n (x) = - \xi^\n R_{\m\n}{}^a (P)\\
\! & + & \!(\d_P (\xi^a) +  \d_M (\xi^\m \o_\m{}^{ab}) + \d_D (b_\m\xi^\m) + \d_Q (\xi^\m \psi_\m^i)) e_\m{}^a \,.\nn
\end{eqnarray}
It is then obvious that we can identify local translations $P$ with g.c.t.~{\it covariantized} with respect to all other transformations except $P$ (see \re{covderg}), at the expense of setting $R(P) = 0$. Hence, we introduce 
\begin{equation}
\delta_{c.g.c.t.}(\xi)=\delta_{g.c.t.}(\xi)-\delta_I(\xi^\mu h_\mu{}^I)\,,
\label{defcgct}
\end{equation}
which works on gauge fields (except the vielbein) as 
\begin{eqnarray}
\delta _{c.g.c.t.}(\xi )h_\mu ^I   & = & -\xi ^\nu \widehat{R}_{\mu \nu }{}^
I-\xi ^\nu  h_\mu ^JM_{\nu J}{}^I
-\xi ^\nu  h_\mu ^Je_\nu ^af_{a J}{}^I\,.
\label{cgctfields}
\end{eqnarray}
By doing so, the commutator of two supersymmetry transformations will also
change. In particular we will find transformations with
field-dependent parameters.

At the same time we have to impose some constraints on the curvatures. One of these constraints was motivated before, while the others are inspired by the need of deriving an irreducible multiplet and the linearities\footnote{We will consider the f\"{u}nfbein as an invertible field.} of some curvatures in certain fields (underlined terms in~\re{fcurv}). The {\it conventional constraints} are
\bea
\label{constraints}
R_{\mu\nu}{}^a(P) = 0 & (50)\nn\,, \\
e^{\nu}{}_b \widehat R_{\mu\nu}{}^{ab} (M) = 0 & (25)\nn\,,\\
\g^\m \widehat R_{\m\n}{}^i (Q) = 0 & (40)\,.
\eea

In parentheses, we denoted the number of restrictions each constraint
imposes. These constraints are similar to those for other Weyl
multiplets in $4$ dimen-\linebreak sions 
\cite{Kaku:1977pa,Ferrara:1977ij,deWit:1980ug,Bergshoeff:1981is} or in $6$ dimensions for the $(1,0)$~\cite{Bergshoeff:1986mz} or $(2,0)$~\cite{Bergshoeff:1999db} Weyl multiplets.


Due to these constraints the fields $\omega_\m{}^{ab}$, $f_\m{}^a$ and
$\phi_\m^i$ are no longer independent, but can be expressed in terms
of the other fields. In order to write down the explicit solutions of
these constraints, it is useful to extract the terms which have been
underlined in~(\ref{fcurv}). We define $\widehat R'$ as the curvatures
without these terms. Formally,
\begin{equation}
\widehat R'_{\mu \nu }{}^I= \widehat R_{\mu \nu }{}^I+ 2h_{[\mu }^J
e_{\nu ]}^a f_{aJ}{}^I\,,
\label{hatRprime}
\end{equation}
where the $F^2(4)$-structure constants $f_{aJ}{}^I$ are defined by the commutators of translations with other gauge
transformations. Then the solutions to the constraints are
\bea 
\widehat\o^{ab}_\m
&=& 2 e^{\n[a} \partial_{[\m} e_{\n]}^{~b]} - e^{\n[a} e^{b]\s} e_{\m c}
\partial_\n e^{~c}_\s
 + 2 e_\m^{~~[a} b^{b]} - \ft12 \bar{\p}^{[b} \g^{a]} \p_\m - \ft14
\bar{\p}^b \g_\m \p^a \,,\nonumber\\
\f^i_\m &=& \ft13\rmi \g^a \widehat{R}^\prime _{\m a}{}^i(Q) - \ft1{24}\rmi
\g_\m \g^{ab} \widehat{R}^\prime _{ab}{}^i(Q)\,, 
 \\
f^a_\m &=& \ft16{\cal R}_\mu {}^a -\ft1{48}e_\mu {}^a {\cal R}\,,\qquad
{\cal R}_{\mu \nu }\equiv \widehat{R}_{\mu \rho }^{\prime~~ab}(M) e_a{}^\rho
e_{\nu b}\,,\qquad {\cal R}\equiv {\cal R}_\mu {}^\mu \,.\nn
\label{transfDepF}
\eea
The constraints imply through Bianchi identities further relations
between the curvatures, \eg, $\widehat R_{\m\n}(D) = 0$. 

\subsection{Matter fields\la{s:matt}}

After imposing the constraints we are left with 21 bosonic and 24
fermionic degrees of freedom. The independent fields are those in the
left upper part of Table~\ref{tbl:fieldsWeyls}. These have to be
completed with matter fields to obtain the full Weyl multiplet. We
have already seen that there are two possibilities for a $D=5$ Weyl
multiplet each with $32+32$ degrees of freedom.

These are obtained by adding either the left or the right
lower corner of Table~\ref{tbl:fieldsWeyls}. To obtain all the extra
transformations we imposed the superconformal algebra, but at the same
time allowing modifications of the algebra by field-dependent
quantities.  The techniques are the same as already used in 4 and 6
dimensions in~\cite{deWit:1980ug,Bergshoeff:1981is}, and were
described in detail in~\cite{Bergshoeff:1986mz}.

For the fields in the upper left corner, we now have to specify the
extra parts $M$ in~(\ref{transfogauge}). This will in fact only apply
to $Q$-supersymmetry. The other transformations are as
in~(\ref{ftr}). The extra terms we can read already from the
linearized rules in~(\ref{lintrDilW}) and~(\ref{lintrStW}). The full
supersymmetry transformations of these fields are
\begin{eqnarray}
\d_Q e_\m{}^a   &=&  \ft 12\bar\e \g^a \psi_\m  \nn\, ,\\
\d_Q \psi_\m^i   &=& {\cal  D}_\m \e^i + \rmi \g\cdot T \g_\m \e^i  \nn\, ,\\
\d_Q V_\m{}^{ij} &=&  -\ft {3}{2}\rmi \bar\e^{(i} \phi_\m^{j)}+ \rmi
\bar\e^{(i} \g\cdot T \psi_\m^{j)} +4 \bar\e^{(i}\g_\m\chi^{j)}
\nn\, ,\\
\d_Q b_\m       &=& \ft 12\rmi\bar\e\phi_\m -2 \bar\e\g_\mu \chi  \,,
\label{modifiedtransfgf}
\end{eqnarray}
where ${\cal D}_\mu \epsilon $ is defined in~(\ref{covder2}). The fields
$T_{ab}$ and $\chi ^i$, together with $D$ that appears below, are independent fields in the Standard
Weyl multiplet, but not in the Dilaton Weyl multiplet. There, they are
given by expressions that are the non-linear extensions of~(\ref{linrelW})
\bea T_{ab}
&=& \ft18 \s^{-2}\bigg(\s \widehat{F}_{ab}
- \ft16 \ve_{abcde} \widehat{H}^{edc} + \ft14\rmi \bar{\p} \g_{ab} \p \bigg)
\,,\label{TchiindilW} \nn \\
\chi^i &=& \ft18\rmi \s^{-1} \slashed{D} \p^i +\ft1{16}\rmi \s^{-2}
\slashed{D} \s \p^i - \ft1{32} \s^{-2} \g \cdot \widehat{F} \p^i 
\nn \\&&
+ \ft14 \s^{-1} \g \cdot \underline{T} \p^i +
\ft1{32}\rmi \s^{-3}  \p_j\bar{\p}^i \p^j \,, \nn \\
D &=& \ft14 \s^{-1} \Box^c \s + \ft18 \s^{-2} (D_a \s) (D^a \s)
-\ft1{16} \s^{-2} \widehat{F}^2
\nn \\&&
- \ft18 \s^{-2} \bar{\p} \slashed{D} \p - \ft1{64} \s^{-4} \bar{\p}^i
\p^j \bar{\p}_i \p_j - 4 \rmi \s^{-1} \bar{\p} \underline{\chi} 
\nn \\&&
+ \left(- \ft{26}3 \underline{T_{ab}} + 2 \s^{-1} \widehat{F}_{ab} +
\ft{1}{4} \rmi \s^{-2} \bar{\p} \g_{ab} \p \right) \underline{T^{ab}}\,,
\label{valueD}
\eea
where the conformal d'Alembertian is defined by
\bea
\Box^c \s \equiv D^a D_a \s
&=& \left( \partial^a  - 2 b^a + \o_b^{~ba} \right) D_a \s - \ft12 \rmi
\bar{\p}_a D^a \p - 2  \s \bar{\p}_a \g^a \underline{\chi}
\nn \\&&
+ \ft12 \bar{\p}_a \g^a \g \cdot \underline{T} \p + \ft12 \bar{\f}_a
\g^a \p + 2 f_a{}^a \s\,,
\eea
and where the underlining indicates that these terms are dependent
fields.  We have not substituted these terms in the expression for $D$
for reasons of brevity.

The modification $M$ in~(\ref{transfogauge}) is the last term of the
transformations of $\psi _\mu ^i$, $V_\mu ^{ij}$ and $b_\mu $. The second
term in the transformation of $V_\mu ^{ij}$ on the other hand is due to
the fact that the structure constants have become structure functions,
and in particular there appears a new $T$-dependent $\SU(2)$
transformation in the anti-commutator of two supersymmetries. We will give
the full new algebra in Sect.~\ref{s:wmul}.
The transformation rules for the matter fields of the
Weyl multiplets are as follows. For the Standard Weyl multiplet we
have ($Q$ and $S$ supersymmetry)
\begin{eqnarray}
\d T_{ab}     &=&  \ft {1}{2}\rmi \bar\e \g_{ab} \chi -\ft
{3}{32}\rmi \bar\e\widehat R_{ab}(Q)    \,,
\nn\\
\d \chi^i     &=&  \ft 14 \e^i D -\ft{1}{64} \g\cdot \widehat R^{ij}(V) \e_j
+ \ft18\rmi \g^{ab} \slashed{D} T_{ab} \e^i
- \ft18\rmi \g^a D^b T_{ab} \e^i  
\nn\\&&
-\ft 14 \g^{abcd} T_{ab}T_{cd} \e^i + \ft 16 T^2 \e^i
+\ft 14 \g\cdot T \eta^i\,,  
\nn\\
\d D &=& \bar\e \slashed{D} \chi - \ft {5 }{3}\rmi \bar\e\g\cdot T
\chi - \rmi \bar\eta\chi \,.
\label{transfosmattStW}
\end{eqnarray}
There are no explicit gauge fields here, as should be the case for
`matter', i.e., non-gauge fields. These are all hidden in the covariant
derivatives and covariant curvatures. The covariant derivatives are for
any matter field given by the rule~\re{covderg} with $A$ restricted to the $I$ range only. 

The covariant curvatures include now matter contributions, e.g.,
\begin{eqnarray}
\widehat{R}_{\m\n}{}^i(Q) &=& R_{\m\n}{}^i(Q) + 2\rmi \gamma \cdot T
\g_{[\m} \p_{\n]}^i\,, 
\label{hatRQV}
\end{eqnarray}
where $R_{\m\n}{}^i(Q)$ 
was computed in~(\ref{fcurv}). 

The knowledge of the covariant curvatures allows  one to calculate the transformations of the dependent fields. Their transformation rules are now determined by their definition due to the constraints. An equivalent way of
expressing this is that their transformation rules are modified
w.r.t.~(\ref{ftr}), due to the non-invariance of the constraints under
these transformations.  We have chosen the constraints to be invariant
under all bosonic symmetries without modifications. Therefore, only
the $Q$- and $S$-supersymmetries of the dependent fields are modified
to get invariant constraints. The new transformation of the spin
connection is
\begin{eqnarray}
\d\widehat\o_\m{}^{ab} &=& \ft12\rmi \bar\e \g^{ab}\phi_\m  - \ft12\rmi \bar
\eta \gamma ^{ab}\psi _\mu 
\nonumber\\&&
- \rmi \bar\e \g^{[a} \g\cdot T \g^{b]} \psi_\m 
\nn\\&&
-\ft 12 \bar\e\g^{[a} \widehat R_\m{}^{b]}(Q) - \ft 14 \bar\e\g_\m
\widehat R^{ab}(Q) - 4 e_\m{}^{[a} \bar\e \g^{b]} \chi \,.
\label{e:var omega}
\end{eqnarray}
The first line is the transformation as implied from the $F^2(4)$
algebra, see~(\ref{ftr}). The second line is due to the modification
of the anti-commutator of two supersymmetries by a $T$-dependent
Lorentz rotation. Finally, the last line contains the terms that go
into the $M$ of~(\ref{transfogauge}). 
Similar thing happens with $\phi _\mu ^i$ and $f_\m{}^a$. For the latter we only need its $M$-dependent $S$-transformation 
\be 
\d_S f_a{}^a = -5 \rmi \bar{\eta} \chi\,.
\ee

The $Q$- and $S$-supersymmetry variations of the matter fields\footnote{As we have already seen, two of the extra fields in the Dilaton Weyl multiplet
are actually gauge fields, rather than matter fields. However, we use
uniformly `matter fields' for them in this context to indicate that
they are not gauging a symmetry of the superconformal algebra.} $A_\m$ and $B_{\m\n}$ in the Dilaton Weyl multiplet are (for full transformations see \re{diltrafo})
\begin{eqnarray}
\d A_\m
&=& -\ft12\rmi \s \bar{\e} \p_\m + \ft12
\bar{\e} \g_\m \p \, , \nonumber\\
 \d B_{\m\n}
&=& \ft12 \s^2 \bar{\e} \g_{[\m} \p_{\n]} + \ft12 \rmi \s \bar{\e} \g_{\m\n} \p
+ A_{[\m} \d(\e) A_{\n]}
\, . 
 \label{transfmatterDilW}
\end{eqnarray}
However, they have also additional symmetries
\bea
\d A_\m &=& \partial_\m \L\,, \nonumber\\
\d B_{\m\n} &=&  2 \partial_{[\m} \L_{\n]} -\ft12 \L F_{\m\n}\,.
\eea
Note that the dependence of the transformation rules for $A_\mu $ and
$B_{\mu \nu }$ on $\psi _\mu $ and $A_\mu $ signals new terms in the
algebra of supersymmetries and $U(1)$ transformations.\footnote{The
$A_{[\mu} \psi_{\nu ]}$ term in $\delta B_{\mu \nu }$ is an extension
of~(\ref{transfogauge}) that occurs for antisymmetric tensor gauge
fields.} This algebra will be written in
Sect.~\ref{s:wmul}. On the other hand, the $F$-term in
$\delta B_{\mu \nu }$ should be interpreted as an $M$ term according
to (\ref{transfogauge}), and modifies the field strength
accordingly. This leads to the following field strengths of these
gauge fields
\bea 
\widehat{F}_{\m\n} &=& 2\partial _{[\mu }A_{\nu
]} + \ft12\rmi \s \bar{\p}_{[\m} \p_{\n]}
 -  \bar{\p}_{[\m} \g_{\n]} \p\, , \nonumber\\
\widehat{H}_{\m\n\r} &=&3\partial _{[\mu }B_{\nu \rho ]} - \ft34 \s^2
\bar{\p}_{[\m} \g_\n \p_{\r]} - \ft32\rmi \s \bar{\p}_{[\m} \g_{\n\r]} \p
+ \ft32 A_{[\m} F_{\n\r]}\, , \label{hatFH}
\eea
with corresponding Bianchi identities for these two curvatures
\bea 
D_{[a} \widehat{F}_{bc]} &=& \ft12 \bar{\p} \g_{[a}
\widehat{R}_{bc]}(Q)\,,
\nn \\
D_{[a} \widehat{H}_{bcd]} &=& \ft34 \widehat{F}_{[ab}
\widehat{F}_{cd]}\,.  \eea

This finishes our derivation of the Standard and Dilaton Weyl
multiplets.  The final results for these multiplets have been
gathered in Sect.~\ref{s:wmul}. 

\subsection{The Weyl multiplets\la{s:wmul}}

For the convenience of the reader we collect in this section the
essential results of the previous sections, and give the supersymmetry
algebra, which is modified by field-dependent terms. The Weyl weight $w$ for each field can be found in Table~(\ref{tbl:fieldsWeyls}), while Lorentz and $\SU(2)$ transformations are evident from the index structure (as explained in \re{Lorrep} and \re{SUtranfo}).

{\bf Standard Weyl multiplet}

The $Q$- and $S$-supersymmetry and $K$-transformation rules for the
independent fields of the Standard Weyl multiplet are
\bea
\d e_\m{}^a   &=&  \ft 12\bar\e \g^a \psi_\m  \nn\, ,\\
\d \psi_\m^i   &=& {\cal  D}_\m \e^i + \rmi \g\cdot T \g_\m \e^i -\rmi
\g_\m
\eta^i  \nn\, ,\\
\d V_\m{}^{ij} &=&  -\ft32\rmi \bar\e^{(i} \phi_\m^{j)} +4
\bar\e^{(i}\g_\m\chi^{j)}
+ \rmi \bar\e^{(i} \g\cdot T \psi_\m^{j)} + \ft32\rmi
\bar\eta^{(i}\psi_\m^{j)} \nn\, ,\\
\d T_{ab}     &=&  \ft12\rmi \bar\e \g_{ab} \chi -\ft
{3}{32}\rmi \bar\e\widehat R_{ab}(Q)    \nn\, ,\\
\d \chi^i     &=&  \ft 14 \e^i D -\ft{1}{64} \g\cdot \widehat R^{ij}(V)
\e_j
+ \ft18\rmi \g^{ab} \slashed{D} T_{ab} \e^i
- \ft18\rmi \g^a D^b T_{ab} \e^i 
\nn\\&&
-\ft 14 \g^{abcd} T_{ab}T_{cd} \e^i + \ft 16 T^2 \e^i
+\ft 14 \g\cdot T \eta^i  \nn\, ,\\
\d D         &=&  \bar\e \slashed{D} \chi - \ft {5}{3} \rmi
\bar\e\g\cdot T
\chi - \rmi  \bar\eta\chi \nn\, ,\\
\d b_\m       &=& \ft 12 \rmi\bar\e\phi_\m -2 \bar\e\g_\mu \chi +
\ft12\rmi \bar\eta\psi_\mu +2\Lambda _{K\mu } \,. \label{modifiedtransf}
 \eea
The covariant derivative ${\cal D}_\mu \epsilon $ is defined
in~(\ref{covder2}) while the general rule follows~(\ref{covderg}). The
covariant curvature\footnote{For $\widehat{R}(V)$ we refer to eq.(3.18) in \cite{Bergshoeff:2001hc}.} $\widehat{R}(Q)$ is given
explicitly in~(\ref{hatRQV}). The expressions for the dependent fields
are displayed in~(\ref{transfDepF}), where the prime indicates the
omission of the underlined terms in~(\ref{fcurv}).

{\bf Dilaton Weyl multiplet}

The Dilaton Weyl multiplet contains two extra gauge transformations:
the gauge transformations of $A_\mu $ with parameter $\Lambda $ and
those of $B_{\mu \nu }$ with parameter $\L _\mu $. The transformation
of the fields are given by:
\bea\la{diltrafo}
\d e_\m{}^a   &=&  \ft 12\bar\e \g^a \psi_\m  \nn\, ,\\
\d \psi_\m^i   &=& {\cal  D}_\m \e^i + \rmi \g\cdot \underline{T} \g_\m
\e^i - \rmi \g_\m
\eta^i  \nn\, ,\\
\d V_\m{}^{ij} &=&  -\ft32\rmi \bar\e^{(i} \phi_\m^{j)} +4
\bar\e^{(i}\g_\m \underline{\chi^{j)}}
  + \rmi \bar\e^{(i} \g\cdot \underline{T} \psi_\m^{j)} + \ft32\rmi
\bar\eta^{(i}\psi_\m^{j)} \nn\, ,\\
 \d A_\m
&=& -\ft12\rmi \s \bar{\e} \p_\m + \ft12
\bar{\e} \g_\m \p +\partial_\m \L \, , \nonumber\\
 \d B_{\m\n}
&=& \ft12 \s^2 \bar{\e} \g_{[\m} \p_{\n]} + \ft12 \rmi \s \bar{\e}
\g_{\m\n} \p + A_{[\m} \d(\e) A_{\n]}
+2 \partial_{[\m} \L_{\n]} -\ft12 \L F_{\m\n}\, , \nonumber\\
\d \p^i &=& - \ft14 \g \cdot \widehat{F} \e^i -\ft12\rmi \slashed{D} \s
\e^i + \s \g \cdot \underline{T} \e^i - \ft14 \rmi \s^{-1} \e_j \bar{\p}^i
\p^j + \s
\eta^i \, , \nonumber\\
\d \s &=& \ft12 \rmi \bar{\e} \p \, ,\nonumber\\
 \d b_\m       &=& \ft12 \rmi \bar\e\phi_\m -2 \bar\e\g_\mu \underline{\chi} +
\ft12\rmi \bar\eta\psi_\mu+2\Lambda _{K\mu } \,.
 \eea
The covariant curvature of $A_\m$ and $B_{\m\n}$ can be found
in~(\ref{hatFH}). The transformation of the dependent fields and the
curvatures have been given in the previous section. We have underlined
the fields $T_{ab}$ and $\chi^i$ to indicate that they are not
independent fields but merely short-hand notations. The explicit
expression for these fields in terms of fields of the Dilaton Weyl
multiplet are given in (\ref{TchiindilW}).

{\bf The new `soft algebra'}

Finally, we present the `soft' algebra that these Weyl multiplets
realize. This is the algebra that all matter multiplets will have to
satisfy, apart from possibly additional transformations under which
the fields of the Weyl multiplets do not transform, and possibly field
equations if these matter multiplets are on-shell (see Ch.~\ref{ch:matter}).

The full commutator of two supersymmetry transformations is
\begin{eqnarray}
\left[\d_Q(\e_1),\d_Q(\e_2)\right] &=&  \d_{cgct}(\xi_3^\m)+
\d_M(\l^{ab}_3) + \d_S(\eta_3)  + \d_U(\l^{ij}_3) 
\nonumber\\ &&
+\d_K(\L^a_{K3})+ \d_{U(1)}(\L_3) + \d_B(\L_{3\m}) \,. \label{algebraQQ}
\end{eqnarray}
The covariant general coordinate transformations have been defined
in~(\ref{defcgct}). The last two terms appear obviously only in the
Dilaton Weyl multiplet formulation. The parameters involved
in~(\ref{algebraQQ}) are
\bea
\xi^\m_3       &=& \ft 12 \bar\e_2 \g_\m \e_1 \,,\nn\\
\l^{ab}_3      &=& - \rmi \bar \epsilon _2\gamma ^{[a}\gamma \cdot T
\gamma ^{b]}\epsilon _1 \,, \nonumber\\
\lambda^{ij}_3 &=& \rmi \bar\e^{(i}_2 \g\cdot T \e^{j)}_1\,, \nn\\
\eta^i_3       &=& - \ft {9}{4}\rmi\, \bar \e_2 \e_1 \chi^i
                   +\ft {7}{4}\rmi\, \bar \e_2 \g_c \e_1 \g^c \chi ^i \nn\\
               &&  + \ft1{4}\rmi\,  \bar \e_2^{(i} \g_{cd} \e_1^{j)}
\left( \g^{cd} \chi_j
                   + \ft 14\, \widehat R^{cd}{}_j(Q) \right) \, ,\nonumber\\
\Lambda^a_{K3} &=& -\ft 12 \bar\e_2\g^a\e_1 D + \ft{1}{96}
\bar\e^i_2\g^{abc}\e^j_1 \widehat R_{bcij}(V)  \nn\\
               && + \ft1{12}\rmi\bar\e_2\left(-5\g^{abcd} D_b T_{cd} +
9 D_b T^{ba} \right)\e_1  \nn\\
               && + \bar\e_2\left(  \g^{abcde}T_{bc}T_{de}
                  - 4 \g^c T_{cd} T^{ad} +  \ft 23  \g^a T^2
                  \right)\e_1 \,,\nonumber\\
\L_3&=& -\ft12\rmi \s \bar{\e}_2 \e_1 \, ,\nonumber\\
\L_{3\m} &=& -\ft12 \s^2 \xi_{3\m} - \ft12 A_\m \L_3\,.
 \eea
For the $Q,S$ commutators we find the following algebra
 \bea\la{QScommu}
 \left[\d_S(\eta),
\d_Q(\e)\right] &=& \d_D( \ft12\rmi \bar\e\eta ) + \d_M( \ft12\rmi \bar\e
\g^{ab} \eta) +      \d_U(  -\ft32\rmi \bar\e^{(i} \eta^{j)} )  +
\delta_K(\L_{3K}^a ) \,,\nn\\
\left[ \d_S(\eta_1), \d_S(\eta_2) \right] &=& \d_K( \ft 12 \bar\eta_2 \g^a
\eta_1 ) \,. \eea
 with
 \be
 \L_{3K}^a= \ft16 \bar{\e} \left(\g \cdot T \g_a - \ft12 \g_a \g \cdot T
\right) \eta \,. \label{algebraQSS}
\end{equation}
The commutator of $Q$ and $U(1)$ transformations is given by
\begin{equation}
 \left[\d(\e), \d(\L) \right] =\d_B\left(-\ft12 \L\d(\e)
A_\m \right) .
 \label{commQU1}
\end{equation}

This ends our description of the Standard and Dilaton Weyl
multiplets.

\section{Conclusions\la{s:concw}}

In this chapter we have taken the first step in the superconformal tensor
calculus by constructing the Weyl multiplets for $N=2$ conformal
supergravity theory in $5$ dimensions.

First, we have applied the standard current multiplet procedure to the
case of the $D=5$ vector multiplet. An unconventional feature is that
the corresponding energy-momentum tensor is neither traceless nor
improvable to a traceless current. However, since one version of the
Weyl multiplet contains a dilaton we could construct this linearized
$32+32$ Dilaton Weyl multiplet \re{lintrDilW} from the current multiplet (Table~\ref{tbl:currentmult}). We also pointed out that there exists a second (`Standard') Weyl multiplet without a dilaton, by comparing with similar Weyl multiplets in~$D=4$ and $D=6$.

\begin{center}
\begin{figure}[ht]
\epsfxsize =13cm
{\hskip 0.5cm \epsffile{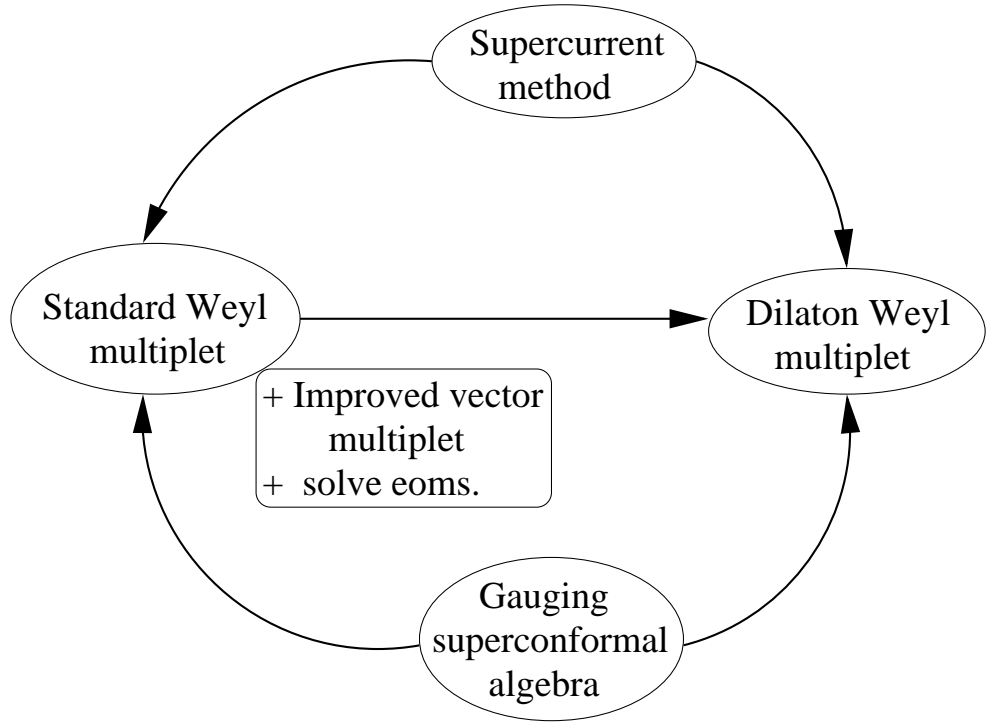}
}
\caption{{\it The two versions of the Weyl multiplet and their connection.}\label{fig:weyl}}\end{figure}
\end{center}

Next, we explained how the non-linear multiplets could be obtained by
gauging the superconformal algebra $F^2(4)$. To that end, we needed some curvature constraints \re{constraints} and the addition of matter fields. The results are collected in Sect.~\ref{s:wmul}. Even if not discussed here, it was shown \cite{Bergshoeff:2001hc} that the coupling
of the Standard Weyl multiplet to an improved vector multiplet leads
to the non-linear relation between the Standard and Dilaton Weyl
multiplets \re{valueD}. Fig.~\ref{fig:weyl} explains the overall picture. 

The fact that there exist two different versions of conformal
supergravity has been encountered before in 6
dimensions~\cite{Bergshoeff:1986mz}. Table~\ref{tbl:compWeyl456} suggests
that the same feature might also occur in $4$ dimensions.
\begin{table}[htbp]
\begin{center}\begin{tabular}{|c|c|c|c|}
\hline
Dimension $D$      &  \# d.o.f.    & Standard Weyl    & Dilaton Weyl \\
\hline
6&10&$T_{abc}^+$&$B_{\mu\nu}$\\
5&10&$T_{ab}$&$A_\mu\, , B_{\mu\nu}$\\
4&6&$T_{ab}$&$A_\mu\, , B_\mu$\\
\hline
\end{tabular}
\caption{{\it The two different formulations of the Weyl multiplet in
$D=4,5,6$.}\label{tbl:compWeyl456}}
\end{center}
\end{table}
It seems plausible that in this case the coupling of a vector
multiplet to the Standard Weyl multiplet will give a Dilaton Weyl
multiplet containing two vectors. It would be interesting to see
whether the Dilaton Weyl multiplet in $4$ dimensions indeed exists, and
how it can be used in matter couplings.


Finally, the results of this chapter constitute the ground on which we are going to build general supergravity/matter couplings in $5$
dimensions. We present that in the next chapter. 
%
\chapter{Conformal supergravity in $D=5$: matter and gauge-fixing \label{ch:matter}}

The Weyl multiplets of the previous chapter play indeed a preveliged role in the SCTC construction, but, for obtaining matter couplings, they ought to be accompanied by matter multiplets. There are in general various representations of the superconformal algebra. In $N=2$, $D=5$ superconformal formalism for instance the vector, vector-tensor, linear, non-linear and the hypermultiplet are the most important ones \cite{Fujita:2001kv,Bergshoeff:2002qk}. Some of them were considered before \cite{Zucker:1999fn,Gunaydin:2000xk,Gunaydin:2000ph,Ceresole:2000jd}, directly in the context of Poincar\'e SUGRA. 

However, our SCTC construction \cite{Bergshoeff:2002qk,Bergshoeff:2003yy} aimed at confirming and challenging the generality of the matter couplings proposed in \cite{Ceresole:2000jd}. In particular, the form of the scalar potential for these matter couplings constitutes the key element when constructing supersymmetric versions of the RS model, which seems not a trivial issue \cite{Kallosh:2000tj,Behrndt:2000tr}. Despite serious progress in that direction \cite{Bergshoeff:2000zn,Behrndt:2000ph,Behrndt:2001qa,Ceresole:2001wi,Anguelova:2002gd}, the problem has not yet been settled. We hope that our results will open new research paths for investigations in that direction.

Due to space limitations, we restrict ourselvesin this work to only some of these representations. Therefore, we will only study vector multiplets, leaving aside their vector-tensor generalization that leads to new contributions to the scalar potential~\cite{Bergshoeff:2003yy}. Although we intentionally ignore this new result, the other new feature of the vector-tensor multiplet, namely the existence of dynamical equations without having an action, is elucidated in the hypermultiplet (the analogous of $N=1$, $D=4$ chiral multiplet) sector where the same phenomenon occurs. Hence, our on-shell hypermultiplet formulation generalizes the (off-shell) discussion in \cite{Fujita:2001kv}. 

The reason to explain the hypermultiplet at the expense of the tensor multiplet originates from geometry. Indeed, the target space parametrized by the hyperscalars describes various geometries in different circumstances. It was known for some time that invariant hypermultiplet actions in rigid SUSY lead to hyperk\"ahler geometry \cite{Alvarez-Gaume:1981hm} , while in SUGRA this becomes quaternionic-K\"ahler \cite{Bagger:1983tt}. Nevertheless, our off-shell formulation of the hypermultiplets emphasizes
the relevance of hypercomplex/quaternionic geometries for SUSY/SUGRA hypermultiplet couplings when no action (metric) is present \cite{Bergshoeff:2002qk,Bergshoeff:2003xx}. 

The rigid SUSY models are presented in Sect.~\ref{s:navec} and Sect.~\ref{s:hyper} and the corresponding actions in Sect.~\ref{s:actr}. For the vector multiplet we use the linear multiplet of Sect.~\ref{s:line}. The coupling to the Standard Weyl multiplet is realized in Sect.~\ref{s:loca}. The SCTC is finally completed in Sect.~\ref{s:gfix}, where the gauge-fixing procedure is carried out. Notice that the geometry of the vector scalars becomes interesting in Poincar\'e SUGRA when it is given by a very special real manifold \cite{Gunaydin:1985ak,deWit:1992cr}. 

The results presented in this chapter are contained in \cite{Bergshoeff:2002qk,Bergshoeff:2003xx,Bergshoeff:2003yy} and part of them overlap with the literature \cite{Kugo:2000af,Gunaydin:2000ph,Ceresole:2000jd,Fujita:2001kv}.

\section{Non-Abelian vector multiplet\la{s:navec}}

In this section, we will discuss non-Abelian, superconformal vector multiplets. From work on $N=2$, $D=5$ Poincar{\'e} matter couplings~\cite{Gunaydin:1999zx} it is known that vector multiplets transforming in representations other than the adjoint have to be dualized to tensor fields. Vector-tensor
multiplets, \ie, vector multiplets transforming in a reducible
representation that contains the adjoint representation as well as
another (arbitrary) representation, were analyzed in the superconformal context in \cite{Bergshoeff:2002qk}. Their field content is presented for the sake of completeness in Table~\ref{tbl:multiplets}. However, apart from some side remarks, they are not included in further considerations. In other words, we limit our discussion for the vector sector to the adjoint representation of some gauge group.
\begin{table}[H]
\begin{minipage}{\linewidth}
\renewcommand{\thefootnote}{\thempfootnote}
\begin{center}
\begin{tabular}{||c|ccc||}
\hline \rule[-1mm]{0mm}{6mm}
Field       & $\SU(2)$ & $w$ & \# d.o.f.  \\
\hline
 & \multicolumn{3}{c||}{off-shell vector multiplet} \\
\rule[-1mm]{0mm}{6mm}
$A_\m^I$      & 1 & 0             & $4 n$ \\
\rule[-1mm]{0mm}{6mm}
$Y^{ij I}$    & 3 & 2             & $3 n$ \\
\rule[-1mm]{0mm}{6mm}
$\s^I$        & 1 & 1             & $1 n$ \\
\hline \rule[-1mm]{0mm}{6mm}
$\p^{i I}$      & 2 & $3/2$         & $8 n$ \\[1mm]
\hline \hline
 & \multicolumn{3}{c||}{on-shell tensor multiplet} \\
\rule[-1mm]{0mm}{6mm}
$B_{\m\n}^M$      & 1 & 0         & $3 m$ \\
\rule[-1mm]{0mm}{6mm}
$Y^{ij M}$    & 3 & 2             & $0$ \\
\rule[-1mm]{0mm}{6mm}
$\sigma ^M$        & 1 & 1             & $1 m$ \\
\hline \rule[-1mm]{0mm}{6mm}
$\psi ^{i M}$      & 2 & $3/2$         & $4 m$ \\[1mm]
\hline \hline
 & \multicolumn{3}{c||}{on-shell hypermultiplet
 }   \\
\rule[-1mm]{0mm}{6mm}
$q^X$      & 2 & $3/2$        & $4 r$ \\
\hline \rule[-1mm]{0mm}{6mm}
$\zeta^A$      & 1 & 2       & $4 r$ \\[1mm]
\hline \hline
 & \multicolumn{3}{c||}{off-shell linear multiplet} \\
\rule[-1mm]{0mm}{6mm}
$L^{ij}$   & 3 & 3             & 3 \\
\rule[-1mm]{0mm}{6mm}
$E_a$      & 1 & 4             & 4 \\
\rule[-1mm]{0mm}{6mm}
$N$        & 1 & 4             & 1 \\
\hline \rule[-1mm]{0mm}{6mm}
$\vf^i$    & 2 & $7/2$ & 8 \\[1mm]
\hline
\end{tabular}
\caption{\it The  $D=5$ matter multiplets. We introduce $n$ vector
multiplets, a linear multiplets and $r$ hypermultiplets. Indicated are
their degrees of freedom, the Weyl weights and the $\SU(2)$
representations, including $m$ tensor multiplets for completeness.}
\label{tbl:multiplets}
\end{center}
\end{minipage}
\end{table}

%
%

Note that, in contrast with the on-shell formulation \re{eq:onshell_maxwell}, we are interested here in the transformation rules for an off-shell non-Abelian vector multiplet~\cite{Fujita:2001kv}. Such an off-shell formulation has $8+8$ real degrees of freedom whose $\SU(2)$ labels and
Weyl weights we have indicated in Table~\ref{tbl:multiplets}.

The gauge transformations that we consider satisfy the commutation\linebreak
relations \re{galgebr}, where we now replace  $A$ by $I=1, \ldots ,n$ and the structure constants appear here multiplied with the coupling constant $g$ of the group G.
The gauge fields~$A_\mu^I$\ ($\mu = 0,1, \ldots ,4$) and general matter fields of the vector multiplet, say $X^I$, transform under gauge transformations with parameters $\Lambda^I$ according to eq.~\re{gtransfor}, respectively
\begin{equation}
\d_G(\L^J) X^I = -g \L^J f_{JK}{}^I X^K \,. \label{eq:gaugecov}
\end{equation}
The expression for the
gauge-covariant derivative of $X^I$  and the field-strengths follow the general rules \re{covderg} and \re{comcovd}, \ie,
\begin{equation}
{\cal D}_\m X^I = \partial_\m X^I + g A_\m^J f_{JK}{}^I X^K \,, \qquad
F_{\mu\nu}^I = 2 \partial_{[\mu} A_{\nu]}^I + g f_{JK}{}^I A_\mu^J
A_\nu^K\,. \label{eq:F}
\end{equation}
The field-strength satisfies the Bianchi identity \re{BIg}

The rigid $Q$- and $S$-supersymmetry transformation rules for the
off-shell Yang-Mills multiplet are given by~\cite{Fujita:2001kv,Bergshoeff:2002qk}
\begin{eqnarray}
\d A_\m^I
&=& \frac 12 \bar{\e} \g_\m \p^I \,, \nonumber\\
\d Y^{ij I}
&=& -\frac 12 \bar{\e}^{(i} \slashed{\cal D} \p^{j) I} -  \frac {1}{2} \rmi g \bar \e^{(i} f_{JK}{}^I \sigma^J \psi^{j) K} + \frac 12 \rmi \bar{\eta}^{(i} \p^{j)I} \,, \nonumber \\
\d \p^{i I}
&=& - \frac 14 \g \cdot F^I \e^i -\frac 12\rmi \slashed{\cal D} \s^I \e^i - Y^{ij I} \e_j + \s^I \eta^i \,, \nonumber\\
\d \s^I &=& \frac 12 \rmi \bar{\e} \p^I \,. \label{ymflat}
\end{eqnarray}
The commutator of two $Q$-supersymmetry transformations yields a
translation with an extra $G$-transformation
\begin{equation}
[ \d(\e_1) , \d(\e_2)] = \d_{P} \left(\frac 12 \bar{\e}_2 \g_\m
\e_1\right) + \d_{G} \left(-\frac 12\rmi \s \bar\e_2 \e_1 \right).
\label{softrigid}
\end{equation}
Note that even though we are considering rigid superconformal symmetry, the al\-ge\-bra~(\ref{softrigid}) contains a field-dependent term on the right-hand side. Such soft terms are commonplace in local superconformal symmetry but here they already appear at the rigid level. In Hamiltonian language, it means that the algebra is satisfied modulo constraints. It should be stressed that when the scalars $\s^I$ acquire some non-vanishing vacuum expectation value the last term of \re{softrigid} can be seen as a central charge extension of the SUSY algebra (an example hereof can be found in \cite{Bergshoeff:2002qk}).

\section{Linear multiplet \la{s:line}}

The second multiplet that we study is the linear multiplet. The significance of the linear multiplet appears in Sect.~\ref{s:vecta} when we introduce an action for the vector multiplet. We will see there that the linear multiplet offers a useful tool in constructing this action via a density formula. Therefore, the eoms that follow from such an action transform precisely as a linear multiplet
in the adjoint representation. In preparation of that section we now present the supersymmetry transformation rules of this multiplet.

The degrees of freedom of the linear multiplet are given in
Table~\ref{tbl:multiplets}. We will consider a linear multiplet in the
background of an off-shell (non-Abelian) vector multiplet. We take the
fields of the linear multiplet in an arbitrary representation of
dimension $m$. The rigid conformal supersymmetry transformation rules for
a linear multiplet in the background of a Yang-Mills multiplet are expressed
as
\begin{eqnarray}
\d L^{ij M}
&=& \rmi \bar{\e}^{(i} \vf^{j) M} \,, \nn \\
\d \vf^{i M} &=& -\frac 12 \rmi \slashed{\cal D} L^{ij M} \e_j - \frac 12
\rmi \g^a E_a^M \e^i + \frac 12 N^M \e^i + \frac 12 g \s^I  t_{IN}{}^M
L^{ij N} \e_j
+ 3 L^{ijM}\eta_j\,, \nn \\
\d E_a^M &=& -\frac 12 \rmi \bar{\e} \g_{ab} {\cal D}^b \vf^M -\frac 12 g
\bar{\e} \g_a t_{IN}{}^M \s^I  \vf^N  + \frac 12 g \bar{\e}^{(i}
t_{IN}{}^M \g_a \p^{j) I}  L_{ij}^N -2 \bar\eta\gamma_a
\varphi^M\,, \nn \\
\d N^M &=& \frac 12 \bar{\e} \slashed{\cal D} \vf^M  + \frac 12 \rmi g
\bar{\e}^{(i} t_{IN}{}^M \p^{j) I}  L_{ij}^N +\frac 32 \rmi \bar\eta
\varphi^M\,. \label{eq:linmultiplet}
\end{eqnarray}
The superconformal algebra closes provided the following differential constraint is satisfied
\begin{equation}
{\cal D}_a E^{a M} + g t_{IN}{}^M \left(Y^{ij I} L_{ij}^N + \rmi \bar
\p^I \vf^N + \sigma^I N^N \right) = 0 \,. \label{eq:constrE}
\end{equation}

Note that the index $I$ refers to the adjoint representation of the
vector multiplet. To obtain the multiplet of equations of motion of the
vector multiplet one should also take for $M$ the adjoint representation
in which case all $t$ matrices become structure constants $f$. If the $g$-dependent terms are separated, one can express \re{eq:constrE} in a more suitable form 
\begin{equation}\la{gconstr}
e^{-1} \del_\m (e G^\m) - g e^{-1} {\cal H}_V =0\,,
\end{equation}
where we restricted ourselves to the case when vector and the linear multiplet components are simple functions, not matrices or column vectors. We also denoted
\begin{eqnarray}\la{notcons}
G^a &\equiv& E^a + \ft 12 \, \rmi  \bar \psi_b \g^{ba} \varphi + \ft 14 \,\bar \psi^i_b \g^{abc} \psi^j_c L_{ij}\,,\nn\\
e^{-1} {\cal H}_V &\equiv& Y^{ij} L_{ij} + \rmi \bar\psi \varphi  - \ft 12 \, 
\bar \psi^i_a \g^a \psi^j L_{ij} + A_a G^a\nn\\
&& + \s (N + \ft 12 \, \bar \psi_a \g^a \varphi + \ft 14 \, \rmi \bar \psi^i_a \g^{ab} \psi_b^j L_{ij})\,. 
\end{eqnarray}
It is crucial to notice that the $E_a$ field that appears in this section transforms under the gauge symmetry.

\section{Hypermultiplet \la{s:hyper}}

In this section, we discuss hypermultiplets in five dimensions. As for
the tensor multiplets \cite{Bergshoeff:2002qk}, there is in general no known off-shell formulation with a finite number of auxiliary fields. Therefore, the supersymmetry algebra already leads to the equations of motion.

A single hypermultiplet contains four real scalars and two spinors subject
to the symplectic Majorana reality condition. For $r$ hypermultiplets, we
introduce real scalars $q^X(x)$, with $X=1,\dots ,4r$, and spinors
$\zeta^A(x)$ with $A=1,\dots ,2r$ satisfying an appropriate symplectic Majorana
condition. 

The scalar fields are interpreted as coordinates of some target space,
and requiring the on-shell closure of the superconformal algebra imposes
certain conditions on the target space, which we derive below.
Superconformal hypermultiplets in four spacetime dimensions were
first discussed in~\cite{deWit:1999fp} and soon after re-investigated in the $D=5$ context \cite{Fujita:2001kv,Bergshoeff:2002qk}. Our presentation follows \cite{Bergshoeff:2002qk} clarifying the terminology ``dynamics without an action", also explained in~\cite{VanProeyen:2001wr}.

\subsubsection{Rigid supersymmetry}

We will show how the closure of the supersymmetry transformation laws
leads to a `hypercomplex manifold'. The closure of the algebra on the
bosons leads to the defining equations for this geometry, whereas the
closure of the algebra on the fermions and its further consistency leads
to equations of motion in this geometry, independent of an action.

The supersymmetry transformations (with $\epsilon^i$ constant parameters)
of the\linebreak bosons $q^X\!(x)$, are parametrized by arbitrary functions
$f^X_{iA}(q)$. Also for the transformation rules of the fermions we write
down the general ansatz 
compatible with the supersymmetry algebra 
\begin{eqnarray}
\delta (\epsilon) q^X
&=& - \rmi \bar\epsilon^i \zeta^A f_{iA}^X \,,\nonumber\\
\delta (\epsilon) \zeta^A &=& \frac 12 \rmi \slashed{\partial} q^X
f_X^{iA} \epsilon_i -\zeta^B \omega_{XB}{}^A \big( \delta (\epsilon) q^X
\big)\,. \label{SUzeta}
\end{eqnarray}
This introduces other general functions $f_X^{iA}(q)$ and
$\omega_{XB}{}^A(q)$ constrained by some reality properties consistent with those of $q^X$ and with symplectic Majorana conditions. 
A priori the functions $f_{iA}^X$ and $f^{iA}_X$ are independent, but the
commutator of two supersymmetries on the scalars only gives a translation
if one imposes
\begin{eqnarray}\label{cov_const}
f^{iA}_Y f^X_{iA} &=& \delta_Y^X \,,\qquad f^{iA}_Xf^X_{jB}=\delta^i_j
\delta^A_B\,,\nonumber\\
\covder_Y f_{iB}^X &\equiv&
\partial_Y f_{iB}^X - \omega_{YB}^{\quad A} f_{iA}^X + \Gamma _{ZY}^{\quad
X} f_{iB}^Z = 0\,,
\end{eqnarray}
where $\Gamma _{XY} {}^Z$ is some object, symmetric in the lower indices.
This means that $f_{iA}^X$ and $f^{iA}_X$ are each other's inverse and are
covariantly constant with connections $\Gamma$ and $\omega$. 
The conditions~(\ref{cov_const}) encode all the constraints on the target
space that follow from imposing the supersymmetry algebra. Below, we show
that there are no further geometrical constraints coming from the fermion
commutator; instead this commutator defines the equations of motion for
the on-shell hypermultiplet. \bigskip

Before doing that, we notice that the supersymmetry transformation rules are covariant with respect to two
kinds of reparametrizations. The first ones are the target space
diffeomorphisms, $q^X\rightarrow {\widetilde q}^X(q)$, under which
$f^X_{iA}$ transforms as a vector, $\omega_{XA}{}^B$ as a one-form, and
$\Gamma_{XY}{}^Z$ as a connection. The second set are the
reparametrizations which act on the tangent space indices $A,B,\ldots$ On
the fermions, they act as
\begin{equation}
\zeta^A \rightarrow {\widetilde \zeta}^A(q)=\zeta^B
U_B{}^A(q)\,,\label{ferm-equiv}
\end{equation}
where $U(q)_A{}^B$ is any invertible matrix. In general, such a
transformation brings us into a basis where the fermions depend on the
scalars $q^X$. In this sense, the hypermultiplet is written in a special
basis where $q^X$ and $\zeta^A$ are independent fields. The supersymmetry
transformation rules~(\ref{SUzeta}) are covariant
under~(\ref{ferm-equiv}) if we transform $f^{iA}_X(q)$ as a vector and
$\omega_{XA}{}^B$ as a connection,
\begin{equation}
\omega_{XA}{}^B\rightarrow {\widetilde \omega}_{XA}{}^B=[(\partial_X
U^{-1})U+ U^{-1}\omega_X U]_A{}^B\,.
\end{equation}
These considerations lead us to define the covariant variation of the
fermions\footnote{A similar argument is applicable on target-space vectors and tensors, \eg, $\widehat \delta \Delta ^X= \delta \Delta ^X + \Delta ^Y \Gamma _{ZY}{}^X\,\delta
  q^Z\,.$}
\begin{equation}
{\widehat \delta} \zeta^A\equiv \delta \zeta^A+\zeta^B\omega_{XB}{}^A
\delta q^X\,, \label{cov-var}
\end{equation}
for any transformation $\delta$ (supersymmetry, conformal
transformations,\ldots). Two models related by either target space
diffeomorphisms or fermion reparametrizations of the
form~(\ref{ferm-equiv}) are equivalent; they are different coordinate
descriptions of the same system. Thus, in  a covariant formalism, the
fermions can be functions of the scalars. 

A second remark concerns the geometry of the target space, which is that of a \emph{hypercomplex} manifold. It is a weakened version of hyperk{\"a}hler geometry where no hermitian covariantly constant metric is defined. We refer the reader to \cite[Appendix B]{Bergshoeff:2002qk} for an introduction to these manifolds,
examples and the mathematical context in which they can be situated.

The crucial ingredient is a triplet of complex structures $\vec{J}_X{}^Y$, the
hypercomplex structure, defined as\footnote{For $\SU(2)$-valued quantities we use alternatively the triplet $\vec{J}_X{}^Y \equiv (J^\alpha)_X{}^Y$ or the doublet notation $J_X{}^Y{}_i{}^j \equiv \rmi
 \vec{J}_X{}^Y\cdot \vec\sigma_i{}^j=2f^{jA}_Xf^Y_{iA}- \delta_i^j\delta_X^Y$.}
\begin{equation}
\J\alpha XY \equiv -\rmi f_X^{iA}(\sigma ^\alpha )_i{}^j f_{jA}^Y\,.
\label{defJf}
\end{equation}
They satisfy the quaternionic algebra
\begin{equation}
J^\alpha J^\beta =- \unity _{4r} \delta^{\alpha \beta} +
\varepsilon^{\alpha \beta \gamma} J^\gamma \,. \label{defJ}
\end{equation}
and, using~(\ref{cov_const}), they are covariantly constant. Under certain restriction (vanishing of the `diagonal' Nijenhuis tensor $N_{XY}{}^Z$), the torsionless connection preserving the $\vec{J}_X{}^Y$ is the Obata connection~\cite{Obata}
\begin{equation}
\Gamma_{XY}{}^Z=-\frac 16\left( 2 \partial_{(X}\vec{J}_{Y)}{}^W +  \vec{J}_{(X}{}^U \times
\partial_{|U|} \vec{J}_{Y)}{}^W\right) \cdot \vec{J}_W{}^Z\,, \label{Obata}
\end{equation}
Moreover, one can show that any torsionless connection that leaves the complex structures invariant is equal to this Obata connection (similar to the fact that a connection that leaves a metric invariant is the Levi-Civita connection).
With this connection one can then construct the
$\Gl(r,\mathbb{H})$-connection
\begin{equation}
  \omega _{XA}{}^B=\frac 12 f_Y^{iB}\left( \partial _X f_{iA}^Y+\Gamma
  _{XZ}^Yf_{iA}^Z\right) \,,
 \label{determineOm}
\end{equation}
such that the vielbeins are covariantly constant. Thus, the only independent quantity remains $f_X^{iA}$ that satisfies \re{cov_const} for \re{Obata} and \re{determineOm}.

The holonomy group of such a space is contained in
$\Gl(r,\mathbb{H})=\SU^*(2r)\times $U$(1)$, the group of transformations
acting on the $A,B$-indices. This follows from the integrability
conditions on the covariantly constant vielbeins $f^{iA}_X$, which
relates the curvatures of the $\omega _{XA}{}^B$ and $\Gamma _{XY}{}^Z$
connections,
\begin{equation}
R_{XYZ}{}^W=f_{iA}^Wf^{iB}_Z{\cal R}_{XYB} {}^A \,,\qquad
\delta_j^i\,{\cal R}_{XYB}{}^A = f^{iA}_W f^Z_{jB}\,R_{XYZ}{}^W\,,
\label{curv-rel}
\end{equation}
such that the Riemann curvature only lies in $\Gl(r,\mathbb{H})$.
From the Riemann tensor one can manufacture an object $W_{ABC}{}^D$
\begin{eqnarray}
&& W_{CDB}{}^A \equiv  f^{iX}_C f^Y_{iD} {\cal R}_{XY}{}_B {}^A = \frac
12f^{iX}_C f^Y_{iD} f_{jB}^Z f_W^{Aj} R_{XYZ}{}^W\,, \label{def-W}
\end{eqnarray}
symmetric in all its three lower indices and whose relevance will become clear shortly. One difference between  hypercomplex and hyperk{\"a}hler manifolds is that the former need not to be Ricci flat. Instead, the Ricci tensor is antisymmetric and defines a closed two-form. 

%

\paragraph{Dynamics.}
Now we consider the commutator of supersymmetry on the fermions, which
will determine the equations of motion for the hypermultiplets.

Using~(\ref{cov_const}),~(\ref{curv-rel}) and~(\ref{def-W}), the commutator on the fermions 
\begin{equation}
[\delta ( \epsilon_1),\delta ( \epsilon_2)] \zeta^A =\frac 12\partial_a
\zeta^A \bar{\epsilon}_2 \gamma^a \epsilon_1 + \frac 14 \Gamma ^A
\bar{\epsilon}_2 \epsilon_1 -\frac 14 \gamma_a \Gamma ^A \bar{\epsilon}_2
\gamma^a \epsilon_1 \,. \label{FinalCommzeta}
\end{equation}
defines, via the non-closure functions, the dynamics of the fermions ($\Gamma^A = 0$)
\begin{equation}
\Gamma^A \equiv \slashed{\covder} \zeta^A + \frac 12 W_{CDB} {}^A
 \zeta^B\bar{\zeta}^D \zeta^C\,, \label{eqmozeta}
\end{equation}
where we have introduced the covariant derivative with respect to the
transformations~(\ref{cov-var})
\begin{equation}
  \covder_\mu  \zeta^A \equiv \partial_\mu  \zeta^A + (\partial_\mu  q^X)
\zeta^B\omega_{XB} {}^A\,.
 \label{defDzeta}
\end{equation}

By varying the functions $\Gamma^A $ under supersymmetry, we derive the
corresponding equations of motion for the scalar fields ($\Delta^X =0$)
\begin{equation}
\widehat \delta(\epsilon ) \Gamma^A =  \frac 12 \rmi f_{X}^{iA}
\epsilon_i \Delta ^X \,,\label{delQGamma3}
\end{equation}
where
\begin{equation}
\label{covscal} \Delta^X \equiv \Box q^X -\frac 12 \bar{\zeta}^B \gamma_a
\zeta^D
\partial^a q^Y f_Y^{iC}f_{iA}^X W_{BCD} {}^A-\frac{1}{4}\covder_Y W_{BCD}{}^A \bar{\zeta}^E \zeta^D
\bar{\zeta}^C \zeta^B f_E^{iY}f_{iA}^X\,,
\end{equation}
and the covariant Laplacian is given by
\begin{equation}
\Box q^X = \partial_a \partial^a q^X+ \left( \partial_a q^Y\right)
\left(\partial^a q^Z \right) \Gamma_{YZ} {}^X\,.
\end{equation}

In conclusion, the supersymmetry algebra imposes the hypercomplex
constraints give by (\ref{cov_const}) and the equations of
motion~(\ref{eqmozeta}) and~(\ref{covscal}). These form a multiplet,
as~(\ref{delQGamma3}) has the counterpart
\begin{equation}
  \widehat \delta (\epsilon ) \Delta^X= -\rmi\bar{\epsilon}^i \slashed\covder \Gamma^A f_{iA}^X+2\rmi\bar{\epsilon}^i \Gamma^B \bar{\zeta}^C \zeta^D f_{Bi}^Y {\cal R}^X
{}_{YCD}\,, \label{susydelta}
\end{equation}
where the covariant derivative of $\Gamma^A$ is defined similar
to~(\ref{defDzeta}). In the following, we will derive more constraints
on the target space geometry from requiring the presence of conformal
symmetry.

\subsubsection{Rigid superconformal symmetry}\label{ss:schyper}

Now we define transformation rules for the hypermultiplet under the full
(rigid) superconformal group. The scalars do not transform under special
conformal transformations and special supersymmetry, but under
dilatations and $\SU(2)$ transformations, we parameterize
\begin{eqnarray}
\delta_D(\Lambda_D) q^X&=& \Lambda_D k^X(q) \,, \nonumber\\
\delta_{\SU(2)}(\vec\Lambda) q^X&=& 2 \vec \L \cdot \vec k^X (q)
\label{SUzetacov}
\end{eqnarray}
for some unknown functions $k^X(q)$ and $k_{ij}^X(q)$.

To derive the appropriate transformation rules for the fermions, we first
note that the hyperinos should be invariant under special conformal
symmetry. This is due to the fact that this symmetry changes the Weyl
weight with one. If we realize the $\left[K,Q\right]$ commutator on the
fermions $\zeta^A$, we read off the special supersymmetry transformation
\begin{equation}
\delta_S(\eta^i)\zeta^A=-k^X f_{X}^{iA} \eta_i\,.
\end{equation}
To proceed, we consider the commutator of regular and special
supersymmetry~(\ref{QScommu}). Realizing this on the scalars, we determine
the expression of the $\SU(2)$ Killing vectors in terms of those corresponding to dilatations and complex structures,
\begin{equation}\label{dil-su2}
\vec k^{X}=\frac 13 k^Y\vec J_X{}^Y\,.
\end{equation}
Realizing~(\ref{QScommu}) on the hyperinos, we determine the covariant
variations
\begin{equation}
\widehat \delta_D \zeta^A = 2 \Lambda_D \zeta^A \,, \qquad \widehat
\delta_{\SU(2)} \zeta^A =0\,,
\end{equation}
and furthermore the commutator~(\ref{QScommu}) only closes if we impose
\begin{equation}\label{conf-constr}
\covder_Y k^X 
= \frac 32 \delta_Y{}^X\,,
\end{equation}
which also implies
\begin{equation}
 \covder_Y \vec k^{ X}=\frac 12\vec J_Y{}^ X \,.
 \label{DkSU2}
\end{equation}
Note that~(\ref{conf-constr}) is imposed by supersymmetry. In a more
usual derivation, where one considers symmetries of the Lagrangian, we
would find this constraint by imposing dilatation invariance of the
action, see~(\ref{exhomK}). Our result, however, does not require the
existence of an action. The relations~(\ref{conf-constr})
and~(\ref{dil-su2}) further restrict the geometry of the target space,
and it is easy to derive that the Riemann tensor has four zero
eigenvectors
\begin{equation}
k^XR_{XYZ}{}^W=0\,,\qquad  \vec k^{ X}\,R_{XYZ}{}^W=0\,, \label{konRis0}
\end{equation}
in analogy with $D=4$ case~\cite{deWit:1980gt,deWit:1999fp}.

%

\subsubsection{Symmetries}\label{ss:symmhyper}

We now consider the action of a symmetry group on the hypermultiplet. We
have no action, but the `symmetry' operation should leave invariant the
set of equations of motion. The symmetry algebra must commute with the
supersymmetry algebra (and later with the full superconformal algebra).
This leads to hypermultiplet couplings to a non-Abelian gauge group $G$.
The symmetries are parametrized by
\begin{eqnarray}
\delta_G q^X&=& -g \Lambda_G^I k_I^X(q)\,, \nonumber\\
\widehat \delta_G \zeta^A &=& -g \Lambda^I_G t_{IB}{}^A(q) \zeta^B\,.
\label{gauge-tr}
\end{eqnarray}
The vectors $k_I^X$ depend on the scalars and generate the algebra of $G$
with structure constants $f_{IJ}{}^K$
\begin{equation} \label{G-alg}
k_{[I|}^Y \partial_Y k^X_{|J]}=-\frac 12 f_{IJ}{}^K k_K^X \,.
\end{equation}
The commutator of two gauge transformations 
on the fermions requires the following constraint on the field-dependent matrices
$t_I(q)$
\begin{equation} \label{tt-comm}
[t_I,t_J]_B{}^A=-f_{IJ}{}^Kt_{KB}{}^A -2k^X_{[I|}\covder_Xt_{|J]B}{}^A
+k^X_Ik^Y_J{\cal R}_{XYB}{}^A\,.
\end{equation}

Requiring the gauge transformations to commute with supersymmetry leads
to further relations between the quantities $k_I^X$ and $t_{IB}{}^A$.
Vanishing of the commutator on the scalars yields
\begin{equation} \label{GG-q}
t_{IB}{}^Af^X_{iA}=\covder_Yk^X_If^{Y}_{iB}\,.
\end{equation}
These constraints determine $t_I(q)$ in terms of the vielbeins $f^{iA}_X$
and the vectors $k^X_I$
\begin{equation} \label{def-t}
t_{IA}{}^B=\frac 12f^Y_{iA}\covder_Yk^X_If^{iB}_X\,,
\end{equation}
and furthermore
\begin{equation} \label{symm-ff}
f^{Y(i}_Af^{j)B}_X \covder_Y k^X_I=0\,.
\end{equation}
The relations~(\ref{symm-ff}) and~(\ref{def-t}) are equivalent
to~(\ref{GG-q}). We interpret~(\ref{def-t}) as the definition for
$t_{IA}{}^B$. The vanishing of an $(ij)$-symmetric part in an equation
as~(\ref{symm-ff}) can be expressed as the vanishing of the commutator of
$\covder_Y k^X_I$ with the complex structures
\begin{equation}
  \left(\covder_X k^Y_I\right) \vec J_Y{}^Z= \vec J_X{}^Y\left(\covder_Y
  k^Z_I\right)\,.
 \label{commDkJ}
\end{equation}
Extracting affine connections from this equation, it can be written as
\begin{equation}
\label{intJ} \left( {\cal L}_{k_I} \vec J\right) _X{}^Y\equiv  k^Z_I
\partial_Z \vec J_X{}^Y -\partial_Z k^Y_I \vec J_X{}^Z+\partial_X
k^Z_I \vec J_Z{}^Y=0\,.
\end{equation}
The l.h.s.~is the Lie derivative of the complex structure in the
direction of the vector~$k_I$.

The vanishing of the gauge-supersymmetry commutator on the fermions requires
\begin{equation}
\covder_Y t_{IA}{}^B= k^X_I {\cal R}_{YXA} {}^B\,. \label{t-id}
\end{equation}
Using~(\ref{GG-q}) this implies a new constraint,
\begin{equation}
\label{DDk} \covder_X\covder_Y k^Z_I=R_{XWY}{}^Zk_I^W\,.
\end{equation}
Note that this equation is in general true for any Killing vector of a
metric. As we have no metric here, we could not rely on this fact, but
here the algebra imposes this equation. It turns out that~(\ref{symm-ff})
and~(\ref{DDk}) are sufficient for the full commutator algebra to hold.
In particular,~(\ref{t-id}) follows from~(\ref{DDk}), using the
definition of $t$ as in~(\ref{def-t}), and~(\ref{curv-rel}).

%

The group of gauge symmetries should also commute with the superconformal
algebra, in particular with dilatations and $\SU(2)$ transformations.
This leads to
\begin{equation}
\label{comDG} k^Y \covder_Y k_I^X=\frac {3}{2} k_I^X\,, \qquad \vec k^{Y}\covder_Y k_I^X=\frac 12 k_I^Y\vec J_Y{}^X\,,
\end{equation}
coming from the scalars, and there are no new constraints from the
fermions or from other commutators. Since $\covder_Yk_I^X$ commutes with
$\J\alpha YX$, the second equation in~(\ref{comDG}) is a consequence of
the first one.
\bigskip

The parameters $\Lambda^I$ were considered constant in the above analysis. 
We now relax that by allowing local gauge transformations. The gauge coupling is done by introducing vector multiplets and defining the covariant derivatives
\begin{eqnarray}
\covder_{\mu} q^X&\equiv& \partial_{\mu} q^X +g A_{\mu}^I k_I^X \,, \nonumber\\
\covder_{\mu} \zeta^A &\equiv& \partial_\mu \zeta^A + \partial_\mu q^X
\omega_{XB}{}^A \zeta^B + g A_{\mu}^I t_{IB} {}^A \zeta^B\,.
\label{covdergene}
\end{eqnarray}
The commutator of two supersymmetries should now also contain a local
gauge transformation similar to~(\ref{softrigid}). This requires an extra term in the supersymmetry transformation law of the fermion
\begin{equation}
\widehat \delta (\epsilon) \zeta^A = \frac 12 \rmi \slashed{\covder} q^X
f_X^{iA} \epsilon_i +\frac {1}{2}g \sigma^I k_I^X f_{iX}^A \epsilon^i\,.
\end{equation}
With this additional term, the commutator on the scalars closes, whereas
on the fermions, it determines the equations of motion
\begin{equation}\la{hyperinoeom}
\Gamma^A\equiv \slashed{\covder} \zeta^A +\frac 12 W_{BCD} {}^A
\bar{\zeta}^C \zeta^D \zeta^B-g(\rmi k_I^X f_{iX}^A \psi^{iI}+ \rmi
\zeta^B\sigma^I t_{IB} {}^A )=0\,,
\end{equation}
with the same conventions as in~(\ref{FinalCommzeta}).

Acting on $\Gamma^A$ with supersymmetry determines the equation of motion
for the scalars
\begin{eqnarray}\la{hyperseom}
\Delta^X &\!\!=\Box q^X -\frac 12 \bar{\zeta}^B \gamma_a \zeta^D \covder^a
q^Y f_Y^{iC}f_{iA}^X W_{BCD} {}^A-\frac {1}{4} \covder_Y W_{BCD}{}^A
\bar{\zeta}^E \zeta^D \bar{\zeta}^C \zeta^B f_E^{iY}f_{iA}^X \nonumber\\
&-\,g \left(  2\rmi \bar{\psi}^{iI} \zeta^B t_{IB}{}^A f_{iA}^X- k_I^Y
J_Y{}^X {}_{ij} Y^{ijI} \right) +g^2 \sigma^I \sigma^J \covder_Y k_I^X
k_J^Y\,.
\end{eqnarray}
The first line is the same as in~(\ref{covscal}), the second line
contains the corrections due to the gauging. The gauge-covariant
Laplacian is here given by
\begin{equation}
\Box q^X=\partial_a \covder^a q^X +g \covder_a q^Y \partial_Y k_I^X
A^{aI}+\covder_a q^Y \covder^a q^Z \Gamma_{YZ}^X\,.
\end{equation}
The equations of motions $\Gamma^A $ and $\Delta ^X$ still satisfy the
same algebra with~(\ref{delQGamma3}) and~(\ref{susydelta}).

\section{Rigid superconformal actions\la{s:actr}}

After determining the relevant multiplets, we continue in this section the SCTC program by presenting rigid superconformal actions for the those
multiplets. We will see that demanding the existence of an action is more restrictive than only considering equations of motion. For the different multiplets, we find that new geometric objects have to be introduced.

\subsection{Vector multiplet action\la{s:vecta}}

The method of invariant density formulae was successfully employed in the past in $D=4$ \cite{deWit:1983tq,Kugo:1983cu} and $D=6$ \cite{Bergshoeff:1986mz} to construct superconformal invariant actions for vector multiplets. We will use it also here to determine the action for the Maxwell version of the off-shell vector multiplet (Sect.~\ref{s:navec}). We focus for simplicity on the rigid case. A density formula provides in general an invariant action up to a total derivative. The desired density formula for the product of a vector and a linear multiplet is given by 
\begin{equation}\la{densM}
\cL_{ \rm VL} = \ft 13 \, \left[ Y^{ij} L_{ij} +\rmi \, \bar \psi \varphi + A_a E^a + \s N\right]\,,
\end{equation}
which indeed has Weyl weight $5$ and, according to \re{eq:gaugecov}-\re{ymflat} and \re{eq:linmultiplet}, it varies to a total derivative under $Q$, $S$ or gauge transformations. 
Notice, however, that $E^a$ appearing in this action and in \re{embM} is not the same with the gauge field denoted also $E^a$ in Sect.~\ref{s:line}, even though their supersymmetry transformations coincide.

The action for the (off-shell) Maxwell multiplet is an immediate consequence of \re{densM} if we can find an embedding of the vector multiplet in the linear one. In other words, we intend to express the fields of the linear multiplet as functions of the vector multiplet components, just like we did with the current multiplet \re{currmvec}. Such an embedding was proposed in \cite{Bergshoeff:2001hc,Fujita:2001kv} and it reads
\begin{eqnarray}\la{embM}
L^{ij} &=& 2 \s Y^{ij} - \ft12 \rmi \bar{\p}^i \p^j\,, \nn\\
\varphi^i &=&\rmi \s \slashed{\del} \p^i + \ft12 \rmi \slashed{\del} \s \p^i
- \ft14 \g \cdot F \p^i + Y^{ij} \p_j\,,\nn\\
E_a &=& -\ft1{8} \ve_{abcde} F^{bc} F^{de} + \del^b
\left(  \s F_{ba} + \ft14 \rmi \bar{\p} \g_{ba}
\p  \right)\,, \nn\\
N &=& -\ft14 F_{ab} F^{ab} - \ft12 \bar{\p}
\slashed{\del}
\p + \s \Box \s + \ft12 \del^a \s \del_a \s + Y^{ij} Y_{ij}\,.
\end{eqnarray}
The substitution of these last relations in \re{densM} yields the improved Lagrangian of the of the rigid Maxwell multiplet 
\bea
\label{fstl}
{\cal L}^{\rm imp}_{\rm Max}
&=& -\ft14 \s F_{\m\n} F^{\m\n} - \ft12 \s \bar{\p} \slashed{\partial} \p -
\ft12 \s (\partial \s)^2 + \s Y^{ij} Y_{ij}
\nn \\&&
-\ft18 \rmi \bar{\p} \g \cdot F \p -\ft1{24} \ve_{\m\n\l\r\s} A^\m
F^{\n\l} F^{\r\s} -\ft12 \rmi \bar{\p}^i \p^j Y_{ij} \,.
\eea

We remark in particular that the action is cubic in the fields of the vector multiplet, enforcing the correct Weyl weight, and the appearance of the CS term. An analogous density formula can be applied for the local superconformal Maxwell action, but we postpone it until Sect.~\ref{s:vectla} and discuss now the non-Abelian action.


The rigidly superconformal invariant action describing $n$ vector
multiplets was obtained in~\cite{Kugo:2000af} from tensor calculus using an intermediate linear multiplet as explained above. The Abelian part can be obtained by just
taking the (cubic) action of one vector multiplet as given
in~\cite{Bergshoeff:2001hc}, adding indices $I,J,K$ on the fields and
multiplying with the symmetric tensor $C_{IJK}$. The resulting action is still supersymmetric. For the non-Abelian
case, we need conditions expressing the gauge invariance of this tensor
\begin{equation}
  f_{I(J}{}^HC_{KL)H}=0\,.
 \label{cyclic}
\end{equation}
Moreover one has to add a few more terms, e.g., to complete the
CS term to its non-Abelian form. This leads to the action
\begin{eqnarray}
{\cal L}_{\rm vector} &\!\!=
 \Bigl[ \left(- \frac 14 F_{\m\n}^{ I} F^{\m\n {
J}} - \frac 12  \bar{\p}^{ I} \slashed{\cal D} \p^{ J} - \frac 12 {\cal
D}_a \s^{ I} {\cal D}^a \s^{ J}
+  Y_{ij}^{ I} Y^{ij { J}}  \right) \s^{ K}  \nn \\
&\!\!\!\!\
     -\frac 1{24} \ve^{\m\n\l\r\s} A_\m^I \left( F_{\n\l}^J F_{\r\s}^K + \frac 12 g [A_\n, A_\l]^J F_{\r\s}^K + \frac 1{10} g^2 [A_\n, A_\l]^J [A_\r, A_\s]^K \right) \nn \\
&
 \!\!\!\!\!\!\!\!\!\!-\frac 18 \rmi \bar{\p}^{ I} \g \cdot F^{ J} \p^{ K} -\frac 12
\rmi \bar{\p}^{i { I}} \p^{j { J}} Y_{ij}^{ K}  + \frac 14 \rmi g
\bar{\p}^{ L}\psi ^H\sigma ^{{I}}\sigma ^{{J}} f_{LH}{}^{K} \Bigr]
C_{IJK} . \label{eq:Vectoraction}
\end{eqnarray}
The eoms for the fields of the vector multiplet generated by the action~(\ref{eq:Vectoraction}) are
\begin{equation}
0 = L^{ij}_I = \varphi^i_I = E^a_I = N_I \,,
\end{equation}
generalizing \re{embM} for non-Abelian vectors as 
\begin{eqnarray}
L^{ij}_I &\equiv& C_{IJK} \left(2 \s^{ J} Y^{ij { K}} - \frac 12 \rmi
\bar{\p}^{i { J}} \p^{j { K}}\right),
\nonumber\\
\varphi^{i}_I &\equiv& C_{IJK} \left(
  \rmi \sigma ^{ J}\slashed{\cal D}\psi ^{i K}
  +\frac 12\rmi(\slashed{\cal D}\sigma ^{ J})
  \psi ^{i K}+Y^{ik J}\psi _k^{ K}
  -\frac 14\gamma \cdot F^{ J}\psi ^{i K}\right) \nonumber \\
  && -\,g C_{IJK} f_{LH}{}^{K} \sigma^{ J} \sigma ^{ L}\psi ^{iH} \,, \nonumber\\
E_{a I} &\equiv&  C_{IJK} \left[{\cal D}^b \left(\s^{ J} F_{ba}{}^{ K} +
\frac 14 \rmi \bar{\p}^{ J} \g_{ba} \p^K \right) - \frac 18 \ve_{abcde}
F^{bc {J}}
F^{de { K}} \right] \nonumber\\
&& -\,\frac 12 g C_{JKL} f_{IH}{}^J \s^K \bar\p^L \g_a \p^H - g C_{JKH}
f_{IL}{}^J \s^K \s^L {\cal D}_a \s^H
\,,\nonumber\\
 N_I
&\!\!\equiv&\!\! C_{IJK} \left(\s^{ J} \Box \s^{ K} + \frac 12 {\cal D}^a \s^{ J} {\cal D}_a \s^{ K}
-\frac14 F_{ab}^{ J} F^{ab  K} - \frac 12 \bar{\p}^{ J} \slashed{\cal D} \p^{ K} + Y^{ij  J} Y_{ij}{}^{ K} \right)  \nonumber \\
&& +\, \frac 12 \rmi g C_{IJK} f_{LH}{}^K \s^{J} \bar{\p}^{ L} \p^H \,.
 \label{eq:vectoreom}
\end{eqnarray}
We labeled these eoms as $L^{ij}_I, \phi^i_I,
E_{a I}, N_I$ since they form a linear multiplet in the adjoint
representation of the gauge group, for which the transformation rules were displayed in~(\ref{eq:linmultiplet}).

\subsection{The hypermultiplet action\la{s:hypa}}
Until this point, the eoms we derived for the hypermultiplet originated from the fact that we had an open superconformal algebra. The non-closure
functions $\Gamma^A$, together with their supersymmetric partners
$\Delta^X$ yielded these equations of motion. We discovered a
hypercomplex scalar manifold $\mathcal{M}$, where $\Gamma_{XY}{}^Z$ was
interpreted as an affine connection. We also needed a $\Gl
(r,\mathbb{H})$-connection $\omega_{XA}{}^B$ on a vector bundle.

Now, we will introduce an action to derive the field equations of the
hypermultiplet. An important point to note is that the necessary data for determining the scalar manifold we had in the previous section, are not sufficient any more. 

This is not specific to our setting, but is a general property
of non-linear sigma models. In such models \re{nonsig}, the kinetic term for the scalars is multiplied by a scalar-dependent symmetric tensor $g_{\alpha\beta}(\phi)$ 
interpreted as the metric on the target space
$\mathcal{M}$. As the field equations for the scalars should now also be
covariant with respect to coordinate transformations on the target
manifold, the connection on the tangent bundle $T \mathcal{M}$ should be
the Levi-Civita connection. Only in that particular case, the field
equations for the scalars will be covariant. It means that in the covariant box $\Box
\phi^\alpha+\dots =0$ the Levi-Civita connection on $T \mathcal{M}$ must be used.

To conclude, we will need to introduce a metric on the scalar manifold,
in order to be able to write down an action. This metric will also
restrict the possible target spaces for the theory.


\subsubsection{No gauged isometries}

To start with, we take the non-closure functions $\Gamma^A$ to be
proportional to the field equations for the fermions $\zeta^A$. In other
words, we ask
\begin{equation}
\frac{\delta S_{\rm hyper}}{\delta \bar{\zeta}^A}=2C_{AB}\Gamma ^B\,.
\end{equation}

In general, the tensor $C_{AB}$ could be a function of the scalars and
bilinears of the fermions. If we try to construct an action with the
above ansatz, it turns out that this tensor  has to be anti-symmetric in
$AB$ and
\begin{eqnarray}
\frac{\delta C_{AB}}{\delta \zeta^C}&=&0\,, \label{Czeta}\\
\covder_X C_{AB}&=&0\,, \label{Cq}
\end{eqnarray}
\ie,  it is independent on the fermions and covariantly constant. 
The tensor $C_{AB}$ will be used to raise and lower indices according to
the NW-SE convention similar to $\varepsilon_{ij}$
\begin{equation}
A_A = A^BC_{BA} \,,\qquad A^A = C^{AB} A_B\,,
 \label{NWSE}
\end{equation}
where, for consistency, it should satisfy $ C_{AC} C^{BC} =   \delta_A{}^B \,.$ The presence of\linebreak these $C_{AB}$ objects breaks the structure group $\Gl (r,\mathbb{H})$ to $\USp(2r-2p,2p)$. It is further possible to show \cite{Bergshoeff:2002qk} that $ C_{AB}$ can be chosen constant, but, for our purposes, the covariant constancy \re{Cq} of $ C_{AB}$ will suffice.

%
We now construct the metric on the scalar manifold as
\begin{equation}
  g_{XY} = f_X^{iA} C_{AB} \varepsilon_{ij}f_Y^{jB}\,.
 \label{defg}
\end{equation}
The above-mentioned requirement that the Levi-Civita connection should be
used (as $\Gamma_{XY}{}^Z$) is satisfied due to~(\ref{Cq}). Indeed, this
guarantees that the metric is covariantly constant, such that the affine
connection is the Levi-Civita one. On the other hand, we have already
seen that for covariantly constant complex structures we have to use the Obata
connection (and $\o_X^{AB}$ defined in \re{determineOm}). Hence, the Levi-Civita and Obata connection should coincide,
and this is obtained from demanding~(\ref{Cq}) using the Obata
connection. This makes us conclude that we can only write down an action
for a hyperk{\"a}hler scalar manifold.

We can now write down the action for the rigid hypermultiplets. It has
the following form
\begin{equation} \label{abaction}
S_{\rm hyper}= \int d^5 x  \left(-\frac 12 g_{XY} \partial_a q^X
\partial^a q^Y + \bar{\zeta}_A \slashed{\covder} \zeta^A-\frac 14 W_{ABCD}
\bar{\zeta}^A \zeta^B \bar{\zeta}^C \zeta^D\right),
\end{equation}
where the tensor $W_{ABCD}$ can be proven to be completely symmetric in
all of its indices. The field
equations derived from this action are
\begin{equation}\label{feh}
\frac{\delta S_{\rm hyper}}{\delta \bar{\zeta}^A} = 2C_{AB}\Gamma^B \,, \qq \frac{\delta S_{\rm hyper}}{\delta q^X} = g_{XY} \Delta^Y - 2\bar{\zeta}_A \Gamma^B \omega_{XB} {}^A\,.
\end{equation}
We remark that the introduction of the metric simplifies the expression of
$\Delta^X$ 
\begin{equation}
\Delta^X=\Box q^X - \bar{\zeta}^A \slashed{\partial} q^Y \zeta^B  {\cal
R}^X {}_{YAB}-\frac {1}{4} \covder^X W_{ABCD}\bar{\zeta}^A \zeta^B
\zeta^C \zeta^D\,.
\end{equation}


\paragraph{Conformal invariance.} Due to the presence of the metric, the
condition for the homothetic Killing vector~(\ref{conf-constr}) implies
that $k_X$ is the derivative of a scalar function.
This scalar function $\chi (q)$, called the hyperk{\"a}hler
potential~\cite{Swann,deWit:1998zg,deWit:1999fp}, determines the
conformal structure and should be restricted to
\begin{equation}
  \covder_X\del_Y\,\chi =\frac 32 g_{XY}\,.
 \label{DDchiisg}
\end{equation}
The relation with the exact homothetic Killing vector is
\begin{equation}
  k_X=\partial_X \chi \,,\qquad \chi =\frac 13 k_Xk^X\,.
 \label{kchi}
\end{equation}
Note that this implies that, when $\chi $ and the complex structures are
known, one can compute the metric with~(\ref{DDchiisg}), using the
formula for the Obata connection~(\ref{Obata}).

\subsubsection{With gauged isometries\label{isoaction}}

With a metric, the symmetries $k_I^X$ in \re{gauge-tr} should be
isometries, i.e.,
\begin{equation}
  \covder_X k_{YI} + \covder_Y k_{XI}=0\,.
 \label{Killingeq}
\end{equation}
This makes the requirement~(\ref{DDk}) superfluous, but we still have to
impose the tri\-holo\-morphicity expressed by either~(\ref{symm-ff})
or~(\ref{commDkJ}) or~(\ref{intJ}).

In order to integrate the equations of motion to an action we have to
define (locally) triples of `moment maps', according to
\begin{equation}
\partial_X \vec P_I =-\frac 12 \vec J{}_{XY} k_I^Y\,. \label{momentmap}
\end{equation}
The integrability condition that makes this possible follows from \re{intJ}.

In the kinetic terms of the action, the derivatives should now be
covariantized with respect to the new transformations. We are also forced
to include some new terms proportional to $g$ and $g^2$
\begin{eqnarray}
S_{\rm hyper}^g&= \int d^5x \biggl(\!-\frac 12 g_{XY} \covder_a q^X
\covder^a q^Y+ \bar{\zeta}_A \slashed{\covder} \zeta^A-\frac 14 W_{ABCD}
\bar{\zeta}^A \zeta^B \bar{\zeta}^C \zeta^D
\label{Sna}\\
& -g\left( 2 \rmi k_I^X f_{iX}^A
\bar{\zeta}_A \psi^{iI}+\rmi \sigma^I t_{IB} {}^A \bar{\zeta}_A \zeta^B-2
P_{Iij} Y^{Iij}\right)-g^2\frac 12 \sigma^I \sigma^J k_I^X k_{JX}\!
\biggr) \,,\quad \nonumber
\end{eqnarray}
[where the covariant derivatives $\covder$ now also include
gauge-covariantization proportional to $g$ as in~(\ref{covdergene})], while
the field equations have the same form as in~(\ref{feh}). Supersymmetry
of the action imposes
\begin{equation}
  k_I^X \vec J^{}_{XY} k_J^Y= 2f_{IJ}{}^K \vec P_K \,.
 \label{constraintfP}
\end{equation}
As only the derivative of $P$ appears in the defining
equation~(\ref{momentmap}), one may add an arbitrary constant to $P$. But
that changes the r.h.s.~of~(\ref{constraintfP}). One should then
consider whether there is a choice of these coefficients such
that~(\ref{constraintfP}) is satisfied. This is the question about the
center of the algebra, which is discussed
in~\cite{D'Auria:1991fj,Andrianopoli:1997cm}. For simple groups there is
always a solution. 
For Abelian theories the constant remains undetermined.
This free constant is the so-called Fayet-Iliopoulos term.

In a conformal invariant theory, the Fayet-Iliopoulos term is not
possible. Indeed, dilatation invariance of the action needs
\begin{equation}
3P^\alpha _I=k^X \partial_XP^\alpha _I\,.
\end{equation}
Thus, $P_I^\alpha $ is completely determined [using~(\ref{momentmap})
or~(\ref{comDG})] as (see also~\cite{Wit:2001bk})
\begin{equation}
  -6P^\alpha _I=k^XJ^\alpha{}_{XY}  k_I^Y= -\frac 23k^Xk^Z\J\alpha ZY \covder_Y k_{IX}\,.
\end{equation}
The proof of the invariance of the action under the complete
superconformal group, uses the equation obtained from~(\ref{comDG})
and~(\ref{momentmap}):
\begin{equation}
k^{X\alpha } \covder_X k_I^Y= \partial^Y P^\alpha _I\,.
\end{equation}
If the moment map $P^\alpha _I$ has the value that it takes in the
conformal theory, then~(\ref{constraintfP}) is satisfied due
to~(\ref{G-alg}). Indeed, one can multiply that equation with
$k_Xk^ZJ^\alpha {}_Z{}^W\covder_W$ and use~(\ref{commDkJ}),~(\ref{DDk})
and~(\ref{konRis0}). Thus, in the superconformal theory, the moment maps
are determined and there is no further relation to be obeyed, i.e., the
Fayet-Iliopoulos terms of the rigid theories are absent in this case.
\bigskip

To conclude, isometries of the scalar manifold that commute with
dilatations can be gauged (see~(\ref{comDG})). The resulting theory has an
extra symmetry group $G$, and its algebra is generated by the corresponding
Killing vectors.

The results~(\ref{eq:Vectoraction}) and~(\ref{Sna}) add up to a total
Lagrangian for matter-coupled rigid superconformal theory 
\begin{equation}
{\cal L}_{\rm total} = {\cal L}_{\rm vector} + {\cal L}^g_{\rm
hyper}\,. \label{Ltotal}
\end{equation}
This implies the following form for our scalar potential\footnote{A more general potential including also tensor multiplet contributions was given in \cite[(3.41)]{Bergshoeff:2002qk}.} 
\begin{equation}
V(\s^{ I}, q^X) = \s^{ K} C_{{I}{J}{ K}} Y_{ij}^{ I} Y^{ij\, {J}}  + \frac
12 g^2 \s^I \s^J  k^X_I k_{JX} \,, \label{potential}
\end{equation}
where $Y^{ij\, I}$ is understood to be the solution of the corresponding field equations $Y^{ijJ}C_{IJ K}\sigma ^{K}=-gP_I^{ij}$. 

This potential reflects the general form in supersymmetry that it is the
square of the transformations of the fermions \re{potentge}. The first term is the square of
the transformations of the gauginos, while the last one is the square of the transformation law of the hyperinos. 

Summarizing, in this section the actions of rigid superconformal
vector-hy\-per\-mul\-ti\-plet couplings have been constructed. The
full answer is~(\ref{Ltotal}). We found that the existence of an action
requires the presence of additional tensorial objects.
Table~\ref{tbl:action} gives an overview of what are the independent
objects to know, either to determine the transformation laws, or to
determine the action. In the next section we generalize our results to the local case.

\landscape
\begin{table}[htb]
\begin{center}
\begin{tabular}{||c||c|c||c|c|}
\hline &
 \multicolumn{2}{c||}{ALGEBRA (no action)}&
 \multicolumn{2}{c|}{ACTION}\\ \hline
 multiplets & objects & Def/restriction & objects & Def/restriction\\
\hline\hline\rule[-1mm]{0mm}{6mm}
 Vect.    & $f_{[IJ]}{}^K$ & Jacobi identities
            & $C_{(IJK)}$ & $f_{I(J}{}^H C_{KL)H} = 0\ \blacktriangle$\\[1mm]
\hline\rule[-1mm]{0mm}{11mm}
 Hyper     &  $f_X{}^{iA}$ &
\begin{tabular}{l}
  invertible and real 
  \\[2mm]
   Nijenhuis condition: $N_{XY}{}^Z=0$
\end{tabular}
 & $C_{[AB]}$ & $\covder_XC_{AB}=0$\\[1mm]
\hline \rule[-1mm]{0mm}{12mm}
 \begin{tabular}{l}Hyper  +\\ gauging \end{tabular}  & $k_I^X$ &
 \begin{tabular}{l} $\covder_X\covder_Y k^Z_I=R_{XWY}{}^Zk_I^W\ \blacktriangleleft$\\[2mm]
 $k_{[I|}^Y \partial_Y k^X_{|J]}=-\frac 12 f_{IJ}{}^K k_K^X $\\[2mm]
 ${\cal L}_{k_I} J^\alpha=0\ \blacktriangleleft $
  \end{tabular}
 & $P_I^\alpha\ \blacktriangle$ &   \begin{tabular}{l}
 $\covder_X k_{YI} + \covder_Y k_{XI}=0$\\[2mm]
 $\partial_X P_I^\alpha =-\frac 12 J^\alpha{}_{XY} k^Y_I\ \blacktriangle$\\[2mm]
 $ k_I^X J_{XY}^\alpha k_J^Y=2f_{IJ}{}^K P_K^\alpha\ \blacktriangle$
 \end{tabular}
 \\[1mm]
\hline
 \begin{tabular}{c}Hyper + \\ conformal \end{tabular}&$k^X\ \blacktriangleleft$ &
  $\covder_Y k^X =\frac32 \delta_Y{}^X\ \blacktriangleleft$&$\chi $ &
  $ \covder_X\covder_Y\, \chi =\frac32 g_{XY}$ \\ \hline
 \begin{tabular}{c}Hyper + \\ conformal +
 gauged \end{tabular} & &
 $k^Y \covder_Y k_I^X=\frac{3}{2} k_I^X$ & & \\ \hline
\end{tabular}
\caption{\it Various matter couplings with or without action. We indicate
which are the geometrical objects that determine the theory and what are
the independent constraints. The symmetries of the objects are already
indicated when they appear first. In general, the equations should also
be valid for the theories in the rows below (apart from the fact that
`hyper+gauging' and `hyper+conformal' are independent, but both are used
in the lowest row). However, the symbol $\blacktriangle$ indicates that
these equations are not to be taken over below. For instance, the moment map
$P_I^\alpha $ itself is completely determined in the conformal theory,
and it should thus not any more be given as an independent quantity. For
the rigid theory without conformal invariance, only constant pieces can
be undetermined by the given equations, and are the Fayet-Iliopoulos
terms. On the other hand, the equations and objects indicated by $\blacktriangleleft$
have not to be taken over for the theories with an action, as they are satisfied due to the Killing equation or are defined by $\chi$.
\label{tbl:action} }
\end{center}
\end{table}
\endlandscape

\section{Local multiplets and actions \la{s:loca}}

We are now sufficiently equipped to extend the rigid multiplets of the previous sections to versions invariant under local superconformal symmetry. More precisely, we will couple the vector- and hyper-multiplets to the Standard Weyl multiplet.

Since in the previous sections we have explained most of the subtleties
concerning the possible geometrical structures, we will be brief here. We
will obtain our results in two steps. First, the realization of the soft superconformal\linebreak algebra \re{algebraQQ}-\re{QScommu} of the Standard Weyl multiplet on the fields of the matter multiplets (modulo transformations that leave the Weyl multiplet invariant or eoms) determines the superconformal transformations of the latter. We will apply next a standard Noether procedure to extend the rigid supersymmetric actions to a locally supersymmetric one. This will introduce the full complications of coupling the matter multiplets to
conformal supergravity.

\subsection{Vector multiplet \la{s:vectla}}

The local supersymmetry rules are given by
\begin{eqnarray}
\d A_\m^I
&=& \frac 12 \bar{\e} \g_\m \p^I -\frac 12 \rmi \s^I \bar\e \p_\mu \,, \nonumber\\
\d Y^{ij I}
&=& -\frac 12 \bar{\e}^{(i} \slashed{D} \p^{j)  I}
+\frac 12 \rmi \bar\e^{(i} \gamma \cdot T \p^{j)  I}
- 4 \rmi \s^{I} \bar\e^{(i} \chi^{j)} \nonumber \\
& &
 -\frac 12 \rmi g \bar \e^{(i} f_{{ J}{K}}{}^{ I} 
\sigma^{ J} \psi^{j) { K}} + \frac 12 \rmi \bar{\eta}^{(i} \p^{j)  I} \,, \nonumber \\
\d \p^{i  I} &=& - \frac 14 \g \cdot \widehat{F}{}^{ I} \e^i -\frac 12\rmi \slashed{D} \s^{ I}
\e^i - Y^{ij  I} \e_j + \s^{ I} \gamma \cdot T \e^i
 + \s^{ I} \eta^i \,, \nonumber\\
\d \s^{ I} &=& \frac 12 \rmi \bar{\e} \p^{ I} \,.
\label{tensorlocal}
\end{eqnarray}
The covariant derivatives and the field strength are now also covariantized w.r.t.~the local $Q$- and $S$-transformations, \eg,
\begin{eqnarray} \label{localderiv-tensor}
D_\mu \psi^{i  I} &=& {\cal D}_\m \psi^{i  I} + \frac
14 \g \cdot \widehat{F}^{ I} \p_\mu^i + \frac 12\rmi
\slashed{D} \s^{ I} \p_\mu^i + Y^{ij  I} \p_{\mu\, j}
 - \s^{ I} \g \cdot T \p_\mu^i 
 - \s^{ I} \phi_\mu^i  \,, \nn \\
{\cal D}_\m \psi^{i  I} &=& (\partial_\mu - \frac 32 \, b_\mu +
\frac 14\, \g_{ab}\, \o_\mu {}^{ab}) \p^{i  I} - V_\mu^{ij}
\p_j^{ I} + g f_{J K}{}^{ I} A_\mu^J
\psi^{i  K} \,,\nn\\ 
\widehat{F}_{\m\n}^I &=& 2 \partial_{[\mu } A_{\nu ]}^I + g f_{JK}{}^I
A_\mu^J A_\nu^K - \bar{\p}_{[\m} \g_{\n]} \p^I + \frac 12 \rmi \s^I
\bar{\p}_{[\m} \p_{\n]}
 \,.
 \end{eqnarray}

%

The action invariant under local superconformal symmetry can be
computed by replacing the rigid covariant derivatives
in~(\ref{eq:Vectoraction}) by the local covariant
derivatives~(\ref{localderiv-tensor}) and adding extra terms proportional
to gravitinos or matter fields of the Weyl multiplet, dictated by
supersymmetry
\begin{eqnarray}\label{conf-VTaction}
&\!\!&\!\!e^{-1}\! {\cal L}_{\rm vector}^{\rm conf} = \biggl[ \bigl(-
\frac14 {\widehat F}_{\m\n}^{ { I}} {\widehat F}^{\m\n
{ { J}}} - \frac 12 \bar{\p}^{ { I}} \slashed{D} \p^{
{ J}} + \frac 13 \s^{ I} \Box^c \s^{ J} +
\frac 16 D_a \s^{ { I}} D^a \s^{ {J}}
\nn\\
&&+  Y_{ij}^{ { I}} Y^{ij { { J}}}  \bigr) \s^{ { K}} 
- \frac 43 \s^{I} \s^{ J}
\s^{ K} \left(D + \frac{26}{3} T_{ab} T^{ab} \right)
+ 4 \s^{ I} \s^{ J} {\widehat  F}_{ab}^{ K} T^{ab}  \nn \\
& & + \!\left(\!{-}\frac 18 \rmi \bar{\p}^{
{ I}} \g \cdot {\widehat F}^{ { J}} \p^{
{ K}}
-\frac 12 \rmi \bar{\p}^{i { { I}}} \p^{j { { J}}} Y_{ij}^{ { K}}  + \rmi \s^{ I} \bar\p^{ J}
\g \cdot T \p^{ K} - 8 \rmi \s^{ I} \s^{ J} \bar\p^{ K} \chi\!\right)\! \nn \\
& &
+ \frac 16 \s^{ I} \bar \p_\m \g^\m \!\left(\!\rmi \s^{ J} \slashed{D} \p^{ K} + \frac 12 \rmi \slashed{D} \s^{ J} \p^{ K} - \frac 14 \g
{\cdot} {\widehat  F}^{ J} \p^{ K}
+ 2 \s^{ J} \g {\cdot} T \p^{ K} - 8 \s^{ J} \s^{ K} \chi \!\right)\! \nn\\
&&
- \frac 16 \bar \p_a \g_b \p^{ I} \left(\s^{ J} {\widehat  F}^{ab { K}} -8 \s^{ J} \s^{ K} T^{ab} \right) -\frac 1{12} \s^{ I} \bar \p_\l \g^{\m\n\l} \p^{ J} {\widehat  F}_{\m\n}^{ K}  \nn\\
&&+ \frac 1{12} \rmi \s^{ I} \bar \p_a
\p_b  \left(\s^{ J} {\widehat  F}^{ab { K}} -8
\s^{ J} \s^{ K} T^{ab} \right) +\frac 1{48} \rmi
\s^{ I} \s^{ J} \bar \p_\l \g^{\m\n\l\r} \p_\r
{\widehat  F}_{\m\n}^{ K}  \nn \\
&& 
- \frac 12 \s^{ I} \bar \p_\m^i \g^\m \p^{j
 J} Y_{ij}^{ K} +\frac 1{6} \rmi \s^{ I}
\s^{ J} \bar\p_\m^i \g^{\m\n} \p_\n^j Y_{ij}^{ K}
-\frac{1}{24} \rmi \bar \p_\m \g_\n \p^{ I} \bar \p^{
J} \g^{\m\n} \p^{ K} \nn \\
&& +
\frac{1}{12} \rmi \bar \p_\mu^i \g^\mu \p^{j { I}} \bar
\p_i^{ J} \p_j^{ K} -\frac{1}{48} \s^{ I}
\bar \p_\m \p_\n \bar \p^{ J} \g^{\m\n} \p^{ K}
+\frac {1}{24} \s^{ I} \bar\p_\m^i \g^{\m\n} \p_\n^j
\bar\p_i^{ J} \p_j^{ K}  \nn\\
&& 
-\frac 1{12} \s^{ I} \bar \p_\l \g^{\m\n\l} \p^{ J} \bar{\p}_{\m} \g_{\n} \p^{ K} + \frac 1{24} \rmi  \s^{ I} \s^{ J} \bar \p_\l \g^{\m\n\l} \p^{ K} \bar{\p}_{\m} \p_{\n} \nn \\
& &
+ \frac 1{48} \rmi \s^{ I} \s^{ J} \bar \p_\l \g^{\m\n\l\r} \p_\r \bar{\p}_{\m} \g_{\n} \p^{ K} + \frac1{96}\s^{ I} \s^{ J} \s^{ K} \bar \p_\l \g^{\m\n\l\r} \p_\r \bar{\p}_{\m} \p_{\n}\biggr]
  C_{{ I}{ J}{ K}}  \nn\\
& &  -\frac 1{24} e^{-1} \ve^{\m\n\l\r\s} C_{IJK} A_\m^I\!
 \left(\!F_{\n\l}^J F_{\r\s}^K + f_{FG}{}^J \!A_\n^F A_\l^G \!  \left(\!{-} \frac 12 g \, F_{\r\s}^K \right. \right.\nn\\
 &&\left.\left. + \frac1{10} g^2 f_{HL}{}^K A_\r^H A_\s^L  \!\right)\!\right)
 + \frac14 \rmi g \bar{\p}^{ I}\psi
^{J}\sigma ^{K}\sigma ^{L}
f_{IJ}{}^{H}C_{H K L}
 \,,
\end{eqnarray}
where the superconformal d'Alembertian is defined as
\begin{eqnarray}
 \Box^c
\s^{ I}
&=& D^a D_a \s^{I} \nn \\[2pt]
&=& \left( \partial^a  - 2 b^a + \o_b^{~ba} \right) D_a \s^{ I}
+ g f_{J  K}{}^{ I} A_a^J D^a \sigma^{ K} -
\frac{\rmi}2 \bar{\p}_\m D^\m \p^{ I} - 2 \s^{I}
\bar{\p}_\m \g^\m \chi 
\nn \\[2pt]
& & + \frac12 \bar{\p}_\m \g^\m \g \cdot T \p^{ I} + \frac12
\bar{\f}_\m \g^\m \p^{ I} + 2 f_\m{}^\m \s^{I}
-\frac 12 g \bar\p_\m \g^\m f_{ J  K}{}^{
I}\p^{ J} \s^{ K}\,.
\end{eqnarray}

\subsection{Hypermultiplet \la{s:hypla}}

Imposing the local superconformal algebra we find the following
supersymmetry rules
\begin{eqnarray}\la{hyperlocal}
\delta q^X
&=& - \rmi \bar\epsilon^i \zeta^A f_{iA}^X \,,\nonumber\\
\widehat \delta \zeta^A &=& \!\!\!\! \frac 12 \rmi \slashed{D} q^X f_X^{iA}
\epsilon_i
  - \frac13 \gamma \cdot T k^X \!\!f^A_{iX} \epsilon^i - \!\!\!\frac 12 g \sigma^I k_I^X f_{iX}^A \epsilon^i +\!\! k^X f^A_{iX} \eta^i .
\end{eqnarray}
The covariant derivatives can be deduced from the general rule \re{covderg} as
\begin{eqnarray}
D_\mu q^X
&=& {\mathcal D}_\mu q^X + \rmi \bar{\psi}_\mu^i \zeta^A f_{iA}^X\, , \nonumber\\
{\mathcal D}_{\mu} q^X &=& \partial_\mu  q^X - b_\mu  k^X - V_\mu^{jk}
k_{jk}^X +
 g A_{\mu}^I k_I^X \,, \nn\\
D_\mu \zeta^A &=& {\mathcal D}_\mu \zeta^A - k^X f_{iX}^A \phi_\mu^i +
\frac 12 \rmi \slashed{D} q^X f_{iX}^A \psi_\mu^i + \frac13 \gamma \cdot
T k^X f_{iX}^A \psi_\mu^i + g \frac 12 \sigma^I k_I^X f_{iX}^A \psi^i_{\mu} \nonumber\\
{\mathcal D}_\mu \zeta^A &=& \partial_\mu  \zeta^A + \partial_\mu  q^X
\omega_{XB} {}^A \zeta^B + \frac14 \omega_\mu {}^{bc} \gamma_{bc} \zeta^A
- 2 b_\mu  \zeta^A + g A_\mu^I t_{IB}{}^A \zeta^B\, .
\end{eqnarray}

Similar to Sect.~\ref{s:hyper}, requiring closure of the
commutator algebra on these transformation rules yields the eoms for the fermions
\begin{eqnarray}
\Gamma_{\rm conf}^A &=& \slashed{D} \zeta^A + \frac 12 W_{CDB} {}^A
\zeta^B \bar{\zeta}^D \zeta^C - \frac 83 \rmi k^X f_{iX}^A \chi^i + 2
\rmi \gamma \cdot T \zeta^A 
\nn\\
&& -g\left(\rmi k_I^X f_{iX}^A \psi^{iI}+ \rmi \sigma^I t_{IB} {}^A
\zeta^B\right). \label{fermion_eom_local}
\end{eqnarray}
The scalar eom can be obtained from
varying~(\ref{fermion_eom_local})
\begin{equation}
\widehat \delta_Q \Gamma^A = \frac 12 \rmi f_X^{iA} \Delta^X \epsilon_i +
\frac14 \gamma^\mu \Gamma^A \bar{\epsilon} \psi_\mu - \frac14 \gamma^\mu
\gamma^\nu \Gamma^A \bar{\epsilon} \gamma_\nu \psi_\mu \,,
\end{equation}
where
\begin{eqnarray}
\Delta^X_{\rm conf}
\!\!&= \Box^c q^X - \bar{\zeta}^B \gamma^a \zeta^C D_a q^Y {\cal R}^X {}_{YBC} + \frac89 T^2 k^X  \nn\\
&\! \!+\frac43 D k^X + 8 \rmi \bar{\chi}^i \zeta^A f_{iA}^X -
 \frac14 \mathcal{D}^X W_{ABCD} \bar{\zeta}^A \zeta^B \bar{\zeta}^C \zeta^D \nn\\
&\!\!\! -g \big( 2\rmi \bar{\psi}^{iI} \zeta^B t_{IB}{}^A f_{iA}^X - k_I^Y J_Y{}^X {}_{ij} Y^{Iij} \big)  + g^2 \sigma^I \sigma^J \mathfrak{D}_Y k_I^X k_J^Y ,
\end{eqnarray}
and the superconformal d'Alembertian becomes here
\begin{eqnarray}
\Box^c q^X
&\equiv& D_a D^a q^X \nn \\
&=& \partial_a D^a q^X - \frac{5}{2} b_a D^a q^X - \frac 12 V_a^{jk}
J_Y{}^X{}_{jk} D^a q^Y + \rmi \bar{\psi}_a^i
D^a \zeta^A f_{iA}^X  \nn\\
&& + 2 f_a{}^a k^X - 2\bar{\psi}_a \gamma^a \chi k^X + 4
\bar{\psi}_a^{(j} \gamma^a \chi^{k)} k_{jk}^X - \bar{\psi}_a^i \gamma^a
\gamma \cdot T \zeta^A f_{iA}^X \nn \\
&&- \bar{\phi}_a^i \gamma^a \zeta^A f_{iA}^X + \omega_a {}^{ab} D_b q^X
-\frac 12 g \bar{\psi}^a \gamma_a \psi^I k_I^X-D_a q^Y \partial_Y k_I^X
A^{aI} \nonumber\\&&+D_a q^Y D^a q^Z \Gamma_{YZ}^X\,.
\end{eqnarray}

Note that, so far, we did not require the presence of an action. Introducing
a metric, the locally conformal supersymmetric action is given by
\begin{eqnarray}\label{conf-hyperaction}
&&e^{-1} \mathcal{L}_{\rm hyper}^{\rm conf}
= - \frac 12 g_{XY} \mathcal{D}_a q^X {\mathcal{D}^a} q^Y +\bar{\zeta}_A \slashed{D} \zeta^A + \frac 49 Dk^2 + \frac {8}{27} T^2 k^2
 \\
&& - \frac{16}{3} \rmi \bar{\zeta}_A \chi^i k^X f_{iX}^A
+ 2 \rmi \bar{\zeta}_A \gamma \cdot T \zeta^A
-\frac14 W_{ABCD} \bar{\zeta}^A \zeta^B \bar{\zeta}^C \zeta^D  \nn\\
&& - \frac29 \bar{\psi}_a \gamma^a \chi k^2 + \frac 13 \bar{\zeta}_A \gamma^a \gamma \cdot T \psi_a^i k^X f_{iX}^A
+ \frac 12 \rmi \bar \zeta_A \gamma^a \gamma^b \psi_a^i \mathcal{D}_b q^X f_{iX}^A  \nn\\
&& + \frac23 f_a {}^a k^2 - \frac16 \rmi \bar{\psi}_a \gamma^{ab} \phi_b k^2
 - \bar{\zeta}_A \gamma^a \phi_a^i k^X f_{iX}^A \nn\\
&& + \frac{1}{12} \bar{\psi}_a^i \gamma^{abc} \psi_b^j \mathcal{D}_c q^Y J_Y {}^X {}_{ij} k_X - \frac 19 \rmi \bar{\psi}^a \psi^b T_{ab} k^2 + \frac {1}{18} \rmi \bar{\psi}_a \gamma^{abcd} \psi_b T_{cd} k^2
\nn\\
&& - g\biggl( \rmi \sigma^I t_{IB} {}^A  \bar{\zeta}_A \zeta^B + 2 \rmi k_I^X f_{iX}^A \bar{\zeta}_A \psi^{iI} +\frac 12 \sigma^I k_I^X f_{iX}^A \bar{\zeta}_A \gamma^a \psi_a^i
 \nn\\
&& 
+ \bar{\psi}_a^i \gamma^a \psi^{jI}
P_{Iij}-\frac{1}{2} \rmi \bar{\psi}_a^i \gamma^{ab} \psi^j_b \sigma^I
P_{Iij}\biggr)
+2g Y^{ij}_I P_{ij}^I  -\frac 12 g^2 \sigma^I \sigma^J k_I^X
k_{JX}\,.\nn
\end{eqnarray}
No further constraints, other than those given in
Sect.~\ref{s:hyper} were necessary in this local case. In
particular, the target space is still hypercomplex or, when an action
exists, hyperk{\"a}hler. This action leads to the following dynamical
equations
\begin{eqnarray}
\frac{\delta \mathcal{S}_{\rm hyper}^{\rm conf}}{\delta \bar{\zeta}^A}
&=& 2\, C_{AB} \Gamma^B_{\rm conf}\, ,\nonumber\\
\frac{\delta \mathcal{S}_{\rm hyper}^{\rm conf}}{\delta q^X} &=& g_{XY}
\left(\Delta^Y_{\rm conf} - 2 \bar{\zeta}_A \Gamma^B_{\rm conf} \omega^Y
{}_B {}^A - \rmi \bar{\psi}_a^i \gamma^a \Gamma^A_{\rm conf}
f_{iA}^Y\right).
\end{eqnarray}

The Lagrangians~(\ref{conf-VTaction}) and~(\ref{conf-hyperaction}) are
the starting point for obtaining matter couplings to Poincar{\'e}
supergravity. This involves a gauge fixing of the local scale and
$\SU(2)$ symmetries, which will be addressed next.

\section{Poincar{\'e} supergravity \la{s:gfix}}

Our initial goal was to derive matter-coupled\footnote{For a more general discussion including tensor multiplets we direct the reader to \cite{Bergshoeff:2003yy}.} Poincar\'e SUGRA that does not require superconformal symmetry. However, the discussion up to this point was carried out within the context of superconformal symmetry where this symmetry is manifest. The last step to be undertaken in this section resides therefore in breaking the superconformal symmetry to the super-Poincar\'e subgroup. As explained in Sect.~\ref{s:SCTC}, this breaking of unwanted symmetries is achieved with the help of compensating multiplets and appropriate gauge-fixing conditions. 

For our particular model, we will use as compensators one hypermultiplet and a vector multiplet. The strategy is illustrated in Fig.~\ref{fig:matter}.  

We start thus in conformal SUGRA with $n+1$ YM vector multiplets\linebreak labeled $ I = (x,1)$, one of which will deliver the graviphoton of the Poincar\'e multiplet. We also need $r+1$ hypermultiplets whose scalars define a $4(r+1)$ hypercomplex/hyperk\"ahler target space. Four of these scalars will compensate the $D$- and $\SU(2)$-symmetries. The `$D$' gauge-fixing condition together with the eom for the field $D$ will lead to a restriction on the $n+1$ scalars in the vector sector. The upshot will then be a very special real manifold parametrized by the $n$ independent scalars. The fermions of the compensating vector- and hyper-multiplets are eliminated by the $S$-gauge corroborated  with the field equation of the auxiliary field $\chi^i$. The other auxiliary fields of the Weyl multiplet, \ie, $T_{ab}$ and $V_\m^{ij}$, are expressed via their eoms in terms of the field strengths of the gauge vectors (dressed by scalars) or the $\SU(2)$ connection of the resulting scalar manifold in the hyper sector, respectively. The process of breaking the superconformal symmetry restricts the hypercomplex/hyperk\"ahler target space to a $4r$-dimensional quaternionic/(-K\"ahler) manifold. Finally, the $b_\m$ field from the Weyl multiplet is removed by the $K$ gauge-fixing condition. 

Hereafter we only sketch this procedure, which will be extensively elaborated in \cite{Bergshoeff:2003yy}. Also the interconnection between quaternionic-like manifolds will be further clarified in \cite{Bergshoeff:2003xx}. 

Since $b_\m$ does not effectively appears in the the action, we will choose as $K$-gauge 
\begin{equation}\la{Kgaug}
b_\m =0 \,.
\end{equation}
Next, the conventional constraints \re{constraints} are solved eliminating the dependent gauge fields $f_\m{}^a$, $\phi^i_\m$ and $\widehat \o_\m{}^{ab}$, where the latter defines the Lorentz covariant derivatives, \eg, ${\cal D}_\m \psi^{iI} = (\del_\m + \ft 14 \widehat \o_\m{}^{ab} \g_{ab})\psi^{iI}$, used throughout this section. For the remaining auxiliary fields $T_{ab}$, $V_\m^{ij}$, $Y^{ij}$, $D$ and $\chi^i$ we write down their field equations, \eg,  
\begin{eqnarray}
D &:&\!\!\!\!\!
\quad C - \ft13 k^2 = 0 \,, \quad \mbox{with} \: C \equiv C_{ I J K} \s^{ I}\s^{ J}\s^{ K}\,, \:  k^2 \equiv k^{\widehat X} g_{\widehat X \widehat Y} k^{\widehat Y}, \label{eq:D-EOM} \\
\chi^i &:& \!\!\!\!\!\quad -8 \rmi C_{ I J K}\s^{ I}\s^{ J}\p^{ K}_i - \ft43 \left( C
-\ft13 k^2 \right) \g^\m \p_{\m i} + \ft{16}{3}\rmi A^{\hat A}_i
\zeta_{\hat A} = 0\,.\la{eq:chi-EOM}
\end{eqnarray}

Collecting terms in the action coming from the $\Box^c \s^{ I}$ and $f_a{}^a k^2$, which are proportional to the Ricci scalar, we notice that a conventional Einstein-Hilbert\linebreak term,  $\ft{1}{2 \kappa^2} R$, is guaranteed if we impose as $D$-gauge 
\begin{equation}\la{Dgaug}
\ft 1{24} \left( C + k^2\right) = - \ft{1}{2 \kappa^2} \,,
\end{equation}
where the constant $\kappa$ introduces the mass scale. The last condition combined with the \re{eq:D-EOM} determines  
\begin{equation}\la{Dgres}
k^2 = - \ft 9{\kappa^2} \,, \qq C = - \ft 3{\kappa^2}  
\end{equation}
whose significance will become evident shortly. In fact, the $D$ gauge-fixing condition is sufficient for a canonical Rarita-Schwinger kinetic component in the action, $- \ft 12 \bar \psi_\m \g^{\m\n\r} {\cal D}_\n \psi_\r$, if we rescale the conformal graviton to its Poincar\'e cousin $(\psi_\m^i)_{\rm conf} \ra \kappa \psi_\m^i$. The supersymmetry variation of \re{Dgaug} can be chosen as the S-gauge condition
\begin{equation}\la{Sgaug}
C_{IJK} \s^I \s^J \psi^{i K} =0\,,
\end{equation}
which actually eliminates one of the gauginos and guarantees the $Q$-invariance of the $D$-gauge.

Before imposing the $\SU(2)$-gauge, it is wiser to make an incursion in the hypermultiplet sector. In Sect.~\ref{s:hypa} we identified the hyperk\"ahler (HK) geometry as the hypermultiplet target space geometry (when a metric exists). The symmetries generated by the homothetic Killing vector $ k^{\widehat X}$ and its $\SU(2)$ partner $ \vec k^{\widehat X}$ can be used to single out four directions on the manifold. We call the coordinates along these directions $z^0$, respectively $z^\a$ such that $q^{\widehat X} = (z^0, z^\a, q^X)$. In this basis, the above Killing vectors simplify 
\begin{equation}\la{Kvectors}
k^{\widehat X} (z,q) = (3 z^0, 0,0)\,, \qq \vec k^{\widehat X} (z,q)=(0, \vec k^\a (z^\b, q), 0)\,.
\end{equation} 
Choosing an appropriate parameterization of the metric $\rmd s^2 = -\frac{(\rmd z^0)^2}{z^0} + \dots $ on the HK cone, one can relate one-to-one all the quantities of the HK geometry (vielbeins, complex structures, connections, etc.) to those of the associated quaternionic-K\"ahler manifold \cite{Bergshoeff:2003xx}.

After this intermezzo, we interpret the first equation of \re{Dgres} as fixing a\linebreak slice $z^0 = \ft 1\kappa$ on the HK cone. It turns out that 
a suitable $\SU(2)$-gauge condition, compatible also with our coordinate choice, is expressed by the vanishing of the compensating hyperinos
\begin{equation}\la{SU2g}
\zeta_i =0\,.
\end{equation}  

Yet another important consequence of the gauge-fixing conditions is to eliminate the parameters of the gauge-fixed symmetries. The $\L_K$ parameter can be computed for instance form $\d_{SC} (b_\m =0)$ as 
\begin{equation}\la{lambdak}
\Lambda_K^a = -\ft12 e^{\mu a} \left( \partial_\mu \Lambda_D + \ft12\rmi \bar\e\phi_\mu - 2 \bar\e\g_\mu\chi + \ft{\kappa}{2}\rmi \bar\eta\psi_\mu \right)\,,
\end{equation}
where $\L_D$ and $\eta$ are determined by other similar decomposition relations.

Let us return to the second eq. \re{Dgres}, which becomes
\begin{equation}\la{vsrm}
{\cal C} \equiv {\cal C}_{IJK} h^I h^J h^K =1\,
\end{equation} 
upon performing suitable rescalings 
\begin{equation}
\s^I \equiv \sqrt{\ft 3{2\kappa^2}}\, h^I\,, \qq C_{IJK} \equiv - \sqrt{\ft {8\kappa^2}3}\, {\cal C}_{IJK}\,.
\end{equation}
The coordinates $h^I$ parametrize in general an $n+1$-dimensional space whose metric 
\begin{equation}\la{metrica}
 a_{IJ} \equiv -2 {\cal C}_{IJK} h^K + ({\cal C}_{ILK} h^L h^K)({\cal C}_{IGH} h^G h^H)\,
\end{equation}
enters the kinetic term for the gauge vector (after eliminating the $T_{ab}$ fields)
\begin{equation}\la{standkin}
 - \ft 14 a_{IJ} \widehat F^I \widehat F^J\,.
\end{equation}
Furthermore, eq. \re{vsrm} determines an $n$-dimensional space known as very special real manifold. We parameterize it here by coordinate $\varphi^x$ such that $h^I(\varphi^x)$ represent now the embedding functions in the ambient $n+1$-dimensional space. 
As a result, the induced metric on the very special geometry is simply $g_{xy} = h^I_x h^J_y a_{IJ}$ and it should be positive. We denoted $h^I_x \equiv - \sqrt{\ft 32}\, \del_x h^I(\varphi)$. The name of very special geometry was coined for manifolds defined by \re{vsrm}  in \cite{Gunaydin:1984bi,deWit:1992nm} and they were later analyzed in \cite{deWit:1992cr}. One can introduce the usual geometrical objects (vielbeins, connections, etc.) for them. In particular, the vielbeins $f_x^{\tilde a}$ are used to determine the unconstrained gauginos of the Poincar\'e vector multiplets, \ie, the solutions to the $S$-gauge \re{Sgaug}
\begin{equation}\la{uncgaug}
\lambda^{i\,\tilde a} \equiv - f_x^{\tilde a} h^x_{I}\,
\psi^{i I} = h^{\tilde a}_{ I}\,
\psi^{i I}\,, \qquad \psi^{i I} = - h^{
I}_a \lambda^{i\, \tilde a} \,.
\end{equation} 

Mixing all these ingredients together we reduce the actions \re{conf-VTaction}, \re{conf-hyperaction} and the superconformal transformation rules \re{tensorlocal}, \re{hyperlocal} to their super-Poincar\'e versions. We present here only the pieces relevant for the next chapter, but the reader is encouraged to consult the full result \cite{Bergshoeff:2003yy}. Thus, the bosonic part of the action characterizing the SUGRA coupling to $n$ non-Abelian vector multiplets and $r$ gauged hypermultiplets has the following form
{\allowdisplaybreaks
\begin{eqnarray}
&&e^{-1}\mathcal{L}=
\ft1{2\kappa^2} R(\omega) -\ft14a_{{I}{J}}
\widehat{F}^{{I}}_{\mu \nu}\widehat{F}^{{J}\mu \nu} -\ft3{2\kappa^2}h_{XY}\mathcal{D}_aq^{X}\mathcal{D}^aq^{Y} \nonumber\\ &&-\ft1{\kappa^2} {\cal C}_{{I}{J}{K}} h^{{K}} h^{{I}}\Box h^{{J}} -\ft1{2\kappa^2} N_{{I}{J}{K}} h^{{K}} \mathcal{D}_ah^{{I}} \mathcal{D}^ah^{{J}} \nonumber\\
&&+ \ft{\kappa}{12}\sqrt{\ft23} e^{-1} \ve^{\m\n\l\r\s} {\cal C}_{IJK} A_\m^I \left[ F_{\n\l}^J F_{\r\s}^K + f_{FG}{}^J A_\n^F A_\l^G \left(- \ft12 g \, F_{\r\s}^K + \ft1{10} g^2 f_{HL}{}^K A_\r^H A_\s^L \right)\right] \nn \\
&&-\ft{g^2}{\kappa^4}\Big( 
+ 2\vec{P}^x\cdot\vec{P}_x - 4\vec{P}\cdot \vec{P}
+ 2\mathcal{N}_{iA}\mathcal{N}^{iA}\Big)\,, \la{Poincact}
\end{eqnarray}
} where the fermionic shifts are denoted ($\vec P_I$ are the Poincar\'e moment maps)
\begin{equation}
\begin{array}{llll}
 {\mathcal N}^{iA} \equiv
\ft{\sqrt6}{4} h^I k_I^X f_X^{iA}\,,\qq 
&\vec P \equiv  h^I \vec P_I \,, \qq & \vec P_x \equiv
 h_x^I \vec P_I  \,.
\end{array}
\end{equation}
By fermionic shifts we mean that, besides derivative terms and higher-order fer\-mio\-nic contributions, the Poincar\'e supersymmetry transformation rules of the fer\-mi\-ons contain  
\begin{eqnarray}\la{fermtra}
\d(\e) \psi_\mu^i &=& \dots  - \ft{1}{\kappa^2\sqrt6} \rmi \g_\m g P^{ij} \e_j \,,\nn\\
\d(\e) \lambda^{i\,\tilde a} &=& \dots  - \ft{1}{\kappa^2} g P^{\tilde a\, ij} \e_j \,,\nn\\
\d(\e) \zeta^A &=& \dots + \ft{1}{\kappa^2} g {\mathcal
N}_i^A \e^i\,.
\end{eqnarray}
A simple check shows that the super-Poincar\'e invariant theory \re{Poincact} derived via SCTC satisfies the generic property of the potential \re{potentge}, \ie, the potential can be written as the sum of squares of the fermionic shifts. Finally, the third line of~\re{Poincact} gives the explicit form of the ${\cal L}_{CS}$ mentioned in \re{genaction}. This concludes the SCTC construction of $D=5$ matter-coupled SUGRA.


%
%

\section{Conclusions\la{s:conmat}}

Matter-couplings of $N=2$, $D=5$ SUGRA were constructed in this chapter with the help of SCTC, completing the discussion we started in Ch.~\ref{ch:weyl}. The main targets were vector- and hyper-multiplets, but we also mentioned in passing the linear multiplet. The latter can be a useful tool in deriving vector multiplet actions  via density formulae similar to \re{densM}. Even though our ignorance of vector-tensor multiplets made the discussion of vector multiplets rather simple, the new feature that appears in that case was exemplified in the hypermultiplet sector. Thus, we noticed that the dynamics of a supersymmetric theory is not necessarily derivable from an action. The SUSY algebra can generate eoms \re{hyperinoeom}-\re{hyperseom} (via non-closure functions) without the need of an action. This is not surprising since the gaugings of $N=8$ supergravity in $5$ dimensions require in some cases an odd number of antisymmetric tensors, which prohibits the construction of an action~\cite{Andrianopoli:2000fi}. Its reduction to~$N=2$ theories should be in the class of theories extending those presented in this chapter by including vector-tensor multiplets. These latter models are also not based on an action~\cite{Bergshoeff:2002qk}.

The existence of an invariant action principle came together with the presence of extra tensorial quantities that restricts the class of allowed models (and their geometries). The geometrical objects demanded by theories with or without actions were displayed in Table~\ref{tbl:action}. After analyzing rigid superconformal theories, we promoted these symmetries to local invariances using the Noether method. However, it turned out that neither extra conditions, nor geometrical objects had to be introduced w.r.t.~the rigid conformal case. The local theories were inferred by coupling their rigid predecessors to the Standard Weyl multiplet. 

\begin{figure}[htbp]
\begin{center}
\input{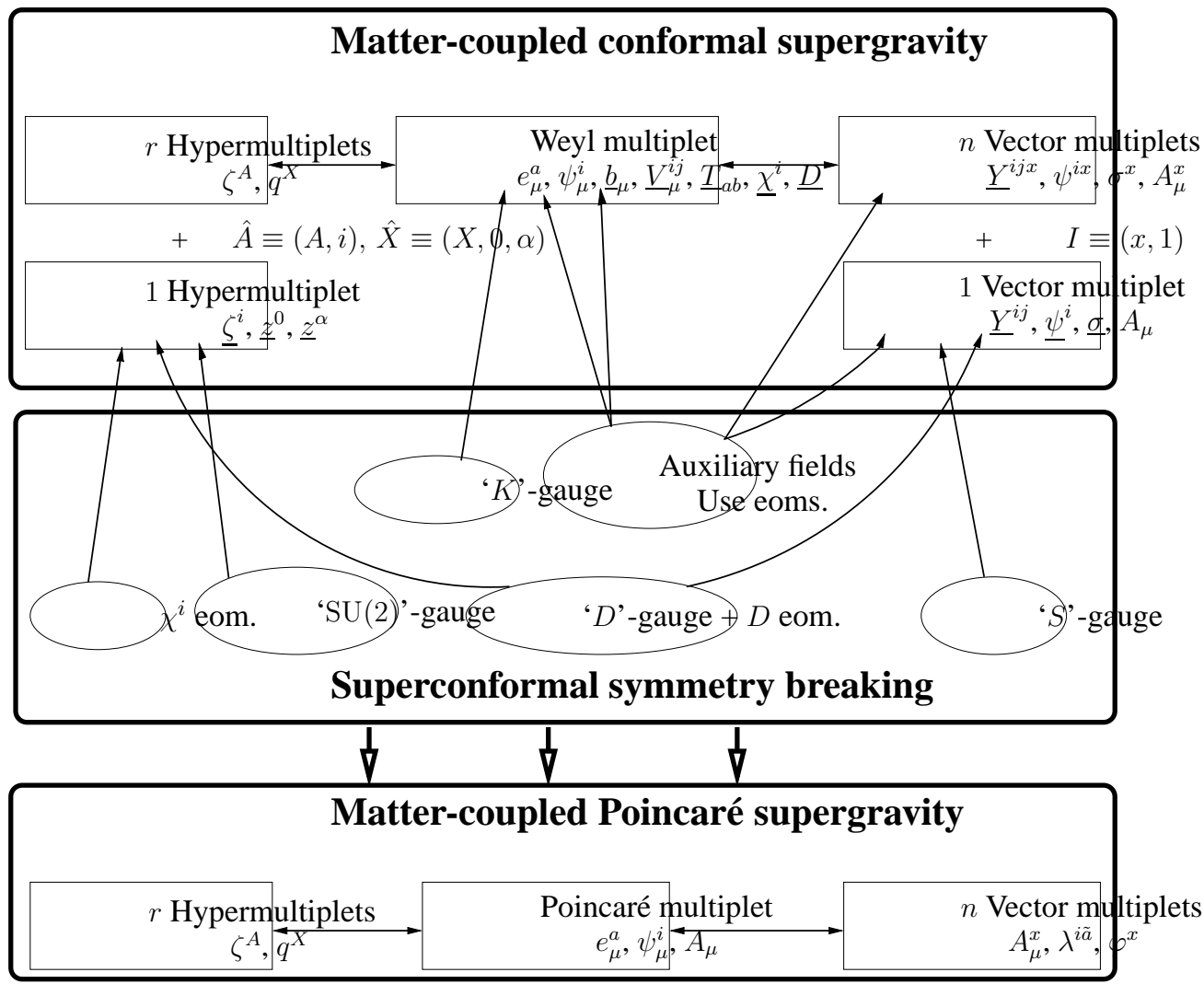}
\caption{{\it The gauge-fixing procedure: the underlined fields are eliminated when passing to Poincar\'e SUGRA.}\label{fig:matter}}
\end{center}
\end{figure}

Althogh superconformal gravity proved to be very rich in geometrical aspects (remember the hypercomplex/hyperk\"ahler geometries of the hypermultiplets), our final destination was Poincar\'e SUGRA. Therefore, we had to make a last step in that direction by eliminating the extra superconformal symmetries: $D$, $K$, $\SU(2)$ and $S$. The gauge-fixing procedure is illustrated in Figure~\ref{fig:matter}. The key elements are a compensating vector- and one compensating hyper-multiplet. The resulting matter-coupled SUGRA action \cite{Bergshoeff:2003yy} was partially given in \re{Poincact} together with the fermionic shifts \re{fermtra}. We remark at the same time that the space parametrized by the remaining hyperscalars is quaternionic/(-K\"ahler) manifold \cite{Bergshoeff:2003xx}, while in the vector sector a very special real manifold \cite{deWit:1992cr} is encountered. 

We will apply a truncated version of our result \re{Poincact} in the next chapter when we present supersymmetric solutions of the theory. Another challenging question is whether the full results \cite{Bergshoeff:2003yy}, in particular the new (off-diagonal) tensor multiplet couplings in the scalar potential, have any relevance in finding a supersymmetric Randall-Sundrum scenario or for studying domain walls and renormalization group flows in the context of the AdS/CFT correspondence.

\landscape
\endlandscape

%
\chapter{Supersymmetric solutions in $D=5$ gauged supergravity\label{ch:sols}}

Supersymmetric branes appear naturally as SUGRA solutions (Sect.~\ref{s:pbra}). They were essential for the discovery and the study of the AdS/CFT correspondence \cite{Maldacena:1998re} as well as for phenomenological model building \cite{Randall:1999ee,Randall:1999vf}. Such supersymmetric solutions of the previously constructed $D=5$ SUGRA are the subject of this chapter. In fact, supersymmetric solutions of $D=5$ pure SUGRA have been classified \cite{Gauntlett:2002nw}. Our main concern in this chapter are solutions to AdS gauged supergravity recently discussed in \cite{Gauntlett:2003fk}. 

AdS geometry is a vacuum solution of Einstein's theory with negative cosmological constant. Its boundary under the Poincar\'e patch is just Minkowski space with one dimension less. In the context of string theory compactifications, it is interesting to find solutions of gauged SUGRAs that asymptotically approach  
\be d\hat s^2_D = e^{2k z}\, ds_{D-1}^2 + dz^2\,,\label{adslike}
\ee
where the boundary ($z\ra \infty$) is generalized from Minkowski space to 
\be
ds_{D-1}^2 = dx^\mu\, dx_\mu + \lambda^{-2}\, d\Omega_2^2\,, \qq \mu = 0, \dots, D-4\,.
\label{boundmetric}
\ee
Here $d\Omega_2^2$ denotes the metric of a two-sphere $S^2$, hyperbolic
two-plane $H^2$ or two-torus $T^2$. We will refer to the metric
(\ref{adslike}), with the boundary given by (\ref{boundmetric}), as
the AdS-type metric.
    
    Brane solutions whose boundary metric is given by
(\ref{boundmetric}) have been studied in various dimensions
\cite{Romans:1992nq,Caldarelli:1998hg,Chamseddine:1999xk,Klemm:2000nj, Chamseddine:2000bk,Maldacena:2000mw,Nunez:2001pt,Gauntlett:2001qs,Sabra:2002xy,Sabra:2002qp,Cacciatori:2003kv}. They are all supported by a
set of $2$-form field strengths, which are proportional to the
volume of $\Omega^2$, together with a set of scalars and a
superpotential. Thus, these solutions can be viewed as the
magnetic duals of the AdS black hole solutions obtained in
\cite{Behrndt:1998ns,Birmingham:1998nr,Klemm:1998in,Behrndt:1998jd,Duff:1999gh,Cvetic:1999xp}. Smooth and supersymmetric
solutions of this type have hitherto been limited to $H^2$, with
the exception of a smeared NS$5$-brane on $S^2$ \cite{Nunez:2001pt}. The AdS-type spaces with spontaneously compactified boundaries (such as the boundary metric~\re{boundmetric}) provide a useful tool for studying branes wrapped on non-trivial cycles, twisted field theories and for describing flows between field theories in different dimensions \cite{Maldacena:2000mw}. 

The standard way of manufacturing supersymmetric solutions for a given \linebreak SUSY/SUGRA theory is to set all fermions to zero and search bosonic configurations for which there exist some non-vanishing (constant) supersymmetry parameters satisfying $\d \p =0$. These are first-order differential equations (called BPS eqs.), the solutions of which determine the amount of preserved supersymmetry. Our alternative way to obtain supersymmetric solutions is based on the presence of a superpotential~\re{superPot}. The superpotential method for constructing supersymmetric solutions is first illustrated in Sect.~\ref{s:MaxEin} for the simple case of $D=5$ AdS Einstein-Maxwell gravity, and then applied several times for more intricate examples: the dilatonic string (Sect.~\ref{s:singstri}) and three-charge solutions (Sect.~\ref{s:thrsol}). It boils down to first making a Kaluza-Klein reduction on $S^2$ (keeping only the singlet of the group action $\SO(3)$) and searching for the eoms of a domain wall in $D=3$. In the presence of a superpotential these are first-order equations that can be lifted to the corresponding equations of a magnetic string in $D=5$. The latter equations are solved either analytically or numerically.

This chapter contains some of the results in \cite{Cucu:2003bm,Cucu:2003yk} restricted to $D=5$ (except Sect.~\ref{s:nonBH} inspired by \cite{Behrndt:1998jd,Cvetic:1999xp}). 

%
%

\section{Non-dilatonic string\la{s:MaxEin}}

         Let us consider the $D=5$ (AdS) Einstein-Maxwell theory with the Lagrangian
\be
\hat e^{-1}\, \hat{\cal L}= \hat R - \fft{1}{4}\, \hat F_{\sst{(2)}}^2 +
\Lambda\,,
\ee
where $\hat e=\sqrt{-\hat g}$, $\hat F_{\sst{(2)}} =d\hat A_{\sst{(1)}}$
and the cosmological constant is given by $-\Lambda < 0$. We perform a
dimensional reduction on a $2$-dimensional space with the ansatz
\bea
d\hat s_{5}^2 &=& e^{2\alpha\, \varphi}\, ds_{3}^2 +
\lambda^{-2}\, e^{2\beta\, \varphi}\, d\Omega_2^2\,,\nn\\
\hat F_{\sst{(2)}}&=&\epsilon\, m\,\lambda^{-2}\,\Omega_{\sst{(2)}}\,, \label{reduce1}
\eea
where $m\,, \, \lambda$ are some constants and $\alpha = - \sqrt{1/3}$ and $\beta = - \alpha /2$.
The reduction is consistent\footnote{Note that a similar discussion for arbitrary dimension $D$ and $n$-dimensional reduction can carried out, \eg, \cite{Cucu:2003yk}.} and the resulting $3$-dimensional
Lagrangian becomes
\be e^{-1}\, {\cal L} = R - \ft12 (\del\varphi)^2 - V(\varphi)\,,
\ee
where the scalar potential is given by \cite{Bremer:1998zp}
\be
V= \ft12 \epsilon^2 m^2 e^{4\,\alpha\,\varphi} -
2 \epsilon\,
\lambda^2\, e^{3\alpha\, \varphi}-\Lambda\,
e^{2\alpha\,\varphi}\,.\label{nondilsp}
\ee
The constant $\epsilon =1$, $-1$ or $0$, corresponds respectively to
$d\Omega_2^2$ being the metric for a unit $2$-sphere, hyperbolic
$2$-plane or $2$-torus. We would like to express the
scalar potential (\ref{nondilsp}) in terms of a superpotential
$W$. Namely\footnote{Up to some overall factor and a rescaling of the scalar it reproduces \re{superPot}.},
\be\la{maxEpot}
V=(\fft{\del W}{\del \varphi})^2 -  W^2\,.
\ee
This is indeed possible if the superpotential has the following form 
\be
W=\sqrt{\fft{3}{2}} \, (\epsilon\, m\, e^{2\,\alpha\, \varphi} +
2 \lambda^2\, m^{-1}\, e^{\alpha\, \varphi})\,, \label{W}
\ee
and provided that the constraint $\Lambda=\fft{4\, \lambda^4}{
m^2}$ is satisfied.  Note that, in the case of~$\epsilon=0$ which
corresponds to a 2-torus, the $2$-form field strength vanishes and we recover Einstein's theory with negative cosmological constant. 

    We can then construct a domain wall solution in $D=3$ dimensions,
with the metric ansatz
\be ds_{3}^2 = e^{2A}\, dx^{\mu}\, dx_\mu+ dy^2\,,\qq \mu =0,1\,,
\ee
and with $A$ and $\varphi$ assumed to depend only on the
transverse coordinate $y$. The equations of motion are given by
\bea\la{maxEeom}
&&\varphi'' + 2\, A'\, \varphi' = \fft{\del V}{\del
\varphi}\,, \qquad A'' + 2\, A'^2 = -V\,,\nn\\
&&\ft12 \varphi'^2 - 2\, A'^2 = V\,,
\eea
where a prime denotes a derivative with respect to $y$. The advantage of the superpotential and its connection to the usual potential \re{maxEpot} is that the second-order equations \re{maxEeom} can be solved by the first-order equations
\be
\varphi'=\sqrt2\, \fft{\del W}{\del \varphi}\,,\qquad
A'=-\fft{1}{\sqrt2}\, W\,. \label{fo}
\ee

    We lift these equations of motion back to five dimensions,
expressed in terms of the coordinate $\rho$ such that the solution \re{reduce1}
takes the form
\bea ds_5^2 &=& e^{2u}\, dx^\mu\, dx_\mu + e^{2v}\, \lambda^{-2}\,
d\Omega_2^2 + d\rho^2\,,\nn\\
F_\2&=& \epsilon m\,\lambda^{-2}\, \Omega_\2\,.
\label{metricansatz} \eea
The first-order equations for the functions $u$ and $v$ are given
by
\be \fft{du}{d\rho}=-\fft{2\lambda^2 - \epsilon\,m^2\,
e^{-2v}}{2 m\, \sqrt{3} }\,,\qquad
\fft{dv}{d\rho}=-\fft{\lambda^2 + \,\epsilon\, m^2\,
e^{-2v}}{ m\,\sqrt{3}}\,. \ee 
These equations can be solved directly and, in terms of  
a new coordinate \newline $r=\lambda^{-1}\, {\rm e}^{v}$, the
magnetic string solution can be expressed as
\bea
ds_5^2&=&(\lambda\, r)^2
H^{\frac{3}{2}}dx_{\mu}^2+\frac{dr^2}{(\lambda\, r)^2 H^2}+r^2
d\Omega_2^2\,,\nn\\
F_\2&=&\epsilon\, m\, \lambda^{-2}\Omega_\2 =
\epsilon\, \sqrt{4\, \Lambda}\, \Omega_\2, \label{sabra}
\eea
where $H=1+ \frac{4 \epsilon}{\Lambda r^2}$.  It is interesting
to note that, once the cosmological constant is fixed, there is no
free parameter associated with the $F_\2$ field strength.  This is
rather different from the standard brane solution, in which the charge
of a supporting field strength is typically an arbitrary integration
constant. In the present case, it is uniquely determined by a
supersymmetric condition, which translates into the condition
presented below (\ref{W}). For vanishing $\epsilon$, the above
solution is purely gravitational with locally AdS$_5$ geometry. The singularity structure of our solutions is such that the $\epsilon=0$ and $\epsilon=-1$ cases are regular while for $\epsilon=1$ the solution is singular.


    These solutions are among those that have already been found in
\cite{Sabra:2002xy} by solving the D-dimensional equations of motion
directly\footnote{To recover the notation of \cite{Sabra:2002xy}, let
$\lambda=\ell$ and $m=\sqrt{3 q}\, \ell^2$.} or in \cite{Cucu:2003yk} with the superpotential approach, and the
corresponding Killing spinors have been determined in
\cite{Sabra:2002qp}. The case of the $D=5$ magnetic string presented in this section, as well as its $D=4$
`cosmic monopole' associate, can be embedded in $N=2$ gauged supergravity
\cite{Romans:1992nq,Chamseddine:1999xk,Chamseddine:2000bk}.

\section{Single-charge string in AdS gauged supergravity\la{s:singstri}}
        
	The non-dilatonic solution analyzed above appears only as a particular situation in gauged supergravities. In general, the $2$-form field strength 
has a dilaton coupling (in $D\ge 6$ gauged supergravities this is always true) as can also be seen from~\re{Poincact}.	


          The relevant Lagrangian can be expressed as
\be \hat e^{-1}\, \hat {\cal L} = \hat R - \ft12(\del\phi)^2 -
\fft14 e^{-a_1\, \phi}\, \hat F_\2^2 - \hat V(\phi)\,, \ee
where $a_1^2 = 8/3$ for $D=5$.\footnote{This particular value of $a_1$ \cite{Lu:1995cs} characterizes the single-charge string. For other binding states of this solution, \eg, the two-equal charge string \cite{Cucu:2003yk}, the constant $a_1$ changes according to \cite{Lu:1995cs}. Also the factor multiplying the second term of the r.h.s.~of the following relation is dimension dependent, \ie, it is $(D-1)/2(D-2)$.} The scalar
potential can be expressed in terms of a superpotential
\be \hat V=(\fft{\del\hat
W}{\del \phi})^2 - \fft{2}{3}\,\hat W^2\,,
\ee
and $\hat W$ is given by
\be \hat W=2g\,\Big(\fft{1}{a_1}\, e^{\ft12 a_1\, \phi} -
\fft{1}{a_2}\, e^{\ft12 a_2\, \phi}\Big)\,.
\ee
The value of $a_2$ determined by examining the
gauged supergravities in $D=5$ is given by $a_2 = - 4/(3 a_1)$ and $g$ is the coupling constant.

We can reduce the theory on $d\Omega_2^2$ with the ansatz \re{reduce1} that yields a $3$-dimensional Lagrangian and the potential 
\bea
e^{-1}\, {\cal L}&=& R-\ft12{\del\phi}^2 -
\ft12(\del \varphi)^2 -V\,,\nn\\
V&=&\ft12\epsilon^2 m^2\, e^{-a_1\, \phi +
4\, \alpha\, \varphi} -
2\epsilon\,\lambda^2\, e^{3\,\alpha\,\varphi} +
\hat V\, e^{2\, \alpha\,\varphi}\,.
\eea
As in the previous discussion, $\epsilon=1$ for $S^2$ and
$\epsilon=-1$ for $H^2$.  Now it amounts to finding a superpotential
$W$ such that
\be V=(\fft{\del W}{\del \phi})^2 + (\fft{\del W}{\del \varphi})^2
- W^2. \ee
The superpotential exists, provided that $\lambda$, $g$ and $m$ satisfy
%
\be
\lambda^2=\sqrt2\, a_1^{-1}\, m\, g\,,
\ee
and if $a_2$ is related to $a_1$ as above. Then, the superpotential reduces to
\be W=\ft{1}{\sqrt2}\,\epsilon\,m\,e^{-\ft12 a_1\, \phi +
2\,\alpha\, \varphi} + \, e^{\alpha\,\varphi}\,\hat W
\,.\label{gensup1} \ee
The first-order equations for the $3$-dimensional system are
therefore given by
\be \varphi'=\sqrt2\, \fft{\del W}{\del \varphi}\,,\qquad
\phi'=\sqrt2\, \fft{\del W}{\del \phi}\,,\qquad
A'=-\fft{1}{\sqrt2 }\, W\,. \label{genfo} \ee

      Having obtained the first-order equations for the
$D=3$ domain walls, it is of interest to lift these
equations to $D=5$ dimensions so that they correspond to magnetic strings.  The string solutions have the same structure \re{metricansatz}
with functions~$u$ and $v$ and the dilaton $\phi$ depending only on
the coordinate $\rho$. They are subject to the following first-order equations
\bea
\fft{d\phi}{d\rho}&=&\sqrt2 \Big(-\ft1{2\sqrt2}\epsilon\, m\,a_1\,
e^{-\ft12a_1\, \phi - 2v} + \fft{d\hat W}{d\phi}\Big)\,,\nn\\
\fft{dv}{d\rho}&=&-\fft{1}{3\sqrt2}\,\Big(\sqrt2 \, \epsilon\,
m\,e^{-\ft12a_1\, \phi - 2v} + \hat W\Big)\,,\nn\\
\fft{du}{d\rho}&=&\fft1{3\sqrt2}\, \Big(\ft1{\sqrt2}\epsilon\,
m\,e^{-\ft12a_1\, \phi - 2v}-\hat W\Big)\,.\label{gauged1}
\eea
Note that since the $(\phi ,v)$ fields form a closed system,
eqs. (\ref{gauged1}) can be solved explicitly by making a coordinate
transformation $d\rho={\rm e}^{-\frac{a_2}{2} \phi}\, dy$ and
defining
\be F\equiv {\rm e}^{\frac{1}{2}(a_2-a_1)\phi}\,,\quad G\equiv
{\rm e}^{\frac{1}{2}(a_1+a_2)\phi +2v}\,. \ee

A straightforward computation determines the metric and the dilaton of\linebreak the $D=5$ single charge magnetic string as
\bea ds_5^2 &= (g\, r)^2\, H^{-\ft{1}{3}}\Big(dx_\mu^2 + (1 -
\fft{\epsilon\, m}{\sqrt3 \, g \, (g\, r)^2})\, \lambda^{-2}\,
d\Omega_2^2\Big) + H^{\ft{2}{3}}\, \fft{4dr^2}{3 g^2\,
r^2}\,,\label{singlec}\\
&{\rm e}^{\sqrt{\frac{3}{2}}\phi} = H \equiv \fft{1 - \fft{\epsilon\,
m}{\sqrt3\, g (g r )^2}}{ 1 + (c_1 - \fft{2\,\epsilon\, m}{\sqrt3\, g }\log
(g\, r))\, (g\, r)^{-2}}\,, \label{Hsinglec}
\eea
where, for convenience, we introduced a new coordinate $g r= {\rm e}^{-\sqrt{3}g\, y /2}$ and we denoted the integration constant $c_1$.  

To understand the solution better, it is
instructive to write the\linebreak function $H=\wtd H/W$, where the
definitions of $\wtd H$ and $W$ can straightforwardly be read off
from (\ref{Hsinglec}). Then, the metric can be expressed as
\be ds_5^2 = \wtd H^{-\ft{1}{3}}\, W^{\ft{1}{3}}\,\Big[ (g\,
r)^2\, (dx^\mu\, dx_\mu + \wtd H\, \lambda^{-2}\, d\Omega_2^2) +
\wtd H\, W^{-1}\, \fft{4dr^2}{3 g^2\, r^2}\Big]\,.
\label{gendmetricsol} \ee
Thus, the solution can be viewed as an intersection of a domain wall,
characterized by the function $H$, and a $1$-brane, characterized
by the function $\wtd H$.

         Clearly, the asymptotic infinite region of the metric is
$r\rightarrow \infty$, in which\linebreak case $H\rightarrow 1$ and the metric
behaves as
\be ds_5^2=(g\,r)^2\, (dx^\mu\, dx_\mu + \lambda^{-2}\,
d\Omega_2^2) + \fft{dr^2}{3 g^2\, r^2}\,.\label{larged1} \ee
If $\epsilon=0$, in which case $d\Omega_2^2$ is the metric of a
2-torus, then the above metric describes locally AdS$_5$
spacetime. For $\epsilon=\pm 1$, the metric \re{larged1} can be viewed as a
domain wall wrapped on $\Omega^2$.

       The full solution with $\epsilon=0$, also obtained in
\cite{Cvetic:1999xx}, describes a domain wall.  This can be viewed as a
distribution of branes from the ten or eleven-dimensional point of
view. Hence, this solution corresponds to the Coulomb branch of the
superconformal Yang-Mills theory on the boundary
\cite{Cvetic:1999xx,Kraus:1998hv,Freedman:1999gk}. When $c_1 =0$ the solution is again locally AdS$_5$ while the sign of a non-vanishing $c_1$ determines whether the horizon (coinciding with the singularity) is at finite or vanishing distance.
 
       For $\epsilon=-1$, the time component of the metric vanishes for some finite $r$, suggesting that the horizon and the singularity coincide, while for $\epsilon=1$, the solution is always singular, but the position and the nature of the singularity depends on the values of the parameters $c_1$, $m$, $g$. 
       


The solution \re{singlec}-\re{Hsinglec} was also obtained in \cite{Maldacena:2000mw} by looking at the BPS equations. Its importance in the AdS/CFT context was analyzed therein. We move to study multiple-charge solutions of $D=5$ gauged SUGRA.


\section{Three-charge solutions\la{s:thrsol}}

It was argued \cite{Schwarz:1983qr,Gunaydin:1985fk,Kim:1985ez} that type IIB SUGRA reduced on $S^5$ gives rise to maximal SUGRA in $D=5$ with an $\SO(6)$ Yang-Mills gauge group. We will be interested below in a consistent truncation of that theory to $N=2$ SUGRA. This truncation breaks $\SO(6)$ to its $\U(1)^3$ Cartan subgroup. The corresponding Abelian gauge fields are the graviphoton and two gauge vectors in two accompanying vector multiplets. According to our discussion in Sect.~\ref{s:gfix}, the bosonic sector must also contain two scalars. The advantage of such a truncation is the simplification of the BPS equations that makes them tractable. The bosonic sector of this $\U(1)^3$ theory can be further truncated as we will explain below. 

The Lagrangian of this $N=2$, $D=5$ gauged supergravity coupled to two vector multiplets is given by
\be e^{-1}{\cal L}_5=\hat R - \ft12 (\del \phi_1)^2 -\ft12 (\del
\phi_2)^2 - \fft14\sum_{i=1}^3 X_i^{-2} (\hat F_\2^i)^2 - \hat V +
e^{-1}\, \ft14 \epsilon^{\mu\nu\rho\sigma\lambda}\, \hat
F^1_{\mu\nu}\,\hat F^2_{\rho\sigma}\, \hat A^3_\lambda\,, \la{Lag2vec}\ee
with the scalar potential 
\be
\hat V=-4g^2 \sum_{i=1}^3
         X_i^{-1}\,.\label{stdscalarpot}
\ee
The quantities $X_i$ (the analogous of $h^I$ in~\re{vsrm}) satisfy $X_1 X_2 X_3 =1$ and are parametrized as
\bea\la{paramX}
X_i=e^{\ft12\vec a_i\cdot\vec \phi}\,, \q \vec a_1=(\sqrt2\,, \fft{2}{\sqrt6})\,, \quad \!\!
\vec a_2=(-\sqrt2\,,\fft2{\sqrt6})\,,\quad\!\! \vec
a_3=(0,- \,\fft4{\sqrt6}).
\eea

An alternative way to derive this truncation is offered by the $K_3 \times S^1$ compactification of the heterotic string \cite{Antoniadis:1996cy}.

\subsection{Non-extremal black hole\la{s:nonBH}}

By studying directly the eoms of the supersymmetric system~\re{Lag2vec} a non-extremal (electrical) three-charge BH solution was found in \cite{Behrndt:1998jd}
\begin{eqnarray}
d s_5^2 &=& - (H_1H_2H_3)^{-2/3} f d t^2 + (H_1H_2H_3)^{1/3} (f^{-1} d r^2 + r^2 d \O^2_{3})\,,\nn\\
X_i &=& H_i^{-1} (H_1H_2H_3)^{1/3}\,, \qq A^i_\1 = \ep\, (1 - H_i^{-1}) \coth{\b_i} \, d t\,,\la{3qBH}
\end{eqnarray}
where the non-extremality parameter $\m$ enters in
\begin{equation}
f= \ep -\frac{\mu}{r^2} + g^2 r^2 H_1 H_2 H_3\,, \qq H_i = 1 + \frac{\mu \sinh^2{\b_i}}{\ep r^2}\,.
\end{equation}
The parameter $\ep$ determines again the constant curvature $1$, $0$ or $-1$ of the foliating spaces (with unit metric $d \O^2_{3}$) in the transverse direction $S^3$, $T^3$ or $H^3$. The\linebreak case $\ep =0$ is a bit tricky since one has first to rescale $\sinh^2{\b_i} \ra \ep \sinh^2{\b_i}$ and then take the limit $\ep \ra 0$. Consequently, the gauge field supported in the $\ep = 0$ case reads $A^i_\1 =  (1 - H_i^{-1})  d t / \sinh{\b_i}$. The existence of horizons and the thermodynamics of this solution were discussed in \cite{Behrndt:1998jd}.

\subsection{String solution\la{s:thrstri}}

We concentrate in this subsection on the magnetically-charged string solution  of~\re{Lag2vec}. Our approach is again based on the superpotential technique of Sect.~\ref{s:MaxEin}. Therefore, we notice that the scalar potential $\hat V$ in (\ref{stdscalarpot}) can be also expressed in terms of the superpotential $\hat W$, given by
\be
\hat W=\sqrt2\, g\, \sum_i X_i\,.
\ee

We now reduce the theory on $S^2$ with the previous metric ansatz \re{reduce1}, while the three $U(1)$ $2$-form are taken of the form
field strengths is given by
\be\la{chargepa}
F_\2^i= \epsilon\,  m_i\, \lambda^{-2}\, \Omega_\2\equiv
\ft12 \epsilon\, q_i\, g^{-1}\, \Omega_\2\,.
\ee
The resulting $D=3$ scalar potential
\be
V=\ft12 \epsilon^2 (\sum_i m_i^2\, X_i^{-2})\,
e^{4\alpha\, \varphi} -
2 \epsilon \lambda^2\, e^{3\alpha\, \varphi} +
\hat V\,e^{2\alpha\,\varphi}
\,\label{stdd3pot}
\ee
can be inferred from a corresponding superpotential 
\be
W=\ft{\epsilon}{\sqrt2}(\sum_im_i\,X^{-1}_i)\,
e^{2\alpha\, \varphi} +
\hat W\, e^{\alpha\, \varphi}\,,
\ee
provided that the following constraint is satisfied
\be
\lambda^2=g\sum_i m_i\,.
\ee
Thus, the charge parameters $q_i$ satisfy $q_1 + q_2 + q_3=2$. After writing down the first-order equations for the three-dimensional system and lifting them back\linebreak to $D=5$, we obtain the first-order equations describing the three-charge magnetic string in $D=5$
\bea
\fft{d\vec \phi}{d\rho} &=&
\sqrt2\,\Big(-\fft{\epsilon}{2\sqrt2}\, (m_1\, \vec a_1\, X_1^{-1}
+ m_2\, \vec a_2\, X_2^{-1} + m_3\, \vec a_3\, X_3^{-1})\,e^{-2v}
+
\fft{d\hat W}{d\vec \phi}\Big)\,,\nn\\
\fft{dv}{d\rho} &=& -\fft{1}{3\sqrt2}\,
\Big(\sqrt2\, \epsilon\,(m_1\, X_1^{-1} +
m_2\, X_2^{-1} + m_3\, X_3^{-1})\, e^{-2v} + \hat W\Big)\,,\nn\\
\fft{du}{d\rho} &=& \fft{1}{3\sqrt2}\, \Big(
\fft{\epsilon}{\sqrt2}\,(m_1\, X_1^{-1} + m_2\, X_2^{-1} + m_3\,
X_3^{-1})\, e^{-2v} - \hat W\Big)\,. \label{5} \eea

A 3-charge string with constant scalars was found in
\cite{Klemm:2000nj}. On the other hand, for $m_1=m_2=m_3$, it is
consistent that $\phi_1$ and $\phi_2$ vanish and this system
reduces to the $D=5$ system of Sect.~\ref{s:MaxEin}, with the solution given
by (\ref{sabra}) \cite{Sabra:2002xy}.\linebreak For $m_1=m_2$ we can consistently
set $\phi_1=0$ while keeping $\phi_2$. In particular,\linebreak if
$m_1=m_2=0$, this system reduces to the $D=5$ system of Sect.~\ref{s:singstri}. After\linebreak taking $g\rightarrow \sqrt{3/4}\, g$, the exact solution
is displayed in formula \re{singlec}-\re{Hsinglec}.

For $m_1=m_2=m_3$, it has recently been found that one can
consistently\linebreak set $\phi_1=\sqrt{3} \phi_2$, with an exact solution
of the form \cite{Cacciatori:2003kv}
\bea
ds_5^2&=&H^{-2/3}\Big[ (g\, r)^2 ({\rm e}^{\frac{\epsilon\,
m}{2g^3\, r^2}}dx_{\mu}^2+H^2 \lambda^{-2} d\Omega_2^2)+H^2
\frac{dr^2}{(g\, r)^2} \Big]\,,\nn\\
{\rm e}^{-\sqrt{6}\phi_2}&=& H \equiv 1-\frac{\epsilon\, m}{2g^3\, r^2}\,, \label{new}
\eea
where we made the coordinate transformation $ d\rho = - {\rm e}^v d r/(g\, r)^2 $.

Because analytical solutions for arbitrary $m_i$ are difficult to find, it
is nevertheless instructive to study the system with a numerical approach.

\subsection{Asymptotic boundary region\la{s:asybou}}

A coordinate transformation $d\rho=dy\, {\rm e}^u$ brings the metric \re{metricansatz} in the form  
\be ds_5^2={\rm e}^{2u} (dx^{\mu} dx^{\nu}\eta_{\mu \nu}
+dy^2)+{\rm e}^{2v}\lambda^{-2}d\Omega_2^2 \,. \label{expansion} \ee
We take as boundary conditions $u(y)\sim v(y)\sim -{\rm log}\, y$ and $\phi
\rightarrow 0$ for small $y$. The leading terms in the Taylor
expansion of the solution of (\ref{5}) read
\bea e^{2u} &=& \fft{1}{g^2\, y^2}\, ( 1 +
\fft{\epsilon\, (m_1 + m_2+m_3)\, g}{9 c^2}\, y^2 + \cdots)\,,\nn\\
e^{2v} &=& \fft{c^2}{g^2\, y^2}\, ( 1 -
\fft{7\epsilon\, (m_1+m_2+m_3)\, g}{18c^2}\, y^2 + \cdots)\,,\nn\\
e^{\ft{\phi_1}{\sqrt2}} &=& 1 - \fft{\epsilon\, (m_1 -
m_2)\,g}{2c^2}\, y^2\, {\rm log}\, y + \cdots\,,\nn\\
e^{\fft{\phi_2}{\sqrt{6}}} &=& 1 - \fft{\epsilon\, (m_1 +
m_2-2m_3)\,g}{6c^2}\, y^2\, {\rm log}\, y + \cdots\,,
\label{3cd5asym} \eea
where the integration constant $c$ measures the relative scale size of $dx^{\mu} dx_{\mu}$\linebreak and $d\Omega_2^2$. The scalar fields correspond in the dual field theory to operators of dimension $\Delta=2$ \cite{Maldacena:2000mw}. Contrary to $D=6,7$ \cite{Cucu:2003yk}, the appearance of the non-normalizable solutions $y^2\, {\rm log}\, y$ for $\phi_1$ and $\phi_2$ imply that
the dual operators are turned on, provided that the values of
$m_i$ are such that the fields are non-vanishing \cite{Maldacena:2000mw}.
Next, we examine how these solutions flow from the above boundary
into the bulk.

\subsection{AdS$_3\times H^2$ and AdS$_3\times S^2$\la{s:fixpoi}}

For $\ep=\pm 1$, there exists a fixed-point solution, given by 
\bea e^{\sqrt2\, \phi_1} &=& \fft{m_1}{m_2}\,
\Big(\fft{m_3+m_2-m_1}{m_3 - m_2 + m_1}\Big)\,,\qquad e^{\sqrt6\,
\phi_2} = \fft{m_1\,m_2\,(m_3^2 - (m_1-m_2)^2)}{
m_3^2\,(m_1 + m_2 - m_3)^2}\,,\nn\\
e^{-2v}&=&-\epsilon\,g\,\Big(\fft{(m_1 + m_2 - m_3)(m_3^2 -
(m_1-m_2)^2)}{m_1^2\,m_2^2\,m_3^2}\Big)^{\ft13}\,,\nn\\
u&=& -g\, e^{\fft{\phi_2}{\sqrt6}}\, \Big(\cosh(\phi_1/\sqrt2) +
\ft12 e^{-\sqrt{\ft32}\,\phi_2}\Big)\, \rho
\equiv \fft{-2}{R}\, \rho\,.\label{ads3omega2}
\eea
The reality condition of the solution implies some restrictions on the parameters~$m_i$. When $m_1=m_2$ for instance, $d\Omega_2^2$ should be the
$H^2$ metric if $2m_1>m_3$ and the $S^2$ metric for $2 m_1< m_3$.\footnote{AdS$_3\times S^2$ solutions
were also recently found in \cite{Cacciatori:2003kv} in the BPS construction.}  If any of the $m_i$ vanish, there is no fixed-point
solution, except when one vanishes with the remaining two being equal.
The AdS$_3$ radius $R$ depends on $g$ and two of the three charge
parameters.

       It is of interest to study whether this fixed-point solutions
lie in the IR or UV region.  If AdS$_{3}\times \Omega^2$ lies in the IR region, then it can smoothly flow to the UV region of the
AdS$_{5}$-type solution (\ref{3cd5asym}).  On the other hand, if
AdS$_{3}\times \Omega^2$ lies in the UV region, then it must instead flow into a singularity.

\subsection{Interpolating from AdS$_3 \times H^2$ to AdS$_5$\la{s:interH}}

We look first at AdS$_3\times H^2$ that can occur either as an
asymptotic geometry or on the horizon. In order to demonstrate
this, we take $m_1=m_2=2$ and $m_3=3$. For this choice of charge
parameters, we can set $\phi_1=0$.  For simplicity, we also set~$g=1$.  As an asymptotic boundary geometry, the Taylor expansions
of the metric components and the remaining scalar are
\bea
e^{2u}&=& \ft{8\, 2^{1/3}}{25}\, y^{-2} (1 + c\, y^n + \cdots)\,,\nn\\
e^{2v} &=& 2\, 2^{1/3}\, (1 -\ft12 c\,y^n
+\cdots)\,,\nn\\
e^{\ft{\phi_2}{\sqrt6}} &=& 2^{1/3}\, (1  -\ft1{16}(11  \pm 3\sqrt{17})
\, y^n + \cdots)\,,\la{ads3h2ads5}
\eea
where the coordinate $y$ is defined to be $d\rho=e^{u}\, dy$ and
$c$ is an arbitrary integration constant.  The constant $n$ can
take two values: $n=\ft15(-3 \pm \sqrt{17})$ according to corresponding signs in \re{ads3h2ads5}.  Since there is no AdS$_5$-like solution in the near region, these solutions will flow into a singularity in the bulk.

\begin{figure}
   \epsfxsize=4.0in \centerline{\epsffile{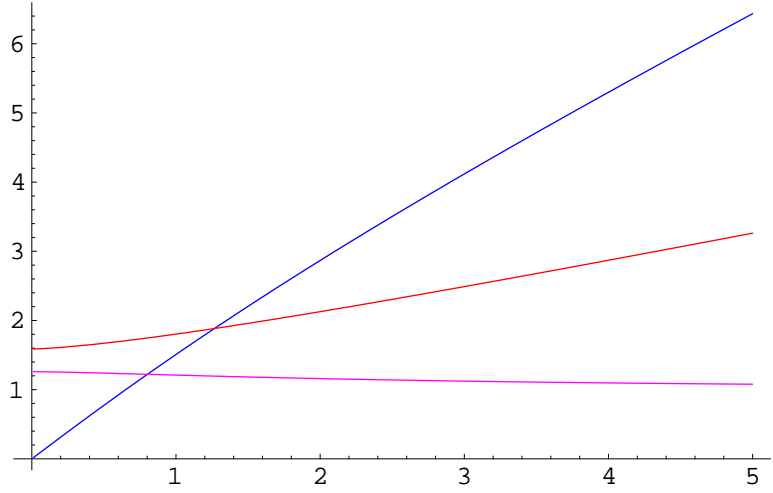}}
   \caption[FIG. \arabic{figure}.]{{\it $e^u$ (blue), $e^{v}$
   (red) and $e^{\phi_2/\sqrt{6}}$
   (purple) for a smooth solution running from AdS$_3\times
   H^2$ at the horizon to an asymptotic AdS$_5$-type geometry.
   $m_1=m_2=2$, $m_3=3$, $g=1$, and $c=-0.05$.} \label{3cd5h2fig}}
\end{figure}

     On the other hand, for the solution with AdS$_3\times H^2$ at
its horizon, we find the Taylor expansion of the next leading
order to be
\bea
e^{2u} &=& \ft{25}{8\, 2^{1/3}}\,r^2\, (1 + c\, r^n + \cdots)\,,\nn\\
e^{2v} &=& 2\, 2^{1/3}\, (1 -\ft1{26}(33 \pm 10\sqrt{17})\, c\, r^n
+ \cdots)\,,\nn\\
e^{\ft{\phi_2}{\sqrt6}} &=& 2^{1/3}\, (1 + \ft{1}{208}\, (147 \mp
11\sqrt{17})\, r^n + \cdots)\,,
\eea
where $c$ is an integration constant, $n= \ft 15 (- 3 \mp \sqrt{17})$ and $r$ is defined\linebreak by
$d\rho=e^{-u}\, dr$.  We can use this (upper sign solution) as the initial conditions
for the numerical calculation. The resulting plots are presented
in Fig.~\ref{3cd5h2fig}, which show that the solution runs
smoothly from AdS$_3\times H^2$ at the horizon (where $\rme^u$ is linear in $r$, while $\rme^v$ is constant) to the AdS$_5$-type
asymptotic behavior given by (\ref{3cd5asym}) (where both $\rme^u$ and $\rme^v$ are linear in $r$). The 3-charge string
with constant scalars also interpolates between AdS$_3\times H^2$
and the AdS$_5$-type geometry \cite{Klemm:2000nj}.

\subsection{Interpolating from AdS$_3 \times S^2$ to AdS$_5$\la{s:interS}}

      The AdS$_3\times S^2$ space can also occur as an asymptotic geometry
for one solution or as a horizon geometry for another.  As a
concrete example, let us consider~$m_1=m_2=-1$ and $m_3=4$, for
which case we have $\phi_1=0$.  For simplicity, we\linebreak set $g=1$. The
boundary solution behaves as
\bea
e^{2u} &=& (\ft{9}{1024})^{1/3}\, y^{-2}\, (1 + c\, y^n + \cdots)\,,
\nn\\
e^{2v} &=& 6^{-1/3}\, (1 -\ft12 c\, y^n + \cdots)\,,\nn\\
e^{\ft{\phi_2}{\sqrt6}} &=& 6^{-1/3}\, (1 + \ft1{20} (11 \pm 3\sqrt{19})
\, c\, y^n + \cdots)\,,
\eea
where $n=\ft12(1 \pm \sqrt{19})$ and $c$ is an integration constant.
Again, the inexistence of the corresponding AdS$_5$-like solution in the near region suggests that these solutions will encounter a singularity as they run away
from the asymptotic region.

\begin{figure}
   \epsfxsize=4.0in \centerline{\epsffile{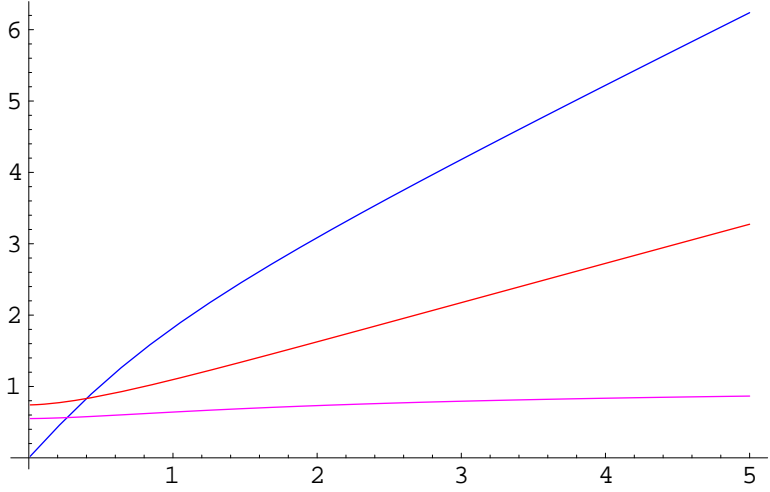}}
   \caption[FIG. \arabic{figure}.]{{\it $e^u$ (blue), $e^{v}$
   (red) and $e^{\phi_2/\sqrt{6}}$
   (purple) for a smooth solution running from AdS$_3\times
   S^2$ at the horizon to an asymptotic AdS$_5$-type geometry.
   $m_1=m_2=-1$, $m_3=4$, $g=1$, and $c=-0.3$.} \label{3cd5s2fig}}
\end{figure}

      For the solution with AdS$_3\times S^2$ at its horizon, the
near-horizon behavior is
\bea
e^{2u} &=& (\ft{1024}{9})^{1/3}\, r^2\, (1 + c\, r^n + \cdots)\,,\nn\\
e^{2v} &=& 6^{-1/3}\, (1 -\ft1{10}(11 \pm 2\sqrt{19})\, c
\, r^n + \cdots)\,,\nn\\
e^{\ft{\phi_2}{\sqrt6}} &=& 6^{-1/3}\, (1 + \ft1{100}(7 \mp 11
\sqrt{19})\, c\, r^n + \cdots)\,,
\eea
where $n=\ft12 (1 \mp \sqrt{19})$ and $c$ is an integration
constant. We can use this (upper sign) as the initial conditions for a
numerical calculation and the results are plotted in
Fig.~\ref{3cd5s2fig}. This shows that the solution runs smoothly
from AdS$_3\times S^2$ at the horizon to the AdS$_5$-type
asymptotic geometry given by (\ref{3cd5asym}).

\section{Type IIB embeddings\la{IIBembedd}}

          Gauged SUGRA in $D=5$ can be obtained
from consistent $S^5$ reduction of type IIB SUGRA.\footnote{Similarly, sphere reductions of M-theory or massive type IIA SUGRAs give rise to $D=4,\, 7$ and $6$ SUGRAs.} Thus, it is straightforward to lift the solutions of the previous sections back to $D=10$ dimensions and study their properties. A
particularly interesting class of solutions which we obtained are
those that smoothly interpolate between AdS$_{3}\times \Omega^2$
at the horizon and AdS$_{5}$-type spacetime in the asymptotic
region. When lifted to higher dimensions, the asymptotic region
becomes a product of AdS$_{5}$-type with an internal space, which is an $S^5$ bundle over $\Omega^2$ boundary of the AdS space. 
On the other hand, the horizon region becomes a
warped product of AdS$_{3}$ with an internal metric that is an $S^5$
bundle over $S^2$.  We shall examine some cases hereafter. For further details
see \cite{Cucu:2003bm,Cucu:2003yk}.

We start by lifting up the BH solution~\re{3qBH} to a type IIB solution. The embedding formula \cite{Cvetic:1999xp} yields 
\begin{eqnarray}\la{BHIIB}
d s_{10}^2 &=& \sqrt{\Delta} d s_5^2 + \frac{1}{g^2 \sqrt \Delta} \, \sum_{i=1}^3 X_i^{-1} (d \mu_i^2 + \mu_i^2 (d \phi_i +  A_\1^i)^2)\,,
\end{eqnarray}
with 
\begin{equation}
\Delta = \sum_{i=1}^3 X_i\, \m_i^2\,, \qq \m_1 = \sin\th \,,\q\! \m_2 = \cos \th \sin \psi\,, \q \! \m_3 = \cos \th \cos \psi\,.
\end{equation}
The authors of~\cite{Cvetic:1999xp} showed that the embedding~\re{BHIIB} for the $\ep=0$ three-charge BH coincides with the decoupling limit of the rotating \br{3} with the angular momenta associated with the charges of the BH. 

        The AdS$_3\times S^2$ and AdS$_3\times H^2$ solutions given by
(\ref{ads3omega2}) can be lifted to ten-dimensional IIB theory
with the reduction ansatz obtained in \cite{Cvetic:1999xp}.  Since
the solution with general $m_i$ is complicated to present, we
consider a simpler case\linebreak with $m_2=m_1$. The ten-dimensional metric
is then
\bea ds_{10}^2&=& \!\!\!\sqrt{\Delta}\Big\{ ds_{\rm{AdS_3}}^2
+\epsilon\,g^{-2}
(\fft{m_1}{m_3-2m_1})^{1/3}\, (\ft12 q_1\, d\Omega_2^2 + d\theta^2)\nn\\
&+& \!\!\!g^{-2}\,\Delta^{-1}\,\Big[ c^{-1/3}\, \cos^2\theta\,\Big
(d\psi^2 + \sin^2\psi\, (d\varphi_1 + \ft12 q_1\, A_\1)^2\nn\\
&+&\!\!\!\!\!\cos^2\psi\, (d\varphi_2 + \ft12 q_1 A_\1)^2\Big) +
c^{2/3} \sin^2\theta\, (d\varphi_3 + \ft12 q_3
A_\1)^2\Big]\Big\}, \eea
where
\bea &&c=\Big|\fft{m_1}{2m_1-m_3}\Big|\,,\qquad \Delta=c^{1/3}\,
\cos^2\theta
+ c^{-2/3} \sin^2\theta>0\,,\quad dA_\1=\Omega_\2\,,\nn\\
&&ds_{\rm{AdS_3}}^2 = e^{-\fft{2\rho}{R}}\, (-dt^2 + dx^2) +
d\rho^2\,, \qquad R=\Big|\fft{2m_1}{g\,(4m_1-m_3)\, c^{1/3}}\Big|
\,, \eea
with $q_i$ defined in \re{chargepa}.
In the above solution, if $|m_3|< 2|m_1|$, we should\linebreak have
$\epsilon=-1$, corresponding to $H^2$; if $|m_3|> 2|m_1|$, we
should have $\epsilon=1$, corresponding to $S^2$.  In general, the
internal metric is an $S^5$ bundle over $S^2$ or $H^2$, depending
the values of the $q_i$ according to the above rules.

It is especially simple to lift the five-dimensional
equal-three-charge string given by (\ref{sabra}), due to the
absence of scalars. The corresponding ten-dimensional metric is
given by
$$
ds_{10}^2=ds_5^2+\frac{1}{g^2} \Big[ d\theta^2+{\rm cos}^2 \theta
\Big( d\psi^2+{\rm sin}^2 \psi (d\phi_1+\frac{1}{3}A_\1)^2
$$
\be +{\rm cos}^2 \psi (d\phi_2+\frac{1}{3}A_\1)^2 \Big)+{\rm
sin}^2 \theta (d\phi_3+\frac{1}{3}A_\1)^2 \Big], \ee
where $dA_\1=\Omega_\2$.

For the five-dimensional three-equal-charged string with a
non-trivial dilaton, given by (\ref{new}), the corresponding
ten-dimensional metric is
$$
ds_{10}^2=\sqrt{\Delta}ds_5^2+\frac{1}{g^2 \sqrt{\Delta}} \Big[
\Delta H^{1/3}d\theta^2+H^{2/3}{\rm sin}^2
\theta(d\phi_1+\frac{1}{3} A_\1)^2
$$
\be + H^{-1/3}{\rm cos}^2 \theta \Big( d\psi^2+{\rm sin}^2 \psi
(d\phi_2+\frac{1}{3} A_\1)^2+{\rm cos}^2 \psi (d\phi_3+\frac{1}{3}
A_\1)^2 \Big) \Big], \ee
where
\be \Delta=H^{1/3}{\rm cos}^2 \theta+H^{-2/3}{\rm sin}^2 \theta >0
\,, \qquad dA_\1=\Omega_\2\,. \label{delta} \ee

Finally, the five-dimensional single-charge string (\ref{singlec}) 
can be lifted to $D=10$ as
\be ds_{10}^2=\sqrt{\Delta}ds_5^2+\frac{1}{g^2 \sqrt{\Delta}}
\Big[ H^{1/3} \Delta\, d\theta^2+H^{-1/3}{\rm cos}^2 \theta\,
d\Omega_3^2 +H^{2/3}{\rm sin}^2 \theta (d\phi_3+A_\1)^2 \Big], \ee
where $\Delta$ and $A_\1$ are given by (\ref{delta}).

In conclusion, we have investigated a large class of supersymmetric solutions of $D=5$ gauged SUGRA. These solutions are supported by $\U(1)$ gauge fields leading to electric or magnetic dual solutions. Using the superpotential technique, we characterized the dynamics by first-order equations. These equations admit (three-charge) stationary solutions \re{ads3omega2} of type ${\rm AdS}_3 \times \Omega^2$.\footnote{In the case of two-equal charge, there exists only the ${\rm AdS}_3 \times H_2$ fixed-point solution \cite{Cucu:2003bm}.} 

A particularly interesting solution interpolates between ${\rm AdS}_3 \times \Omega^2$ at the horizon and ${\rm AdS}_5$-type geometry in the asymptotic region. Such a solution indicates the existence of a class of superconformal field theories whose renormalization group flow runs from the
IR to the UV fixed-point. The dual field theory description of such SUGRA solutions was realized by wrapping \br{3}s on an $H^2$ inside a $K3$ and looking at the field theory on the brane \cite{Maldacena:2000mw}. It turns out that the usual (UV) $D=4$ SYM is flowing in the IR to a twisted two-dimensional CFT. In the language of \cite{Maldacena:2000mw}, there is a `flow across dimensions' from CFT$_4/S^2$ to some CFT$_2$. Thus, our solution (for $S^2$) offers a new example of AdS/CFT correspondence where the boundary of the AdS$_5$-type geometry compactifies spontaneously. Some steps towards the dual $D=2$ twisted field theory were taken in \cite{Imaanpur:1998rc}, but the full understanding of this correspondence needs further investigations. The dS analog of this solution was recently examined and possible connections with cosmology were clarified \cite{Lu:2003dn}.  

We have shown that the same ${\rm AdS}_3 \times \Omega^2$ solution may also appear in the\linebreak far-away region, and flow to a singularity at small distance. This implies that the stationary AdS$_{3}\times
\Omega^2$ solutions typically lie on the inflection points of the
modulus space. Finally the type IIB origin of our solutions was elucidated. The internal spaces of these embeddings belong to geometries with generalized holonomy advocated in \cite{Duff:2003ec}.

Another open question is related to the hypermultiplet sector. Due to \cite{Cacciatori:2002qx} we know that the inclusion of hypermultiplets can modify considerably the structure of $D=5$ SUGRA supersymmetric solutions. Obviously, clear statements in this respect require further research.

%
\addcontentsline{toc}{chapter}{Appendix}


%
\appendix

\renewcommand{\chaptername}{Appendix}


\typeout{:<+ appa.tex}
\chapter{Notations and conventions\la{a:appa}}
\section{Antisymmetric tensor fields}\la{a:form}

This appendix explains our notation and conventions regarding antisymmetric tensor fields. For the Minkowski metric we use the `mostly plus' signa-\linebreak ture $\eta_{ab} = {\rm diag}(-,+,\dots, +)$. Unless otherwise specified we distinguish between curved indices $\m,\n, \dots$ and (local) flat indices $a,b, \dots$. 

Symmetrization and antisymmetrization is always performed with weight 1, \ie, \begin{equation}\la{antisymdef}
A_{[ab]} = \ft 12 (A_{ab} - A_{ba})\,, \qq   A_{(ab)} = \ft 12 (A_{ab} + A_{ba})\,.
\end{equation}
The Levi-Civita tensor is defined as
\begin{equation}\la{LeviCivi}
\ve_{01\ldots (D-1)} = - \ve^{01\ldots (D-1)} = 1
\end{equation}
and it has the following contraction
\begin{equation}\la{Levicontr}
\ve_{a_1 \ldots a_n b_1 \ldots b_p} \ve^{a_1 \ldots a_n c_1 \ldots c_p} = - n ! p!\d_{[b_1}^{[c_1} \ldots \d_{b_p]}^{c_p]}
\,.\end{equation}
This tensor appears in the definition of the dual objects, which specifically in five dimensions reads
\begin{equation}\la{dualin5}
{\tilde   A}{}^{a_1\ldots a_{5-n}}=\ft1{n!}\rmi \varepsilon _{a_1\ldots
a_{5-n}b_1\ldots b_n} A^{b_n\ldots b_1}\,.
\end{equation}
That implies the following properties
\begin{equation}\la{tildetilde}
\tilde {\tilde A}=A\,, \qquad \frac{1}{n!}A^{a_1\ldots a_n}B_{a_1\ldots
a_n} \equiv \frac{1}{n!}A\cdot B=\frac{1}{(5-n)!}\tilde A\cdot \tilde B\,.
\end{equation}

The corresponding Levi-Civita tensor in curved spacetime is introduced with the help of the vielbeins $e^\mu _a$ and their determinant $e = \sqrt{|g|}$
\begin{equation}\la{Levicurved}
\ve_{\m_1 \ldots \m_D} = e^{-1} e_{\mu_1} ^{a_1} \dots e_{\mu_D} ^{a_D} \ve_{a_1 \ldots a_D} \,, \qq \ve^{\m_1 \ldots \m_D} = e e^{\mu_1} _{a_1} \dots e^{\mu_D} _{a_D} \ve^{a_1 \ldots a_D}\,.
\end{equation}

Furthermore, to avoid combinatoric factors and a plethora of indices, it is sometimes preferable to use differential forms. To each rank-$p$ skew-symmetric object one associates a $p$-form
\begin{equation}\la{pform}
A_{(p)} = \ft 1{p!} A_{\m_1 \ldots \m_p} d x^{\mu_1} \wedge \dots \wedge d x^{\mu_p}\,,
\end{equation}
where $\wedge$ denotes the exterior (wedge) product. Applying the Hodge-duality operation on $A_{(p)}$, one infers a $(D-p)$ dual form $ B_{(D - p)} = * A_{(p)}$ whose components are
\begin{equation}\la{Hodged}
B_{\m_1 \ldots \m_{D-p}} = \ft 1{p!} \, e \, \ve_{\m_1 \ldots \m_{D-p}}{}^{\n_1 \ldots \n_p}A_{\n_1 \ldots \n_p}\,.
\end{equation}
In particular, the Hodge dual of a zero-form, say 1, is a $D$-form, nothing else than the invariant volume element
\begin{equation}\la{volform}
* 1 = \ft 1{D!} \, e  \,\ve_{\m_1 \ldots \m_{D}} d x^{\mu_1} \wedge \dots \wedge d x^{\mu_D} = e\, d^D x\,. 
\end{equation}
The Hodge dual \re{Hodged} has two important consequences
\begin{eqnarray}
* * A_{(p)} &=& (-)^{p (D-p) + 1} A_{(p)}\,,\la{HodgeH}\\
*A_{(p)}\wedge B_{(p)} &=& *B_{(p)}\wedge A_{(p)} = A \cdot B *1 \,. \la{kinform}
\end{eqnarray}
The latter property allows a differential form of the kinetic terms in the action. 

There are two differential operations on $p$-forms that will play an important role in this work. The first one is the {\it exterior derivative} $d$ that takes the derivative of the components of $A_{(p)}$ and, at the same time, it raises with one the rank of the differential form. More precisely, we define
\begin{equation}\la{extddef}
d A_{(p)} = \ft 1{p!} \del_{[\n} A_{\m_1 \ldots \m_p]} d x^\n \wedge d x^{\mu_1} \wedge \dots \wedge d x^{\mu_p}\,.
\end{equation}
Besides of being a differential ($dd=0$), one additional property of the exterior derivative, namely the way it works on wedge products \begin{equation}\la{extd}
d (A_{(p)} \wedge B_{(q)}) = d A_{(p)} \wedge B_{(q)} + (-)^p A_{(p)} \wedge d B_{(q)}\,
\end{equation}
is necessary when performing partial integrations within the action.

The second relevant operation is the {\it interior product} $i_v$ that can be defined with the help of a vector field $v= v^\m \del_\m$ as 
\begin{equation}\la{intpdef}
(i_v A_{(p)})_{\m_1 \ldots \m_{p-1}} = v^\n A_{\n \m_1 \ldots \m_{p-1}}\,.
\end{equation}
The interior product is also a differential that lowers with one the rank of the differential form it works upon. We will need the following properties of this product
\begin{eqnarray}\la{intppro}
i_v v + v i_v = v^2\,,\nn\\
* i_v = (-)^{D-p} v*\,, \qq i_v * =  (-)^{D-p + 1} * v\,,
\end{eqnarray}
where the last relations hold when applied on a $p$-form in $D$ dimensions.

\section{ Supersymmetry conventions\la{a:hyp}}

In this appendix we explain our conventions for fermions and Dirac matrices in $N=2$,  $D=5$ supergravity. We give some tips for manipulating this quantities and the properties of the ($R$-symmetry)-valued objects.

The generators $U_{ij}$ of the $R$-symmetry group $\SU(2)$ are defined to
be anti-hermitian and symmetric, i.e.,
\be 
(U_i{}^j)^*=-U_j{}^i\,, \qq U_{ij} = U_{ji}\,, \q i=1,2\,.
\ee 
A symmetric traceless $U_i{}^j$ corresponds to a symmetric $U^{ij}$
since we lower or raise $\SU(2)$ indices using the $\ve$-symbol, in
NW--SE convention:
\begin{equation}
X^i=\varepsilon ^{ij}X_j\,,\qquad X_i=X^j\varepsilon _{ji}\,,\qquad
\varepsilon _{12}=-\varepsilon _{21}= \varepsilon ^{12}=1\,.
\label{NWSEconv}
\end{equation}
The actual value of $\varepsilon $ is here given as an example. It is
in fact arbitrary as long as it is antisymmetric, $\varepsilon
^{ij}=(\varepsilon _{ij})^*$ and $\varepsilon _{jk}\varepsilon
^{ik}=\delta_j{}^i$.

The charge conjugation matrix ${\cal C}$ and ${\cal C}\gamma _a$ are
antisymmetric. The matrix ${\cal C}$ is unitary and $\gamma _a$ is
hermitian apart from the timelike one, which is anti-hermitian. The
bar is the Majorana bar\footnote{When there is no possibility of confusion, spinor indices are mostly omitted.}:
\begin{equation}
\bar \lambda ^i =(\lambda ^i) ^T {\cal C}\,.
\label{barlambda}
\end{equation}
We define the charge conjugation operation on spinors as
\begin{equation}
( \lambda ^i)^C\equiv \alpha^{-1}B^{-1}\varepsilon ^{ij}(\lambda
^j)^*\,, \qquad \bar \lambda ^{iC}\equiv \overline {( \lambda ^i)^C}=
\alpha ^{-1}\left( \bar \lambda{}^k\right) ^*B\varepsilon ^{ki}\,,
\label{defCspinors}
\end{equation}
where $B={\cal C}\gamma _0$, and $\alpha = \pm 1$ when one uses the
convention that complex conjugation does not interchange the order of
spinors, or $\alpha = \pm\rmi$ when it does.  Symplectic Majorana
spinors satisfy $\lambda =\lambda ^C$. Charge conjugation acts on
gamma matrices as $(\gamma _a)^C=-\gamma _a$, does not change the
order of matrices, and works on matrices in $\SU(2)$ space as
$M^C=\sigma _2 M^*\sigma _2$. Complex conjugation can then be replaced
by charge conjugation, if for every bispinor one inserts a factor
$-1$. Then, e.g., the expressions
\begin{equation}
\bar \lambda ^i \gamma _\mu \lambda^j \,,\qquad \rmi \bar{\l}^i \l_i
\label{realexpr}
\end{equation}
are real for symplectic Majorana spinors. For more details, see
e.g.,\cite{VanProeyen:1999ni}.

When the $\SU(2)$ indices on spinors are omitted, northwest-southeast
contraction is understood, e.g.
\be
{\bar\l} \g^{(n)} \psi = {\bar\l}^i \g^{(n)} \psi_i\,,
\ee
where we have used the following notation
\be
\g^{(n)} = \g^{a_1\cdots a_n} =\g^{[a_1}\g^{a_2}\cdots \g^{a_n]}\,.
\ee

Antisymmetrizations are done according to \re{antisymdef}. Changing the
order of spinors in a bilinear leads to the following signs
\be
{\bar \psi}^{(1)} \g_{(n)} \chi ^{(2)} = t_n \ {\bar \chi }^{(2)}
\g_{(n)} \psi^{(1)}\ ,\qquad \left\{ \begin{array}{c} t_n=-1\mbox{ for
}n=2,3 \\ t_n=+1\mbox{ for }n=0,1
\end{array}\right.
\ee
where the labels $(1)$ and $(2)$ denote any $\SU(2)$ representation.

We frequently use the following Fierz rearrangement formulae \cite{Fierz:1937}
\be
\psi_j \bar{\l}^i = - \ft 14 \bar{\l}^i \psi_j -\ft14 \bar{\l}^i \g^a \psi_j
\g_a + \ft18 \bar{\l}^i \g^{ab} \psi_j \g_{ab}\,,\qquad \bar{\psi}^{[i}
\l^{j]} = - \ft 12 \bar{\psi} \l \ve^{ij}\, .
\ee

\begin{table}[htbp]
\begin{center}
\begin{tabular}{|l|rr|} \hline
$D=5$  &  $m=1$  &  $m=2$ \\ \hline\rule[-1mm]{0mm}{6mm}
$n=0$  &  $5$    &  $-20$ \\\rule[-1mm]{0mm}{6mm}
$n=1$  &  $-3$   &  $-4$  \\\rule[-1mm]{0mm}{6mm}
$n=2$  &  $1$    &  $4$   \\\rule[-1mm]{0mm}{6mm}
$n=3$  &  $1$    &  $4$   \\ \hline
\end{tabular}
\caption{The coefficients $c_{n,m}$
in~(\ref{contractions}).\label{tbl:contractions}}
\end{center}
\end{table}

When one multiplies three spinor doublets, one 
should be
able to write the result in terms of $(8\times 7\times 6)/3!=56$
independent structures.  {}From analyzing the representations, one can
obtain that these are in the $(4,2)+(4,4)+(16,2)$ representations of
$\overline{\SO(5)}\times \SU(2)$.  They are
\begin{eqnarray}
&& \xi _j\bar \xi ^j\xi ^i= \gamma ^a\xi _j\bar \xi ^j\gamma _a\xi
^i=\ft18\gamma ^{ab} \xi ^i \bar \xi \gamma
_{ab}\xi\,, \nonumber\\
&& \xi ^{(k} \bar \xi ^{i}\xi ^{j)}\,,\nonumber\\
&& \xi _j\bar \xi ^j\gamma _a\xi ^i\,.
 \label{cubicfermions}
\end{eqnarray}

The product of all gamma matrices is proportional to the unit matrix
in odd dimensions. We use
\begin{equation}  
\gamma ^{abcde}=\rmi\varepsilon^{abcde}\,.
\label{gammaepsilon}
\end{equation}
This implies that the dual of a $(5-n)$-antisymmetric gamma matrix is the
$n$-antisymmetric gamma matrix given by 
\be 
\label{gamma_dual}
\g_{a_1\ldots a_{n}} = \ft1{(5-n)!} \rmi \ve_{a_1\ldots a_{n} b_1
\ldots b_{5-n}} \g^{b_{5-n} \ldots b_1}\,.
\ee

For convenience we will give a rule for calculating gamma-contractions
like 
\be 
\label{contractions} 
\g^{(m)} \g_{(n)} \g_{(m)} = c_{n,m} \g_{(n)}\, ,
\ee 
where the constants $c_{n,m}$ are given for the most frequently used
cases in Table~\ref{tbl:contractions}.

\section{ A promenade in SS-Maxwell equivalence\label{a:Max}}

The generating functional for the Maxwell theory in the Hamiltonian
approach is well known (see e.g.,\cite{Girotti:1997kn,Henneaux:1992ig,Gomis:1995he})
\begin{equation}
Z= \int {\cal D} A_m \,{\cal D}\pi_m \,{\cal D}c \,{\cal D} 
\bar {\cal P}\,{\cal D}\bar c\, {\cal D}{\cal P} \, \exp iS^M_\P\,.
\end{equation}
As usual, $\pi_m$ is the conjugate momentum of $A_m$, $c$ and $\bar c$
are ghosts, and $\bar {\cal P}$\linebreak and ${\cal P}$ are their respective
conjugate momenta.  The Hamiltonian gauge-fixed action in the Coulomb
gauge is given by
\begin{equation}
S^M_\Psi= \int d^4 x(\pi_m\dot A^m + \dot c \bar {\cal P} - {\cal H}_0
+ A_0 \partial_i \pi^i + \pi_0\partial^i A_i + i \bar {\cal P}{\cal P}
- i\bar c \square c)\,,
\end{equation}
where $i$, $j$, $\dots$ stand for spatial indices in the 3 dimensional
hyperplane $x^0$ constant.  The Hamiltonian density is equal to
\begin{equation}
{\cal H}_0 = \frac{1}{2}  (\pi^i\pi_i + B^i B_i)\,.
\end{equation}
The magnetic field is $B^i=F^*_{0i}$.  We can easily integrate the
fields $A^0$, $\pi_0$, the ghosts $c$, $\bar c$ as well as their
conjugate momenta ${\cal P}$, $\bar {\cal P}$.  Then, we obtain that
\begin{equation}
Z= \int {\cal D} A_i \,{\cal D}\pi_i \,{\rm
det}(\square)\,\delta(\partial_i \pi^i)\,\delta(\partial^i A_i) \, \exp
i{\tilde S}^M_\P
\end{equation}
with
\begin{equation}
{\tilde S}^M_\Psi= \int d^4 x(\pi_i\dot A^i - {\cal H}_0)\,.
\end{equation}
The determinant of $\square$ comes from the integration on the fermionic
ghosts.  The integration on $A^0$ and $\pi_0$ gives the
delta-functions enforcing, respectively, the Gauss law and the Coulomb
gauge.

In order to make the connection with the gauge-fixed Schwarz-Sen
action, we have to move to a two-potential formulation, that is we
have to solve the Gauss constraint $\partial_i \pi^i=0$ by introducing
a potential $Z^i$ such that
\begin{equation}
\pi^i=\epsilon^{ijk}\partial_jZ_k\,.
\label{e:pi}
\end{equation}
The potential $Z_i$ can be decomposed into a sum of a longitudinal and
a transverse part: $Z_i=Z_i^L+Z_i^T$, where $\del^i Z_i^T = 0$. 
When $Z_i$ is transverse
($Z_i=Z^T_i$), the equation~(\ref{e:pi}) is invertible (with
appropriate boundary conditions).  More precisely, in that case one
expresses $Z_i$ as
\begin{equation}
Z_i=-\bigtriangleup^{-1}\epsilon_{ijk}\partial^j \pi^k\,.
\label{e:zed}
\end{equation}
We can introduce the field $Z^i$ in the path integral in the following
way
\begin{equation}
Z= \int {\cal D} A_i \,{\cal D}\pi_i \,{\cal D}Z_i\, {\rm
det}(\square)\,\delta(\partial_i \pi^i)\,\delta(\partial^i
A_i)\,\delta(Z^i+\bigtriangleup^{-1}\epsilon^{ijk}\partial_j \pi_k) \,
\exp i{\tilde S}^M_\P\,,
\end{equation}
where we used 
\begin{equation}
1= \int \cD Z_i^{L}\cD Z_i^{T} \delta(Z^{L\,i})\delta(Z^{T\,i}+\bigtriangleup^{-1}\epsilon^{ijk}\partial_j \pi_k)
= \int \cD Z_i \delta(Z^i+\bigtriangleup^{-1}\epsilon^{ijk}\partial_j \pi_k) 
\end{equation}
with $\delta(Z^{L\,i})=\delta(\partial_iZ^i)$. We also notice that
\begin{eqnarray}
\!\!\delta(Z^{T\,i}+\bigtriangleup^{-1}\epsilon^{ijk}\partial_j
\pi^T_k)\,=\,\underbrace{{\rm det}^{-1}\,(\bigtriangleup^{-1}{\rm
curl})}_{={\rm det}({\rm
curl})}\,\delta(\pi^{T\,i}-\epsilon^{ijk}\partial_jZ^T_k)\,,
\end{eqnarray}
where ``${\rm curl}$" stands for the operator
$\epsilon^{ijk}\partial_j$ and $\partial_i\pi^{T\,i}=0$. In order to make the connection with the SS path-integral, we identify the two potentials as follows
\begin{equation}
A^{1}_{i} =  A_i \,, \qquad A^{2}_{i} =  Z_i\,. 
\end{equation}
Putting all these remarks together and integrating out the $\pi_i$'s we obtain
\begin{equation}
\label{e:genfunc}
Z= \int {\cal D} A^\a_i \, {\rm det}(\square)\,{\rm det}({\rm
curl})\,\delta(\partial^i A^\a_i) \, \exp iS^{S-S}_\P\,,
\end{equation}
where $S^{S-S}_\P$ is the Schwarz-Sen gauge-fixed action
\begin{equation}\label{e:ScSe}
S^{S-S}_\Psi= \int d^4 x\,\fr12 (\cL^{\a\b}\dot{A}^\a_i
-\delta^{\a\b}B^\a_i)B^{\b\, i}\,.
\end{equation}

\section{The $F^2(4)$ superconformal algebra \la{a:alg}}

There exist many varieties of superconformal algebras, when one allows
for central charges~\cite{vanHolten:1982mx,D'Auria:2000ec}. However,
so far a suitable superconformal Weyl multiplet has only been
constructed from those superconformal algebras\footnote{One notable
case is the $10$-dimensional Weyl multiplet~\cite{Bergshoeff:1983az},
that is not based on a known algebra.} that appear in the Nahm's
classification~\cite{Nahm:1978tg}. In that classification appears one
exceptional algebra, which is $F(4)$. The particular real form that we
need here is denoted by $F^2(4)$, see Tables 5 and 6
in~\cite{VanProeyen:1999ni}.

The commutation relations defining the $F^2(4)$ algebra are given by (the bosonic subalgebra $\SO(5,2)$ is a particular case of \re{confAlg})
\begin{equation}
\begin{array}{rclrcl}
\left[P_a   , M_{bc}\right]  & = & \eta_{a[b}P_{c]}\,,\qquad  &  \left[K_a   , M_{bc} \right] & = & \eta_{a[b} K_{c]} \, ,  \\
\left[D     , P_a \right]    & = & P_a\, , \qquad &
\left[D     , K_a \right]    & = & -K_a\, ,  \\
\left[M_{ab}, M^{cd}\right]  & = & -2 \d_{[a}{}^{[c} M_{b]}{}^{d]}\,
,\qquad
  & \left[P_a   , K_b  \right]   & = & 2(\eta_{ab} D + 2 M_{ab})\, , \\
  &  &  &  &  &  \\
\left[M_{ab}, Q_{i\a} \right] & = & \displaystyle -\ft 14 (\g_{ab} Q_i)_\a \, ,\qquad  &
\left[M_{ab}, S_{i\a} \right] & = & \displaystyle -\ft 14 (\g_{ab} S_i)_\a \, , \\
\left[D     , Q_{i\a} \right] & = &  \displaystyle \ft 12 Q_{i\a}\, , &
\left[D     , S_{i\a} \right] & = & \displaystyle -\ft 12 S_{i\a}\, ,   \\
\left[K_a   , Q_{i\a} \right] & = &  \rmi(\g_a S_i)_\a \, , \qquad &
\left[P_a   , S_{i\a} \right] & = & -\rmi(\g_a Q_i)_\a \, ,   \\
  &  &  &  &  &  \\
\left\{ Q_{i\a}, Q_{j\b} \right\} & = & \displaystyle -\ft 12 \ve_{ij} (\g^a)_{\a\b} P_a
\, ,\qquad &\left\{ S_{i\a}, S_{j\b} \right\} & = & \displaystyle -\ft 12 \ve_{ij}
(\g^a)_{\a\b} K_a
\, ,   \\
\left\{ Q_{i\a}, S_{j\b} \right\} & = &\multicolumn{4}{l}{\displaystyle -\ft 12 \rmi
\left(\ve_{ij} C_{\a\b}
 D +
      \ve_{ij} (\g^{ab})_{\a\b} M_{ab}  + 3 C_{\a\b} U_{ij} \right)
\, ,}  \\
  &  &  &  &  &  \\
\left[Q_{i\a}, U_{kl} \right] & = & \ve_{i(k} Q_{l)\a} \, , \qquad &
\left[S_{i\a}, U_{kl} \right] & = & \ve_{i(k} S_{l)\a} \, , \\
\left[U_{ij}, U^{kl}  \right] & = & 2 \d_{(i}{}^{(k} U_{j)}{}^{l)} \, .
 &  &  &
\end{array}
\label{suf2(4)}
\end{equation}


\typeout{:<+ appbned.tex}
\chapter{Samenvatting\la{a:appb}}

\section{Het kader: M-theorie\la{a:Mth}} 

De zoektocht naar een ge\"unificeerde theorie van alle krachten en materie vormt de niet waargemaakte droom van Einstein. Sindsdien hebben natuurkundigen zich veel moeite getroost om deze moeilijke taak te voltooien maar, ondanks de vooruitgang die ondertussen geboekt is, is het einde nog niet in zicht. De niet-gravitationele krachten (elektrozwak en sterk) worden door het Standaard Model beschreven\footnote{Hoewel er geen twijfel bestaat over de geldigheid van dit model, moeten sommige stukken van de puzzel nog gevonden worden.}, een kwantumveldentheorie gebaseerd op de  $\SU(3) \times\SU(2)\times\U(1)$ ijkgroep.

Het concept van ijksymmetrie, dat essentieel is voor de kwantificatie van niet-gravitationele interacties, is niet meer voldoende voor het beschrijven van de\linebreak zwaartekracht. Omwille van zijn gedrag bij hogere energie\"en ontsnapt de gravitatie aan het standaard formalisme van kwantumveldentheorie. In een volgende stap werd het begrip supersymmetrie ingevoerd, een symmetrie (met een fermionische parameter $\epsilon$) die bosonen (b) en fermionen (f) uitwisselt,
\begin{equation}\la{gensusy}
\d  b = \epsilon f \,, \qq \d f = (\del b) \epsilon\,.
\end{equation} 
Maar de resulterende (lokale) supersymmetrische theorie, supergravitatie (SUGRA) genoemd, kan de problemen van kwantumgraviatie niet oplossen. Dit duidt de nood  aan van nieuwe, radikale idee\"en.

Zo een idee kwam in de jaren '80 met {\it snaartheorie} en in de jaren '90 met zijn verfijnde versie {\it M-theorie}. In snaartheorie, om hiermee te beginnen, worden de elementaire deeltjes als verschillende excitatietoestanden van hetzelfde fundamentele voorwerp (de snaar) voorgesteld. Net zoals de snaren van een viool die met verschillende tonen kunnen trillen, kan de fundamentale snaar op vele manieren oscilleren. Elke trillingstoestand stemt overeen met een bepaalde massa, lading, spin, enz., dus met een deeltje. Een merkwaardige eigenschap van het spectrum van massaloze deeltjes is dat het de gravitonen (voor gesloten snaren, d.w.z.~samenvallende eindpunten) en fotonen (voor open snaren, d.w.z.~snaren met vrije eindpunten) bevat. Bijgevolg zouden, in principe, de zeer kleine snaren (verondersteld van de orde van de Planck schaal $\sim 10^{-33} {\rm cm}$) zowel de niet-gravitationale als de graviationale krachten moeten kunnen beschrijven.

Wanneer de snaren in de tijd evolueren, spannen ze een twee-dimensionale oppervlakte op die een wereldvlak wordt genoemd. Op dezelfde wijze als de puntdeeltjes, interageren de snaren volgens twee-dimensionale veralgemeningen van Feynman diagrammen. Deze zijn opgebouwd uit samenvoegingen en opsplitsingen van wereldvlakken (Fig.~\ref{fig:stringinter}). Een andere vernieuwing van snaartheorie is dat de snaarkoppelingskonstante een dynamische grootheid wordt en bepaald is door de vacu\"um verwachtingswaarde van een scalair dilaton veld.

Een bijzonder verrassende ontdekking van de (perturbatieve) snaartheorie is het feit dat de theorie enkel consistent is in $D=10$ ruimtetijd dimensies. Men veronderstelt dat de extra $6$ dimensies klein zijn (van de orde van de Planckschaal) en compact (zoals cirkels of meer ingewikkelde ruimtes zoals Calabi-Yau-vari\"eteiten (CY) of $K3$). Bijgevolg zijn ze niet `zichtbaar' aan versnellers de dag van vandaag, maar ze zullen detekteerbaar worden indien de versnellingsenergie gevoelig wordt opgedreven. Men kan van $D=10$ naar $D=4$ of een andere tussenliggende dimensie gaan door de niet-gewenste dimensies compact te nemen en hun afmetingen naar nul te laten gaan. Dit wordt Kaluza-Klein (KK) reduktie genoemd. Er is in feite niet \'e\'en consistente snaartheorie, maar vijf: type IIA, IIB, I, heterotische $\SO(32)$ en heterotische $\E_8 \times \E_8$. Zij worden gekarakteriseerd door het type snaren (gesloten en/of open) die ze beschrijven, het aantal supersymmetrie\"en en de aard van de ijksymmetrie\"en die ze bevatten. De complexiteit van de theorie bij hoge energie\"en, waar een hele toren van massieve toestanden verschijnt, suggereert dat men zich in de eerste instantie tot  de lichtste modes beperkt. In deze lage-energie benadering reduceert elke snaartheorie tot de overeenkomstige SUGRAtheorie. De laatstgenoemde genereert via KK reductie verschillende SUGRA's in lagere dimensies, zoals we hieronder beschrijven voor de $D=5$ SUGRA. Dit motiveert eerder uitgevoerd SUGRA onderzoek en moedigt verdere exploratie in deze richting aan.

In de jaren '80 zat men verveeld met het bestaan van vijf snaartheorie\"en voor een unificerende theorie. De oplossing hiervoor kwam door het opgeven van de perturbatieve benadering en door de volledige theorie te beschouwen: hierin zijn de vijf theorie\"en met elkaar verwant door de dualiteiten. Type II theorie\"en bijvoorbeeld zijn door T-dualiteit met elkaar verbonden, in de zin dat IIA op een cirkel met straal $R$ equivalent is met IIB op een cirkel met straal $1/R$ en omgekeerd. Deze T-dualiteit verwisselt de KK modes $n/R$ in de ene theorie met de windingstoestanden $m R$ in de duale theorie en omgekeerd. Bovendien, de zwak gekoppelde ($g_S<1$) type I theorie is ook equivalent met de sterk gekoppelde ($g_S>1$) heterotische $\SO(32)$ snaartheorie en omgekeerd. In de taal van snaartheorie noemt men ze S-duaal met elkaar. Anderzijds het type IIB is zelfduaal onder de S-dualiteit \ie de beschrijving van de theorie voor $g_S<1$ zou moeten samenvallen met de overeenstemmende  $g_S>1$.

De vijf snaartheorie\"en zijn daarom gewoon verschillende `gezichten' (gebieden van de parameterruimte) van een enkele, omvattende theorie. Deze mysterieuze en nog onbekende theorie werd M-theorie gedoopt. Hoewel M-theorie vandaag de dag niet volledig begrepen is, wordt algemeen aangenomen dat zij elf dimensies heeft en dat  zij niet enkel snaren bevat maar ook andere multidimensionale voorwerpen zoals membranen. De hogerdimensionale veralgemeningen van membranen worden {\it $p$-branen} genoemd.

M-theorie heeft een lage-energie benadering, de $D=11$ SUGRA, die geen plaats vond in de snaartheorie voor bijna twintig jaar. De elfde dimensie is een natuurlijke niet-perturbatieve eigenschap van de snaartheorie. De type IIA snaar op het perturbatieve niveau wordt een twee-dimensionale `fietsband' als men \'e\'en dimensie hoger gaat, wanneer de koppelingsconstante grote waarden ($>1$) aanneemt. Meer precies, de straal van de elfde dimensie is recht evenredig met de koppelingsconstante van de IIA snaar, zodanig dat de extra dimensie perturbatief afwezig is, terwijl zij belangrijk wordt in het $g_S>1$ regime. Van niet-perturbatieve snaaranalyse leren we bijgevolg dat de fundamentele voorwerpen van M-theorie twee-dimensionale membranen zijn. Een gelijkaardige discussie kan gevoerd worden voor de heterotische  $\E_8$ snaar.

Branen zijn oorspronkelijk als SUGRA oplossingen uitgevonden. Vanuit dit oogpunt kunnen ze opgevat worden als hoger-dimensionale veralgemeningen van zwarte gaten (ZG) en analoog kunnen ze elektrische of magnetische fluxen dragen. 
Het geval van een $3$-braan in $D=10$ is speciaal omdat zijn vier-dimensionale wereldvolume een vier-vorm potentiaal draagt met een zelf-duale vijf-vorm veldsterkte  $F_{(5)} = d A_{(4)}$. Bijgevolg zal zo een $3$-braan zowel elektrisch als magnetisch geladen zijn. Polchinski besefte later het belang van deze braanoplossingen voor de snaartheorie in zijn niet-perturbatief regime.  Hij verklaarde de branen van de snaartheorie als topologische defecten waarop open snaren eindigen. In deze context zijn branen gekend als D$p$-branen. 
In tegenstelling tot de gesloten snaren (gravitatie) die ongehinderd in de bulk kunnen bewegen, zijn de open snaren vastgemaakt aan D$p$-branen. 
Vermits de open snaren ijkvelden dragen zijn de D$p$-branen geladen onder de overeenkomstige  $F_{(p+2)}$ veldsterkte. Om een lang verhaal kort te maken, het kan aangetoond worden dat type IIA snaartheorie D$p$-branen bevat met even $p$, terwijl type IIB de branen bevat met oneven $p$. De studie van braanoplossingen is duidelijk een noodzakelijke stap in het begrip van\linebreak snaar-/M-theorie.

Aangezien branen SUGRA oplossingen zijn, kan men verwachten dat ze verbonden zijn met een overeenstemmende braan in de lagere-dimensionale of duale theorie wanneer men dimensionale reductie of een dualiteit toepast. Dit is inderdaad het geval. Het M$2$-braan van M-theorie reduceert dimensioneel hetzij tot de fundamentele type IIA snaar (zoals hierboven uitgelegd) of tot de D$2$-braan van dezelfde theorie. Dit hangt af van het feit of de dimensionele reductie evenwijdig aan of loodrecht op het braan gebeurt. Een ingewikkelder voorbeeld is de toro\"idale compactificatie langsheen het  M$5$-braan, hetgeen T-duaal is tot een cirkel (transverse) reductie van het type IIB  D$3$-braan. In dit geval is de  $\Sl(2,\Z)$ symmetrie van de torus geassocieerd met de geconjectureerde  $\Sl(2,\Z)$-symmetrie (een veralgemening van de (S) zelf-dualiteit aanwezig in type IIB) op het D$3$-braan zelf. Bijgevolg bekomt men een mooie geometrische interpretatie van een fysische dualiteit.

Ge\"inspireerd door de ontwikkelingen in de M-theorie opperden Randall en Sundrum (RS) in 1999 de volgende fenomenologische idee\footnote{De RS scenario's zouden ook ontdekt en besproken kunnen worden onafhankelijk van M-theorie.}: we `leven' op een $3$-braan terwijl de wereld vijf-dimensionaal is. De deeltjes van het Standaard Model zijn beperkt tot het braan terwijl de zwaartekracht in de vijfde dimensie kan bewegen. Ze stelden twee mogelijke scenario's voor. Het eerste is dat waarin er twee gescheiden branen zijn (\'e\'en waarop we leven en het ander van de zwarte materie) en de extra dimensie compact is (meer precies een orbifold). Deze opstelling boekte successen in het verklaren van het hierarchieprobleem maar kwam tekort in het stabiliseren van de afstand tussen de branen. 

Een tweede interessantere RS-configuratie bestaat uit \'e\'en enkel braan en een grote extra dimensie. Deze extra dimensie kan oneindig zijn, zolang als de gravitatie maar snel genoeg afvalt wanneer we ons verwijderen van het braan zodanig dat de afwijkingen van de vier-dimensionale wet van Newton klein zijn. Experimentele metingen van de standaard Newtonwet $1/r^2$ geven een belangrijke test voor de geldigheid van de braanwereld scenario's (BWs). De vernieuwing in de tweede (RS2) opstelling is het alternatief dat ze biedt voor de compactificatie. De extra dimensie moet niet langer klein of compact zijn. Veeleer een metriek met een sterke warp factor
\begin{equation}\la{RSsc}
d s_5^2 = \rme^{-2 k \vert z \vert } d x_4^2 + d z^2
\end{equation} 
is voldoende om de gravitationele interactie die doorlekt in de $z$ richting, te onderdrukken. Het braan splitst de ruimtetijd langsheen de $z$ co\"ordinaat in twee gescheiden anti-de-Sitter \footnote{Een AdS ruimte is een vacu\"um oplossing voor Einstein's algemene relativiteit met een negatieve kosmologische constante.} (AdS) stukken. Vanuit theoretisch standpunt is het belangrijk om deze configuratie te implementeren in M-theorie. Het onderzoek\linebreak in $D=5$ supergravitatie en zijn materiekoppelingen gedurende de afgelopen jaren was gemotiveerd door het vinden van een supersymmetrische versie van RS2. Het vinden van een supersymmetrische RS2, hetgeen geen gemakkelijke taak bleek te zijn, impliceert een oplossing voor het $D=5$ SUGRA systeem gekoppeld aan de materie op het braan. Deze oplossing interpoleert tussen twee stabiele ADS vacua van de theorie.

We concentreren ons nu op de materie op het $3$-braan. Het is gekend sinds Maxwell dat de elektromagnetische vergelijkingen (in de afwezigheid van bronnen) invariant zijn onder de omwisseling van elektrische en magnetische velden. Algemener gesteld, men kan een  $\Sl(2,\R)$ transformatie uitvoeren op een vector samengesteld uit de veldsterkte $F_{(2)}$ en zijn duale tensor $* F_{(2)}$, wat leidt tot een menging van veldvergelijkingen en Bianchi identiteiten. Dit is nu begrepen in type IIB snaartheorie als een klassieke zelf-dualiteit van het D$3$-braan (hetgeen in de volledige kwantumtheorie gebroken is tot $\Sl(2,\Z)$ zoals hierboven vermeld). Nocthans in geen enkele van deze contexten (zuivere Maxwell of D$3$-braan wereldvolume beschrijving) is het triviaal om een actie te formuleren zodanig dat ze de dualiteits-symmetrie bevat. Dit werd enkel gedurende de laatste tien jaar covariant gerealizeerd.

Het probleem wordt nog ingewikkelder in het geval van coincidente branen (bovenop elkaar). Alvorens verder te gaan, vermelden we dat de lage-energie effectieve actie die de wereldvolume theorie van een enkele D$p$-braan beschrijft, de $\U(1)$ super-Maxwell is in de $p+1$ ruimtetijd dimensies. Aan de andere kant, de ijksymmetrie in het geval van $n$ coincidente D$p$-branen is uitgebreid tot een $\U(n)$ super-Yang-Mills (SYM) ijkgroep. Een interessant geval is de wereldvolume theorie van de samenvallende D$3$-branen die de (conforme) $D=4$, $N=4$ SYM is. Naast het feit dat ze divergentievrij is, bevat deze veldtheorie dezelfde symmetrie\"en als de dicht bij de horizon geometrie van een stapel  D$3$-branen, namelijk een AdS$_5\times S^5$ geometrie. Dit feit leidde Maldacena in 1997 tot het vermoeden van een dualiteit tussen type IIB snaartheorie in een AdS$_5\times S^5$-achtergrond en de conforme veldentheorie  (CFT) op de Minkowski-rand van de AdS-ruimte. Sindsdien werd deze gravitatie/veldentheorie correspondentie succesvol getest op verschillende manieren en werd ze ook uitgebreid tot andere dimensies. Dit is opnieuw een prachtig voorbeeld van hoe snaartheorie gravitationele en niet-gravitationele krachten samenbrengt. In het vervolg bekijken we hoe deze dualiteit verandert wanneer een AdS-achtige rand niet langer vlak is. 

Terugkomend op het probleem van de dualiteit-symmetrische actie voor de Maxwell theorie of het D$3$-braan wereldvolume, wordt de situatie ingewikkelder wanneer men $n$ co\"incidente D$3$-branen beschouwt. Zoals hierboven vermeld, ver\-wacht men dat zulk een theorie een (supersymmetrische) Yang-Mills-versie van dualiteit-symmetrische \'e\'en-vorm ijkpotentialen nodig heeft. De uitdaging om zo'n niet-Abelse actie voor dualiteit-symmetrische vectorvelden kan men vergelijken met de moeilijkheden bij het zoeken van een niet-Abelse theorie van twee-vorm ijkpotentialen. Deze laatste wordt verondersteld het wereldvolume van $n$ samenvallende M$5$-branen te beschrijven. 

Met deze uitgekozen onderwerpen uit M-theorie in ons achterhoofd, zullen we in het vervolg de inhoud en de belangrijkste resultaten van deze thesis bespreken.

\section{Overzicht van de thesis\la{thout}}

We beginen deze thesis met een inleidend hoofdstuk waarin de achtergrond en de motivatie van het onderzoek uitgelegd wordt. De idee\"en van Hfdst.~\ref{intro} zijn samengevat in Appendix~\ref{a:Mth}.

Als voorbereiding voor de technische ontwikkelingen in de volgende hoofdstukken, geven we in Hfdst.~\ref{ch:kit} een overzicht van sommige technische en conceptuele aspecten van snaar- en veldentheorie. Dit hoofdstuk kan gezien worden als een inleiding voor de onervaren lezer maar ook als de bron van fysische methoden die we later gebruiken. Na een korte uitleg van perturbatieve snaartheorie (Sect.~\ref{string}), bespreken we de wereldvolumebeschrijving van D-branen. Dit onderwerp is gekoppeld aan sommige resultaten van Hfdst.~\ref{ch:defo}. Behalve de D-branenclassificatie in termen van de theorie waarin ze verschijnen, trekken we een belangrijke conclusie over de veldentheorie die op een braan leeft. Deze laaste is gegeven door een actie die twee stukken bevat: een Dirac-Born-Infeld (DBI) term en een Wess-Zumino (WZ) term. We vullen de discussie over $p$-branen aan met enkele standaard $p$-branenoplossingen van $D=10$ en $D=11$ supergravitatietheorie\"en in Sect.~\ref{s:pbra}. Deze oplossingen worden veralgemeend in Hdstk.~\ref{ch:sols}. Het beeld dat men dient te onthouden is:
\begin{equation}
{\rm D-braan} = \left\{ 
\begin{array}{l}
\ra{\rm SUGRA} \,{\rm oplossing}\, + \:{\rm fluxen}\\
\ra{\rm wereldvolume} \: {\rm veldentheorie} \q S= S_{DBI} + S_{WZ}
\end{array}\right.\,.\la{dbr}
\end{equation}
We vatten de KK-compactificatiemethode samen, en daarna leggen we zorgvuldig de begrippen supersymmetrie en supergravitatie uit, met nadruk op het $5$-dimen\-sio\-nale geval. De generieke actie \re{genaction}, de supersymmetrie transformatie\-regels~\re{genertr} en de vorm van de scalaire potentiaal \re{superPot} zijn de essenti\"ele resulaten. 

In Sect.~\ref{confSugra} en Sect.~\ref{s:BV} beschrijven we de superconforme methode voor de constructie van SUGRA en de Batalin-Vilkovisky (BV) kwantisatietechniek van ijksystemen. De idee achter de superconforme tensorcalculus (SCTC) bestaat in het gebruik van een (grotere) superconforme symmetrie om uiteindelijk een super-Poincar\'e invariante theorie af te leiden. Voor het compenseren van de extra symmetrie, die gebruikt wordt in de tussenstappen voor het vereenvoudigen van de berekeningen en een beter begrip van de struktuur van het uiteindelijk model, worden sommige compenserende multipletten ingevoerd. Deze multipletten hebben geen fysische vrijheidsgraden: hun waarde wordt vastgesteld door een ijkkeuze op het einde van de ganse superconforme constructie.  De SCTC methode is expliciet uitgewerkt in Hfdst.~\ref{ch:weyl} en \ref{ch:matter}. 
In de BV methode die we gebruiken in Hfdst.~\ref{ch:dual} wordt de verzameling van velden vergroot met extra variabelen, ghosts genaamd, die overeenstemmen met de ijkinvarianties van het systeem. De kracht van de BV methode (Sect.~\ref{s:BV}) bestaat in het coderen van de volledige informatie over het ijksysteem in de oplossing van de mastervergelijking
\begin{equation}
(S,S) =0\,.
\end{equation} 
De haakjes tonen de `canonische' struktuur aan over de ruimte van velden en hun canonische toegevoegden die als antivelden gekend zijn. Van zodra de oplossing $S$ (met inbegrip van de oorspronkelijke actie als het eerste deel en andere termen die ghosts en antivelden bevatten) is gevonden, is de padintegraal quantisatie rechtstreeks.

De nieuwe resultaten van dit proefschrift zijn bevat in de Hfdst.~\ref{ch:dual}-\ref{ch:sols}. Ze zijn gebaseerd op verschillende samenwerkingen \cite{Bekaert:2000rh,Bekaert:2001wa,Bekaert:2001vq,Bergshoeff:2001hc,Bergshoeff:2002qk,Cucu:2001cs,Cucu:2003bm,Cucu:2003yk,Bergshoeff:2003xx,Bergshoeff:2003yy}.

Het eerste onderwerp van de thesis gaat over {\it dualiteitssymmetrische theorie\"en}. In Hfdst.~\ref{ch:dual} bestuderen we de verschillende formuleringen van de vrije theorie, terwijl in Hfdst.~\ref{ch:defo} we hun consistente interacties onderzoeken. Theorie\"en van chirale bosonen of zelfduale theorie\"en verwijzen naar die theorie\"en die $p$-vorm ijkvelden bevatten en voor dewelke de veldsterktes voldoen aan 
\begin{equation}\label{sdual}
F_{(p+1)} = * F_{(p+1)}\,.
\end{equation}
Omwille van de realiteitseigenschappen van de Hodge duale operator $*$ heeft deze vergelijking enkel zin in $2$, $6$ en $10$ dimensies. De laatste twee situaties zijn ingebed in de M-theorie door de zelf-duale $2$-vorm die leeft op het M$5$-braan wereldvolume en de zelf-duale $4$-vorm van het type IIB SUGRA. Het concept kan uitgebreid worden tot andere even dimensies ($4$ en $8$) ten koste van een veldcomplexificatie. Equivalent kan men enkel het aantal velden $(A^\a)_{\a =1,2}$ verdubbelen en hun veldsterktes relateren via een veralgemeende zelf-dualiteitsvoorwaarde 
\begin{equation}\label{dualsyme}
{\cal L}^{\b\a} F^\b_{(p+1)} = * F^\a_{(p+1)}\,.
\end{equation}
Deze laatste theorie\"en worden dualiteits-symmetrische theorie\"en genoemd. In deze thesis beperken we onszelf tot $D=4$ wat equivalent is met de Maxwelltheorie. Het probleem met chirale bosonen is juist hun zelf-duale veldvergelijking \re{sdual}-\re{dualsyme} die een eerste-orde vergelijking is. Bijgevolg is het een vrij moeilijke taak om ze af te leiden van een actieprincipe, hetgeen de quantisatie van zulke systemen niet triviaal maakt. Vandaag de dag zijn er verschillende manieren om \re{dualsyme} te implementeren voor Maxwell theorie:
\begin{itemize}
\item{{\bf De Schwarz-Senformulering}} (SS) leidt tot een niet-covariante actie;
\item{{\bf De Pasti-Sorokin-Toninformulering}} (PST) bevat een hulpscalair die niet-polynomisch voorkomt in de actie;
\item{{\bf De McClain et al. formulering}} gebaseerd op oneindig veel hulpvelden.
\end{itemize} 
Klassiek gezien zijn al deze manieren equivalent aan de gewone Maxwellformulering, zoals besproken in Hfdst.~\ref{ch:dual}. Verder wordt daar besproken dat dezelfde conclusie ook op het kwantumniveau stand houdt. Als een bonus, geven we een overzicht van de manier waarop de covariante PST methode een oplossing geeft voor het probleem van de complete actie van het type IIB SUGRA. 

Consistente vervormingen van de verschillende formuleringen van het \linebreak Maxwellmodel worden geanalyseerd in Hfdst.~\ref{ch:defo}. {\it Consistente vervorming}  van een theorie verwijst naar een aanpassing van de vorm van de (ijk)symmetrie\"en en de actie van een gegeven systeem, op zodanige manier dat de bekomen theorie een zelfde aantal ijkinvarianties bevat en hetzelfde aantal van fysische vrijheidsgraden als het oorspronkelijke model. 
In Sect.~\ref{s:defPST}  voeren we zulk een vervorming uit voor de PST formulering, door de actie en een speciale ijksymmetrie te veranderen. We vinden dat zulk een vervorming mogelijk is gegeven dat aan de Courant-Hilbert (CH) vergelijking voldaan is $f_+ \cdot f_- =1$. Dezelfde CH-vergelijking treedt op als een consistentievoorwaarde wanneer men Lorentz invariantie van een zelf-vervormde SS formulering of dualiteitssymmetrie van een zelf-interagerende Maxwell theorie (vervormde Maxwell) vereist. Wanneer we de CH vergelijking (Sect.~\ref{s:solC-H}) oplossen, vinden we tot onze verbazing dat, naast de DBI oplossing die leidt tot de dualiteits-symmetrische formulering van de D$3$-braan wereldvolumetheorie (Sect.~\ref{s:D3br}), er nog een oneindige klasse van oplossingen bestaat. Van deze dienen de fysische interpretatie en toepassingen nog onderzocht te worden. Het algemeen beeld is bijgevolg
\begin{equation}
\left.\begin{array}{l}
{\rm vervorming}\: {\rm SS} + {\rm Lorentz}\: {\rm invariantie}\\
{\rm vervorming}\: {\rm PST} + {\rm ijk}\: {\rm invariantie}\\
{\rm vervorming}\: {\rm Max.} + {\rm dualiteit}\: {\rm symmetrie}
\end{array}\right\} \ra f_+ \cdot f_- =1 \ra
\left\{\begin{array}{l}
{\rm DBI}\: {\rm actie,} \\
{\rm andere}\\  
{ \rm oplossingen (?)} 
\end{array}\right.
\end{equation}

We benadrukten in de vorige sectie dat de wereldvolume theorie van samenvallende D$3$-branen een niet-Abelse formuleringen van de dualiteit-symmetrische Maxwell theorie zou moeten zijn. De mogelijkheid voor zulk een niet-Abelse vervorming van de som van de dualiteit-symmetrische SS acties wordt geanalyseerd in het tweede deel van Hfdst.~\ref{ch:defo}. Het antwoord is negatief. Daarin bewezen we het volgende no-go theorema: er bestaan geen consistente, lokale interacties van een set van vrije Abelse ijkvectoren die de Abelse ijktransformaties kunnen vervormen indien we vereisen dat de (lokale) vervormde actie continu reduceert tot de som van SS acties, wanneer de koppelingsconstante naar nul gaat. Dit theorema suggereert dat zulke niet-Abelse beschrijvingen van samenvallende D$3$-branen mogelijks niet-lokaal zijn, of vereist het gebruik van niet-perturbatieve methoden. Verder onderzoek zal ons vertellen welke richting te nemen.  

Een tweede onderwerp van de thesis behandelt de {\it constructie van de $D=5$ materie-gekoppelde SUGRA met behulp van SCTC}. Dit werd bereikt in Hfdst.~\ref{ch:weyl} en~\ref{ch:matter} voor het geval van een $D=5$ SUGRA met $8$ superladingen (soms $N=2$ genoemd). Vijf-dimensionale SUGRA werd hierboven gemotiveerd door toepassingen op RS scenario's en AdS/CFT correspondentie. Bovendien treedt deze theorie ook op als $D=11$ SUGRA CY compactificaties en $S^5$ reductie van type IIB of een lage-energie limiet $K3 \times S^1$  compactificatie van een heterotische snaar, hetgeen de link maakt met M-theorie. Bovendien is het $N=2$ geval belangrijk wegens geometrische redenen. Zoals bij $D=4$ heeft de ruimte, die geparameteriseerd is door zijn scalairen (scalaire manifold of targetruimte), een heel rijke geometrie zoals we verder verklaren. 

De SCTC gebruikt conforme symmetrie als een hulpmiddel om theorie\"en op te stellen die eigenlijk deze symmetrie niet vertonen. De groep van de conformele transformaties bevat, naast de gebruikelijke Poincar\'esymmetrie\"en, schaaltransformaties (dilataties) en sommige speciale conforme symmetrie\"en. In het geval van supersymmetrie worden ze aangevuld met $8$ supersymmetrie\"en $Q$ en $8$ andere speciale superladingen $S$ die worden voortgebracht door de commutator van gewoonlijke supersymmetrie\"en en speciale conforme transformaties. Daar bovenop vervolledigt een extra bosonisch stuk de superconforme groep. Dit laatse is ondergebracht onder de naam van R-symmetrie en in het geval van $D=5$, $N=2$ wordt het gegeven door $\SU(2)$. De strategie van SCTC kan men samenvatten als:  
\begin{itemize}
\item construeer verschillende superconforme invariante multipletten (representaties van de groep) en de overeenkomstige acties;
\item kies compenserende multipletten;
\item breek de superconforme symmetrie tot super-Poincar\'esymmetrie en herschrijf de acties en de transformatieregels.
\end{itemize}

In Hfdst.~\ref{ch:weyl} stellen we de SCTC voor door het Weyl multiplet op te bouwen, \ie het multiplet dat de ijkvelden van de superconformele groep bevat. Er zijn twee versie van dit multiplet: het dilaton en het standaard Weyl multiplet afhankelijk of ze al dan niet het dilaton (een scalair veld met geen massadimensie maar met een Weyl gewicht) tussen hun materievelden bevatten. De laatstgenoemde velden verwijzen naar de extra velden die men moet toevoegen aan de ijkvelden van de Weyl multipletten om irreducibele representaties te bekomen. We construeren beide multipletten met behulp van twee technieken. In de supercurrent methode bepalen we eerst de Noetherstroom die overeenstemt met de actie van het vector multiplet. Daarna koppelen we het aan het dilaton Weyl multiplet en zo bekomen we zijn supersymmetrietransformaties. De standaardversie van het Weyl multiplet is afgeleid door het maken van een consistente herdefini\"ering van de velden in het dilaton multiplet. De tweede benadering bestaat in het rechtstreeks ijken van de superconforme algebra. Het blijkt dat, om een verbinding te maken met algemene co\"ordinaattransformaties en om overbodige vrijheidsgraden te elimineren in de representatie, men een aantal conventionele beperkingen moet opleggen op de krommingen, \eg $R_{\m\n} (P) =0$, en men materievelden moet toevoegen. De twee versies van het Weyl multiplet (gegeven in \re{modifiedtransf} en \re{diltrafo})  zijn equivalent zoals men kan zien na de koppeling van het standaard Weyl aan een verbeterd vectormultiplet en na het oplossen van de bewegingsvergelijkingen 
\begin{equation}
\left\{\begin{array}{l}
{\rm Standard}\: {\rm Weyl} \: {\rm multiplet}\\
+ {\rm verbeterd}\: {\rm vector} \: {\rm multiplet}
\end{array}\right.\stackrel{{\rm los}\: {\rm bwvgln.}\: {\rm op}}{\longrightarrow} 
{\rm dilaton}\: {\rm Weyl} \: {\rm multiplet}\,.
\end{equation}   

De overige relevante multipletten (vector-, lineaire-, hyper-multipletten) en de overeenkomstige acties zijn opgebouwd in Hfdst.~\ref{ch:matter}. Het lineaire multiplet is een handig hulpmiddel om een actie voor het vectormultiplet (Sect.~\ref{s:vecta}) te bekomen via de densiteitsformule \re{densM}. Voor de duidelijkheid van de presentatie bespreken we eerst de rigide conforme multipletten en acties (Sect.~\ref{s:navec}-\ref{s:actr}). Daarna voegen we de complicatie van de koppeling tot de Weyl multiplet achtergrond toe (Sect.~\ref{s:loca}). Uiteindelijk, door het kiezen van de gepaste ijkcondities en door het elimineren van hulpvelden via hun veldvergelijkingen (Sect.~\ref{s:gfix}), werden de super-Poincar\'e invariante actie en de supersymmetrie transformaties afgeleid. De details van deze procedure zijn ge\"illustreerd in Fig.~\ref{fig:matter} en kunnen samengevat worden als
\begin{equation}
\begin{array}{l}
{\rm Materiegekoppelde} \\
 conforme \; {\rm SUGRA} 
\end{array}
\stackrel{{\rm ijkfixing}\: {\rm condities}}{\longrightarrow} 
\begin{array}{l}
{\rm Materiegekoppelde}  \\
Poincare \; {\rm SUGRA}
\end{array}\,.
\end{equation} 
Een belangrijk resultaat besproken in dit hoofdstuk is het concept van {\it theorie\"en zonder een actie}. Het verwijst naar het feit dat de sluiting van de supersymmetriealgebra al bewegingsvergelijkingen oplegt voor de dynamische velden, zonder de nood om die vergelijkingen af te leiden uit een actie. In feite, vraagt het bestaan van een actie de aanwezigheid van extra geometrische objecten, hetgeen de klasse van mogelijke supersymmetrische theorie\"en beperkt. 

De geometrie van de hypermultipletscalairen loopt over verschillende types: een hyperk\"ahler manifold in het conforme geval (met actie) en een quaternionisch-K\"ahler manifold op het Poincar\'eniveau. Maar, in de afwezigheid van een actie wordt het gegeven door zwakkere versies: een hypercomplexe ruimte in conforme theorie\"en en een quaternionische ruimte in de super-Poincar\'e modellen. De vectormultipletscalairen anderzijds parameteriseren de zogenoemde heel speciale reele manifolds. 

Uiteindelijk, in Hfdst.~\ref{ch:sols} stellen we enkele oplossingen voor voor de $D=5$ geijkte SUGRA. Deze theorie is een bijzonder geval van de modellen afgeleid in de vorige hoofdstukken met enkel  $\U(1)$ vector multipletten. De gebruikte techniek is gebaseerd op een superpotentiaal $hat W$ in termen waarvan men de potentiaal $\hat V$ van de (ijk)theorie kan schrijven, \eg als in 
\begin{equation}
\hat V=(\fft{\del\hat
W}{\del \phi})^2 - \fft{2}{3}\,\hat W^2\,.
\end{equation}
Dit leidt tot een vervanging van de dynamische, tweede-orde, vergelijkingen van het systeem door eerste-orde vergelijkingen die de superpotentiaal bevatten. In praktijk, om $D=5$ oplossingen te bekomen, maken we eerst een dimensionele reductie  (op $\O_{(2)}=$ $S^2$ of $H^2$) naar $D=3$. De dynamische vergelijkingen van\linebreak de $D=3$ domeinmuur worden dan teruggebracht tot vergelijkingen voor de $D=5$ snaar. 

De oplossingen getoond in Hfdst.~\ref{ch:sols} kan men onderbrengen in drie categorie\"en: niet-dilatonische, dilatonsiche enkel-geladen en drie-geladen oplossingen (ZG en snaren), \ie oplossingen met drie niet-triviale fluxen. Het complexe geval van de drie-geladen snaar is onderzocht met zowel analytische als numerische methoden. We vinden AdS$_3 \times \O^{2}$ fixed-puntoplossingen, zowel als twee klassen van interpolerende oplossingen. De eerste interpoleert continu van AdS$_3 \times \O^{2}$ in het horizongebied naar  AdS$_5$-type (\ie met de Minkowski grens aangepast aan een ruimte met gecompactificeerde rand) in de asymptotische limiet, terwijl de andere loopt van  AdS$_3 \times \O^{2}$ in de asymptotische limiet tot een singularitiet in de bulk. De $D=10$ type IIB oorsprong van de oplossingen wordt uitgelegd. 


\addcontentsline{toc}{chapter}{Glossary of Terms}

\typeout{:<+ glosst.tex}
\section*{Glossary of Terms}

\vskip 3cm

\begin{description}

\item[(A)dS] a constant-curvature spacetime describing a positive (negative) cosmological constant.

\item[AdS/CFT]  the conjecture of the equivalence between the gravity (string theory) on an AdS space and a CFT on its boundary. 

\item[Antifield] the `canonical conjugate' of an ordinary field or ghost with respect to the antibracket structure.

\item[Algebra]
    \begin{description}
    \item[YM] (Yang-Mills) ordinary Lie algebra with true structure constants.
    \item[soft] a Lie algebra with the structure constants replaced by structure functions.
    \item[open] an algebra closing modulo trivial transformations of the action (in the context of field theory).
    \end{description}

\item[Antibracket] a Poisson-like structure that pairs, in the BV antifield formalism, the fields with their conjugate antifields.

\item[Axion] the RR scalar field of type IIB string theory that combines with the dilaton into a complex scalar controlling the $\Sl (2,\R)$ symmetry of the theory.

\item[$\b$-function] a function giving the running of the coupling constant with the scale of the theory. 

\item[Black hole] a particular solution of (super)gravity theory characterized by a time-like (one-dimensional) `worldvolume' and (macroscopically) by only a few parameters (mass, charge, etc.). This solution becomes singular at some particular points.

\item[Braneworld scenarios] models in which matter fields are confined to a hypersurface within a higher-dimensional geometries. The gravitational force can feel all dimensions.

\item[BV formalism] (Batalin-Vilkoviski) a quantization procedure applied to to systems with dynamical constraints, in particular to gauge theories. 

\item[BRST symmetry or differential] (Becchi-Rouet-Stora-Tyutin) a (rigid) fermionic invariance of the extended action. It is usually represented by a differential $s$.

\item[BRST cohomology group] the group of BRST-closed  ($s A=0$) functions where functions are identified when they differ by a BRST-exact quantity ($A \sim A +s B$).

\item[CFT] (Conformal Field Theory) a conformally-invariant field theory.

\item[Cohomological approach] a reformulation (in terms of BRST-cohomologies) of the old Noether procedure for computing consistent deformations of a gauge theory. 

\item[Compactification] a procedure to reduce the number of dimensions by considering some of them to be compact and very small.

\item[Compensating multiplet] a non-physical multiplet used to compensate (fix) some extra symmetry. 

\item[Conformal symmetry] the group of transformations that leaves angles invariant.

\item[Conventional constraints] constraints imposed in SCTC to transform local translations into general coordinate transformations and to express the dependent gauge fields (\eg, the spin connection) in terms of the independent ones (\eg, the vielbeins).

\item[CY] (Calabi-Yau) a geometrical space with special properties (\ie, a complex structure and vanishing Ricci tensor) normally used for compactification of\linebreak string/M-theory down to four/five dimensions.

\item[$D=11$ SUGRA] eleven-dimensional supergravity theory considered as low-energy limit of M-theory.

\item[DBI theory] (Dirac-Born-Infeld) generalization of non-linear electrodynamics to curved spaces. Nowadays, the theory is believed to describe the worldvolume dynamics of the D-branes.

\item[D-brane] a special case of a $p$-brane on which open strings can end.

\item[Deformation of a theory] the procedure of modifying the original action (and possibly the original symmetries) such that the resulting (interacting) theory reduces in some limit to the original one. 

\item[Density formula] relation providing an invariant action formula.

\item[Dilaton] a scalar field in string theory whose vacuum expectation value controls the string coupling constant $g_S$.

\item[Domain wall] topological defect of co-dimension one, \ie, an object separating the space (along one coordinate) into two disjoint regions. 

\item[Duality] the property of two (apparently) different theories which describe the same physics for different values of their parameters.

\item[Duality-symmetry] property of some $p$-forms of having `generalized' self-dual (under Hodge duality) field strength, realized in $D=4$, $D=8$ (for spaces with Minkowski signature).

\item[Extended action] an action derived through the BV method that encodes the whole information about the gauge structure of the system.

\item[Fixed-point solution] SUGRA solution with constant scalars.

\item[Gauged SUGRA] theory of SUGRA containing (at least) some gauge vectors that serve to gauge some rigid symmetry of the ungauged version.

\item[Gauge fixing] procedure followed when eliminating undesired gauge degrees of freedom from a theory. 

\item[Gauge-fixing fermion] fermionic object used in the BV formalism to carry out the gauge-fixing procedure.

\item[GR] (General Relativity) Einstein's theory of gravitation.

\item[Hadrons] strong interacting particles (e.g., quarks, protons, neutrons, etc.).

\item[Heterotic string] consistent closed string theory supporting $16$ supercharges and gauge group $\SO(32)$ or $\E_8 \times \E_8$.

\item[Hypercomplex manifold] a geometrical space that admits a triplet of complex structures satisfying the quaternionic algebra and covariantly constant w.r.t.\linebreak some affine connection. It has antisymmetric Ricci tensor.

\item[Hyperk\"ahler manifold] a geometrical space that admits a triplet of complex structures satisfying the quaternionic algebra and covariantly constant w.r.t.~the Levi-Civita connection. It has vanishing Ricci tensor.

\item[Hypermultiplet] representation of the super-Poincar\'e/conformal algebra\linebreak for $N=2$ containing spinor and scalar fields.

\item[Hodge duality] generalization to $p$-forms of the operation interchanging (in Max\-well's theory) the electric and magnetic fields. In general, it interchanges a $p$-form with the dual $(D-p)$-form. 

\item[Improved multiplet] supersymmetric multiplet containing a dilaton field that can compensate the lack of scale invariance for certain terms in the action.

\item[IR region] (infrared) describes the behavior of a theory at large distances (small energies).

\item[Killing vector] a vector field satisfying the Killing equation, indicating an isometry of a manifold.
    \begin{description}
    \item[conformal] a vector field satisfying the conformal Killing equation, \ie, the Killing equation up to a term proportional to the metric.
    \item[homothetic] a vector field satisfying the conformal Killing equation for some constant value of the scale parameter.
    \item[exact] a homothetic Killing vector that can be expressed as the derivative of another quantity.
    \end{description}

\item[Linear multiplet] in $N=2$, $D=5$ SUGRA is the multiplet of the eoms of the vector multiplet.
    
\item[M2-brane] fundamental (electrically charged) object of M-theory extended in two spatial directions.

\item[M5-brane] the magnetic dual of a \mbr{2}.

\item[Master equation] the central equation of the (BV) antifield method, whose minimal solution encodes all the information about the gauge system.

\item[Matter fields in Weyl multiplet] fields of the Weyl multiplet that do not correspond to gauge fields of the superconformal symmetries.

\item[Matter multiplets] representations of the the super-Poincar\'e/conformal algebra\linebreak whose field content can describe matter particles (electrons, quarks, etc.) and their supersymmetric partners (selectrons, squarks, etc.). 

\item[Minimal solution] the solution to the master equation that carryies the full (non-trivial) gauge information of the system. It is sometimes extended to a {\it non-minimal} solution that includes also the trivial gauge symmetries.

\item[Moduli space] the space parametrized by the scalars (moduli) of the theory (see also Target space).

\item[M spinors] (Majorana) spinors constrained by a reality condition.

\item[M-theory] name given to a (not yet completely known) theory believed to describe all five string theories and $D=11$ SUGRA as different regions of its parameter space. This theory is expected to provide the unification of GR with SM.

\item[MW spinors] (Majorana-Weyl) spinors with both Majorana and Weyl properties.

\item[Noether method] a procedure of constructing local invariant actions by adding perturbatively to the rigid ones extra terms containing the corresponding gauge field and Noether current. 

\item[NSNS sector] (Neveu-Schwarz) universal part of the string  (bosonic) spectrum including the graviton, the dilaton and the Kalb-Ramond $2$-form.

\item[Off-shell formulation] realization of a multiplet/algebra without the necessity of eoms. 

\item[On-shell formulation] realization of a multiplet/algebra up to eoms, \ie, when the fields satisfy on-shell conditions. 

\item[$p$-brane] a SUGRA solution describing a $p$-dimensional generalization of a black hole. 

\item[$p$-form] a field described by a skew-symmetric tensor of rank $p$.

\item[PST action] (Pasti-Sorokin-Tonin) action principle that implements all on-shell conditions (in particular the self-duality or duality-symmetry) as eoms. 

\item[QCD] (Quantum Chromodynamics) quantum field theory of the strong interactions, based on the gauge group $\SU(3)$.

\item[QED] (Quantum Electrodynamics) unifying theory of weak and electromagnetic interactions, based on the gauge group $\SU (2) \times \U (1)$.

\item[Quaternionic manifold] a manifold whose complex structures are preserved w.r.t.~an affine connection and an $\SU(2)$-connection.

\item[Quaternionic-K\"ahler manifold] a quaternionic manifold with a metric such that the affine connection is the Levi-Civita connection.

\item[R-symmetry] automorphism group of extended SUSY that rotates supercharges into each other.

\item[RR sector] (Ramond-Ramond) part of the type II string (bosonic) spectrum containing $p$-forms that couple to D-branes.

\item[RS scenario] (Randall-Sundrum) a particular realization of braneworlds with one (or two) $3$-brane(s) embedded in a five-dimensional space. Our world is supposed to live on such a brane.

\item[SCTC] (Superconformal Tensor Calculus) method based on the superconformal group used to construct SUGRAs.

\item[S-duality] a duality relating the strong coupling regime of a theory with the weak coupling description of another, or the same, theory.

\item[Self-duality] property of some $p$-forms of having self-dual (under Hodge duality) field strength, realized in $D=2$, $D=6$ and $D=10$ (for spaces with Minkowski signature).

\item[Self-coupling] see `self-interaction'.

\item[Self-interaction] a deformation of a theory where the interactions contain the same fields as the undeformed theory. For example, one self-interacting model associated to the Maxwell theory is the DBI action.

\item[SM] (Standard Model) (still incomplete) theory unifying all non-gravitational forces (strong and electro-weak).

\item[SM spinors] (symplectic-Majorana) spinors constrained by a generalized reality condition. 

\item[SS theory] (Schwarz-Sen) a non-covariant formulation of a duality-symmetric action for Maxwell theory. 

\item[String solution] a solution of SUGRA theory with a $1+1$-dimensional worldvolume.
     \begin{description}
     \item[non-dilatonic] the scalar fields of the theory vanish in this solution.  
     \item[dilatonic] the scalar fields are non-trivial.
     \item[three-charged] the solution allows three non-trivial gauge fields.
     \end{description}

\item[String theory] a theory of elementary particles where the fundamental constituents (e.g., the electron, the photon, etc.) are described as different vibration modes of a fundamental string.

\item[SUGRA] (Supergravity) a supersymmetric version of general relativity (local supersymmetry includes gravity).

\item[Superconformal symmetry] supersymmetric extension of the conformal group.

\item[Supercurrent multiplet] the multiplet of Noether currents and their supersymmetric partners for a given supersymmetric theory.

\item[Superpotential] function whose square and derivative squared determines the potential of a theory. It only exists for some theories.

\item[SUSY] (Supersymmetry) a symmetry connecting bosons to fermions and vice versa. Its main consequence resides in the existence of a superpartner for each known elementary particle. 

\item[SYM] (Super Yang-Mills) a supersymmetric version of the Yang-Mills theory.

\item[Target space] in string theory it is the embedding ten-dimensional spacetime; in SUSY or SUGRA the space described by the scalars of the theory (see also Moduli space).

\item[Twisted field theory] a SUSY field theory on a curved background where some supersymmetry is saved by coupling the theory to an external gauge field.

\item[Type I] string theory of closed and open strings supporting $16$ supercharges.

\item[Type IIA] string theory of closed strings containing $N=2$ MW spinors ($32$ supercharges) of opposite handedness.

\item[Type IIB] string theory of closed strings containing $N=2$ MW spinors ($32$ supercharges) with the same handedness.

\item[UV region] (ultraviolet) describes the behavior of a theory at small distances (large energies).

\item[Vector multiplet] a representation of the super-Poincar\'e/conformal algebra containing spin-$1$, spin-$1/2$ and spin-$0$ states, and possibly some auxiliary fields, all in the adjoint representation of some gauge group.

\item[Vector-tensor multiplet] a generalization of the vector multiplet containing rank-$2$ tensorial objects in an arbitrary (reducible) representation of some gauge group.

\item[Very special real geometry] the geometry of the scalars of the vector multiplets in $N=2$, $D=5$ SUGRA characterized by a totally symmetric tensor $C_{IJK}$.

\item[Weyl multiplet] representation of the superconformal algebra containing the corresponding (independent) gauge fields and possibly other `matter' fields.
    \begin{description}
    \item[dilaton version] formulation of the Weyl multiplet including as matter fields: a dilaton field, a fermion and a gauge vector together with a gauge tensor.
    \item[standard version] formulation of the Weyl multiplet including as matter fields: a tensor, a fermion and a scalar field with the wrong properties for a dilaton (non-vanishing mass dimension).
    \end{description}

\item[W spinors] (Weyl) spinors restricted via a chirality projection.

\item[WZ action] (Wess-Zumino) piece of a D-brane action including the charged character of the brane under the RR fields. 

\end{description}

\addcontentsline{toc}{chapter}{Glossary of Symbols}

\typeout{:<+ glosss.tex}
\section*{Glossary of Symbols}

\vskip 3cm

\begin{description}

\item[Action] $S$, $S_0$, $S_\Psi$, etc.
\item[Antifields] $\phi^{\dots\, *}$, $A^{\dots\, *}$, $B^{\dots\, *}$, $C^{\dots\, *}$ etc.
\item[Antisymmetric rank-2 tensors] $B_{mn}^{(M)}$
\item[Auxiliary scalar] for PST $a$, its gradient $u_m$ and its normed gradient $v_m$
\item[Beta-function] $\b (g)\equiv E \,\frac{\del g(E)}{\del E}$
\item[BRST differential] $s$
\item[Charge parameters] $m(_i)$, $q(_i)$ 
\item[Conformal boosts generators] $K_\m$ $\ra$ parameters $\L_K^\m$ $\ra$ gauge field $f_\m{}^a$
\item[Complex stuctures] $\vec J_X{}^Y \equiv J_X{}^{\a Y}$, $J_X{}^Y{}_i{}^j$
\item[Connection] $\Gamma_{XY}^Z$
\item[Cosmological constant] $\L$
\item[Coupling constant] $g$ or, for string theory, $g_s$
\item[Covariant derivative] $\nabla_\m$, $D_\m$, ${\cal D}_\m$, $\mathfrak{D}_\m$
\item[Curvatures]
    \begin{description}
    \item[Weyl multiplet] $R_{\m\n}^{\dots}$
    \item[hypermultiplet] $R_{XYZ}{}^W$, ${\cal R}_{XY A}{}^B$
     \end{description}
\item[Dilatation generator] $D$ $\ra$ parameter $\l_D$ $\ra$ gauge field $b_\m$
\item[Dilaton] $\phi$ or $\Phi$
\item[Electric field] $E_i^{(\a A)}$
\item[Field strength] $F_{(p+1)} = d A_{(p)}$, $H_{(3)} = d B_{(2)}$, $F^\a_{mn}$, etc.\hfill
    \begin{description}
    \item[contracted] $\cH^\a_{m}$ and contracted dual $\tilde \cH^\a_{m}$
    \item[dual] $ *F_{(p+1)}$, $F^{*\a}_{mn}$
    \item[self-dual] $\cF^\a_{mn}$, ${\cal F}_{(5)}$
    \end{description}
\item[Gauge-fixing fermion] $\Psi[\dots]$
\item[Gauge vector field or photon] $A^{(I)}_\m$, $A_\m^x$, $h_\m^{(A)}$, $V_\m$, $A_\m^{\a}$ $\ra$ parameter $\L^{(A)}$ 
\item[Gaugino] $\p^i$, $\p^{iI}$, $\l^{i \tilde a}$
\item[Ghosts] $c^{\dots}$, $C^\a$
\item[Gravitino] $\p_\m^i$
\item[Graviton] $e_\m^a$, $g_{\m\n}$, $h_{\m\n}$
\item[Hamiltonian] $H$ and Hamiltonian density $\cH$
\item[Hyperinos] $\zeta^A$, $\zeta^i$, $\zeta^{\widehat A}$ 
\item[Hyperk\"ahler potential] $\chi(q)$
\item[Hypermultiplet eoms.] $\Delta^X$, $\G^A$
\item[Hyperscalars] $q^X$, $q^{\widehat X}$
\item[Killing vector] $k_I^X$
\item[Lagrangian density] $\cL$
\item[Linear multiplet components] $E_a^{(M)}$, $L^{ij (M)}$, $N^{(M)}$, $\varphi^{i (M)}$
\item[Lorentz generators] $M_{\m\n}$ $\ra$ parameters $\l_M^{\m\n}$
\item[Magnetic field] $B_i^{(\a A)}$
\item[Moment maps] $\vec P_I \equiv P_I^{\alpha}$
\item[Noether current] $J_\m$
     \begin{description}
     \item[energy-momentum tensor] $\theta_{\m\n}$
     \item[supercurrent] $J_\m^i$
     \item[$\SU(2)$ current] $v_\m^{ij}$ 
     \end{description}
\item[Path integral] $Z$
\item[p-form] $A_{(p)}$ $\ra$ gauge parameter $\L_{(p-1)}$
\item[Planck]\hfill
    \begin{description}
    \item[length] $l_p \sim 10^{- 33} {\rm cm}$
    \item[mass] $M_p \sim 10^{19} {\rm GeV}$
    \end{description}
\item[Scalar potential] $V(\s)$
\item[Scalars] $\phi$, $\s^{(\dots)}$, $q^X$
\item[Sign parameters] $\ep= 1,\, 0, \, -1$ for $S^2$, $T^2$, $H^2$
\item[$\SO(2)$-antisymmetric matrix] $\cL^{\a\b}$
\item[Special supercharges] $S^i_\a$ $\ra$ parameters $\eta^{i \,\a}$ $\ra$ gauge field $\phi_\m^i$
\item[Structure constants] $f_{IJ}{}^K$
\item[Supercharges] $Q^i_\a$ $\ra$ parameters $\e^{i \,\a}$ $\ra$ gauge field $\p_\m^i$
\item[Superpotential] $W$
\item[Symmetries]
     \begin{description}
     \item[conformal] $k^X$, $\vec k^X$
     \item[target space hypermultiplet] $k_I^X$
     \end{description}
\item[Translation generators] $P_\m$$\ra$ parameters $\xi^\m$ $\ra$ gauge field $e_\m^a$
\item[Vector scalars] $\s$, $\s^I$, $\varphi^{(x)}$, $\phi$
\item[Vielbeins] $e_\m^a$, $f_{X}^{iA}$, $f_{\widetilde x}^{\widetilde a}$
\item[Volume form] (two-dimensional) $\O_{(2)}$
\item[Weyl multiplet `matter' fields] \hfill
    \begin{description}
    \item[dilaton] $B_{\m\n}$, $A_\m$, $\p^i$, $\varphi$
    \item[standard] $T_{ab}$, $\chi^i$, $D$
    \end{description}
\item[Weyl weight] $w$
\end{description}





\cleardoublepage
\addcontentsline{toc}{chapter}{Bibliography}


\footnotesize

\providecommand{\href}[2]{#2}\begingroup\raggedright\endgroup

\end{document}